\documentclass[superscriptaddress,prc,twocolumn,showpacs,preprintnumbers,
amsmath,amssymb,bibnotes,secnumarabic,altaffilletter,floatfix]{revtex4}
\usepackage{graphicx}
\usepackage{rotating}
\usepackage{xspace}
\usepackage{multirow}

\newcommand{\bof}{$\beta_{14}$\xspace}
\newcommand{\nit}{$^{16}$N }
\newcommand{\nitnosp}{$^{16}$N}
\newcommand{\lit}{$^{8}$Li }

\newcommand{\cf}{$^{252}$Cf }

\newcommand{\dto}{D$_2$O\xspace}
\newcommand{\dtonosp}{D$_2$O}
\newcommand{\hto}{H$_2$O\xspace}
\newcommand{\htonosp}{H$_2$O}
\newcommand{\bet}{$\beta_{14}$ }
\newcommand{\betnosp}{$\beta_{14}$}

\newcommand{\gam}{$\gamma$}
\newcommand{\gamsp}{$\gamma$ }
\newcommand{\cts}{$\cos\theta_{\odot}$ }
\newcommand{\ctsnosp}{$\cos\theta_{\odot}$}
\newcommand{\bi}{$^{214}$Bi\xspace}
\newcommand{\tl}{$^{208}$Tl\xspace}
\newcommand{\bisp}{$^{214}$Bi }
\newcommand{\tlsp}{$^{208}$Tl }
\newcommand{\degreesp}{$^{\circ}$ }
\newcommand{\bo}{$\beta_{14}$}
\newcommand{\nhits}{N$_{\rm hit}$ }
\newcommand{\nhitsnosp}{N$_{\rm hit}$}
\newcommand{\NS} {\chem{16}{N}}

\newcommand{\chem}[2]{\mbox{$\rm ^{#1}{#2}$}}

\begin{document}

\title{Low Energy Threshold Analysis of the Phase I and Phase II Data Sets
of the Sudbury Neutrino Observatory}

\newcommand{\alta}{Department of Physics, University of 
Alberta, Edmonton, Alberta, T6G 2R3, Canada}
\newcommand{\ubc}{Department of Physics and Astronomy, University of 
British Columbia, Vancouver, BC V6T 1Z1, Canada}
\newcommand{\bnl}{Chemistry Department, Brookhaven National 
Laboratory,  Upton, NY 11973-5000}
\newcommand{\carleton}{Ottawa-Carleton Institute for Physics, Department of 
Physics, Carleton University, Ottawa, Ontario K1S 5B6, Canada}
\newcommand{\uog}{Physics Department, University of Guelph,  
Guelph, Ontario N1G 2W1, Canada}
\newcommand{\lu}{Department of Physics and Astronomy, Laurentian 
University, Sudbury, Ontario P3E 2C6, Canada}
\newcommand{\lbnl}{Institute for Nuclear and Particle Astrophysics and 
Nuclear Science Division, Lawrence Berkeley National Laboratory, Berkeley, 
CA 94720}
\newcommand{\lanl}{Los Alamos National Laboratory, Los Alamos, NM 87545}
\newcommand{\lanla}{Los Alamos National Laboratory, Los Alamos, NM 87545}
\newcommand{\oxford}{Department of Physics, University of Oxford, 
Denys Wilkinson Building, Keble Road, Oxford OX1 3RH, UK}
\newcommand{\penn}{Department of Physics and Astronomy, University of 
Pennsylvania, Philadelphia, PA 19104-6396}
\newcommand{\queens}{Department of Physics, Queen's University, 
Kingston, Ontario K7L 3N6, Canada}
\newcommand{\uw}{Center for Experimental Nuclear Physics and Astrophysics, 
and Department of Physics, University of Washington, Seattle, WA 98195}
\newcommand{\uta}{Department of Physics, University of Texas at Austin, Austin,
 TX 78712-0264}
\newcommand{\triumf}{TRIUMF, 4004 Wesbrook Mall, Vancouver, BC V6T 2A3, Canada}
\newcommand{\ralimp}{Rutherford Appleton Laboratory, Chilton, Didcot OX11 0QX, 
UK}
\newcommand{\uci}{Department of Physics, University of California, Irvine, 
CA 92717}
\newcommand{\snoi}{SNOLAB, Sudbury, ON P3Y 1M3, Canada}
\newcommand{\lbla}{Lawrence Berkeley National Laboratory, Berkeley, CA}
\newcommand{\llnl}{Lawrence Livermore National Laboratory, Livermore, CA}
\newcommand{\iusb}{Department of Physics and Astronomy, Indiana University, 
South Bend, IN}
\newcommand{\fnal}{Fermilab, Batavia, IL}
\newcommand{\uo}{Department of Physics and Astronomy, University of Oregon, 
Eugene, OR}
\newcommand{\hu}{Department of Physics, Hiroshima University, Hiroshima, Japan}
\newcommand{\slac}{Stanford Linear Accelerator Center, Menlo Park, CA}
\newcommand{\mac}{Department of Physics, McMaster University, Hamilton, ON}
\newcommand{\doe}{US Department of Energy, Germantown, MD}
\newcommand{\lund}{Department of Physics, Lund University, Lund, Sweden}
\newcommand{\mpi}{Max-Planck-Institut for Nuclear Physics, Heidelberg, Germany}
\newcommand{\uom}{Ren\'{e} J.A. L\'{e}vesque Laboratory, Universit\'{e} de 
Montr\'{e}al, Montreal, PQ}
\newcommand{\cwru}{Department of Physics, Case Western Reserve University, 
Cleveland, OH}
\newcommand{\pnnl}{Pacific Northwest National Laboratory, Richland, WA}
\newcommand{\uc}{Department of Physics, University of Chicago, Chicago, IL}
\newcommand{\mitt}{Laboratory for Nuclear Science, Massachusetts Institute of 
Technology, Cambridge, MA} 
\newcommand{\ucsd}{Department of Physics, University of California at 
San Diego, La Jolla, CA }
\newcommand{	\lsu	}{Department of Physics and Astronomy, Louisiana State 
University, Baton Rouge, LA}
\newcommand{\imp}{Imperial College, London, UK} 
\newcommand{\ucia}{Department of Physics, University of California, Irvine, CA}
\newcommand{\suss}{Department of Physics and Astronomy, University of Sussex, 
Brighton, UK} 
\newcommand{	\lifep	}{Laborat\'{o}rio de Instrumenta\c{c}\~{a}o e 
F\'{\i}sica Experimental de
Part\'{\i}culas, Av. Elias Garcia 14, 1$^{\circ}$,  Lisboa, Portugal} 
\newcommand{\hku}{Department of Physics, The University of Hong Kong, 
Hong Kong.}
\newcommand{\aecl}{Atomic Energy of Canada, Limited, Chalk River Laboratories, 
Chalk River, ON , Canada} 
\newcommand{\nrc}{National Research Council of Canada, Ottawa, ON, Canada} 
\newcommand{\princeton}{Department of Physics, Princeton University, 
Princeton, NJ} 
\newcommand{\birkbeck}{Birkbeck College, University of London, Malet Road, 
London, UK} 
\newcommand{\snoia}{SNOLAB, Sudbury, ON, Canada} 
\newcommand{\uba}{University of Buenos Aires, Argentina}
\newcommand{\hvd}{Department of Physics, Harvard University, Cambridge, MA}
\newcommand{\pny}{Goldman Sachs, 85 Broad Street, New York, NY}
\newcommand{\pnv}{Remote Sensing Lab, PO Box 98521, Las Vegas, NV} 
\newcommand{\psis}{Paul Schiffer Institute, Villigen, Switzerland}
\newcommand{\liverpool}{Department of Physics, University of Liverpool, 
Liverpool, UK}
\newcommand{\uto}{Department of Physics, University of Toronto, Toronto, ON, 
Canada}
\newcommand{\uwisc}{Department of Physics, University of Wisconsin, 
Madison, WI}
\newcommand{\psu}{Department of Physics, Pennsylvania State University,
     University Park, PA}
\newcommand{\anl}{Deparment of Mathematics and Computer Science, Argonne
     National Laboratory, Lemont, IL}
\newcommand{\cornell}{Department of Physics, Cornell University, Ithaca, NY}
\newcommand{\tufts}{Department of Physics and Astronomy, Tufts University, 
Medford, MA}
\newcommand{\ucd}{Department of Physics, University of California, Davis, CA}
\newcommand{\unc}{Department of Physics, University of North Carolina, Chapel 
Hill, NC}
\newcommand{\dresden}{Institut f\"{u}r Kern- und Teilchenphysik, Technische 
Universit\"{a}t Dresden,   Dresden, Germany} 
\newcommand{\isu}{Department of Physics, Idaho State University, Pocatello, ID}
\newcommand{\qmul}{Dept. of Physics, Queen Mary University, London, UK}
\newcommand{\ucsb}{Dept. of Physics, University of California, Santa Barbara, 
CA}
\newcommand{\cern}{CERN, Geneva, Switzerland}
\newcommand{\utah}{Dept. of Physics, University of Utah, Salt Lake City, UT}
\newcommand{\casa}{Center for Astrophysics and Space Astronomy, University
of Colorado, Boulder, CO}  
\newcommand{\aasu}{Present address: Department of Chemistry and Physics, 
Armstrong Atlantic State University, Savannah, GA}
\newcommand{\susel}{Sanford Laboratory at Homestake, Lead, SD}  
\newcommand{\uwa}{Center for Experimental Nuclear Physics and Astrophysics, 
and Department of Physics, University of Washington, Seattle, WA} 
\newcommand{\queensa}{Department of Physics, Queen's University, 
Kingston, Ontario, Canada} 

\affiliation{\alta}
\affiliation{\ubc}
\affiliation{\bnl}
\affiliation{\carleton}
\affiliation{\uog}
\affiliation{\lu}
\affiliation{\lbnl}
\affiliation{\lifep}
\affiliation{\lanl}
\affiliation{\lsu}
\affiliation{\mitt}
\affiliation{\oxford}
\affiliation{\penn}
\affiliation{\queens}
\affiliation{\ralimp}
\affiliation{\snoi}
\affiliation{\uta}
\affiliation{\triumf}
\affiliation{\uw}

\author{B.~Aharmim}\affiliation{\lu}
\author{S.\,N.~Ahmed}\affiliation{\queens}
\author{A.\,E.~Anthony}\altaffiliation{Present address: \casa}
\affiliation{\uta}
\author{N.~Barros}\affiliation{\lifep}
\author{E.\,W.~Beier}\affiliation{\penn}
\author{A.~Bellerive}
\affiliation{\carleton}
\author{B.~Beltran}\affiliation{\alta}
\author{M.~Bergevin}\affiliation{\lbnl}\affiliation{\uog}
\author{S.\,D.~Biller}\affiliation{\oxford}
\author{K.~Boudjemline}
\affiliation{\carleton}
% ...
			
\collaboration{SNO Collaboration}
\noaffiliation

\begin{abstract}
	Results are reported from a joint analysis of Phase~I and
Phase~II data from the Sudbury Neutrino Observatory.  The effective
electron kinetic energy threshold used is $T_{\rm eff}=3.5$~MeV, the
lowest analysis threshold yet achieved with water Cherenkov detector
data.  In units of $ 10^6$ cm$^{-2}$ s$^{-1}$, the total flux of
active-flavor neutrinos from $^8$B decay in the Sun measured using the
neutral current (NC) reaction of neutrinos on deuterons, with no
constraint on the $^8$B neutrino energy spectrum, is found to be
$\Phi_{\textrm{NC}} = 5.140 \,^{+0.160}_{-0.158} \textrm{(stat)}
\,^{+0.132}_{-0.117} \textrm{(syst)}.$ These uncertainties are more
than a factor of two smaller than previously published results.  Also
presented are the spectra of recoil electrons from the charged current
reaction of neutrinos on deuterons and the elastic scattering of
electrons. A fit to the SNO data in which the free parameters directly
describe the total $^8$B neutrino flux and the energy-dependent
$\nu_e$ survival probability provides a measure of the total $^8$B
neutrino flux $\Phi_{^8{\rm B}} = 5.046\,^{+0.159}_{-0.152}
\textrm{(stat)} \,^{+0.107}_{-0.123} \textrm{(syst)}$.  Combining
these new results with results of all other solar experiments and the
KamLAND reactor experiment yields best-fit values of the mixing
parameters of $\theta_{12}=34.06\,^{+1.16}_{-0.84}$ degrees and
$\Delta m^2_{21}=7.59\,^{+0.20}_{-0.21}\times 10^{-5}$~eV$^2$.  The
global value of $\Phi_{^8{\rm B}}$ is extracted to a precision of
$^{+2.38}_{-2.95}$\%.  In a three-flavor analysis the best fit value
of $\sin^2\theta_{13}$ is $2.00^{+2.09}_{-1.63}\times 10^{-2}$.
This implies an upper bound of
$\sin^2\theta_{13}< 0.057$ (95\% C.L.).
\end{abstract}

\maketitle

\section{Introduction \label{sec:intro}}

	It is by now well-established that neutrinos are massive and
mixed, and that these properties lead to the oscillations observed in
measurements of neutrinos produced in the
Sun~\cite{home2}--\cite{bor}, in
the atmosphere~\cite{SKatm}, by accelerators~\cite{minos,k2k}, and by
reactors~\cite{kam}.  The mixing model predicts not only neutrino
oscillations in vacuum, but also the effects of matter on the
oscillation probabilities (the `MSW' effect)~\cite{wolf,msw}.  To
date, the effects of matter have only been studied in the solar
sector, where the neutrinos' passage through the core of both the Sun
and the Earth can produce detectable effects.  The model predicts
three observable consequences for solar neutrinos: a suppression of
the $\nu_e$ survival probability below the average vacuum value of
$1-\frac{1}{2}\sin^22\theta_{12}$ for high-energy ($^8$B) neutrinos, a
transition region between matter-dominated and vacuum-dominated
oscillations, and a regeneration of $\nu_e$s as the neutrinos pass
through the core of the Earth (the day/night effect).  In addition to
improved precision in the extraction of the total flux of $^8$B
neutrinos from the Sun, an advantage of the low energy threshold
analysis (LETA) presented here is the enhanced ability to explore the
MSW-predicted transition region and, in addition, more stringent
testing of theories of non-standard interactions that affect the shape
and position of the predicted rise in survival
probability~\cite{solstatus}--\cite{nsiagain}.

We present in this article a joint analysis of the data from the first
two data acquisition phases of the Sudbury Neutrino Observatory (SNO),
down to an effective electron kinetic energy of $T_{\rm eff}=3.5$~MeV,
the lowest analysis energy threshold yet achieved for the extraction
of neutrino signals with the water Cherenkov technique.  The previous
(higher threshold) analyses of the two data sets have been documented
extensively elsewhere~\cite{longd2o,nsp}, and so we focus here on the
improvements made to calibrations and analysis techniques to reduce
the threshold and increase the precision of the results.

We begin in Section~\ref{sec:detector} with an overview of the SNO
detector and physics processes, and provide an overview of the data
analysis in Section~\ref{sec:anal_overview}.  In
Section~\ref{sec:dataset} we briefly describe the SNO Phase~I and
Phase~II data sets used here.  Section~\ref{sec:montecarlo} describes
changes to the Monte Carlo detector model that provides the
distributions used to fit our data, and Section~\ref{sec:hitcal}
describes the improvements made to the hit-level calibrations of PMT
times and charges that allow us to eliminate some important
backgrounds.

Sections~\ref{sec:recon}-~\ref{sec:beta14} describe our methods for
determining observables like position and energy, and estimating their
systematic uncertainties. Section~\ref{sec:cuts} describes the cuts we
apply to our data set, while Section~\ref{sec:treff} discusses the
trigger efficiency and Section~\ref{sec:ncap} presents the neutron
capture efficiency and its systematic uncertainties.  We provide a
detailed discussion of all background constraints and distributions in
Section~\ref{sec:backgrounds}.

Section~\ref{sec:sigex} describes our `signal extraction' fits to the
data sets to determine the neutrino fluxes, and
Section~\ref{sec:results} gives our results for the fluxes and mixing
parameters.

\section{The SNO Detector\label{sec:detector}}

	SNO was an imaging Cherenkov detector using heavy water
($^2$H$_2$O, hereafter D$_2$O) as both the interaction and detection
medium~\cite{snonim}.  SNO was located in Vale Inco's Creighton Mine,
at $46^{\circ} 28^{'} 30^{''}$ N latitude, $81^{\circ} 12^{'} 04^{''}$
W longitude.  The detector was 1783~m below sea level with an
overburden of 5890 meters water equivalent, deep enough that the rate
of cosmic-ray muons passing through the entire active volume was just
3 per hour.

	One thousand metric tons (tonnes) of D$_2$O was contained in a
12~m diameter transparent acrylic vessel (AV).  Cherenkov light
produced by neutrino interactions and radioactive backgrounds was
detected by an array of 9456 Hamamatsu model R1408 20~cm
photomultiplier tubes (PMTs), supported by a stainless steel geodesic
sphere (the PMT support structure or PSUP).  Each PMT was surrounded
by a light concentrator (a `reflector'), which increased the effective
photocathode coverage to nearly $55$\%.  The channel discriminator
thresholds were set to 1/4 of a photoelectron of charge.  Over seven
kilotonnes (7$\times 10^6$~kg) of H$_2$O shielded the D$_2$O from
external radioactive backgrounds: 1.7~kT between the AV and the PSUP,
and 5.7~kT between the PSUP and the surrounding rock. Extensive
purification systems were used to purify both the D$_2$O and the
H$_2$O.  The H$_2$O outside the PSUP was viewed by 91 outward-facing
20~cm PMTs that were used to identify cosmic-ray muons.  An additional
23 PMTs were arranged in a rectangular array and suspended in the
outer H$_2$O region to view the neck of the AV.  They were used
primarily to reject events not associated with Cherenkov light
production, such as static discharges in the neck.

	The detector was equipped with a versatile calibration-source
deployment system that could place radioactive and optical sources
over a large range of the $x$-$z$ and $y$-$z$ planes (where $z$ is the
central axis of the detector) within the D$_2$O volume.  Deployed
sources included a diffuse multi-wavelength laser that was used to
measure PMT timing and optical parameters (the
`laserball')~\cite{laserball}, a $^{16}$N source that provided a
triggered sample of 6.13~MeV $\gamma$s~\cite{n16}, and a $^8$Li source
that delivered tagged $\beta$s with an endpoint near
14~MeV~\cite{li8}.  In addition, 19.8~MeV $\gamma$s were provided by a
$^3{\rm H}(p,\gamma)^4{\rm He}$ (`pT') source~\cite{pt_nim} and
neutrons by a $^{252}$Cf source.  Some of the sources were also
deployed on vertical lines in the H$_2$O between the AV and PSUP.
`Spikes' of radioactivity ($^{24}$Na and $^{222}$Rn) were added at
times to the light water and D$_2$O volumes to obtain additional
calibration data.  Table~\ref{tbl:cal_sources} lists the primary
calibration sources used in this analysis.  \begingroup \squeezetable
\begin{table*}[ht!]
\begin{center}
 \begin{tabular}{lllcc}
\hline \hline Calibration source & Details & Calibration & Deployment
Phase & Ref.  \\ \hline Pulsed nitrogen laser & 337, 369, 385, &
Optical \& & I \& II & \cite{laserball} \\ \qquad(`laserball') & 420,
505, 619~nm & \hspace{0.1in} timing calibration & & \\ \NS & 6.13~MeV
$\gamma$ rays & Energy \& reconstruction & I \& II & \cite{n16} \\
$^8$Li & $\beta$ spectrum & Energy \& reconstruction & I \& II &
\cite{li8} \\ \cf & neutrons & Neutron response & I \& II &
\cite{snonim} \\ Am-Be & neutrons & Neutron response & II only & \\
$^3$H$(p,\gamma)^4$He (`pT') & 19.8~MeV $\gamma$ rays & Energy
linearity & I only & \cite{pt_nim} \\ Encapsulated U, Th &
$\beta-\gamma$ & Backgrounds & I \& II & \cite{snonim} \\ Dissolved Rn
spike & $\beta-\gamma$ & Backgrounds & II only & \\ \textit{In-situ}
$^{24}$Na activation & $\beta-\gamma$ & Backgrounds & II only & \\
\hline \hline
\end{tabular}
\caption{\label{tbl:cal_sources} Primary calibration sources.}
\end{center}
\end{table*}
\endgroup

	SNO detected neutrinos through three processes~\cite{herb}:

\begin{center}
  \begin{tabular}{lcll}
$ \nu_x + e^-$ & $\rightarrow$ & $\nu_x + e^-$ & (ES)\\
$\nu_e + d$ & $\rightarrow$ & $p + p + e^-$\hspace{0.5in} & (CC)\\
$ \nu_x + d$ & $\rightarrow$ & $p + n + \nu_x'$ & (NC)\\ \\  \end{tabular}
 \end{center}

	For both the elastic scattering (ES) and charged current (CC)
reactions, the recoil electrons were detected directly through their
production of Cherenkov light.  For the neutral current (NC) reaction,
the neutrons were detected via de-excitation \gam s following their
capture on another nucleus.  In SNO Phase~I (the `D$_2$O phase'), the
detected neutrons captured predominantly on the deuterons in the
D$_2$O.  Capture on deuterium releases a single 6.25~MeV $\gamma$ ray,
and it was the Cherenkov light of secondary Compton electrons or
$e^+e^-$ pairs that was detected.  In Phase II (the `Salt phase'), 2
tonnes of NaCl were added to the D$_2$O, and the neutrons captured
predominantly on $^{35}$Cl nuclei, which have a much larger neutron
capture cross section than deuterium nuclei, resulting in a higher
neutron detection efficiency.  Capture on chlorine also releases more
energy (8.6~MeV) and yields multiple $\gamma$s, which aids in
identifying neutron events.

The primary measurements of SNO are the rates of the three neutrino
signals, the energy spectra of the electrons from the CC and ES
reactions, and any asymmetry in the day and night interaction rates
for each reaction.  Within the Phase~I and II data sets, we cannot
separate the neutrino signals on an event-by-event basis from each
other or from backgrounds arising from radioactivity in the detector
materials.  Instead, we `extracted' the signals and backgrounds
statistically by using the fact that they are distributed differently
in four observables: effective kinetic energy ($T_{\rm eff}$), which
is the estimated energy assuming the event consisted of a single
electron, cube of the reconstructed radial position of the event
($R^3$), reconstructed direction of the event relative to the
direction of a neutrino arriving from the Sun (\cts), and a measure of
event `isotropy' ($\beta_{14}$), which quantifies the spatial
distribution of PMT hits in a given event (Sec.~\ref{sec:beta14}).
Low values of $\beta_{14}$ indicate a highly isotropic distribution.

\begin{figure}
\begin{center}
\includegraphics[width=0.42\textwidth]{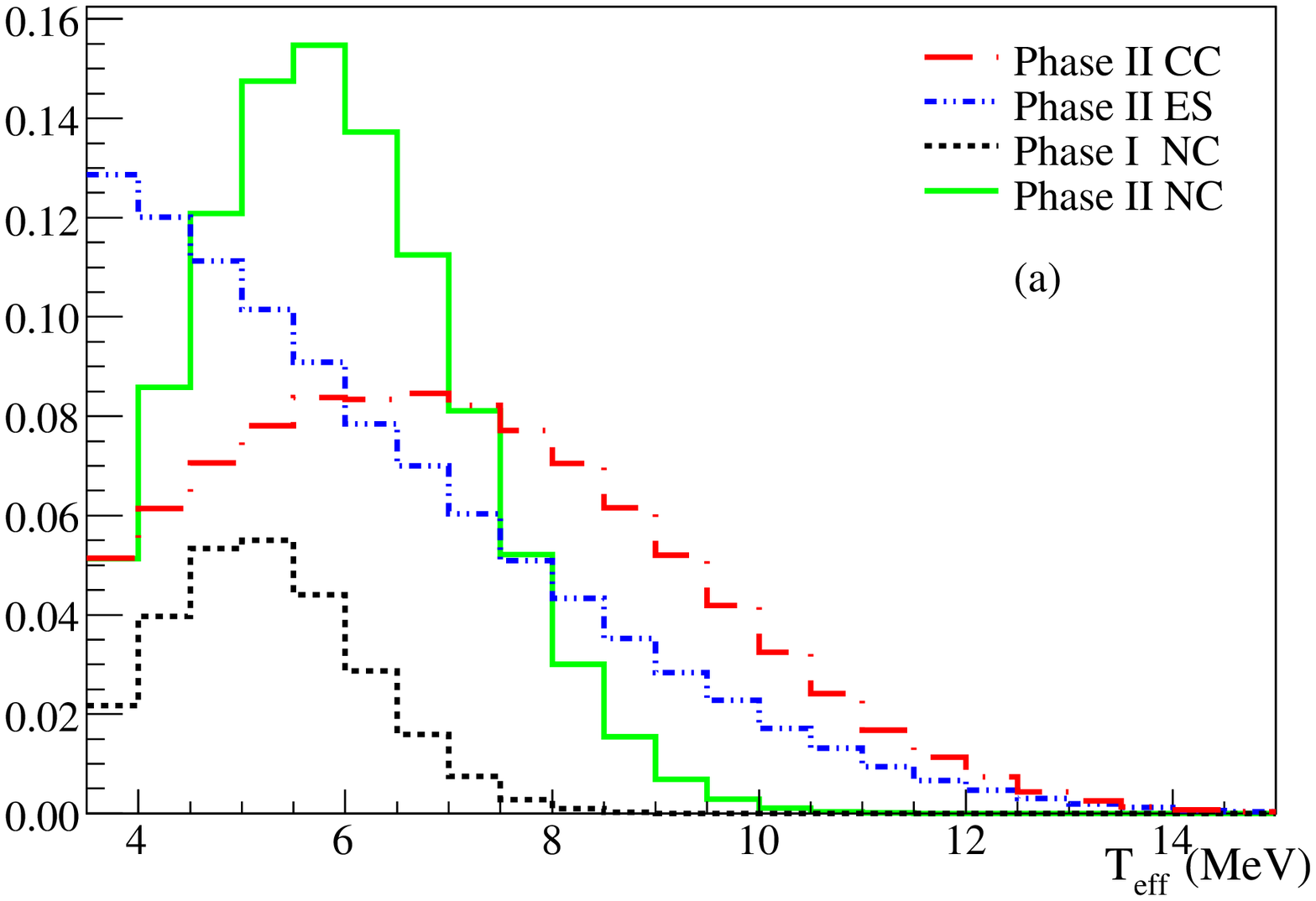}
\includegraphics[width=0.42\textwidth]{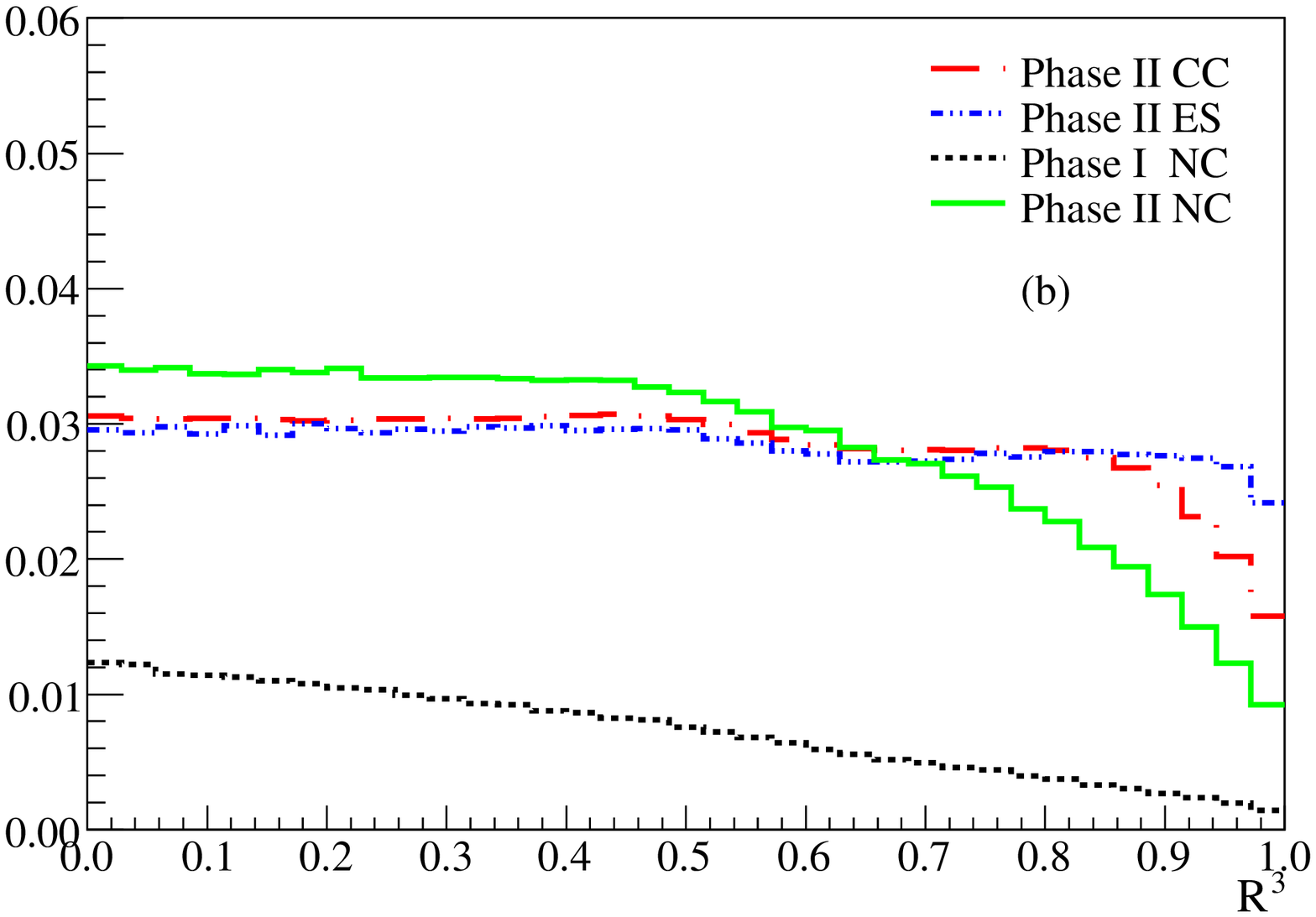}
\includegraphics[width=0.42\textwidth]{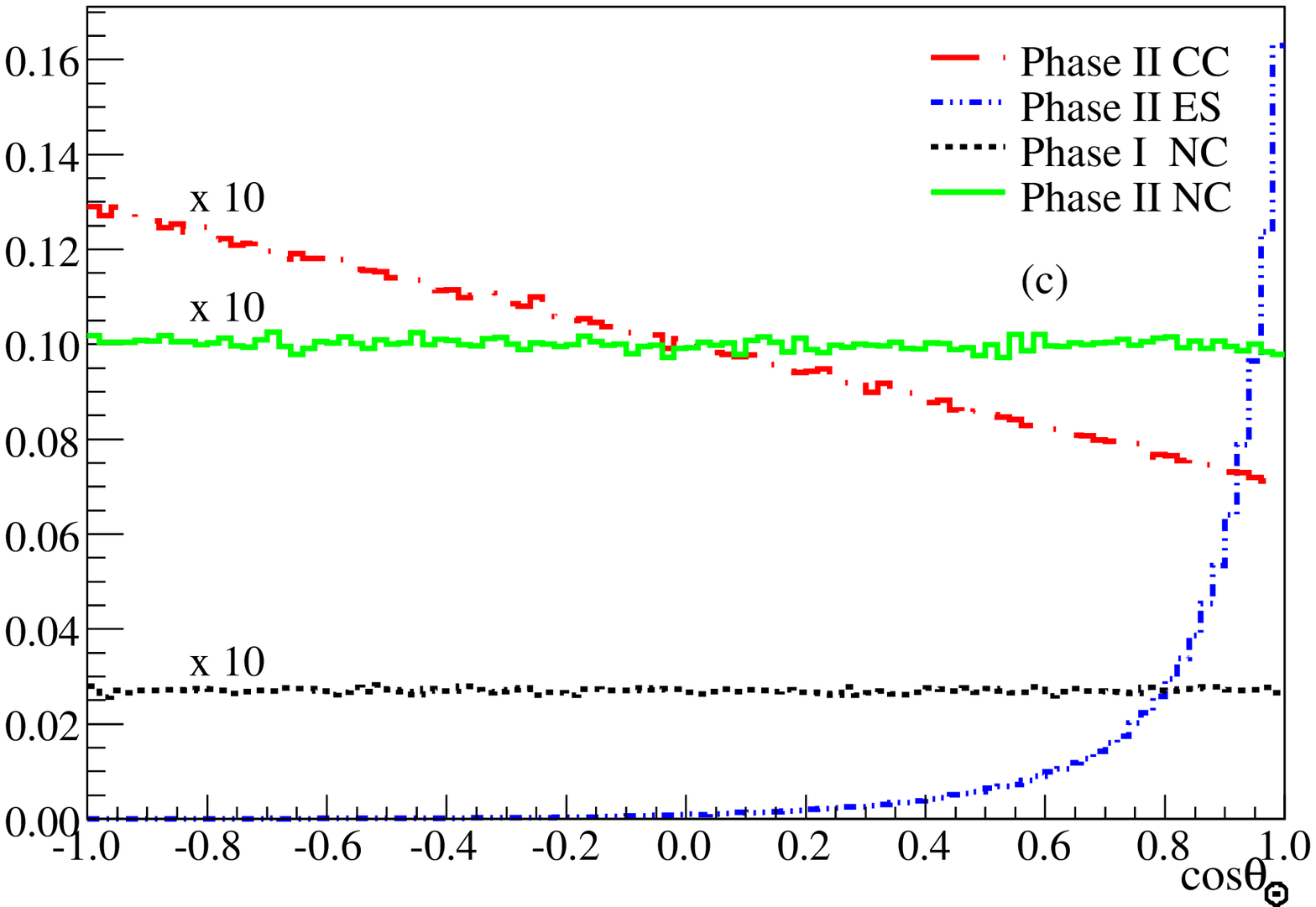}
\includegraphics[width=0.42\textwidth]{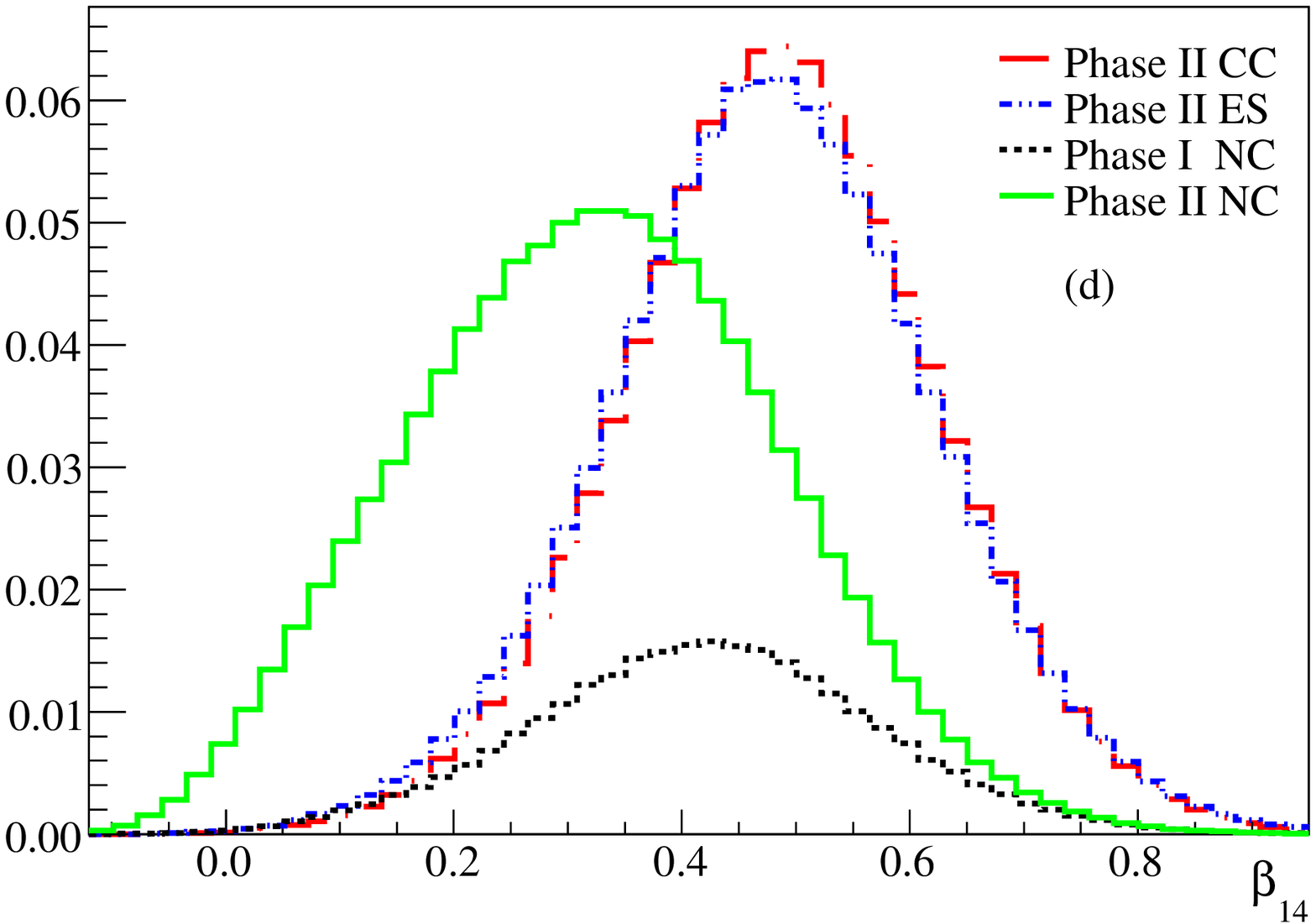}
\caption{(Color online) The Monte Carlo-generated distributions of (a)
energy ($T_{\rm eff}$), (b) radius cubed ($R^3$), (c) direction
(\ctsnosp), and (d) isotropy (\betnosp) for signal events.  The same
simulation was used to build multi-dimensional PDFs to fit the
data. In calculating $R^3$, the radius $R$ is first normalized to the
600~cm radius of the AV.  The CC and NC \cts distributions are scaled
by a factor of 10 for clarity against the ES
peak. \label{fig:pdfsnus}}
\end{center}
\end{figure}

\begin{figure}
\begin{center}
\includegraphics[width=0.42\textwidth]{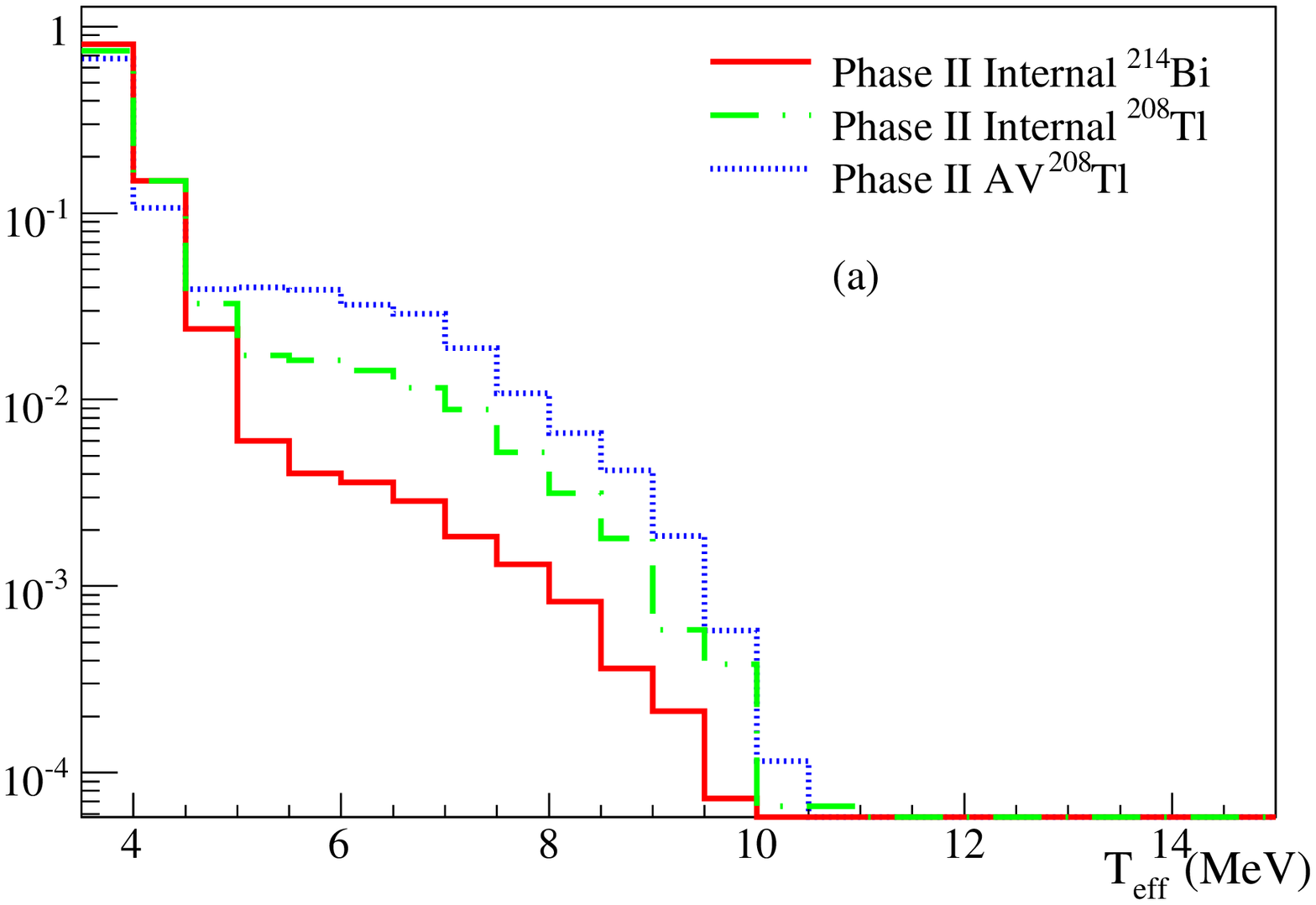}
\includegraphics[width=0.42\textwidth]{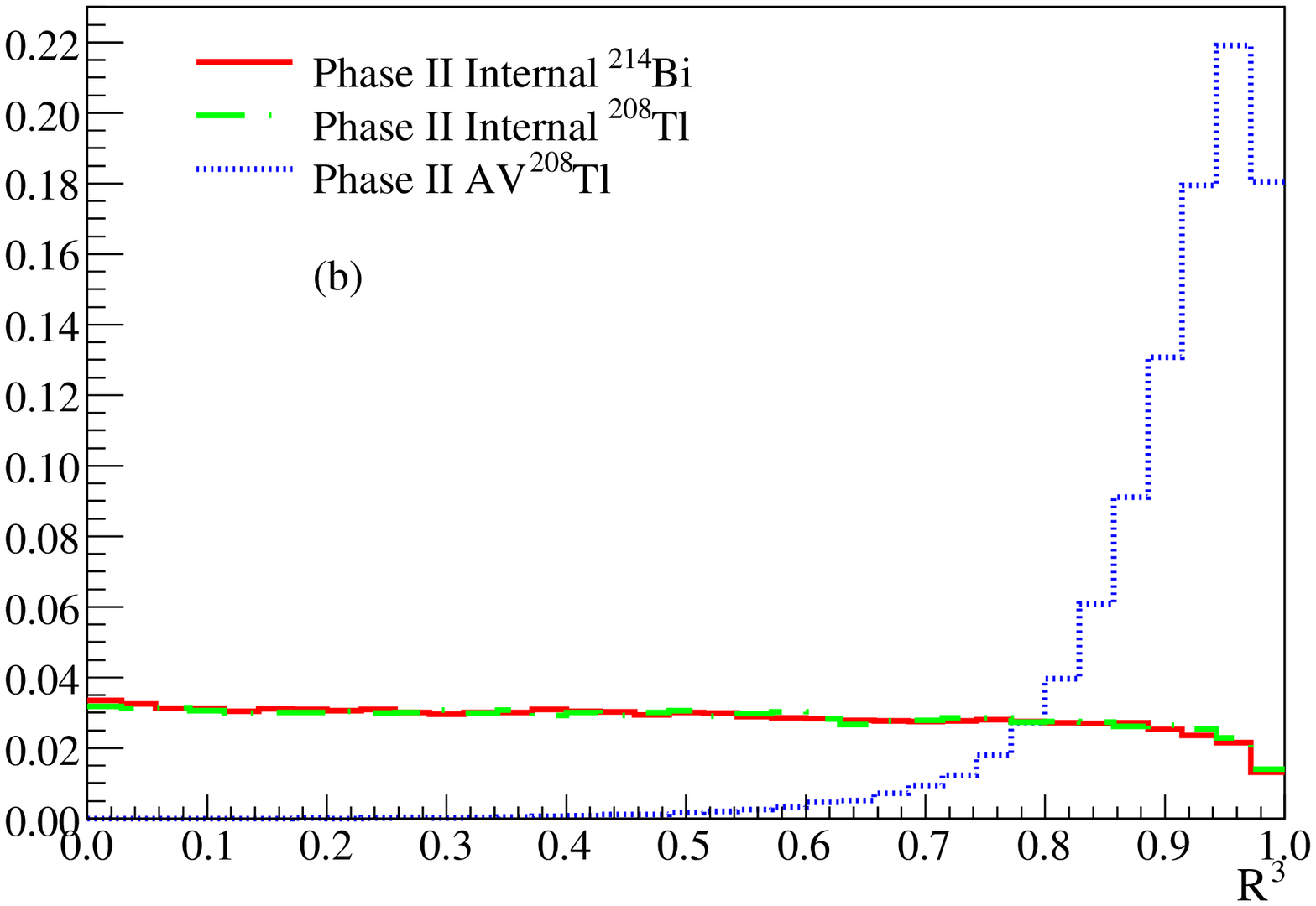}
\includegraphics[width=0.42\textwidth]{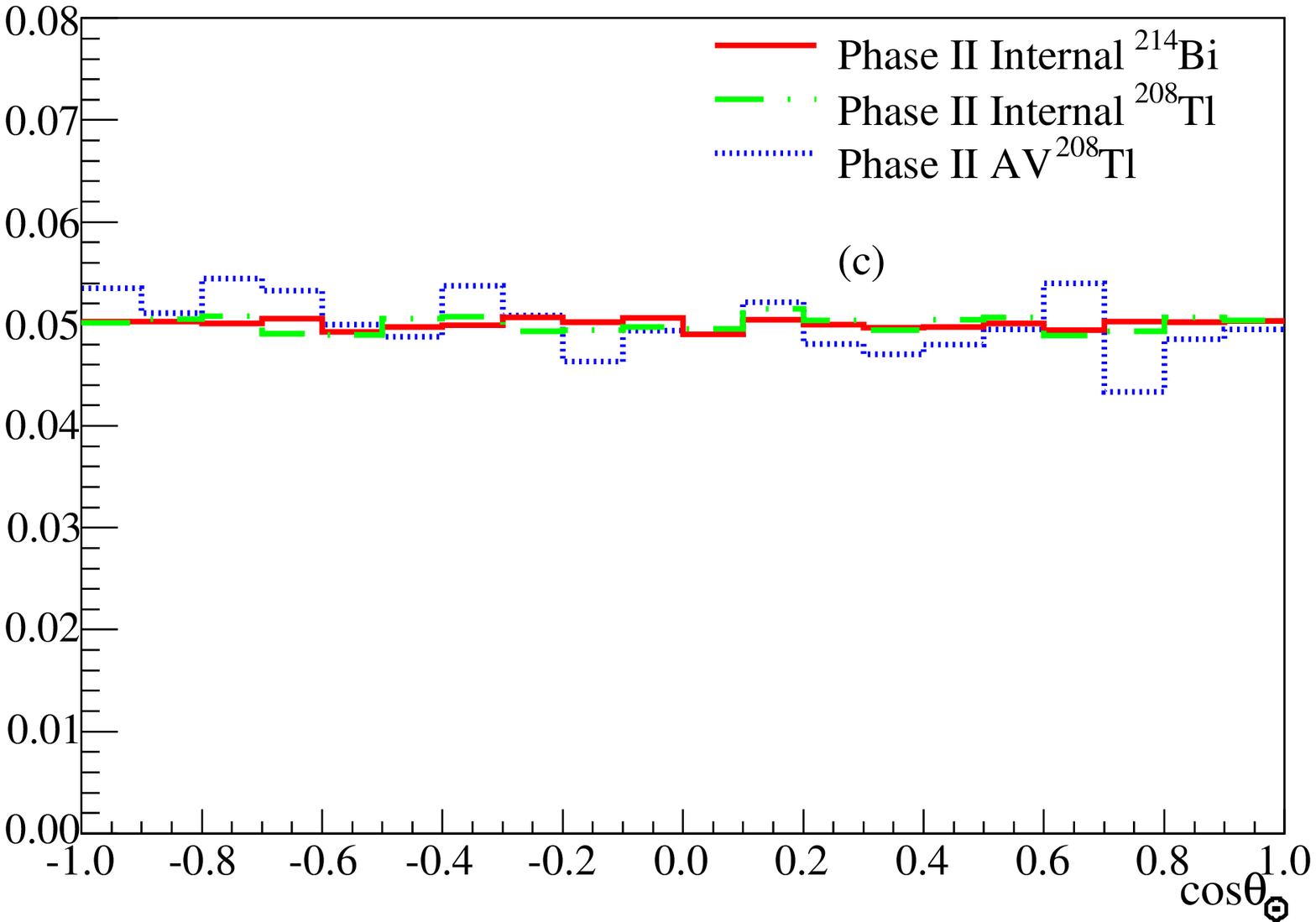}
\includegraphics[width=0.42\textwidth]{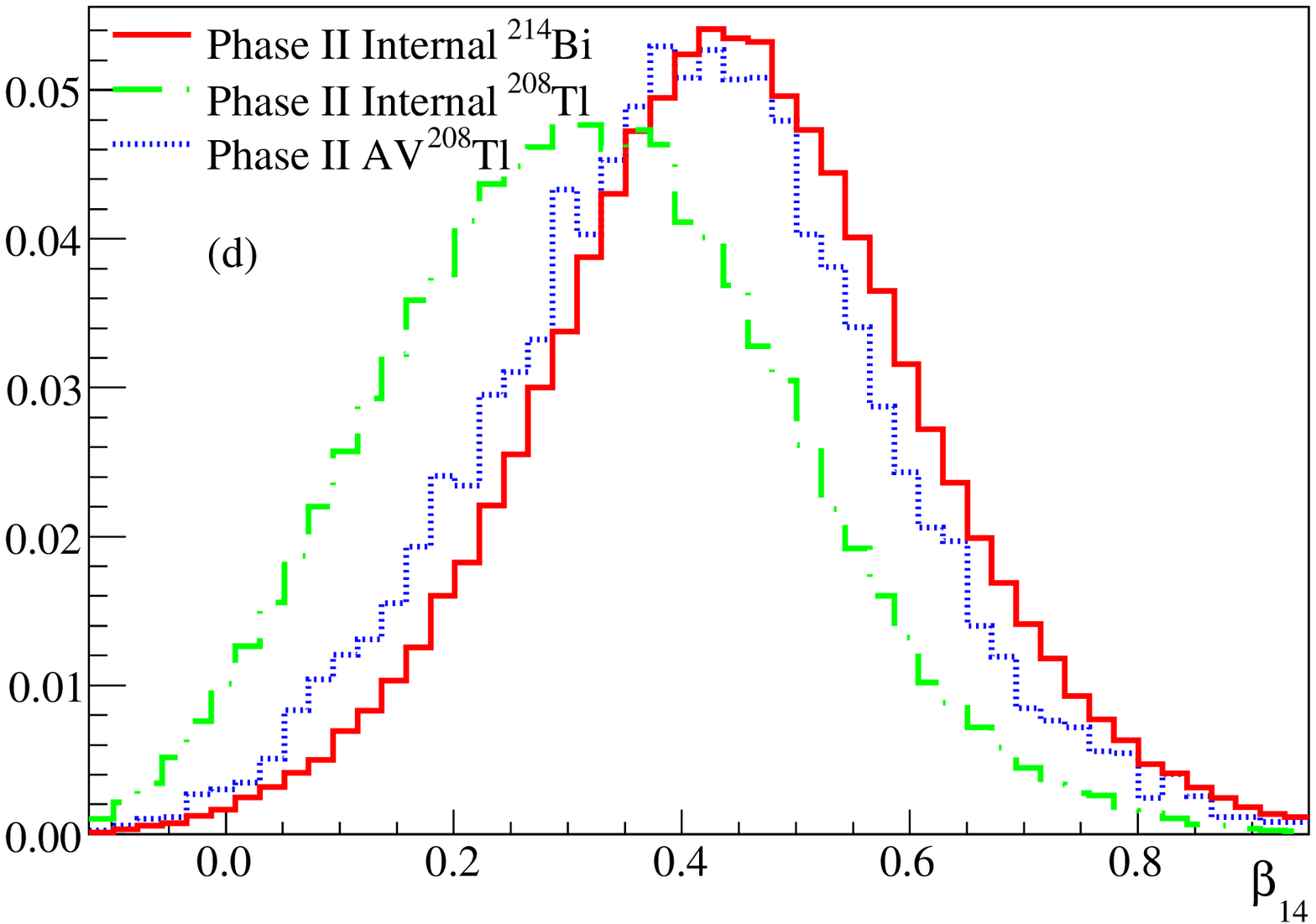}
\caption{(Color online) The Monte Carlo-generated distributions of (a)
energy ($T_{\rm eff}$) on a log scale, (b) radius cubed ($R^3$), (c)
direction (\ctsnosp), and (d) isotropy (\betnosp) for background
events.  The same simulation was used to build multi-dimensional PDFs
to fit the background events. The backgrounds shown are internal
$^{214}$Bi, internal $^{208}$Tl, and AV
$^{208}$Tl. \label{fig:pdfsbkds}}
\end{center}
\end{figure}

Figure~\ref{fig:pdfsnus} shows the one-dimensional projections of the
distributions of these observables for the three neutrino signals,
showing CC and ES in Phase~II and NC for both data sets.  The Phase~II
distributions are normalized to integrate to 1 except in
Fig.~\ref{fig:pdfsnus}(c), in which the CC and NC distributions are
scaled by a factor of 10 relative to ES for the sake of clarity.  The
Phase~I NC distributions are scaled by the ratio of events in the two
phases, to illustrate the increase in Phase~II.  In the figure, and
throughout the rest of this article, we measure radial positions in
units of AV radii, so that $R^3 \equiv (R_{\rm fit}/R_{AV})^3$.
Figure~\ref{fig:pdfsbkds} shows the same distributions for some of the
detector backgrounds, namely `internal' $^{214}$Bi and $^{208}$Tl
(within the D$_2$O volume) and `AV' $^{208}$Tl (generated within the
bulk acrylic of the vessel walls). While some of the $^{214}$Bi nuclei
came from decays of intrinsic $^{238}$U, the most likely source of
$^{214}$Bi was from decays of $^{222}$Rn entering the detector from
mine air. The $^{208}$Tl nuclei came largely from decays of intrinsic
$^{232}$Th.  Near the $T_{\rm eff}=3.5$~MeV threshold the dominant
signal was from events originating from radioactive decays in the
PMTs.  These events could not be generated with sufficient precision
using the simulation, and so were treated separately from other event
types, as described in Sec.~\ref{s:pmtpdf}.  There were many other
backgrounds; these are described in Sec.~\ref{sec:backgrounds}.

The energy spectra provide a powerful method for separating different
event types.  The CC and ES spectra depend on the shape of the
incident neutrino spectrum.  We treated the CC and ES spectra in two
different ways: in one fit we made no model assumptions about the
underlying spectral shape, allowing the CC and ES spectra to vary in
the fit, and in a second fit we assumed that the underlying incident
neutrino spectrum could be modeled as a smoothly distorted $^8$B
spectrum.  The shapes of NC and background spectra do not depend on
neutrino energy and so were fixed in the fit, to within the systematic
uncertainties derived later.  Decays of $^{214}$Bi and $^{208}$Tl in
the detector both led to $\gamma$ rays above the deuteron binding
energy of 2.2~MeV, which created higher energy events when the
photodisintegration neutron was subsequently captured on either
deuterium (Phase~I) or predominantly $^{35}$Cl (Phase~II).  A
significant fraction of $^{214}$Bi decays produce a 3.27~MeV-endpoint
$\beta$.  These background events are therefore characterized by
steeply falling energy spectra with a photodisintegration tail, as
shown in Fig.~\ref{fig:pdfsbkds}(a).

CC and ES events produced single electrons and, hence, the observed
light from these events was fairly anisotropic, yielding a
correspondingly high value for the isotropy parameter, \betnosp.  The
\bet distributions show small differences due to the different energy
spectra of the two event types, which affects \bet through the known
correlation between energy and isotropy of an event.  The isotropy of
Phase~I NC events looks similar to that of CC and ES events, because
the $\gamma$ ray tended to produce light dominated by that from one
Compton electron.  By contrast, the isotropy distribution of Phase~II
NC events is peaked noticeably lower because neutron capture on
$^{35}$Cl atoms nearly always resulted in multiple \gam s, which could
each scatter an electron and, hence, produce a more isotropic PMT hit
pattern.  Therefore, \bet provides a sensitive method for separation
of electron-like events from neutron capture events in this phase,
without requiring a constraint on the shapes of the CC and ES energy
spectra, thus providing an oscillation-model-independent measurement
of the flux of solar neutrinos.  The isotropy distributions for
$^{214}$Bi events and $^{208}$Tl events inside the heavy water are
noticeably different because, above the $T_{\rm eff}=$ 3.5~MeV
threshold, Cherenkov light from $^{214}$Bi events was dominated by
that from the ground state $\beta$ branch while that from $^{208}$Tl
events was from a $\beta$ and at least one additional Compton
electron.  The difference allowed these events to be separated in our
fit, as was done in previous SNO {\it in-situ} estimates of detector
radioactivity~\cite{longd2o,nsp}.

The \cts distribution is a powerful tool for distinguishing ES events
since the scattering of $\nu_e$ from the Sun resulted in electron
events whose direction is strongly peaked away from the Sun's
location.  The direction of CC events displays a weaker correlation of
$\sim (1 - \frac{1}{3}$\ctsnosp) relative to the direction of the Sun.
The NC distribution is flat since the \gam s generated by neutron
capture carried no information about the incident neutrino direction.
Background events had no correlations with the Sun's location and,
thus, also exhibit a flat distribution, as shown in
Fig.~\ref{fig:pdfsbkds}(c).

The radial position of events within the detector yields a weak
separation between the three neutrino interaction types, but a much
more powerful level of discrimination from external background events.
CC and ES events occurred uniformly within the detector and hence have
relatively flat distributions.  NC events occurred uniformly, but
neutrons produced near the edge of the volume were more likely to
escape into the AV and \hto regions, where the cross section for
neutron capture was very high due to the hydrogen content.  Neutron
capture on hydrogen produced 2.2~MeV \gam s, below the analysis
threshold and thus less likely to be detected.  Therefore, the radial
profile of NC events falls off at the edge of the volume.  This effect
is more noticeable in Phase~I, since the neutron capture efficiency on
deuterium is lower than on $^{35}$Cl and, hence, the neutron mean-free
path was longer in Phase~I than in Phase~II.

\section{Analysis Overview \label{sec:anal_overview}}

The `LETA' analysis differs from previous SNO analyses in the joint
fit of two phases of data, the much lower energy threshold, (which
both result in increased statistics) and significantly improved
systematic uncertainties.

The neutrino signal rates were determined by creating probability
density functions (PDFs) from distributions like those in
Figs.~\ref{fig:pdfsnus} and~\ref{fig:pdfsbkds} and performing an
extended maximum likelihood fit to the data.  The CC and ES spectra
were determined by either allowing the flux to vary in discrete energy
intervals (an `unconstrained fit') or by directly parameterizing the
$\nu_e$ survival probability with a model and fitting for the
parameters of the model.

	There were three major challenges in this analysis: reduction
of backgrounds, creation of accurate PDFs (including determination of
systematic uncertainties on the PDF shapes), and extracting the
neutrino signals, energy spectra, and survival probabilities from the
low-threshold fits.

Three new techniques were applied to reduce backgrounds compared to
previous SNO analyses~\cite{longd2o,nsp}.  First, we made substantial
improvements to energy reconstruction by developing a new algorithm
that included scattered and reflected light in energy estimation.  The
inclusion of `late light' narrowed the detector's effective energy
resolution by roughly 6\%, substantially reducing the leakage of
low-energy background events into the analysis data set by $\sim$60\%.
Second, we developed a suite of event-quality cuts using PMT charge
and time information to reject external background events whose
reconstructed positions were within the fiducial volume.  Third, we
removed known periods of high radon infiltration that occurred during
early SNO runs and when pumps failed in the water purification system.

	Creation of the PDFs was done primarily with a Monte Carlo
(MC) simulation that included a complete model of physics processes
and a detailed description of the detector.  We made substantial
improvements to the Monte Carlo model since our previous publications,
and we describe these improvements in detail in
Sec.~\ref{sec:montecarlo}.

Our general approach to estimating systematic uncertainties on the
Monte Carlo-simulated PDF shapes was based on a comparison of
calibration source data to Monte Carlo simulation, as in previous SNO
analyses.  In cases where the difference between calibration data and
simulation was inconsistent with zero, and we had evidence that the
difference was not caused by a mis-modeling of the calibration source,
we corrected the PDF shapes to better match the data.  For example, we
applied corrections to both the energy (Sec.~\ref{sec:energy}) and
isotropy (Sec.~\ref{sec:beta14}) of simulated events.  Any residual
difference was used as an estimate of the uncertainty on the Monte
Carlo predictions.  Corrections were verified with multiple
calibration sources, such as the distributed `spike' sources as well
as encapsulated sources, and additional uncertainties were included to
account for any differences observed between the various measurements.
Uncertainties were also included to take into account possible
correlations of systematic effects with the observable parameters.
So, for example, we allowed for an energy dependence in the fiducial
volume uncertainty, and the uncertainty on the energy scale was
evaluated in a volume weighted fashion to take into account possible
variations across the detector.

	The final extraction of signal events from the data was a
multi-dimensional, many-parameter fit.  Although marginal
distributions like those shown in Figs.~\ref{fig:pdfsnus}
and~\ref{fig:pdfsbkds} could be used as PDFs, in practice there are
non-trivial correlations between the observables that can lead to
biases in the fit results. We therefore used three-dimensional PDFs
for most of the backgrounds and for the NC signal, factoring out the
dimension in \ctsnosp, which is flat for these events.  The CC and ES
events had PDFs whose dimensionality depended on the type of fit.  For
the unconstrained fit, we used three-dimensional PDFs in
$(R^3,\beta_{14},\cos\theta_{\odot})$, factoring out the $T_{\rm eff}$
dimension because the fit was done in discrete intervals, within which
the $T_{\rm eff}$ spectrum was treated as flat. For the direct fit for
the $\nu_e$ survival probability, we used fully four-dimensional PDFs
for the CC and ES signals.

	The parameters of the `signal extraction' fits were the
amplitudes of the signals and backgrounds, as well as several
parameters that characterized the dominant systematic uncertainties.
\textit{A priori} information on backgrounds and systematic
uncertainties was included.  To verify the results, we pursued two
independent approaches, one using binned and the other unbinned
PDFs. We describe both approaches in Sec.~\ref{sec:sigex}.

We developed and tuned all cuts using simulated events and calibration
source data.  Signal extraction algorithms were developed on Monte
Carlo `fake' data sets, and tested on a 1/3-livetime sample of data.
Once developed, no changes were made to the analysis for the final fit
at our analysis threshold on the full data set.

	In treating systematic uncertainties on the PDF shapes, we
grouped the backgrounds and signals into three classes:
`electron-like' events, which include true single-electron events as
well as those initiated via Compton scattering from a single \gam;
neutron capture events on chlorine that produced a cascade of many
$\gamma$s with a complex branching table; and PMT $\beta$-$\gamma$
decays, which occurred in the glass or envelope of the PMT assembly and
support structure.  The PMT $\beta$-$\gamma$ events were treated 
separately from other
$\beta$-$\gamma$ events because they were heavily influenced by local
optical effects near the PMT concentrators and support structure, and
are therefore hard to model or simulate.
	
The analysis results presented here have substantially reduced
uncertainties on the neutrino interaction rates, particularly for
SNO's signature neutral current measurement.  Although there are many
sources of improvement, the major causes are:
\begin{itemize}
\item The lower energy threshold increased the statistics of the CC
and ES events by roughly 30\%, and of the NC events by $\sim 70$\%;
\item In a joint fit, the difference in neutron detection sensitivity
in the two phases provided improved neutron/electron separation,
beyond that due to differences in the isotropy distributions;
\item Significant background reduction due to improved energy
resolution, removal of high radioactivity periods, and new event
quality cuts;
\item Use of calibration data to correct the PDF shapes.
\end{itemize}

\section{Data Sets \label{sec:dataset}}

     The Phase~I and Phase~II data sets used here have been described
in detail elsewhere~\cite{longd2o,nsp}.  We note only a few critical
details.

     SNO Phase~I ran from November 2, 1999 to May 31, 2001.  Periods
of high radon in Phase~I were removed for this analysis based on the
event rate.  To minimize bias, we used Chauvenet's criterion to
eliminate runs in which the probability of a rate fluctuation as high
or higher than observed was smaller than 1/(2$N$), where $N$ is the
total number of runs in our data set ($\sim500$).  With this cut, we
reduced the previously published 306.4 live days to 277.4.  Most of
the runs removed were in the first two months of the phase, or during
a period in which a radon degassing pump was known to have failed.
This $\sim$9\% reduction in livetime removed roughly 50\% of all
$^{214}$Bi events from the Phase~I data set.  SNO Phase~II ran from
July 2001 to August 2003, for a total of 391.4 live days.

	SNO had several trigger streams, but the primary trigger for
physics data required a coincidence of $N_{\rm coinc}$ or more PMT
hits within a 93~ns window.  From the start of Phase~I until December
20, 2000, $N_{\rm coinc}$ was set to 18; it was subsequently lowered
to 16 PMT hits.  This hardware threshold is substantially below the
analysis threshold, and no efficiency correction was required, even at
3.5~MeV (see Sec.~\ref{sec:treff}).

\section{Monte Carlo Simulation \label{sec:montecarlo}}

	SNO's Monte Carlo simulation played a greater role here than
in previous publications, as we used it to provide PDFs of not only
the neutrino signals, but for nearly all backgrounds as well.  The
simulation included a detailed model of the physics of neutrino
interactions and of decays of radioactive nuclei within the detector.
Propagation of secondary particles was done using the EGS4 shower
code~\cite{egs}, with the exception of neutrons, for which the
MCNP~\cite{mcnp} neutron transport code developed at Los Alamos
National Laboratory was used.  Propagation of optical photons in the
detector media used wavelength-dependent attenuations of \dto and \hto
that were measured {\it in situ} with laserball calibrations, and
acrylic attenuations measured {\it ex situ}.  The simulation included
a detailed model of the detector geometry, including the position and
orientation of the PSUP and the PMTs, the position and thickness of
the AV (including support plates and ropes), the size and position of
the AV `neck', and a full three-dimensional model of the PMTs and
their associated light concentrators.  SNO's data acquisition system
was also simulated, including the time and charge response of the PMTs
and electronics.  Details of the simulation have been presented
in~\cite{longd2o,nsp}; we describe here the extensive upgrades and
changes that were made for this analysis.

	Ultimately, SNO's ability to produce accurate PDFs depends on
the ability of the Monte Carlo simulation to reproduce the low-level
characteristics of the data, such as the distributions of PMT hit
times and charges.  We therefore improved our timing model to more
correctly simulate the `late pulsing' phenomenon seen in the Hamamatsu
R1408s used by SNO. We also added a complete model of the PMT single
photoelectron charge distribution that includes PMT-to-PMT variations
in gain. Gain measurements were made monthly with the laserball source
at the center of the detector, and the simulation uses different
charge distributions for each PMT according to these gain
measurements.

	Addition of the more complete charge spectrum also allowed us
to add a detailed model of each electronics channel's discriminator.
On average, the threshold voltage was near 1/4 of that for a single
photoelectron, but there were large variations among channels because
of variations in noise level.  Over time, the channel thresholds were
adjusted as PMTs became quieter or noisier; these settings were used
in the simulation for each run.  The discriminator model also provided
for channel-by-channel efficiencies to be included, thus improving
simulation of the detector's energy resolution.

	We made several important changes to the optical model as
well.  The first was a calibration of PMT efficiencies, which
accounted for tube-to-tube variations in the response of the
photomultipliers and light concentrators.  These efficiencies are
distinct from the electronics discriminator efficiency described
above, as they depended on the PMT quantum efficiency, local magnetic
field, and individual concentrator reflectivity, while the
discriminator efficiency depended upon PMT channel gain and threshold
setting.  The PMT efficiencies were measured using the laserball, as
part of the detector's full optical calibrations, which were performed
once in Phase~I and three times in Phase~II.  The efficiencies in the
simulation were varied over time accordingly.

\begin{figure} 
\begin{center}
\includegraphics[width=0.48\textwidth]{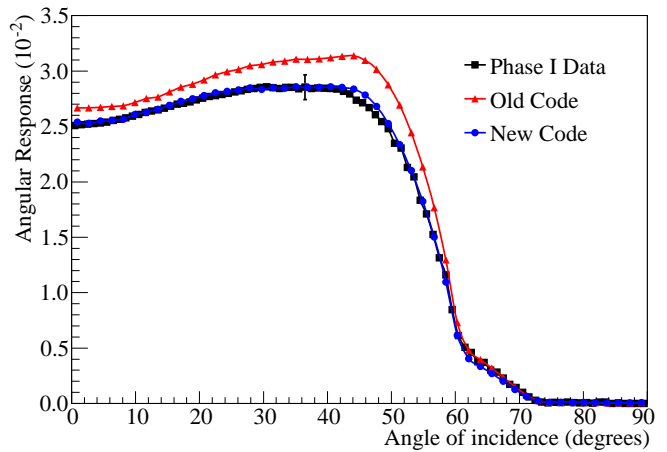}
\caption{(Color online) Comparison of new model of photomultiplier
angular response to data and the old model for Phase~I at
365$\,$nm.\label{fig:3dpmt}}
\end{center} 
\end{figure}

	The light concentrators themselves are known to have degraded
over time and the three-dimensional model of the collection efficiency
of the PMT-concentrator assembly used in previous analyses had to be
modified.  We developed for this analysis a phenomenological model of
the effects of the degradation to the concentrator efficiency. Rather
than modifying the concentrator model itself, we altered the PMT
response as a function of the position at which the photon struck the
photocathode.  In effect, this produced a variation in the response of
the concentrator and PMT assembly as a function of photon incidence
angle.  A simultaneous fit was performed to laserball calibration data
at six wavelengths, with each wavelength data set weighted by the
probability that a photon of that wavelength caused a successful PMT
hit.  The extraction of optical calibration data was extended to a
larger radius than in previous analyses, in order to extract the PMT
response at wider angles.  {\it Ex-situ} data were also included in
the fit to model the response at $>\,$40\degreesp for events in the
light water region.  Time dependence was accommodated by performing
separate fits in time intervals defined by the available calibration
data: one interval in Phase~I and three in Phase~II.  This change
improved the modeling of any position-dependence of the energy
response but did not affect the overall energy scale, which was
calibrated using the \nit source.  We also made a global change to the
light concentrator reflectivity based on measurements with the
$^{16}$N source.  Figure~\ref{fig:3dpmt} compares the new model of the
PMT-concentrator response as a function of incidence angle to that
used in earlier publications.

The laserball calibration data were used as a direct input to the
energy reconstruction algorithms, providing media attenuations, PMT
angular response measurements, and PMT efficiencies.  For wavelengths
outside the range in which data were taken, the Monte Carlo simulation
was used to predict the response.

\section{Hit-level Calibrations \label{sec:hitcal}}

The accuracy with which we know the charge and time of each PMT hit
directly affects event position and energy uncertainties.  To
calibrate the digitized charges and time, we performed pulser
measurements twice weekly, measuring pedestals for the charges and the
mapping of ADC counts to nanoseconds for the times.  The global
channel-to-channel time offsets and the calibration of the pulse
risetime corrections were done with the laserball source deployed near
the center of the detector.  These calibrations have been described
elsewhere~\cite{longd2o}.

	Four significant changes were made to the calibration of PMT
charges and times.  The first was the removal of hits associated with
channel-to-channel crosstalk.  Crosstalk hits in the SNO electronics
were characterized by having low charges, slightly late times, and
being adjacent to a channel with very high charge.

The second change was a correction to the deployed positions of the
laserball source to ensure that the time calibrations were consistent
between calibration runs.  Prior to this correction, the global PMT
offsets had been sensitive to the difference between the nominal and
true position of the source, which varied from calibration run to
calibration run.  The new correction reduced the time-variations of
the PMT calibrations noticeably, but there was a residual 5~cm offset
in the reconstructed $z$-position of events, for which a correction
was applied to all data.

        There were a variety of ways in which PMTs could fail, and we
therefore applied stringent criteria for a PMT to be included in
position and energy reconstruction.  The criteria were applied to both
calibration and `neutrino' data sets as well as to run simulations.

	The last improvement was a calibration to correct for a
rate-dependence in the electronics charge pedestals.  Crosstalk hits
were used to monitor the pedestal drift and a time-varying correction
was applied.  With this correction we could use the PMT charge
measurements to remove certain types of background events, and to
substantially reduce systematic uncertainties on the energy scale
associated with variations in PMT gain, which affected the photon
detection probability.

\begin{figure} 
\begin{center}
\includegraphics[height=0.2\textheight]{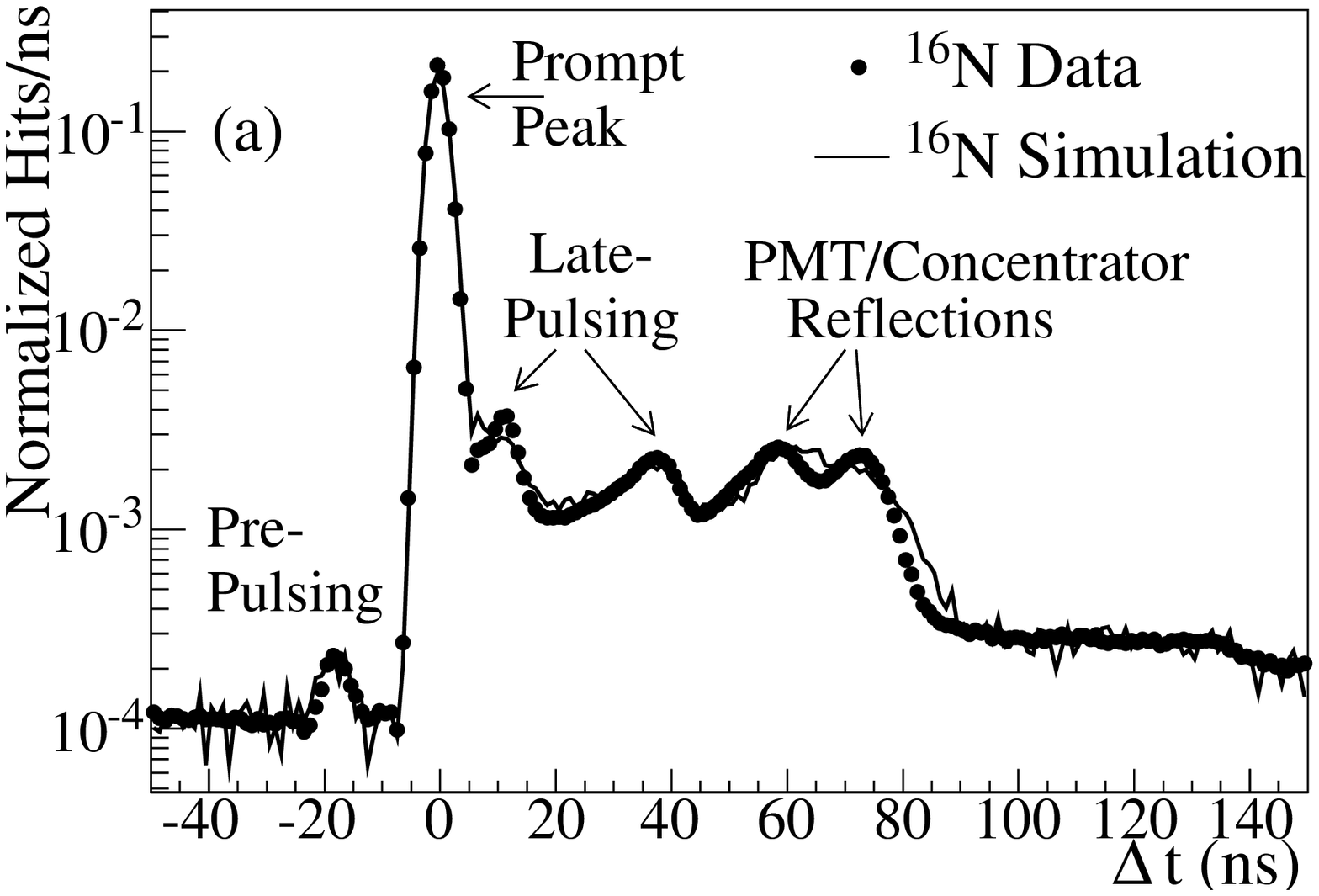}
\includegraphics[height=0.2\textheight]{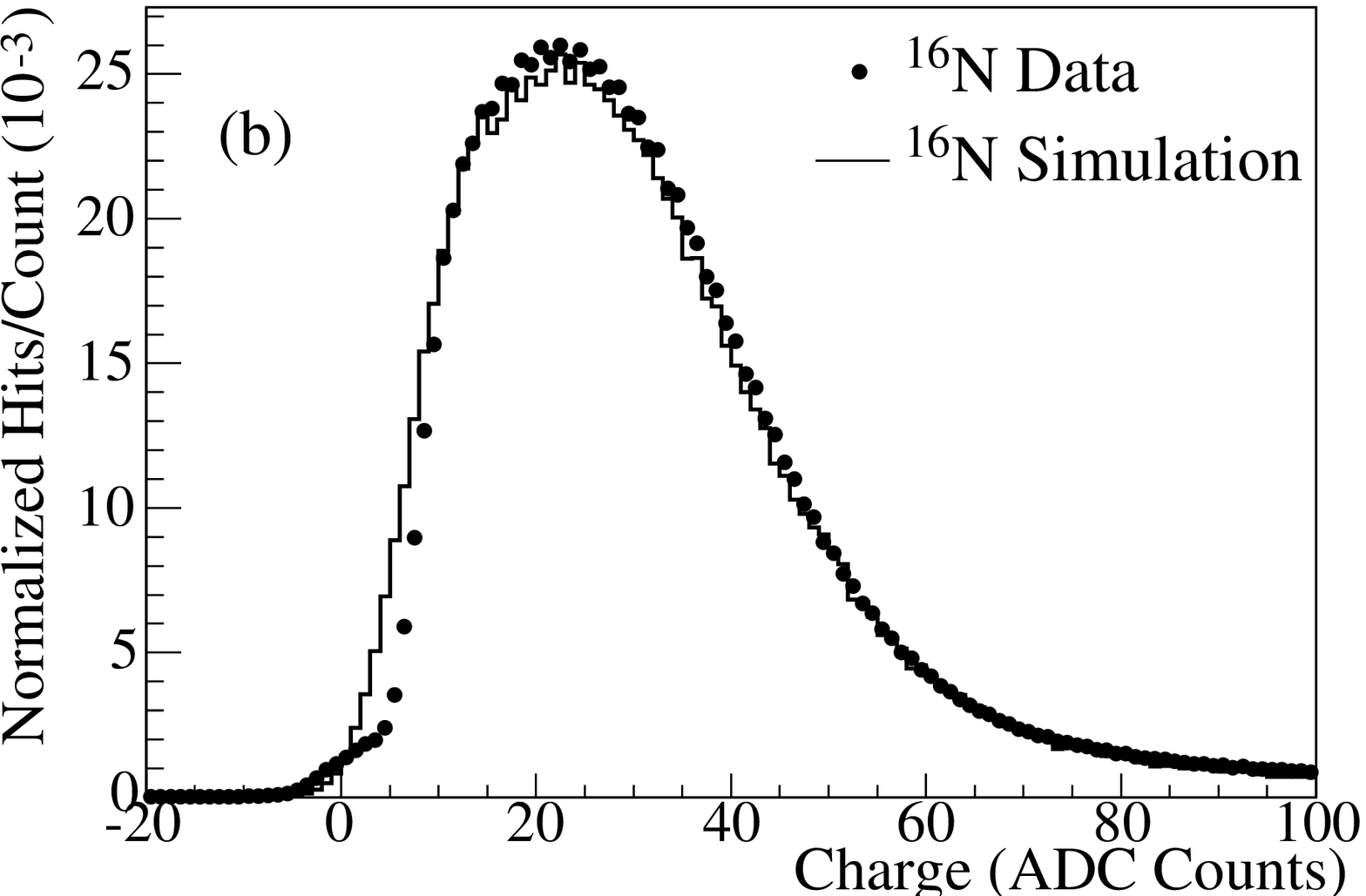}
\caption{Comparison of $^{16}$N simulation to data for (a) PMT hit
time-of-flight residuals and (b) photoelectron charge
spectra.\label{fig:qt}}
\end{center} 
\end{figure}

	Figure~\ref{fig:qt} shows the distributions of PMT
time-of-flight residuals and measured photoelectron charges for a
$^{16}$N calibration run at the center of the detector compared to a
simulation of that run.  The simulation includes the upgrades
discussed in Sec.~\ref{sec:montecarlo}.  The time residuals show
excellent agreement in the dominant prompt peak centered near $\Delta
t=0$~ns, as well as good agreement for the much smaller pre-pulsing
($\Delta t\sim -20$~ns) and late-pulsing ($\Delta t \sim 15$~ns and
$\Delta t \sim 35$~ns) features. For the charge distribution, the
agreement is also excellent above 10 ADC counts or so, which
corresponds to the majority of the charges used in the analysis.
Thus, we are confident that the simulation models the behavior of
reconstruction and cuts with sufficient accuracy.

\section{Position and Direction Reconstruction\label{sec:recon}}

        The primary reconstruction algorithm used in this analysis was
the same as in previous Phase~I publications.  We used reconstructed
event position and direction to produce the PDFs shown in
Figs.~\ref{fig:pdfsnus} and~\ref{fig:pdfsbkds}, and to reject
background events originating outside the AV.  Knowledge of event
position and direction was also used in the estimation of event energy
(see Sec.~\ref{sec:energy}).  Below we outline the reconstruction
method, and then discuss the uncertainties in our knowledge of event
positions and directions.

\subsection{Reconstruction Algorithm}

The vertex and direction reconstruction algorithm fitted event position,
time, and direction simultaneously using the hit times and locations
of the hit PMTs.  These values were found by maximizing the
log-likelihood function,
\begin{equation}
{\log\cal L}(\vec{r}_e,\vec{v}_e,t_e) = \sum_{i=1}^{N_{\rm hit}} \log
{\cal P}(t^{\rm res}_i,\vec{r}_i;\vec{r}_e,\vec{v}_e,t_e),
\end{equation}
with respect to the reconstructed position ($\vec{r}_e$), direction
($\vec{v}_e$), and time ($t_e$) of the event.  ${\cal
P}(t^{\rm res}_i,\vec{r}_i;\vec{r}_e,\vec{v}_e,t_e)$ is the
probability of observing a hit in PMT $i$ (located at $\vec{r}_i$)
with PMT time-of-flight residual $t^{\rm res}_i$
(Eq.~\eqref{eqn:ftp-tresid}), given a single Cherenkov electron track
occurring at time $t_e$ and position $\vec{r}_e$, with direction
$\vec{v}_e$.  The sum is over all good PMTs for which a hit was
recorded.  The PMT time-of-flight residuals relative to the
hypothesized fit vertex position are given by:
\begin{equation}
\label{eqn:ftp-tresid} 
t^{\rm res}_i = t_{i} - t_{\rm e} - |\vec{r}_{\rm e} -
\vec{r}_{i}|\frac{{n_{\rm eff}}}{c},
\end{equation} 
where $t_i$ is the hit time of the $i$th PMT.  The photons are assumed
to travel at a group velocity $\frac{{c}}{n_{\rm eff}}$, with ${n_{\rm
eff}}$ an effective index of refraction averaged over the detector
media.

The probability ${\cal P}$ contains two terms to allow for the
possibilities that the detected photon arrived either directly from
the event vertex (${\cal P}_{\rm direct}$) or resulted from
reflections, scattering, or random PMT noise (${\cal P}_{\rm other}$).
These two probabilities were weighted based on data collected in the
laserball calibration runs.

        The azimuthal symmetry of Cherenkov light about the event
direction dilutes the precision of reconstruction along the event
direction.  Thus, photons that scattered out of the Cherenkov cone
tended to systematically drive the reconstructed event vertex along
the fitted event direction.  After initial estimates of position and
direction were obtained, a correction was applied to shift the vertex
back along the direction of the event so as to compensate for this
systematic drive.  The correction varied with the distance of the
event from the PSUP as measured along its fitted direction.

The reconstruction algorithm returned a quality-of-fit statistic
relative to the hypothesis that the event was a correctly
reconstructed single electron.  This statistic was used later in the
analysis to remove backgrounds and reduce tails on the reconstruction
resolution.  Details of the reconstruction algorithm can be found
in~\cite{longd2o}.

\subsection{Uncertainties on Position and Direction}

Many effects that could produce systematic shifts in reconstructed
positions were modeled in the simulation.  Data from calibration
sources deployed within the detector were compared to Monte Carlo
predictions, and the differences were used to quantify the uncertainty
on the simulation.  The observed differences were not deemed significant
enough to warrant applying a correction to the Monte Carlo-generated
positions, and so the full size of the difference was taken as the
magnitude of the uncertainty.  The differences between data and Monte
Carlo events were parameterized as four types:
\begin{itemize}
\item vertex offset: a constant offset between an event's true and
reconstructed positions;
\item vertex scale: a position-dependent shift of events either inward
or outward;
\item vertex resolution: the width of the distribution of
reconstructed event positions;
\item angular resolution: the width of the distribution of
reconstructed event directions relative to the initial electron
direction.
\end{itemize}

These uncertainties can have an impact upon the flux and spectral
measurements in two ways: by altering the prediction for the number of
events reconstructing inside the fiducial volume and by affecting the
shape of the PDFs used in the signal extraction.

Reconstruction uncertainties were determined primarily from \nit
source data.  In previous analyses \cite{longd2o}, the volume density
of Compton-scattered electrons relative to the source location was
modeled with the analytic function $S(r) \sim
\exp(\frac{-r}{\lambda})/(r^2)$.  Model improvements for this analysis
allowed us to extract this distribution for each \nit source run from
the Monte Carlo simulation of that run, and take into account the
exact source geometry, effect of data selection criteria on the
distribution, and any time-dependent detector effects.

\begin{figure}[!ht]
\begin{center}
\includegraphics[width=0.48\textwidth]{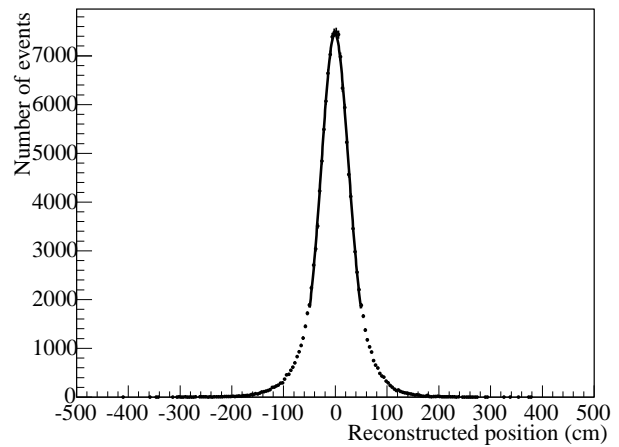}
\caption{\label{f:n16fit}Fit of the \nit Compton-electron position
distribution convolved with a Gaussian to the reconstructed $z$
position of \nit data events for a typical central run in Phase~II.}
\end{center} 
\end{figure}

The distribution of electron positions was convolved with a Gaussian,
representing the detector response, and the resulting function was fit
to the one-dimensional reconstructed position distribution along each
axis, allowing both the mean and standard deviation of the Gaussian to
vary for each orthogonal axis independently.  An example of such a fit
is shown in Figure~\ref{f:n16fit}.  This fit was done separately for
the \nit data and the Monte Carlo simulation of each \nit run.  The
difference in the Gaussian means gives the vertex offset for that run
and the square root of the difference in the variances represents the
difference in vertex resolution.

\subsubsection{Vertex Offset}
\label{s:voff}
Analysis of the differences between the reconstructed and true event
vertex positions at the center of the detector, or `central vertex
offset', was done using runs with the source within 25$\,$cm of the
center, where the source position is known most accurately.  This
avoids confusion with any position-dependent effects, which are taken
into account in the scale measurement (Sec.~\ref{s:vscale}).  A
data$-$MC offset was determined for each run, along each detector
axis.  The offsets from the runs were combined in weighted averages
along each axis, with the uncertainty for each run offset increased to
include the uncertainty in source position.  Although the results
showed a small mean offset along each axis, the magnitude was
comparable to the source position uncertainty and therefore we did not
correct the PDFs based on this difference.  Instead, asymmetric
double-sided uncertainties were formulated by using the uncertainty in
the weighted average, and increasing it by the magnitude of the
weighted average itself on the side on which the offset was measured.
The effects of these uncertainties were determined during signal
extraction by shifting the position of each event by the positive and
negative values of the uncertainty along each axis independently, and
recomputing the PDFs.  The values of the uncertainties are given in
Table~\ref{t:recunc} in Sec.~\ref{s:recsum}.

\subsubsection{Vertex Scale}
\label{s:vscale}
A potential position-dependent bias in the reconstructed position
that can be represented as being proportional to the distance of the
event from the center of the detector is defined as a vertex scale
systematic.

In previous SNO analyses, uncertainty in the position of the
calibration source was a major contribution to reconstruction
uncertainties, especially away from the $z$-axis of the detector,
where sources were deployed in a less accurate mode.  A new method was
derived for this analysis to reduce sensitivity to this effect.
Although the absolute source position was known only to $\sim2\,$cm on
the $z$-axis and $\sim5\,$cm away from this axis, changes in position
once the source was deployed were known with much greater precision.
By comparing the result from each \nit run to a run at the center of
the detector from the same deployment scan, possible offsets between
the recorded and true source position were removed, thus reducing
source position uncertainties.  In addition, any constant offset in
vertex position, such as that measured in Sec.~\ref{s:voff}, was
inherently removed by this method, thus deconvolving the measurement
of scale from offset.  This method allowed data from different scans
to be combined, providing a more representative sampling across the
time span of the data set and improving the statistics of the
measurement.

\begin{figure}[!ht]
\begin{center}
\includegraphics[width=0.48\textwidth]{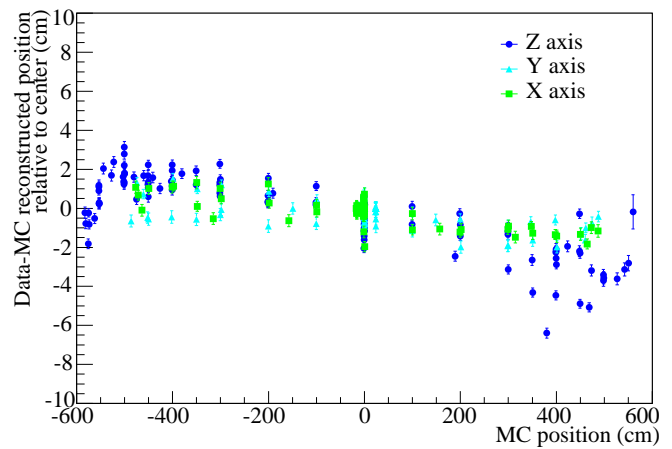}
\caption{\label{f:recxyz}(Color online) Vertex offset along the three
detector axes as a function of position within the detector.}
\end{center} 
\end{figure}

Vertex scale was investigated by using the data$-$MC
reconstructed position offset along each detector axis,
as shown in Figure~\ref{f:recxyz}, using only runs within 50~cm of
that axis to minimize correlations among the three.  The runs were
grouped into 50$\,$cm bins along each axis by source position, and
the weighted average of the offsets for the runs within each bin was
found.  A linear function was fit to the bins as a function of
position along that axis.  Since the method was designed to remove any
central vertex offset, the function was defined to be zero at the
center of the detector.
	
	The slope from the fit provides the scaling required to bring
the simulation into agreement with data. We did not apply a
correction, but instead treated it as an asymmetric uncertainty on the
reconstructed positions of all events.  The effects observed along the
$x$ and $y$ axes were of a very similar magnitude and, therefore, were
assumed to be due to a radial effect, possibly caused either by small
errors in the modeling of the wavelength-dependent refractive index or
residual PMT timing calibration errors.  Conservatively, the larger of
the $x$ and $y$ values was used to bound this effect.  The resulting
uncertainty was applied in our signal extraction fits by multiplying
the $x$, $y$ and $z$ position of each event in our PDFs by the value
of the scale uncertainty, thus shifting events either inwards or
outwards in the detector, and taking the difference from the nominal
fit.  Since the effect observed along the $z$-axis was larger, the
difference of this from the radial effect was treated as an additional
uncertainty, applied only to the $z$ position of events.  The values
used for each uncertainty are listed in Table~\ref{t:recunc} in
Sec.~\ref{s:recsum}.

\begin{figure}[!ht] 
\begin{center}
\includegraphics[width=0.48\textwidth]{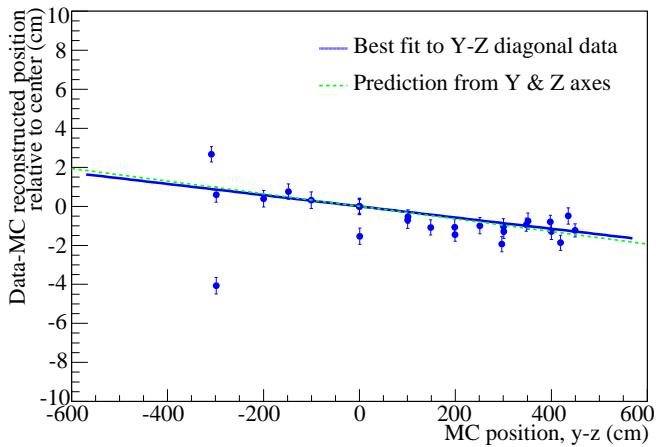}
\caption{\label{f:yztest}(Color online) Vertex offset along the
$y$-$z$ diagonal as a function of position along that diagonal.  The
dashed line shows the prediction from the $y$- and $z$-axis values and
the solid line shows the best fit scaling value for these data points.
Observed variations at negative positions are likely associated with
systematics in source position.}
\end{center} 
\end{figure}

Since only runs within 50$\,$cm of each Cartesian axis were used to
determine vertex scale, diagonal axis runs could be used for
verification.  The method described measured the scale for each
Cartesian axis independently. The values obtained for the $y$ and $z$
axes, for example, could therefore be combined to predict the scaling
for runs on the $y$-$z$ diagonal.  The prediction was shown to agree
very well with the data, as illustrated in Figure~\ref{f:yztest},
demonstrating the robustness of the analysis and its applicability to
events everywhere in the fiducial volume.

A similar analysis was performed using \cf source data in Phase~II.
The results were consistent with those shown here, verifying that the
same uncertainties could be applied to both electron-like and neutron
capture events.

We investigated several other potential causes of variation in
reconstruction accuracy.  The \nitnosp-source event rate during most
calibration runs was high in comparison to our expected neutrino event
rate, so the results were checked using low-rate \nit data.  The
stability over time was determined by comparing runs across the span
of the two phases.  As in previous analyses~\cite{longd2o},
calibration-source dependence was investigated by verifying \nit
results using the \lit source.  This also provides a check on the
energy dependence because the \lit data extended to higher energies
than the \nit data.  The results were all consistent within the
uncertainties presented here.

\subsubsection{Vertex Resolution}
\label{s:vresn}
The position resolution achieved in this analysis was $\sim$20~cm for
data events.  The difference in resolutions between data and Monte
Carlo events was modeled as a Gaussian of standard deviation (or
`width') $\sigma_{\rm extra}$, by which the Monte Carlo distribution
should be smeared to reproduce the data.  $\sigma_{\rm extra}^2$ was
given by $(\sigma_{\rm Data}^2 - \sigma_{\rm MC}^2)$ for each \nit
run.  This procedure is only valid for $\sigma_{\rm MC} < \sigma_{\rm
Data}$, which was the likely scenario since any minor detector
non-uniformities tend to cause a broader resolution in the data.  In
some cases, the simulation and data were close enough to one another
that statistical variation caused $\sigma_{\rm Data}$ to appear to be
less than $\sigma_{\rm MC}$.  In these cases, $ |(\sigma_{\rm Data}^2
- \sigma_{\rm MC}^2)|$ was taken to represent the uncertainty in the
comparison.  The results from the runs were combined in a weighted
average, independently for each detector axis.  The resulting values
for $\sigma_{\rm extra}$ are listed in Table~\ref{t:recunc} in
Sec.~\ref{s:recsum}.  These were applied during the signal extraction
by smearing the positions of all Monte Carlo events by a Gaussian of
the appropriate width.  This was achieved for the binned signal
extraction (Sec.~\ref{s:mxf}) by generating a random number for each
event from a Gaussian of the correct width and adding the result to
the event's position and, for the unbinned method, by a direct
analytic convolution (Sec.~\ref{s:kernel}).

\subsubsection{Angular Resolution}

The \nit source was used for this measurement by relying on the high
degree of colinearity of Compton scattered electrons with the initial
$\gamma$ direction.  The mean of the distribution of reconstructed
event positions was used to estimate the source position.  The
reconstructed event position was used as an estimate for the
scattering vertex.  To reduce the effect of reconstruction errors,
only events reconstructing more than 120$\,$cm from the source were
used.  The angle between the initial $\gamma$ direction (taken to be
the vector from the source position to the fitted scattering vertex)
and the reconstructed event direction was found and the distributions
of these angles were compared for data and Monte Carlo events.

The same functional form used in previous analyses~\cite{nsp} was fit
to the distributions for data and Monte Carlo events within each run.
The weighted average of the differences in the fitted parameters was
computed across the runs and the resulting value used as an estimate
of the uncertainty in angular resolution (given in
Table~\ref{t:recunc}, Sec.~\ref{s:recsum}).

\subsubsection{Energy Dependent Fiducial Volume}

\begin{figure}[!ht]
\begin{center}
\includegraphics[width=0.48\textwidth]{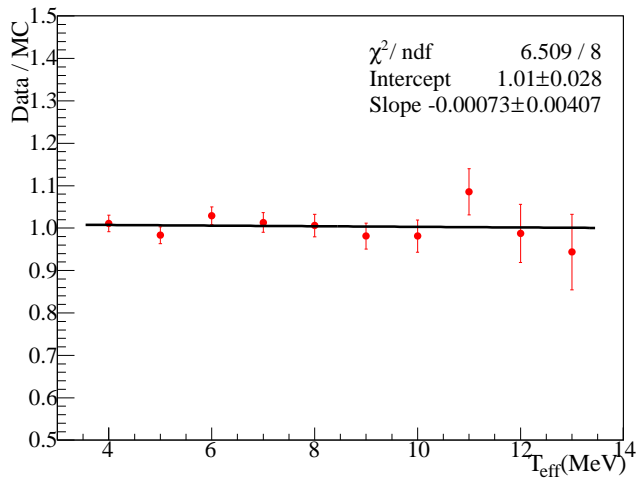}
\caption{\label{f:efv}(Color online) Ratio of the fraction of events
reconstructing inside the source position for data and Monte Carlo
events, as a function of effective electron energy, for \lit source
runs.}
\end{center} 
\end{figure}

The energy dependence of the vertex scaling is of particular
importance since it could affect the number of events that reconstruct
within the fiducial volume as a function of energy and, hence, distort
the extracted neutrino spectrum.  Because the \nit source provided
monoenergetic $\gamma$s, giving rise to electrons around 5~MeV,
whereas the \lit source sampled the full range of the neutrino energy
spectrum, the \lit source was used for this measurement.  The fraction
of events reconstructing inside the source's radial position, closer
to the detector center, was used as a measure of the number of events
reconstructing inside the fiducial volume to take into account both
vertex shift and resolution effects.  Absolute offsets between data
and Monte Carlo events have already been characterized in Sections
\ref{s:voff}--\ref{s:vresn}, so a differential comparison of this
parameter between data and Monte Carlo events was used to evaluate any
energy dependence.  A fit from Phase~II is shown in
Figure~\ref{f:efv}.  The energy dependence is given by the slope of a
straight line fit to the ratio of the data and Monte Carlo parameters,
averaged across calibration runs.  The final uncertainty is quoted as
an asymmetric, double-sided uncertainty to account for the non-zero
value of the slope and its uncertainty.  The values for each phase are
given in Table~\ref{t:recunc}.  The absolute shift, indicated in
Fig.~\ref{f:efv} by an intercept different from one, is a measure of
the global vertex scaling.  This effect has already been evaluated in
Sec.~\ref{s:vscale}.  It does not impact the energy dependence and
therefore is not relevant to this present measurement.

An additional check was performed using neutrino data from outside the
fiducial volume.  All standard analysis cuts were applied, as
described in Sec.~\ref{s:cuts}, as well as a 5.5~MeV threshold to
select a clean sample of neutrino events.  A Hill function was fit to
the radial distribution of the events, with the half-point of the
function representing the position of the AV.  Statistics in the data
were limited, so the fit was performed in just three energy bins.
Monte Carlo simulation of the three types of neutrino interactions was
combined in the signal ratios found in a previous SNO analysis
\cite{nsp} and the same fit was performed.  The ratio of the resulting
fitted AV position in the data and simulation is a measure of the
radial scaling and, therefore, the energy dependence of this ratio is
a check on the analysis described above.  The results were in good
agreement.  In Phase~II the energy dependence was $0.8\pm2.1\%$/MeV,
in comparison to $-0.07\pm0.41\%$/MeV measured using the \lit source.

\subsubsection{Summary of Reconstructed Position Uncertainties}
\label{s:recsum}
Table~\ref{t:recunc} summarizes the uncertainties in reconstructed
position and direction.

\begin{table}[!ht]
\begin{center}
\begin{tabular}{lccc} \hline \hline
 & \multicolumn{2}{c}{Uncertainty, $\delta_i$}& Transformation \\
Parameter & Phase~I & Phase~II &of observables \\ \hline $x$ Offset
(cm) & $^{+1.15}_{-0.13}$ & $^{+0.62}_{-0.07}$ & $x+\delta_i$ \\ $y$
Offset (cm) & $^{+2.87}_{-0.17}$ & $^{+2.29}_{-0.09}$& $y+\delta_i$\\
$z$ Offset (cm) & $^{+2.58}_{-0.15}$ & $^{+3.11}_{-0.16}$&
$z+\delta_i$\\ $R$ Scale (\%)& $^{+0.10}_{-0.57}$ & $^{+0.04}_{-0.34}$
& $(1 + \frac{\delta_i}{100})x_i$ \\ $z$ Scale (\%)& $^{+0.40}_{-0.0}$
& $^{+0.03}_{-0.25}$ & $(1 + \frac{\delta_i}{100})z$\\ $x$ resn
(cm)&$+3.3$ & $+3.1$ & $x + \mathcal{N}(0,\delta_i)$\\ $y$ resn (cm)&
$+2.2$ &$+3.4$ & $y + \mathcal{N}(0,\delta_i)$\\ $z$ resn (cm)& $+1.5$
& $+5.3$ &$z + \mathcal{N}(0,\delta_i)$\\ Angular resn & $\pm 0.11$ &
$\pm 0.11$&$1 + (\cos\theta_{\odot}-1)(1 + \delta_i)$\\ EFV (\%/MeV)&
$^{+0.85}_{-0.49}$ & $^{+0.41}_{-0.48}$ &
$W=1+\frac{\delta_i}{100}(T_{\rm eff}-5.05)$\\ \hline \hline
\end{tabular}
\caption{\label{t:recunc}Systematic uncertainties in the reconstructed
position and direction of events.  EFV is the energy dependent
fiducial volume uncertainty.  The column labeled ``Transformation of
observables'' refers to the formulae used to propagate these
uncertainties into the signal extraction fits.
$\mathcal{N}(0,\delta_i)$ refers to a convolution with a Gaussian
distribution of mean 0.0 and standard deviation $\delta_i$. Events
that are pushed past \cts $=\pm1.0$ are randomly assigned a \cts value
in the interval [-1.0, 1.0].  $W$ is an energy-dependent fiducial
volume factor applied around the midpoint of the \nit energy, where
$T_{\rm eff}$ is the reconstructed effective electron kinetic energy
and 5.05~MeV is the central $T_{\rm eff}$ value for the \nit data.
This was applied as a weight for each event when creating the PDFs.
(``Resolution'' is abbreviated as ``resn'').}
\end{center}
\end{table}

It is worth noting that in previous analyses~\cite{nsp} the radial
scaling uncertainty was evaluated at $\pm\,$1\%, which translates to a
3\% uncertainty in fiducial volume.  The improved analysis presented
here has reduced the scale uncertainty to a little over 0.5\% at its
maximum and significantly less in most dimensions.  The resolution
differences observed previously were on the order of 9$\,$cm
\cite{longd2o}, whereas the differences measured here are roughly one
third of that in most dimensions.  The angular resolution uncertainty
of 11\% is an improvement over the 16\% measured in previous work
\cite{nsp}.

\section{Energy Reconstruction}
\label{sec:energy}

We estimated the kinetic energy of an event after its position and
direction were reconstructed.  The energy estimate was used both to
reject background events and to produce the PDFs shown in
Figs.~\ref{fig:pdfsnus} and \ref{fig:pdfsbkds}.  Improving the
resolution of the energy estimation algorithm was critical because of
the low energy threshold of the analysis -- a 6\% improvement in
energy resolution reduces the number of background events
reconstructing above threshold by $\sim$60\%.

\subsection{Total-Light Energy Estimator}
\label{sec:ftk}

A new algorithm, called ``FTK'', was designed to use all the detected
PMT hits in the energy estimate, including scattered and reflected
light~\cite{dunfordthesis}.  The look-up table approach of the
prompt-light fitter used in previous publications was abandoned in
favor of a maximum likelihood method, in which photon detection
probabilities were generated based on the reconstructed event position
and direction.  The best value of the effective kinetic energy,
$T_{\rm eff}$, was found by maximizing the likelihood given the
observed number of hit PMTs, $N_{\rm hit}$, and taking into account
optical effects due to the reconstructed position and direction of the
event.  In principle, one could consider a more sophisticated approach
in which both the number and distribution of all hit PMTs are used
along with the recorded time of each hit, but such an approach is much
more time intensive and was judged to be impractical for the present
analysis.

We considered five sources of PMT hits in an event, defined by the
following quantities:
\begin{itemize}
	\item $n^{\rm dir}_{\rm exp}$ - the expected number of
	detected photons that traveled directly to a PMT, undergoing
	only refraction at the media boundaries;
	\item $n^{\rm scat}_{\rm exp}$ - the expected number of
	detected photons that were Rayleigh scattered once in the \dto
	or \hto before detection (scattering in the acrylic is
	neglected);
	\item $n^{\rm av}_{\rm exp}$ - the expected number of detected
	photons that reflected off the inner or outer surface of the
	acrylic vessel;
	\item $n^{\rm pmt}_{\rm exp}$ - the expected number of
	detected photons that reflected off the PMTs or light
	concentrators;
	\item $n^{\rm noise}_{\rm exp}$ - the expected number of PMT
	noise hits, based on run-by-run measurements.
\end{itemize}
FTK computed the probabilities of a single photon being detected by
any PMT via the four event-related processes: $\rho_{\rm dir}$,
$\rho_{\rm scat}$, $\rho_{\rm av}$, $\rho_{\rm pmt}$.  The direct
light probability was found by tracing rays from the event vertex to
each PMT, and weighting each ray by the attenuation probability in
each medium, transmittance at each boundary, solid angle of each PMT,
and detection probability given the angle of entry into the light
concentrator.  Scattering and reflection probabilities were found
using a combination of ray tracing and tables computed from Monte
Carlo simulation of photons propagating through the detector.

If $N_\gamma$ is the number of potentially detectable Cherenkov
photons produced in the event given the inherent PMT detection
efficiency, then the expected number of detected photons given these
probabilities is:
\begin{equation}
	n_{\rm exp}(N_\gamma) = N_\gamma \times (\rho_{\rm dir} +
	\rho_{\rm scat} + \rho_{\rm av} + \rho_{\rm pmt}).
\end{equation}
To be able to compare $n_{\rm exp}$ to the observed $N_{\rm hit}$, we
need to account for noise hits and convert from detected photons to
PMT hits, since multiple photons in the same PMT produced only one
hit.  Given the rarity of multiple photons in a single PMT at solar
neutrino energies, FTK made a correction only to the dominant source
term, $n^{\rm dir}_{\rm exp}=N_\gamma\rho_{\rm dir}$.  Letting $N_{\rm
MPC}(n^{\rm dir}_{\rm exp})$ be the multi-photon corrected number of
direct PMT hits, the total expected number of hits is:
\begin{eqnarray}
	N_{\rm exp}(N_\gamma) & \approx & N_{\rm MPC}(n^{\rm dir}_{\rm
	exp}) \nonumber \\ & & + N_\gamma \times (\rho_{\rm scat} +
	\rho_{\rm av} + \rho_{\rm pmt}) + n^{\rm noise}_{\rm exp}.
\end{eqnarray}
The probability of observing $N_{\rm hit}$ hits when $N_{\rm exp}$ are
expected is given by the Poisson distribution:
\begin{equation}
	P(N_{\rm hit}\,|\,N_\gamma) = \frac{(N_{\rm exp})^{N_{\rm
	hit}} e^{-N_{\rm exp}}}{N_{\rm hit}!}.
\end{equation}
To obtain a likelihood function for $T_{\rm eff}$, rather than
$N_\gamma$, we integrate over the distribution of $N_\gamma$ given an
energy $T_{\rm eff}$:
\begin{equation}
	\mathcal{L}(T_{\rm eff}) = \int \frac{(N_{\rm
	exp}(N_\gamma))^{N_{\rm hit}} e^{-N_{\rm
	exp}(N_\gamma)}}{N_{\rm hit}!}\times P(N_\gamma\,|\,T_{\rm
	eff})\,dN_\gamma,
\end{equation}
where $P(N_\gamma\,|\,T_{\rm eff})$ is the probability of $N_\gamma$
Cherenkov photons being emitted in an event with energy $T_{\rm eff}$.
The negative log-likelihood was then minimized in one dimension to
give the estimated energy of the event.

\subsection{Energy Scale Corrections and Uncertainties}
\label{sec:ecorr}

We measured the energy scale of the detector by deploying the tagged
$^{16}$N $\gamma$ source at various locations in the $x$-$z$ and
$y$-$z$ planes within the \dto volume.  Although $^{16}$N was a nearly
monoenergetic $\gamma$ source, it produced electrons with a range of
energies through multiple Compton scattering and $e^+e^-$ pair
production.  As a result, the single 6.13~MeV $\gamma$ produced an
`effective electron kinetic energy' ($T_{\rm eff}$) distribution that
peaked at approximately 5~MeV.

	Using the $^{16}$N $\gamma$-ray source to determine the
detector's energy scale is complicated by its broad spectrum of
electron energies. To separate the detector's response from this
intrinsic electron energy distribution, we modeled the reconstructed
energy distribution with the integral
\begin{equation}
	P(T_{\rm eff}) = N\int P_{\rm
	source}(E_{e^{-}})\frac{1}{\sqrt{2\pi}\, \sigma}
	e^{\frac{(T_{\rm eff} - E_{e^{-}} -
	p_3)^2}{2\sigma^2}}dE_{e^{-}},
\end{equation}
where $N$ is a normalization constant, $\sigma(E_{e^{-}}) = p_1 +
p_2\sqrt{E_{e^{-}}}$ is the detector resolution, and $P_{\rm source}$
is the apparent electron energy distribution from the $^{16}$N
$\gamma$ rays without including the detector optical response.  $p_3$
sets the displacement of the $^{16}$N peak, and therefore the offset
in energy scale at that source location.  The $P_{\rm source}$
distribution was computed from a Monte Carlo simulation of $\gamma$
propagation through the source container and production of Cherenkov
photons from Compton-scattered $e^-$ and pair-produced $e^+e^-$.  We
translated the number of Cherenkov photons in each simulated event to
a most probable electron (MPE) kinetic energy with the same tables
that were used in the FTK energy estimation algorithm, and generated
the distribution, $P_{\rm source}$, of event
values~\cite{dunfordthesis}.  Given this fixed distribution for the
$^{16}$N calibration source, we fit for $N$, $p_1$, $p_2$, and $p_3$
in each source run, for both data and for Monte Carlo simulation of
the same source position and detector state.  The parameter
differences between data and Monte Carlo, run-by-run, determined the
energy corrections and uncertainties.  Parameters $p_1$ and $p_2$
measure the detector energy resolution, and are discussed further in
Sec.~\ref{sec:eres}.  Parameter $p_3$ was used here to define the
spatial energy scale correction and uncertainties.

The Monte Carlo was initially tuned by adjusting a global collection
efficiency parameter in the simulation to minimize the difference
between data and Monte Carlo energy scales for $^{16}$N runs at the
center of the detector.  A series of additional corrections were then
applied to the estimated energy of all the data and Monte Carlo
events, to remedy known biases.

	Approximations in FTK's handling of multiple hits on a single
tube lead to a small energy non-linearity, and we derived a correction
for this by comparing the reconstructed energy for Monte Carlo events
to their true energies.  Similarly, the simple PMT optical model used
by FTK produced a small radial bias in event energies and, again,
comparison of reconstructed energies of Monte Carlo events to their
true values were used to provide a correction.

	Two additional corrections were based on evaluations of
data. The first was to compensate for the degradation of the PMT light
concentrators, which changed the detector's energy response over time
during Phase I.  The degradation affected the fraction of light that
was reflected off the PMT array.  We tracked the variation using
$^{16}$N runs taken at the center of the detector, and created a
time-dependent correction to event energies that shifted their values
by up to 0.4\%~\cite{dunfordthesis}.

\begin{figure}[!ht]
\begin{center}
\includegraphics[width=0.48\textwidth]{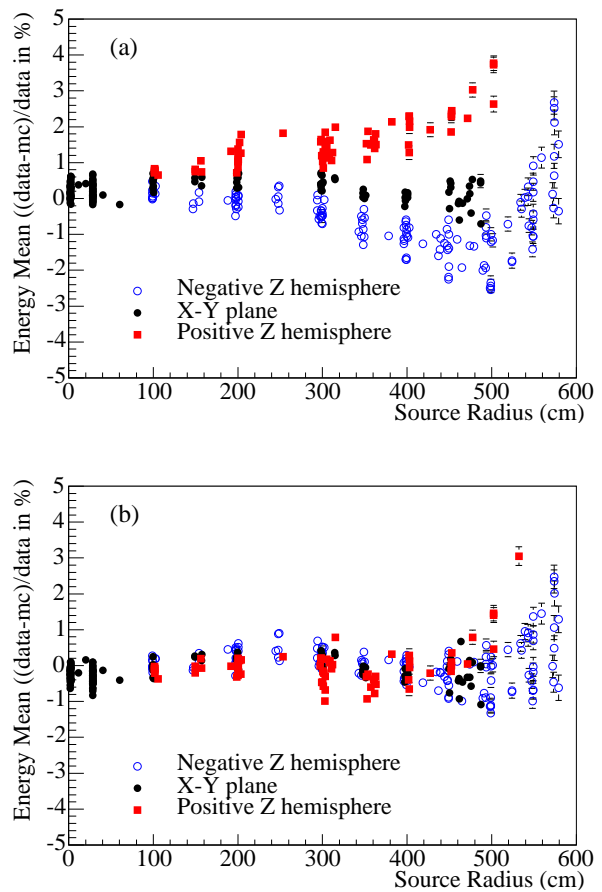}
\caption{\label{f:saltescale}(Color online) Difference between
$^{16}$N data and Monte Carlo energy scales as a function of radius
for Phase~II $^{16}$N source runs in the upper hemisphere, on the
equatorial plane, and in the lower hemisphere.  Panel (a) shows the
significant variation in these three regions before the spatial energy
correction.  Panel (b) shows the same runs after the spatial energy
scale correction is applied.  (The fiducial volume cut is at 550~cm).}
\end{center} 
\end{figure}

	The final correction was applied to remove a variation in
energy with the detector $z$-coordinate.  Figure~\ref{f:saltescale}(a)
shows the difference between the average reconstructed energies of
events from the $^{16}$N source for each calibration run, and the
Monte Carlo simulation of the run, as a function of the radial
position of the source.  As can be seen, for events in the top
(positive $z$) hemisphere of the detector, the Monte Carlo
underestimated the event energies by as much as 3\% and, in the bottom
hemisphere, it overestimated the energies by almost the same amount.
The cause of the former was the simulation's poor optical model of the
acrylic in the neck of the AV.  The latter was likely caused by
accumulation of residue at the bottom of the acrylic vessel and
variations in the degradation of the PMT light concentrators.

To correct for the $z$-dependence of the energy scale, we first split
the $^{16}$N calibration runs into two groups.  One group contained
runs on the $x$-$z$ plane along with half of the runs on the $z$-axis,
and was used to construct the correction function.  The second group
contained runs on the $y$-$z$ plane along with the other half of the
$z$-axis runs, and was used later to independently evaluate the
spatial component of the energy scale uncertainty.

	We found that the variation in the energy scale best
correlated with the vertical position of the event ($z$) and the
direction cosine of the event relative to the $z$-axis ($u_z$).  All
of the $^{16}$N events in the first group were binned in the $(z,
u_z)$ dimensions and the peak of the $^{16}$N energy distribution was
found for data and Monte Carlo events separately.  We fit a
second-order polynomial in $z$ and $u_z$ to the ratio of the data and
Monte Carlo peak energies.  This smooth function provided the spatial
energy correction for data events.  Fig.~\ref{f:saltescale}(b) shows
the spatial variation after this energy correction.

To evaluate the spatial component of the energy scale uncertainty, we
assumed azimuthal symmetry in the detector, and divided the second
group of $^{16}$N calibration runs into regions based on radius and
polar angle.  Within each region, the RMS of the individual run
differences between the corrected data and Monte Carlo energy scales
defined the uncertainty on the energy scale in that volume.  All
regions were then combined into a volume-weighted measure of the
uncertainty on the overall energy scale in the detector due to spatial
variation and non-uniform sampling of the detector volume.  As a
verification of the procedure, we reversed the roles of the two
calibration groups (using the $y$-$z$ plane to construct the
calibration function and the $x$-$z$ plane to evaluate the
uncertainties) and found very similar corrections and uncertainties.

The energy scale uncertainty of the detector also includes uncertainty
in modeling of energy loss in the \nit source itself, uncertainties in
the online status of PMTs, variation in the channel response between
high-rate calibration data and low-rate neutrino data, and
uncertainties in the data acquisition channel gains and thresholds,
which affect the photon detection probability.  Many of these
uncertainties have been substantially reduced compared to previous
publications by the improvements to the Monte Carlo model described in
Sec.~\ref{sec:montecarlo} and the rate-dependent correction to the
channel pedestals described in Sec.~\ref{sec:hitcal}.

	The components of the energy scale uncertainties are
summarized in Table \ref{tab:escale_uncert}.  We take the source
uncertainty as 100\% correlated between phases, and the other
uncertainties as uncorrelated.  To verify the validity of the
$^{16}$N-derived energy corrections and uncertainties over a wider
range of energies, we compared the data and Monte Carlo energy
distributions for $^{252}$Cf neutron source runs and the \dto-volume
radon spike, for both of which events are more widely distributed in
the detector than for the $^{16}$N source.  In both cases, the
agreement between the data and Monte Carlo was well within the
uncertainties stated in Table~\ref{tab:escale_uncert}.

\begin{table}[!h]
\begin{center}
\begin{tabular}{lcc}
\hline \hline Uncertainty & Phase I & Phase II \\ \hline PMT Status &
$\pm 0.01$\% & $\pm 0.01$\% \\ Threshold/Gain & $+0.18\; -0.31$\% &
$+0.13\; -0.07$\% \\ Rate & $\pm 0.3$\% & $\pm 0.05$\% \\ Source &
$\pm 0.4$\% & $\pm 0.4$\% \\ Spatial Variation & $\pm 0.18$\% & $\pm
0.31$\% \\ \hline Total & $+0.56\; -0.62$\% & $+0.52\; -0.51$\% \\
\hline \hline
\end{tabular}
\caption{Summary of energy scale uncertainties.}
\label{tab:escale_uncert}
\end{center}
\end{table}

\subsection{Energy Resolution}
\label{sec:eres}

Energy resolution was a significant systematic uncertainty because of
its impact on background acceptance above the 3.5~MeV energy
threshold.  Due to differing event topologies in the two phases, the
resolution uncertainties were treated as three independent,
uncorrelated systematic parameters: Phase~I events (both electron-like
and neutron capture events), Phase~II electron-like events, and
Phase~II neutron capture events.  In all cases, the resolution was
found to be slightly broader in the data than for Monte Carlo events.
The difference was parameterized as a Gaussian of width $\sigma_{\rm
extra}$, with which the Monte Carlo distribution was convolved to
reproduce the data.  The width of the Gaussian was given by the
quadrature difference of the data and Monte Carlo resolutions:
$\sigma_{\rm extra} = \sqrt{ (\sigma_{\rm Data}^2 - \sigma_{\rm
MC}^2)}$.  A resolution correction was formulated using calibration
source data and applied to the Monte Carlo events used in PDF
generation.  The uncertainties on this correction were then taken from
the spread of the calibration data.

\subsubsection{Energy Resolution Uncertainties for Phase~II Electron-like
Events} \label{s:eres:saltelec}

The \nit source was the primary source for this measurement.  We
evaluated the uncertainties in two ways by measuring the resolution
for the spectrum of Compton electrons differentially and integrally.

The MPE fit described in Sec.~\ref{sec:ecorr} unfolds source effects
from the event distribution, allowing the extraction of the intrinsic
monoenergetic electron resolution as a function of energy.  The fit
was performed for both data and Monte Carlo simulation of \nit runs
and the resulting resolutions were compared differentially in energy.
The energy resolution at threshold is the dominant concern for
electron-like events, due to the exponential rise of the backgrounds,
and the value at 3.5~MeV was therefore used as representative of the
detector resolution.  $\sigma_{\rm extra}$ at threshold was found to
be 0.152 $\pm$ 0.053~MeV.  In terms of the fractional difference:
\begin{equation}
       \sigma_{\rm frac} = \frac{(\sigma_{\rm Data} - \sigma_{\rm
       MC})}{\sigma_{\rm MC}}
\end{equation}
this translates to $\sigma_{\rm frac}=$2.4 $\pm$ 1.6\% at threshold.

To measure the integrated Compton electron resolution using the
monoenergetic $\gamma$ rays produced by the \nit source, the
reconstructed energy distribution for Monte Carlo-simulated $\gamma$s
was convolved with a smearing Gaussian and the result was fit directly
to the data, allowing the mean and width of the smearing Gaussian to
vary.  The resulting $\sigma_{\rm extra}$ of the smearing Gaussian was
$0.0\pm 0.046$~MeV.  This measurement represents a higher average
energy than the `unfolded' MPE value since the \nit provides $\gamma$s
at 6.13~MeV.  The value of $\sigma_{\rm frac}$ from this $\gamma$-ray
measurement is $0.00\pm 0.08$\%.

Two $^{222}$Rn spikes were deployed during Phase~II, one in the \dto
and one in the \hto volume.  These provided a low energy source of
$\beta$s and \gam s, below the analysis threshold and, therefore, all
observed decays appeared due to the detector energy resolution, making
the spikes particularly sensitive to this effect.  The unbinned signal
extraction code (Sec.~\ref{s:kernel}) was used in a simplified
configuration to fit the data from each spike.

The internal spike was fit with 3 PDFs in two dimensions: energy and
isotropy.  The PDFs were $^{214}$Bi electron-like events (primarily
$\beta$s) in the \dto volume, $^{214}$Bi photodisintegration neutrons,
and a `quiet' data set drawn from neutrino runs near the date of the
spike.  The latter provides the energy distribution of all
`background' events to the spike measurement, including other
radioactive decays such as PMT $\beta$-$\gamma$s as well as neutrino
interactions.  An analytic convolution parameter was also floated,
defining the width of the convolving Gaussian applied to the Monte
Carlo electron-like events.  The resulting $\sigma_{\rm extra}$ was
0.139 $^{+0.023}_{-0.036}$~MeV, which is equivalent to $\sigma_{\rm
frac}=$2.0 $\pm$ 1.0\% at threshold.  Floating the $^{214}$Bi
electrons and neutrons independently also allowed a verification of
the Monte Carlo prediction for the photodisintegration rate.  The
results were in good agreement, giving 0.91 $\pm$ 0.13 times the Monte
Carlo predicted rate.

The external spike was fit with two PDFs in just the energy dimension,
due to lower statistics.  The electron to neutron ratio in the
$^{214}$Bi PDF was fixed to the Monte Carlo prediction and the overall
normalization of this PDF was taken as a free parameter, along with
the quiet data normalization.  The Monte Carlo events were again
convolved with a Gaussian, whose width was allowed to vary in the fit.
The resulting value for $\sigma_{\rm extra}$ was
$0.273^{+0.030}_{-0.035}$~MeV, which gives $\sigma_{\rm frac}=$$7.6\pm
1.9$\% at threshold.  The broader resolution for external events,
which were generated in the \hto region but either traveled or were
misreconstructed into the \dtonosp, is not unexpected since the
detector's energy response was modeled less well in the outer detector
regions.

\begin{figure}[!ht]
\begin{center}
\includegraphics[width=0.48\textwidth]{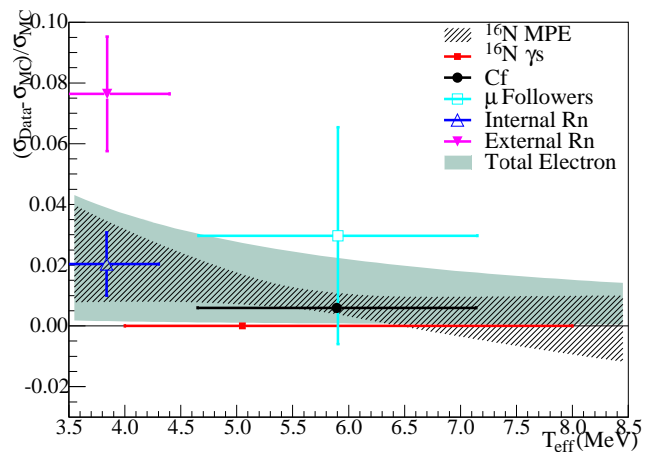}
\caption{\label{f:salteres}(Color online) Measurements of energy
resolution in Phase~II.  The solid area shows the one sigma band on
the energy resolution correction applied to Phase~II electron-like
events.  The \cf and muon follower points show the measurements of the
energy resolution for neutron capture events, and were not used to
evaluate the total shift for electron-like events.}
\end{center} 
\end{figure}

These four measures were combined to give the resolution correction
and associated uncertainty for electron-like events in Phase~II.
Since the two \nit measurements are not independent, they were not
used together.  The weighted mean of the MPE fit and the two spike
points was used to give the correction, with an associated
uncertainty.  The difference of that value from the weighted mean of
the \nit $\gamma$ point and the two spike points was then taken as an
additional one-sided (negative) uncertainty, to take into account the
difference in the two \nit measurements.  This results in a final
value of $\sigma_{\rm extra} = 0.168 ^{+0.041}_{-0.080}$~MeV, which
was applied as a constant smearing across the energy range.  The four
measurements and the resulting one sigma band on the final correction
value for Phase~II electron-like events are shown in
Figure~\ref{f:salteres}.

The MPE fit was also applied to the \lit source but this was not
included in the calculation due to the low statistics of the
measurement.  However, the energy dependence of both the \lit and the
\nit MPE fits were used to demonstrate that the use of a constant
$\sigma_{\rm extra}$ across the energy spectrum was consistent with
the data available.

\subsubsection{Energy Resolution Uncertainties for Phase~II 
Neutron Capture Events}

The energy resolution for neutron capture events in Phase~II was
measured using the \cf source, with a verification performed using a
`muon follower' data set, consisting of neutron capture events
occurring within a defined time window after a muon passed through the
detector.

There are fewer uncertainties associated with the neutron measurement
since the \cf source produced neutrons whose captures on $^{35}$Cl and
deuterium resulted in the same $\gamma$ cascades as those from NC
events.  The measurement was performed by numerically convolving a
spline-fit of the Monte Carlo energy distribution with a Gaussian and
fitting the resulting form to the data.  The mean and width of the
convolving Gaussian were allowed to vary, in order to take into
account possible correlations between energy scale and resolution
effects.  The result was $\sigma_{\rm extra} = 0.153 \pm 0.018$~MeV.
The observed energy scale from this measurement agreed very well with
that evaluated in Sec.~\ref{sec:ecorr}.

The statistics of the muon follower data set were low, and the
resulting uncertainty on the measurement was therefore relatively
large.  Nevertheless, a similar analysis was performed, giving a
$\sigma_{\rm extra}$ of $0.237 \pm 0.144$~MeV.

The weighted mean of the two points was used for the final correction
to the energy resolution of neutron capture events in Phase~II, with
its associated uncertainty, with the value dominated by the \cf
measurement: $\sigma_{\rm extra} = 0.154 \pm 0.018$~MeV.  Both points
are also shown on Fig.~\ref{f:salteres}.

\subsubsection{Energy Resolution 
Uncertainties for Phase~I Electron-like Events}

No radon spikes were deployed in Phase~I, and so only the two \nit
measurements were available.  Both the MPE fit and the Gaussian
convolution to the $\gamma$-ray energy distribution were performed for
Phase~I \nit runs, in the same manner as for Phase~II
(Sec.~\ref{s:eres:saltelec}).  The central correction value was taken
from the MPE fit directly, giving $\sigma_{\rm extra} = 0.155 \pm
0.036$~MeV.  The small number of energy resolution measurements in
Phase I provides fewer handles on the uncertainty than the much-better
calibrated Phase~II.  The uncertainties in Phase I were therefore
chosen to match those of Phase~II.  The width of the convolving
Gaussian for Phase~I events was therefore taken as $\sigma_{\rm extra}
= 0.155 ^{+0.041}_{-0.080}$~MeV.  This was also applied to neutron
capture events in Phase~I, since the event topologies were similar.

\subsection{Energy Linearity}
\label{sec:enonlin}
The corrections derived in Sec.~\ref{sec:ecorr} were done primarily
using the \nit source, and therefore the uncertainty in the energy
scale at the \nit energy is very small.  An additional uncertainty was
included to account for possible differential changes in the energy
scale that were not correctly modeled in the Monte Carlo simulation.
Such changes could be caused by residual crosstalk hits, or
mis-modeling of the multi-photon PMT hit probabilities in the energy
reconstruction algorithm.  The differential changes were determined
relative to the $^{16}$N point, and used calibration sources whose
energies were substantially higher.

The pT source provided $\gamma$s roughly 14~MeV higher in energy than
those from \nitnosp, resulting in a good lever-arm on any non-linear
effects.  This source was only deployed in Phase~I since deployment in
Phase~II would have resulted in an overwhelming neutron signal.  The
difference between data and Monte Carlo-reconstructed event energies
was measured to be $-$1.36 $\pm$ 0.01\% at the energy of the pT
source.

The MPE fit described in Sec.~\ref{sec:ecorr} was applied here to the
\lit source, including an additional term in the parameterization to
model first-order differential changes in the energy scale.  The fit
was done to both data and Monte Carlo events, and a difference of just
$-0.011\pm 0.004$\% was found, evaluated at the same energy as the pT
source $\gamma$ rays.

Giving the pT and \lit sources equal weight, the average shift in
energy scale at the energy of the pT source was found to be $-$0.69\%.
Using this as a measure of the degree by which the Monte Carlo energy
scale could vary differentially from the data and assuming a linear
interpolation between the \nit and pT energies, the linearity
uncertainty was parameterized in terms of the difference of an event's
energy from the \nit source ($\sim$5.05~MeV).  This results in a
scaling factor that can be applied to the energy of each Monte Carlo
event used to build the PDFs in the signal extraction procedure.
Conservatively, this was applied as a two-sided uncertainty:

\begin{eqnarray}
\label{e:enonlin}
T'_{\rm eff} = \left [1.0 \pm 0.0069 \times \left ( \frac{T_{\rm
eff}-5.05}{19.0-5.05}\right ) \right ] T_{\rm eff},
\end{eqnarray}

\noindent where 19$\,$MeV is the effective energy of the pT source,
$T_{\rm eff}$ is the original effective kinetic energy of an
individual event and $T'_{\rm eff}$ is the modified energy.

Tests using both the \lit and \cf sources suggested no evidence for
any linearity shift in Phase~II.  We expect any source of linearity
shift to be common across the two phases, however, and therefore the
results from Phase~I were conservatively taken to apply to both phases
in a correlated fashion.

\section{Event Isotropy}
\label{sec:beta14}

	As discussed in Sec.~\ref{sec:detector}, we used a measure of
event `isotropy' as one dimension of our PDFs to help distinguish
different types of events.  By `isotropy' we mean the degree of
uniformity in solid angle of the hit PMTs relative to the fitted event
location.

	Single electron events, like those created in neutrino CC and
ES reactions, had a Cherenkov cone that, at solar neutrino energies,
was somewhat diffuse due to electron multiple scattering in the water.
Nevertheless, even with the multiple scattering, these events were
characterized by a fairly tight cluster of PMT hits in a cone aligned
with the forward direction of the electron.

Neutron capture events on deuterium in Phase~I led to a single
6.25~MeV $\gamma$ ray.  Although these events could produce multiple
Compton electrons and, hence, a number of Cherenkov cones that
distributed hits more widely than single electrons, Phase~I neutron
capture events in the data set were dominated by single Compton
scatters and, thus, isotropy was not useful in distinguishing them
from CC or ES events.

	In contrast, in Phase~II neutrons captured primarily on
$^{35}$Cl, which typically led to a $\gamma$ cascade that looks very
different from single electrons.  Neutron capture on $^{35}$Cl
typically produced several $\gamma$ rays, with energies totaling
$8.6$~MeV, which distributed PMT hits more uniformly in solid
angle.  The isotropy distribution for these events is thus a
convolution of the isotropy distribution of single $\gamma$-ray events
with the directional distribution of the $\gamma$ rays emitted in the
possible $\gamma$-decay cascades.

	The isotropy of background events can also be significantly
different from that of single electron and neutron events.  Decays of
$^{208}$Tl, for example, produce both a $\beta$ and a 2.614~MeV
$\gamma$ ray and, thus, resulted in a different distribution of hit
PMTs than either single electrons or single $\gamma$s.

	The measure of isotropy was therefore critical to the
analysis, helping us to separate CC and ES events from NC events, and
both of these from low-energy background events.

	We examined several measures of isotropy, including a full
correlation function, the average angle between all possible pairwise
combinations of hit PMTs, and constructions of several variables using
Fisher discriminants. We found that, for the most part, they all had
comparable separation power between the single electron (CC and ES)
and the neutron (NC) signals.  As in our previous Phase~II
publications~\cite{nsp}, we opted to use a linear combination of
parameters, $\beta_{14}\equiv\beta_1+4\beta_4$, where:
\begin{equation}
\beta_l = \frac{2}{N(N-1)}\sum_{i=1}^{N-1} \sum_{j=i+1}^N
P_l(\cos\theta_{ij}).
\end{equation}
In this expression, $P_l$ is the Legendre polynomial of order $l$,
$\theta_{ij}$ is the angle between triggered PMTs $i$ and $j$ relative
to the reconstructed event vertex, and $N$ is the total number of
triggered PMTs in the event.  Very isotropic events have low (even
negative) values of $\beta_{14}$.

\subsection{Uncertainties on the Isotropy Measure}

	We parameterized the difference between the predicted \bof PDF
and the true PDF by a fractional shift in the mean,
$\bar{\beta}_{14}$, and a broadening of the width,
$\sigma_{\beta_{14}}$.  We also allowed for an energy dependence in
the shifts.

\begin{figure}
\begin{center}
\includegraphics[width=3.4in]{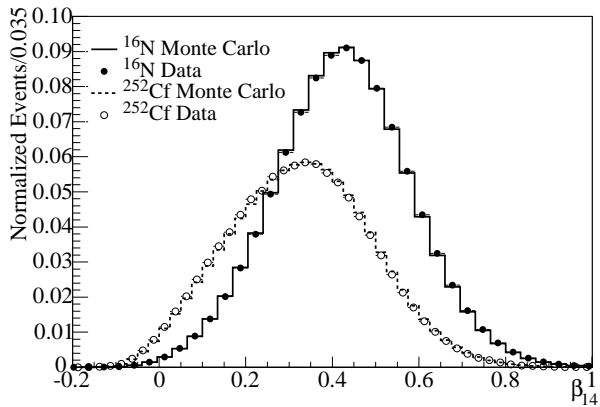}
\caption{\label{fig:isotropy1} $\beta_{14}$ isotropy distributions for
\cf data and MC and \NS~data and MC. There is a very small shift of
the Monte Carlo $^{252}$Cf $\beta_{14}$ distribution toward higher
(less isotropic) values.}
\end{center}
\vspace{-4ex}
\end{figure}

Figure~\ref{fig:isotropy1} shows \bof distributions of Phase~II data
from \cf and \NS~sources and from corresponding MC simulations.  The
\NS~source emitted a single 6.13~MeV $\gamma$ ray, which usually
underwent Compton scattering and produced one or more electron tracks,
while neutrons from the \cf source were typically captured in Phase~II
by the chlorine additive, leading to a cascade of several $\gamma$
rays.  It is clear from the figure that the \NS~data and Monte Carlo
agree very well, while the Monte Carlo simulation of the \cf source
shows a very small shift toward higher $\beta_{14}$ values (less
isotropic events than in the data).  This shift is discussed in
Sec.~\ref{sec:piib14neut}.

	Errors in the simulated distributions of $\beta_{14}$ can have
several sources: incorrect modeling of the detector optics or
photomultiplier tubes, unmodeled event vertex reconstruction errors,
errors in the model of the production of Cherenkov light (including
the interactions of $\gamma$ rays and electrons in the detector) and,
for neutrons captured on $^{35}$Cl, uncertainties in our knowledge of
the $\gamma$ cascade sequences and correlations between the directions
of the multiple $\gamma$ rays.

	Except for the last item, these errors affect all event types.
For Phase~I, in which neutrons were captured on deuterons, we allowed
for correlations among the uncertainties on all signals and most
backgrounds. For Phase~II, we treated the uncertainties on the mean
and width of the $\beta_{14}$ distribution for NC events and
photodisintegration neutrons separately from the other event types.
Uncertainties on the $\beta_{14}$ distributions of $\beta$s and
$\gamma$s from radioactive background events were treated the same as
for CC and ES events.  The one exception to this was PMT
$\beta$-$\gamma$ events, whose location at the PMT array led to
effects on the $\beta_{14}$ distribution that are not present in the
other signals.  The $\beta_{14}$ distribution and associated
uncertainties for PMT $\beta$-$\gamma$s are discussed in
Sec.~\ref{s:pmtpdf}.

	As usual in this analysis, we derived uncertainties on the
mean, width, and energy dependence of the $\beta_{14}$ distribution by
comparing calibration source data to Monte Carlo simulations of the
calibration source runs.  When we found a difference that was
corroborated by more than one source, or was caused by known errors in
the simulation, we adjusted the simulated distribution by shifting the
mean of the distribution and/or convolving the distribution with a
smearing function to better match the calibration data.  In such
cases, additional uncertainties associated with the correction were
included.

\subsubsection{$\beta_{14}$ Uncertainties for Phase~II Electron-like Events}
\label{sec:b14iie}
\begin{figure}
\begin{center}
\includegraphics[width=3.4in]{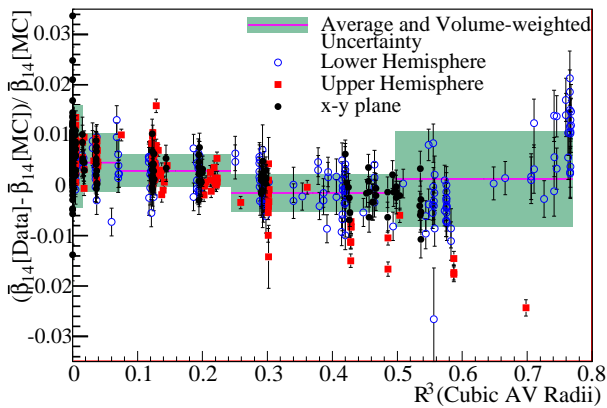}
\caption{\label{fig:piin16}(Color online) Fractional differences in
the mean of the $\beta_{14}$ distributions for data and Monte Carlo,
for the Phase~II $^{16}$N calibration source.  Also shown in the
figure are the averages in each radial bin, with the bands indicating
the volume-weighted uncertainty in each bin. }
\end{center}
\end{figure}

	The primary measure of isotropy uncertainties for Phase~II
electron-like events comes from comparisons of $^{16}$N calibration
source data to Monte Carlo simulation.  We fit Gaussians to both the
data and simulated events for each run, and calculated the fractional
difference between the fitted parameters.  Figure~\ref{fig:piin16}
shows the fractional difference in the means as a function of $R^3$.
Each point shown is the fractional difference for a single run, with
the error bar evaluated as the combination of the uncertainty on the
fit parameters for data and Monte Carlo events.  The detector region
in which the source was deployed has been identified for each run.

Also shown in Fig.~\ref{fig:piin16} are the 
averages of these differences, in several radial bins.
The uncertainty on each average is the standard deviation of the
points in that bin, weighted by the volume represented by the bin
(smaller volumes have larger uncertainties).  The overall weighted
average within the entire 550~cm radius fiducial volume is consistent
with zero, with an uncertainty of $\pm 0.21$\%.  The calibration data
were collected at a high rate relative to normal neutrino data runs
and, so, we added to this an uncertainty to account for the difference
in \bof between high rate and low rate data ($\pm 0.1$\%) by comparing
low-rate and high-rate \nit source runs, as well as a small
uncertainty of $\pm 0.002$\% associated with a possible un-modeled
time-dependence obtained by comparing data and Monte Carlo differences
over time.  The quadrature combination of these uncertainties on the
mean of the $\beta_{14}$ distribution totals $\pm 0.24$\%.  A similar
analysis was performed for the width of the $\beta_{14}$ distribution,
yielding a total fractional uncertainty of $\pm 0.54$\%.

\subsubsection{$\beta_{14}$ Uncertainties for Phase~I Electron-like Events}
\label{sec:pib14}
\begin{figure}
\begin{center}
\includegraphics[width=3.4in]{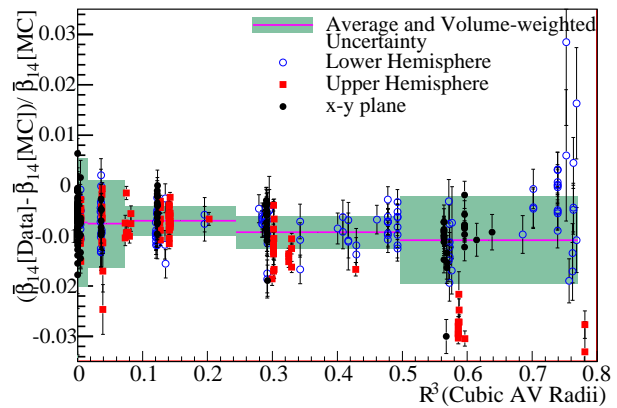}
\caption{\label{fig:pin16}(Color online) Fractional differences in the
mean of the $\beta_{14}$ distributions for data and Monte Carlo, for
the Phase~I $^{16}$N calibration source.  Also shown in the figure are
the averages in each radial bin, with the bands indicating the
volume-weighted uncertainty in each bin. }
\end{center}
\end{figure}

	We applied an identical analysis to the Phase~I $^{16}$N data
but, as shown in Figure~\ref{fig:pin16}, we found a difference of
$-0.81\pm 0.20$\% between the means of the $\beta_{14}$ distributions
for source data and source simulations.  Comparison of $^{16}$N data
between Phase~I and Phase~II showed them to be consistent, and the
(data-Monte Carlo) difference seen in Fig.~\ref{fig:pin16} to be due
to a shift in the simulated events.  Further investigation showed that
the difference was caused by the value of the Rayleigh scattering
length used in the Phase~I simulation.  Explicit measurements of the
Rayleigh scattering had been made and used in the simulation for
Phase~II but no such measurements existed for Phase~I.  Use of the
Phase~II Rayleigh scattering length in Phase~I simulations was found
to produce the desired magnitude of shift, and we therefore corrected
the $\beta_{14}$ values of all simulated Phase~I events by a factor of
$(1-0.0081)=0.9919$.

We included three uncertainties associated with this correction.  The
first was 0.20\% on the correction itself, evaluated from the
volume-weighted average of the data and Monte Carlo differences for
Phase~I, as shown in Fig.~\ref{fig:pin16}.  To take into account the
fact that we used the consistency in the \nit data between the two
phases to support the correction of $-0.81$\%, we added in quadrature
the uncertainty on the difference between the means of the Phase~ I
and Phase~II $^{16}$N $\beta_{14}$ distributions, which was 0.34\%.
Finally, because we used the consistency of the Phase~II data with the
Monte Carlo simulation as evidence that the Phase~I $\beta_{14}$
distribution was correct, aside from the Rayleigh scattering
correction, we included the volume-weighted Phase~II uncertainty on
the offset of the mean (0.21\% from Fig.~\ref{fig:piin16} in
Sec.~\ref{sec:b14iie}).

	The evaluations of the uncertainties associated with rate
dependence and time dependence in Phase~I were 0.08\% and 0.03\%,
respectively, and the overall uncertainty on the mean of the
$\beta_{14}$ distribution in Phase~I thus totaled 0.42\%.

	We evaluated the uncertainty on the width of the $\beta_{14}$
distribution for Phase~I in the same way as for Phase~II, finding a
fractional uncertainty which also totaled 0.42\%.

\subsubsection{$\beta_{14}$ Uncertainties for Phase~II Neutron Capture Events
\label{sec:piib14neut}}
\begin{figure}
\begin{center}
\includegraphics[width=3.4in]{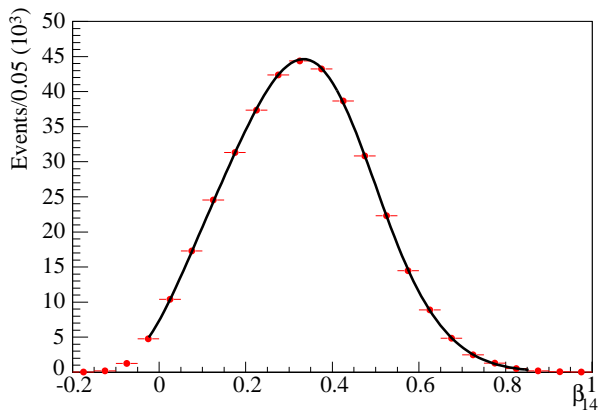}
\caption{\label{fig:cffit}(Color online) Fit of Monte Carlo simulated
$\beta_{14}$ distribution for neutron capture events from $^{252}$Cf
to data taken with the $^{252}$Cf source.  The fitted shift for this
sample is $-$1.2\%, and the additional smear is 1.8\%, before any
corrections for bias.}
\end{center}
\end{figure}

	Neutron capture events in Phase~II were distinct from other
neutrino-induced events and backgrounds in that the $\gamma$ cascade
was more isotropic than a single electron or $\gamma$ ray.  The
primary measurement of the uncertainty on the mean of the $\beta_{14}$
distribution comes from deployments of the $^{252}$Cf source, which
produced several neutrons per fission decay.  The $\beta_{14}$
distribution of the resulting neutron capture events was noticeably
non-Gaussian, and we therefore derived uncertainties on the mean and
width by fitting the $\beta_{14}$ distributions from simulated
$^{252}$Cf runs directly to the distributions of data. The fit allowed
for scaling as well as convolution with a Gaussian smearing function.
Figure~\ref{fig:cffit} shows the fit of a simulated $^{252}$Cf run to
data, in which the fitted scaling was -1.2\% and the smearing was an
additional 1.8\% of the width of the Monte Carlo distribution.

\begin{figure}
\begin{center}
\includegraphics[width=3.4in]{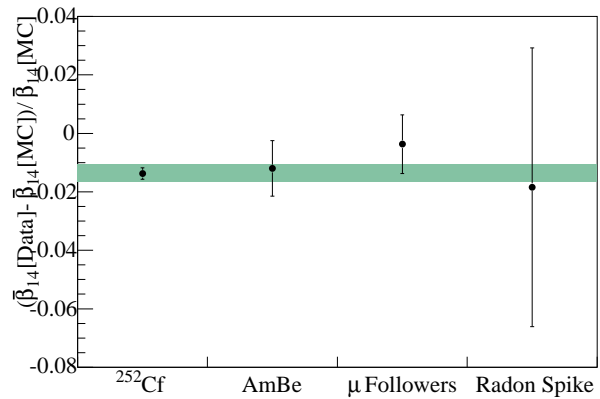}
\caption{\label{fig:b14src}(Color online) Fractional difference in
mean $\beta_{14}$ between data and Monte Carlo events for several
neutron sources.  The horizontal band indicates the error on the
overall $-$1.44\% correction.}
\end{center}
\end{figure}

	We derived scaling factors from fits like that in
Fig.~\ref{fig:cffit} for all $^{252}$Cf runs, and then volume-weighted
them in the same way as for the $^{16}$N data.  The average of the
volume-weighted differences showed an overall offset between the means
of the $\beta_{14}$ distributions for data and Monte Carlo of $\sim
-1.4$\%.  This result was not consistent with that from the $^{16}$N
data for Phase II (which, as discussed above, had no significant
offset), which indicated that the shift was not due to a detector
effect.  To check whether the shift was caused by mis-modeling of the
$^{252}$Cf source in the simulation, we performed the same analysis on
several types of neutron capture events: neutrons produced by passage
of a muon through the detector (`muon followers'), neutrons from a
tagged Am-Be source, and neutrons produced by deuteron
photodisintegration during the deployment of a radon spike in the
detector.  Figure~\ref{fig:b14src} shows results from these sources.
An energy-dependent fit to all sources except $^{252}$Cf showed an
offset of $-1.12\pm0.31$\%, consistent with the data from the
$^{252}$Cf source.  This indicated that the offset was likely not a
source effect but was instead associated with the simulation of the
$\gamma$ cascade from neutron captures on chlorine, possibly with some
contribution from the energy-dependent correction of the Monte Carlo
value for \bof presented in Sec.~\ref{s:bofenergy}.  All sources taken
together gave an overall offset of $-1.44$\%, and we therefore
corrected the $\beta_{14}$ PDF by multiplying each simulated event's
$\beta_{14}$ value by $(1 + \delta_{\beta_{14}}) = (1-0.0144)=0.9856$.

	The uncertainties on this correction came first from the
uncertainty on the overall average, which was 0.17\%.  To this we
added in quadrature the same rate- and time-dependent uncertainties as
were calculated for the Phase~II $^{16}$N sources.  We also added an
uncertainty associated with the multiplicity of neutrons from the
$^{252}$Cf source of 0.09\% (neutrons produced by either
photodisintegration of deuterons or the NC reaction are singles,
whereas the $^{252}$Cf source produces multiple neutrons/decay), and
0.03\% uncertainty to account for the relatively sparse sampling of
the detector, giving a total of 0.22\%.  Conservatively, we included a
further uncertainty based on the difference between $^{252}$Cf and the
other neutron-source data, a one-sided uncertainty of 0.31\%.  The
total uncertainty on the mean of the $\beta_{14}$ distribution for
Phase~II neutron captures was therefore $^{+0.38}_{-0.22}$\%.

As well as a measure of any required shift, the fit described above
also allowed for the widths of the data and Monte Carlo distributions
to differ.  A resolution parameter was varied in the fit, as the
standard deviation of the Gaussian by which the Monte Carlo
distribution was analytically convolved.  The results for each \cf run
were volume-weighted using the procedure described above to result in
an average overall smearing value.  The same fit was performed on a
sample of Monte Carlo-generated data, and the bias determined from
these fits was subtracted from the overall average.  The result was a
fractional smearing correction to be applied to the PDFs of 0.43\%,
with an uncertainty (including all sources described above: time,
rate, multiplicity, and sampling) of 0.31\%.

\subsubsection{$\beta_{14}$ Uncertainties for Phase~I Neutron Capture Events}

Neutrons created in Phase~I captured on deuterons, releasing a
single 6.25~MeV $\gamma$ ray. The uncertainties on the mean
and width of the $\beta_{14}$ distribution were therefore
well-estimated by the measurements made with the $^{16}$N
6.13~MeV $\gamma$-ray source, already discussed in
Sec.~\ref{sec:pib14}.  We therefore used the same
uncertainties for both event types, applied in a correlated
fashion.

\subsubsection{Energy Dependence of $\beta_{14}$ Uncertainties}
\label{s:bofenergy}
\begin{figure}
\begin{center}
\includegraphics[width=3.4in]{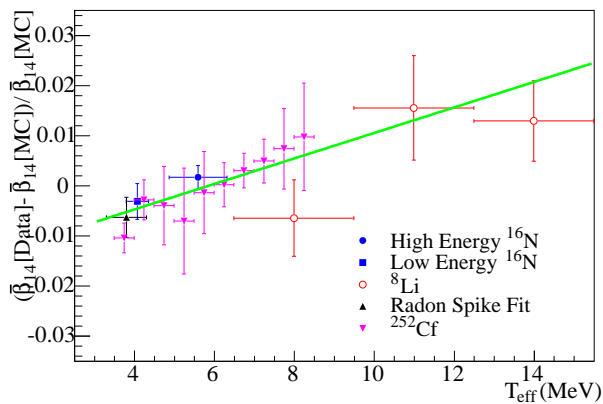}
\caption{\label{fig:b14vE}(Color online) Fractional shift in mean
$\beta_{14}$ between data and Monte Carlo simulation in Phase~II for
several calibration sources as a function of kinetic energy, with the
fit to Eq.~\eqref{eq:b14vE} shown. }
\end{center}
\end{figure}

	A final systematic uncertainty on the $\beta_{14}$
distributions is their energy dependence.  In Figure~\ref{fig:b14vE}
we show the energy
dependence of the fractional difference between Monte Carlo
predictions of the mean of the $\beta_{14}$ distribution and data from
several different sources: the Phase~II radon spike, low and high
energy $^{16}$N source events, the $^{252}$Cf source (with the data
corrected by the 1.44\% shift discussed above), and $^8$Li-source
$\beta$ events in three energy bins.  There clearly is an energy
dependence in the data, which we fit with a function of the form:
\begin{equation}
f = \delta_{\beta_{14}}+m_{\beta_{14}}(T_{\rm eff}-5.6\rm ~MeV),
\label{eq:b14vE}
\end{equation}
where $T_{\rm eff}$ is kinetic energy, and 5.6~MeV is the kinetic
energy at the high-energy $^{16}$N point (the point used to determine
the offset in the mean of the Phase~II electron $\beta_{14}$
distribution).  With this parameterization, the offset
($\delta_{\beta_{14}}$) and the slope ($m_{\beta_{14}}$) are
uncorrelated. Given that all the sources exhibited the same trend, we
applied the same slope to all event types, but used the different
offsets and uncertainties for $\delta_{\beta_{14}}$ described in the
previous sections.  We performed a similar analysis for Phase~I,
although less calibration data were available, and found that the same
slope fit the $^{16}$N and $^8$Li data in this phase.

	We found no energy dependence in the broadening of the width
of the $\beta_{14}$ distributions.  These uncertainties were therefore
treated as independent of energy.

The corrections and uncertainties to the $\beta_{14}$ distributions
are listed in Tables~\ref{tbl:b14sum1} and~\ref{tbl:b14sum2}.

\begin{table}[!ht]
\centering
\begin{tabular}{lcc}
\hline Phase/Particles & $\delta_{\beta_{14}}$ & $m_{\beta_{14}}$
 (10$^{-3}$ MeV$^{-1}$)\\ \hline \hline II/electrons &
 0.0$\,\pm\,$0.0024 & 2.76$\,\pm\,0.696$ \\ II/neutrons &
 $-0.0144\,^{+0.0038}_{-0.0022}$ & 2.76$\,\pm\,0.696$ \\ I/electrons &
 $-$0.0081$\,\pm\,$0.0042 & 2.76$\,\pm\,0.696$ \\ I/neutrons &
 $-$0.0081$\,\pm\,$0.0042 & 2.76$\,\pm\,0.696$ \\ \hline
\end{tabular}
\caption{\label{tbl:b14sum1} Summary of uncertainties on the
$\beta_{14}$ scale.  The \bof of each event was corrected by:
$\beta_{14}\rightarrow\beta_{14}
(1+(\delta_{\beta_{14}}+m_{\beta_{14}}(T_{\rm eff}-5.6{\rm ~MeV})))$.}
\end{table}

\begin{table}[!ht]
\centering
\begin{tabular}{lcc}
\hline Phase/Particles & Correction (\%) & Uncertainty (\%) \\ \hline
 \hline II/electrons & 0.0 & $\pm\,$0.42 \\ II/neutrons & 0.43 &
 $\pm\,$0.31 \\ I/electrons & 0.0 & $\pm\,$0.42 \\ I/neutrons & 0.0 &
 $\pm\,$0.42 \\ \hline
\end{tabular}
\caption{\label{tbl:b14sum2} Summary of uncertainties on the
$\beta_{14}$ width.}
\end{table}

\section{Cuts and Efficiencies\label{sec:cuts}}
\label{s:cuts}

The data set contains two main types of background events: physics
backgrounds, due to radioactive decays, and instrumental backgrounds,
caused by the detector itself.  Two sets of cuts were developed to
remove these events, described in Sections~\ref{s:cutdescdamn}
and~\ref{s:cutdeschlc}.  Each set of cuts had an associated level of
signal loss, which was taken into account in the measurement of
neutrino flux and spectra as described in Sec.~\ref{s:cutacc}.

\subsection{Low-Level (Instrumental) Cuts}
\label{s:cutdescdamn}

	There were many sources of instrumentally-generated events in
the SNO detector, which produced hits originating either in the PMTs
or in the electronics channels.  Static discharges in the nitrogen in
the neck of the acrylic vessel and `flasher' PMTs, in which discharges
occurred within a photomultiplier tube itself, produced light in the
detector.  Electronic pickup generated by noise on the deck above the
detector or by high-voltage breakdown could produce hits in
electronics channels.  We removed these instrumental backgrounds with
a suite of loose `low-level' cuts that rejected events before event
reconstruction.  The cuts were based on event characteristics such as
the distribution of PMT hit times, the presence of unusually low or
high PMT charges, or unusual time correlations between events (such as
bursts of events with large numbers of hits).  More details on these
low-level cuts can be found in~\cite{longd2o,nsp}.  We used the same
cuts and cut criteria here, with the exception that the simple burst
cut used in~\cite{longd2o} was not used in this analysis because it
was redundant with other burst cuts.

	The acceptance of these cuts was re-evaluated for this
analysis, particularly in the low-threshold region (below $T_{\rm
eff}=5.0$~MeV) where the cuts had not previously been examined in
detail. We discuss the results of these cut acceptance measurements in
Sec.~\ref{s:inssac}.
	 
\subsection{High-Level Cuts} \label{s:cutdeschlc}

Background radioactivity events were produced primarily by the decays
of $^{214}$Bi and $^{208}$Tl. Lower energy ($T_{\rm eff}<3$~MeV)
decays of these nuclei in the heavy water could appear above our
$T_{\rm eff}=3.5$~MeV threshold because of the broad energy resolution
intrinsic to a Cherenkov detector. Decays within the walls of the
acrylic vessel, the light water surrounding the vessel, and the
photomultiplier tube array could pass the energy cut and have
misreconstructed vertex positions which falsely placed them within the
fiducial volume.  The PMT array was, by far, the radioactively hottest
component of the SNO detector and, consequently, the largest source of
background events.  We designed a suite of 13 loose cuts that used
`high-level' information (reconstructed event position, direction, and
energy) to remove events whose likely origin was either outside the
fiducial volume or whose true energy was below our threshold.  All of
the cuts were adjusted based exclusively on simulated events and
calibration data.  Several of the cuts had a high degree of redundancy
in order to maximize background rejection. The acceptance of the cuts
was therefore evaluated collectively, as described in
Sec.~\ref{s:cutacc}.

	 Five of the high-level cuts removed backgrounds using
Kolmogorov-Smirnov (KS) tests of the hypothesis that the event had a
single Cherenkov-electron track.  Two of these tests compared
azimuthal and two-dimensional (polar vs azimuthal) angular
distributions to those expected for Cherenkov light produced by an
electron, and two others did the same for hits restricted to a narrow
prompt time window.  The fifth of these KS tests was a comparison of
the distribution of fitted PMT time residuals (see
Eq.~\eqref{eqn:ftp-tresid}) with the expected distribution for direct
Cherenkov light.

	Three more of the cuts applied event `isotropy' to remove
misreconstructed events. Events whose true origins were well outside
the fiducial volume but which reconstructed inside tend to appear very
anisotropic.  For one of these cuts we used the mean angle between
pairs of PMTs, ($\theta_{ij}$), and for another the isotropy parameter
$\beta_{14}$, which is described in Sec.~\ref{sec:beta14}.  Both of
these have been used in previous SNO analyses~\cite{longd2o,nsp}. The
third of these cuts was based on the charge-weighted mean pair angle,
$\theta_{ij}$, in which each pair angle is weighted by the product of
the detected charges of the two PMTs in the pair.

	Further cuts used information from the energy reconstruction
algorithm discussed in Sec.~\ref{sec:ftk}.  Two cuts removed events
whose reported energy uncertainty was well outside of the range
expected from the known energy resolution.  These are referred to in
Sections~\ref{sec:hlcsac}--\ref{sec:desac} as the `energy-uncertainty'
cuts.  The third was a comparison of the energy estimated with FTK
(which used all hits) with that from a prompt-light-only energy
estimator.  Events whose origins were outside the acrylic vessel and
which pointed outward often had a larger fraction of prompt hits
because the direct light was not attenuated by the acrylic vessel.
Such an event would have a higher energy as measured by a prompt-light
energy estimator than by the total-light energy reconstruction of FTK.
We normalized the ratio of these two energy estimates by the ratio of
prompt to total hits in the event.  The cut itself was
two-dimensional: events were removed if the normalized ratio of energy
estimates was unusually large and the charge-weighted $\theta_{ij}$
was unusually low (the latter indicating an outward-pointing event
with a tight cluster of hits).

	The last two high-level cuts were also used in determining the
PDFs for radioactive backgrounds from the PMTs.  The first of these,
the in-time ratio (ITR) cut, removed events based on the ratio of the
prompt hits to the total hits.  The prompt time window for the ITR cut
extended from 2.5~ns before the reconstructed event time to 5.0~ns
after, and the full-event window was roughly 250~ns long. The mean of
the ITR distribution for SNO events is at 0.74. Events that were
reconstructed at positions far from their true origin tend to have
small ITR values, because the PMT hits were spread across the entire
time window.  In previous analyses~\cite{longd2o,nsp,snoncd} we used
the ITR cut with a fixed threshold, rejecting events with an in-time
ratio smaller than 0.55.  For the lower-energy events included in this
analysis, the lower number of hits caused the distribution of ITR to
broaden and introduced a large, energy-dependent bias in the
acceptance of the cut.  We therefore changed the cut threshold to
scale with the number of hits ($N_{\rm hit}$) in an event.  The fixed
value of 0.55 used in earlier publications corresponded to cutting
events that fell more than 2.7$\sigma$ below the mean of the
distribution, and we retained this criterion, so that the new version
of the ITR cut rejected events that were more than 2.7$\sigma$ below
the mean of 0.74, where now $\sigma = 0.43/\sqrt{N_{\rm hit}}$.

	The last cut was aimed directly at removing events produced by
radioactive decays in the PMTs themselves.  Such events produced light
either in the PMT glass or in the light water, just in front of the
PMTs.  Although only a tiny fraction of such events were
misreconstructed inside the fiducial volume, the PMT array was
relatively hot, with a total decay rate from uranium and thorium chain
daughters of a few kHz.  Because of their origin within or near the
PMTs, these events were characterized by a large charge in one PMT (or
distributed over a few nearby PMTs) with hit times that preceded the
reconstructed event time.  The `early charge' (EQ) cut therefore
examined PMT hits in a window that ran from $-$75~ns to $-$25~ns
before the event time. If a PMT hit in this window had an unusually
high charge, or there was an unusually large number of hits in this
window, then the event was cut.  To account for variations in PMT
gain, `unusually high charge' was defined by using the known charge
spectrum of the PMT in question to calculate the probability of
observing a charge as high as observed or higher.  If more than one
hit was in the window, a trials penalty was imposed on the tube with
the lowest probability, and an event was cut if this trials-corrected
probability was smaller than 0.01.  We defined `unusually large number
of hits' in a similar way, by comparing the number of hits observed in
the early time window to the expected number, given the total number
of hits in the event.  If the Poisson probability of having the
observed number in the early time window was below 0.002, the event
was cut.

\subsection{Burst Removal}
\label{sec:bursts}

	Atmospheric neutrinos, spontaneous fission, and cosmic-ray
muons could all produce bursts of events that were clearly not due to
solar neutrinos.  Most of these bursts had a detectable primary event
(like a high-energy atmospheric-neutrino event) followed by several
neutron events.  In addition, many instrumentally-generated events
came in bursts, such as those associated with high-voltage breakdown
in a PMT.

	We therefore applied several cuts to the data set to remove
most of these time-correlated events.  Four of these were part of the
suite of instrumental cuts described in Sec.~\ref{s:cutdescdamn}.  The
first removed events that were within 5~$\mu$s of a previous event
and, therefore, eliminated events associated with PMT afterpulsing or
Michel electrons from decays of stopped muons.  The second removed all
events within 20 seconds of an event that had been tagged as a muon.
Most of these `muon followers' were neutrons created by passage of a
cosmic-ray muon through the heavy water, which captured either on
deuterons or, in Phase II, on $^{35}$Cl, but the cut also removed
longer-lived cosmogenic activity.  The muon follower cut resulted in a
very small additional overall detector deadtime because of the very
low rate of cosmic rays at SNO's depth.  Atmospheric neutrinos could
also produce neutrons, either directly or by creating muons which, in
turn, disintegrated deuterons.  We therefore removed any event within
250~ms of a previous event that had $N_{\rm hit}>60$ (Phase~I) or
$N_{\rm hit}>150$ (Phase~II).  The fourth cut was aimed primarily at
residual instrumental bursts, and removed events that were part of a
set of six or more with $N_{\rm hit}>40$ that occurred within an
interval of six seconds.

	Because of the relatively loose criteria used, after these
cuts were applied there were still time-correlated events in the SNO
data set that were very unlikely to be solar neutrinos, but were
primarily low-multiplicity neutrons created by atmospheric neutrino
interactions.  We therefore applied a final `coincidence cut' that
removed events if two or more occurred within a few neutron capture
times of each other.  For Phase~I this window was 100~ms; a shorter
window of 15~ms was used for Phase~II because of the shorter neutron
capture time on chlorine compared to deuterium.  The cut was
`retriggerable', in that the window was extended for its full length
past the last event found.  If a new event was thus `caught', the
window was again extended.  We calculated that this cut removed less
than one pair of events from each data set due to accidental
coincidences.

\subsection{Cut Summary}

The numbers of events in the data sets after successive application of
each set of cuts are shown in Table~\ref{t:cuts}.  The burst cuts
described in Sec.~\ref{sec:bursts} are included in instrumental cuts,
except for the final coincidence cut, which appears in the last line
of the table.

\begin{table}[!ht]
\begin{center}
\begin{tabular}{lrr} 
\hline \hline Events & \multicolumn{1}{c}{Phase~I} &
\multicolumn{1}{c}{Phase~II} \\ \hline Full data set &128421119 &
115068751\\ Instrumental &115328384 & 102079435\\ Reconstruction
&92159034 & 77661692\\ Fiducial volume ($<$550~cm) &11491488 &
8897178\\ Energy range (3.5--20$\,$MeV) &25570 & 40070\\ High-level
cuts &9346 & 18285\\ Coincidence cut & 9337 & 18228 \\ \hline \hline
\end{tabular}
\caption{\label{t:cuts}Number of events remaining in the data set 
after successive application of each set of cuts.}
\end{center}
\end{table}

\subsection{Cut Acceptance}
\label{s:cutacc}

As in previous analyses \cite{longd2o}, the fraction of signal events
expected to pass the full set of analysis cuts (the `cut acceptance')
was determined by separating the cuts into three groups: instrumental,
reconstruction, and high-level cuts.  Correlations between these
groups had been shown to be minimal \cite{nsp}, and it was verified
that this was still true after the addition of new high-level cuts for
this analysis.

The \nit and \lit calibration sources were used for the primary
measurements of cut acceptance and the \cf source was used for neutron
capture events in Phase~II.  Neutron events in Phase~I are
well-modeled by \nit events since capture on deuterium resulted in a
single $\gamma$ at 6.25~MeV and \nit was a source of 6.13~MeV
$\gamma$s.

\subsubsection{Instrumental Cut Acceptance}
\label{s:inssac}

The instrumental cuts were not simulated in the Monte Carlo code and,
therefore, we could not make a relative estimate of their acceptance
by comparing simulation to data. Instead, an absolute measure of their
acceptance was made using calibration data and applied as a correction
(with uncertainties) to the PDFs.

\begin{figure}[!ht]
\begin{center}
\includegraphics[width=0.48\textwidth]{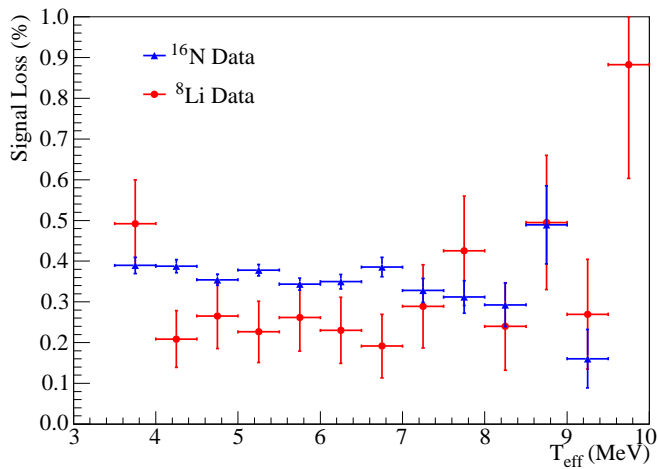}
\caption{\label{f:n16li8inssac}(Color online) Signal loss due to the
instrumental cuts for the \nit and \lit calibration sources as a
function of reconstructed kinetic energy, in Phase~I.}
\end{center} 
\end{figure}

Being a near-perfect source of CC-like electron events, the \lit
source was used to evaluate the signal loss for electron-like events,
and \cf was used for Phase~II neutron capture events.  The \nit source
was used as a check and any difference in the values obtained was
conservatively taken as a two-sided systematic uncertainty.
Figure~\ref{f:n16li8inssac} shows the \nit and \lit measurements in
Phase~I.  The weighted mean of the \lit signal loss shown in the
figure was taken as the correction to the PDFs, and the median
deviation of the points from this value was used to represent the
energy-dependent uncertainty.
 
The \nit source, which was deployed more frequently and at more
positions than $^8$Li, was used to determine time- and
position-dependent uncertainties.  Runs were binned by position and
date, and the median deviation of the bin values from the best-fit
value was taken as the measure of systematic uncertainty.

After combination of the systematic uncertainties in quadrature, 
the final estimates of signal loss due to the instrumental cuts were:
\begin{itemize}
\item Phase~I: $0.214$\% $\pm 0.026$ (stat) $\pm 0.094$ (syst)
\item Phase~II $e^-$: $0.291$\% $\pm 0.028$ (stat) $\pm 0.202$ (syst)
\item Phase~II $n$:  $0.303$\% $\pm 0.003$ (stat) $\pm 0.186$ (syst)
\end{itemize}
where ``$e^-$'' refers to electron-like events and ``$n$'' to neutron
captures.  The acceptance is given by one minus the fractional signal
loss and was applied as an adjustment to the normalization of the
PDFs.

\subsubsection{Acceptance of Reconstruction}

Occasionally, the reconstruction algorithm failed to converge, and
returned no vertex for an event.  In past analyses, an upper bound was
placed on the resulting signal loss, by using calibration source data,
but a different approach was used in this analysis.  What is important
is how well the effect is reproduced in the simulation.  Therefore, a
comparison was made of the acceptance of data and Monte Carlo events
and the difference of the ratio from unity was taken as a systematic
uncertainty on the PDF normalization.

Results from the \nit source, and the \cf source for Phase~II
neutrons, demonstrated that the signal loss in the data was reproduced
by the simulation to within the statistical uncertainties.  Analysis
of runs taken during the two phases showed no significant deviation
with time.  A position-dependent uncertainty was evaluated by taking
the ratio of the acceptance of \nit data and Monte Carlo events as a
function of source deployment position.  The difference of the
weighted average of the points from 1.0 was taken as the value of the
uncertainty.  The \lit source was used to investigate energy
dependence.  As expected, the signal loss decreased at higher
energies, where more information was available to reconstruct an
event.  The simulation was shown to reproduce this effect very
accurately and the uncertainty was therefore treated in the same
manner as the position-dependent uncertainty.

Combining the systematic uncertainties in quadrature, we obtained the
final uncertainties associated with reconstruction acceptance:
\begin{itemize}
\item Phase~I:    $\pm 0.034\%$ (stat) $\pm 0.060 \%$ (syst)
\item Phase~II $e^-$: $\pm 0.037\%$ (stat) $\pm 0.090 \%$ (syst)
\item Phase~II $n$: $\pm 0.000\%$ (stat) $\pm 0.009 \%$ (syst)
\end{itemize}

\subsubsection{High-Level Cut Acceptance}
\label{sec:hlcsac}

To take into account the acceptance of the high-level cuts, the ratio
of the cut acceptance for data and Monte Carlo events was calculated
and applied to the PDFs as a normalization correction.  This ratio was
evaluated as a function of energy, position and time.

The energy-uncertainty cuts described in Sec.~\ref{s:cutdeschlc} were
observed to have much stronger variations in signal loss as a function
of position and energy than the other high-level cuts and were
therefore treated separately.  It was verified that the correlations
between the two resulting subsets of high-level cuts were minimal, so
that treating them independently was a valid approach.  The following
sections describe the analysis for each subset of cuts, where `reduced
high-level cuts' refers to the subset that does not include the
energy-uncertainty cuts.

\subsubsection{Reduced High-Level Cut Acceptance}
\label{s:redhlcsac}

The data/Monte Carlo acceptance ratio and its uncertainty were
calculated for each calibration source run.  The runs were divided
into radial bins, and the error-weighted mean and standard deviation
were calculated in each bin.  Finally, the volume-weighted average of
the bin values was calculated.

The energy dependence of the acceptance ratio was investigated using
\nit and \lit data for electron-like events and \cf for Phase~II
neutron capture events.  The \nit data were restricted to the energies
below 9~MeV to avoid complications associated with event pile-up
caused by the high rate of the calibration source.

The measurements from \nit and \lit were in very good agreement, and
were both consistent with the acceptance ratio having no dependence on
energy.  The normalization correction for the PDFs was therefore
evaluated using the \nit source data by taking the weighted mean of
the values in each energy bin.  The median deviation of the \lit
points from the best-fit was taken as a systematic uncertainty on the
energy-dependence.

	The acceptance ratio for Phase~II neutron capture events was
evaluated using \cf data. To avoid pile-up of fission $\gamma$s, the
events were required to have energies in the interval 4.5-9.5~MeV. An
energy-dependent uncertainty was included to account for any variation
of individual energy bins from the overall average.

The stability of the acceptance as a function of time was studied
using \nit runs taken in the center of the detector.  No trend was
observed, but the time variability was incorporated as an additional
systematic uncertainty.

The \nit source was also used to evaluate a systematic uncertainty
associated with a possible position dependence of the acceptance
ratio.  Runs were binned by position in the detector, the
volume-weighted average of the bins was found and the mean deviation
of the ratio in each bin from this value was calculated.  A comparison
of \nit and \cf source data showed that they exhibited statistically
equivalent position dependences, so the more widely deployed \nit
source was used to quantify this effect for both electron-like and
neutron capture events.

The acceptance corrections and associated uncertainties derived from
the difference between the high-level cut acceptances for data and
Monte Carlo events are summarized in Table~\ref{t:hlcsac}.

\begin{table}[!ht]
\begin{center}
\begin{tabular}{lccc} \hline
\hline & Phase~I & Phase~II $e^-$ & Phase~II $n$ \\ \hline Correction
& 0.9945 & 0.9958 & 0.9983 \\ \hline Stat uncert (\%) & 0.0273 &
0.0159 & 0.0196 \\ Energy dep (\%) & 0.1897 & 0.1226 & 0.0005--2.3565
\\ Position dep (\%) & 0.1630 & 0.3144 & 0.3144 \\ Time dep (\%) &
0.0805 & 0.0130 & 0.0130 \\ \hline \hline
\end{tabular}
\caption{\label{t:hlcsac}Correction and associated uncertainties for
the high-level cut acceptance ratio.  The Phase~II neutron
energy-dependent uncertainty was treated differentially with energy;
the quoted range covers the value across the energy spectrum.}
\end{center}
\end{table}

\subsubsection{Energy-Uncertainty Cut Acceptance}
\label{sec:desac}

We expect that the effect of placing cuts on the uncertainty on the
estimate of an event's energy reported by the energy reconstruction
algorithm should be the same for data and Monte Carlo events.
Nevertheless, uncertainties on this assumption were evaluated using
the \nit and \cf source data, applying the same energy ranges as in
the reduced high-level cut analysis (Sec.~\ref{s:redhlcsac}).

Differential uncertainties were evaluated using the same method as for
the reduced high-level cuts.  The stability over time was measured
using \nit data.  The acceptance ratio was observed to be stable, but
an additional uncertainty was included based on the spread of the
points.

	The \nit and \cf data showed statistically equivalent
position-dependent behavior in the acceptance of the
energy-uncertainty cuts, and we therefore evaluated position-dependent
uncertainties using the more widely-deployed \nit source.  \nit source
data were divided into 50~cm slices along the $z$-axis, and the
acceptance ratios calculated in the slices were combined in a
volume-weighted average.  The uncertainty on this average was derived
from the deviation of the points from unity.

The energy-uncertainty cuts were even more sensitive to the effects of
pile-up than were the other high-level cuts.  Therefore, to evaluate
an energy-dependent uncertainty on the acceptance ratio for
electron-like events, events from the \nit source were restricted to
energies below 7~MeV, and the lower rate \lit source was used for
measurements at higher energies.  \cf data were used for Phase~II
neutron capture events, with the deviations from unity measured in the
8.5--9$\,$MeV bin also applied to higher energy events.  This resulted
in energy-dependent uncertainties for both electron-like and neutron
capture events.

The uncertainties in acceptance were applied as uncertainties in
normalization of the PDFs.  The values are summarized in
Table~\ref{t:desac}.

\begin{table}[!ht]
\begin{center}
\begin{tabular}{lrrr} \hline
\hline & \multicolumn{1}{c}{Phase~I} & \multicolumn{1}{c}{Phase~II
$e^-$} & \multicolumn{1}{c}{Phase~II $n$} \\ \hline Stat uncert (\%) &
0.0377 & 0.0668 & 0.0322 \\ Position dep (+) (\%) & +0.0750 & +0.0838
& +0.0838 \\ Position dep ($-$) (\%) & $-$1.0760 & $-$0.9897 &
$-$0.9897 \\ Time dep (\%) & 0.0834 & 0.0531 & 0.0531 \\ \hline \hline
\end{tabular}
\caption{\label{t:desac}Uncertainties on the energy-uncertainty cut
acceptance ratio.  Energy-dependent uncertainties were treated
differentially with energy and are not shown.  The uncertainty in
position is asymmetric.}
\end{center}
\end{table}

\subsubsection{Overall Cut Acceptance}
The final correction to the PDF normalization comes from combination
of the high-level cut correction (Table~\ref{t:hlcsac}) and the
instrumental cut correction (Sec.~\ref{s:inssac}).  The various
contributions to uncertainty on signal loss were treated as
uncorrelated and combined in quadrature to give the final uncertainty
on the cut acceptance correction.  Table~\ref{t:finalsac} lists the
final corrections and uncertainties.

\begin{table}[!ht]
\begin{center}
\begin{tabular}{lccc} \hline
\hline & Phase~I & Phase~II $e^-$ & Phase~II $n$ \\ \hline Correction
& 0.9924 & 0.9930 & 0.9954 \\ Pos uncertainty (\%) & 0.34--0.45 &
0.41--0.80 & 0.38--2.70 \\ Neg uncertainty (\%) & 1.12--1.17 &
1.07--1.08 & 1.06--1.65 \\ \hline \hline
\end{tabular}
\caption{\label{t:finalsac}Corrections applied to the Monte
Carlo-generated PDFs due to cut acceptance.  The uncertainties were
evaluated differentially with energy; the quoted range covers their
values across the energy spectrum.}
\end{center}
\end{table}

\begin{figure}[!ht]
\begin{center}
\includegraphics[width=0.48\textwidth]{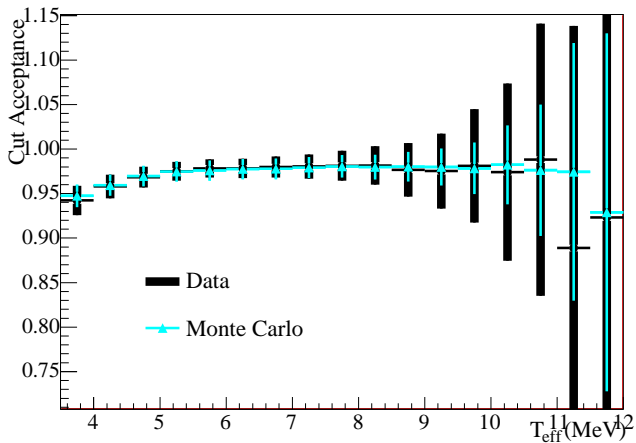}
\caption{\label{f:sacrifice}(Color online) Acceptance of the full set
of analysis cuts for both data and Monte Carlo events from a single
\cf run in Phase~II, as a function of kinetic energy.  }
\end{center} 
\end{figure}

Figure~\ref{f:sacrifice} shows a comparison of the cut acceptance for
data and Monte Carlo events from a single \cf run in Phase~II.  The
full set of analysis cuts was applied to both data and simulation, and
the Monte Carlo-predicted acceptance was corrected by the value from
Table~\ref{t:finalsac}.  As the figure shows, the Monte Carlo
simulation reproduces the shape of the data distribution very closely.

\section{Trigger Efficiency \label{sec:treff}}

\begin{figure}
\begin{center}
\includegraphics[height=0.26\textheight]{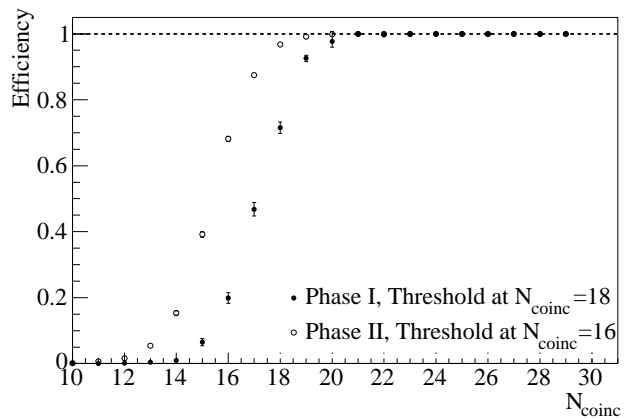}
\caption{\label{fig:trigturn} Comparison of the trigger efficiencies
in the two data-taking phases and for the two different thresholds
used.}
\end{center}
\end{figure}

	As discussed in Sec.~\ref{sec:dataset}, the primary trigger
for SNO was a coincidence of PMT hits within a 93~ns time window, set
to $N_{\rm coinc}=18$ hits for the early part of Phase~I and to
$N_{\rm coinc}=16$ hits for the remainder of Phase~I and all of
Phase~II.  We define the `efficiency' of the trigger as the
probability that an event with $N_{\rm coinc}$ hits actually triggered
the detector.  Small shifts in the analog (DC-coupled) baseline,
noise, and disabled trigger electronics channels could all lead to a
non-unity efficiency.  We measured the efficiency using the isotropic
laser source, by triggering on the laser pulse and comparing an
offline evaluation of the trigger (by counting hits in a sliding 93~ns
window) to the output of the hardware trigger.  We found that for the
$N_{\rm coinc}=18$ hit threshold, events with 23 or more hits in
coincidence triggered the detector with an efficiency greater than
99.9\% and, for the $N_{\rm coinc}=16$ hit threshold, the efficiency
reached 99.9\% at 21 hits.  Figure~\ref{fig:trigturn} shows the
efficiency measured as a function of $N_{\rm coinc}$, for Phase I at
the higher $N_{\rm coinc}=18$ threshold, and for Phase~II at the lower
$N_{\rm coinc}=16$ hit threshold.

	For events at our $T=3.5$~MeV analysis threshold, the mean
number of hits in an event over the full 400~ns event window was
$\sim$ 30 for Phase I and $\sim$ 27 for Phase~II, with RMS's of 1.8
hits and 1.7 hits, respectively.  The numbers of hits in the 400~ns
event window and in the 93~ns trigger coincidence window differed
primarily in the contribution from random PMT noise which, for both
phases, contributed on average roughly 1 additional hit in the 400~ns
event window.  Thus, for both phases, the trigger efficiency was above
99.9\% for all but a negligible fraction of events with a high enough
$N_{\rm coinc}$ to pass the analysis cuts.

	Because our PDFs and overall normalization were derived from
simulation, we compared the trigger-efficiency estimate from the data
to the simulation's prediction. We also compared the idealized
simulated trigger to a simulation that included variations in the
trigger baseline as measured by an online monitor.  We found that the
Monte Carlo simulation's prediction of trigger efficiency was in
excellent agreement with our measurement for both SNO phases, and that
the measured variations contributed a negligible additional
uncertainty to our overall acceptance.

\section{Uncertainties on the Neutron Capture Efficiencies \label{sec:ncap}}

	In Phase~I, neutrons produced through the NC reaction and
background processes were captured on deuterons within the heavy
water, releasing a single 6.25~MeV $\gamma$ ray. In Phase~II, the
neutrons were captured primarily on $^{35}$Cl, releasing a $\gamma$
cascade of total energy 8.6~MeV.  The absolute cross sections for
these capture reactions, along with detector acceptance, determined
the rate of detected neutron events.  The uncertainty on the neutron
capture efficiency for Phase~II overwhelmingly dominates that for
Phase~I in the final flux determinations because of the larger capture
cross section.

In this analysis, we used the Monte Carlo simulation to define the
central values of the neutron capture efficiencies.  Included in our
simulation were the measured isotopic purity of the heavy water, as
well as its density and temperature and, for Phase~II, the measured
density of salt added to the \dtonosp.

	To assess the systematic uncertainties on the neutron capture
efficiencies, we used data taken with the $^{252}$Cf source deployed
at many positions throughout the detector, and compared the observed
counting rates to simulations of the source runs.  The differences
between data and simulated events provide an estimate of the
simulation's accuracy.  The Phase~I and Phase~II data sets are
noticeably different in their neutron detection efficiency because of
the much larger capture cross section in Phase~II, and the higher
energy $\gamma$ cascade from neutron capture on chlorine.  We
therefore assessed the uncertainties in the two phases slightly
differently, as discussed below.  We also compared the results of this
`direct counting' approach with a `time series analysis', in which the
relative times of events were used to extract the capture efficiency.
The two methods were in excellent agreement for both phases.  Our
capture efficiency uncertainty for Phase~II is $\pm 1.4$\%, and for
Phase~I it is $\pm$2\%.

\subsection{Phase~II Neutron Capture Efficiency Uncertainties}

\begin{figure}
\begin{center}
\includegraphics[width=0.48\textwidth]{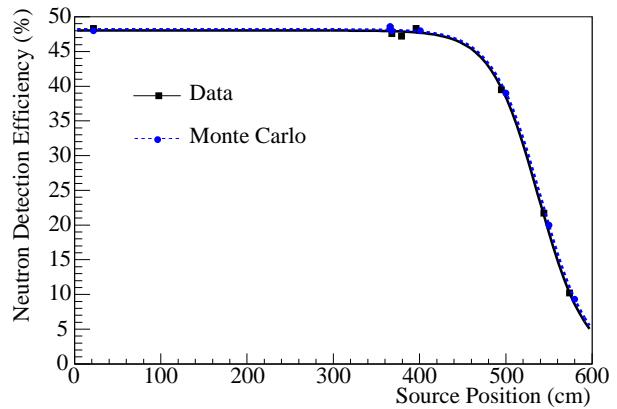}
\caption{\label{fig:funcomps}(Color online) Data and Monte Carlo
neutron detection efficiencies in Phase~II fitted to the
phenomenologically-motivated neutron detection efficiency function.  }
\end{center}
\vspace{-4ex}
\end{figure}

	For the Phase~II analysis, neutron events from the $^{252}$Cf
source were selected using the same burst algorithm that was used in
previous SNO publications~\cite{nsp}.  Neutrons were identified by
looking for prompt fission $\gamma$ events from the $^{252}$Cf decay,
and tagging subsequent events that occurred within 40~ms.
Figure~\ref{fig:funcomps} plots the neutron detection efficiency for
each source run as a function of radial position of the source in the
detector, for both data and Monte Carlo simulated events.  The source
position for a run was determined by finding the mean reconstructed
position of the prompt fission $\gamma$ events, to eliminate the large
positioning uncertainties of the source deployment mechanism.  The
efficiencies shown in Fig.~\ref{fig:funcomps} were each fitted to a
phenomenologically-motivated neutron detection efficiency function:
\begin{equation}
\epsilon(s) = A ({\rm tanh}(B(s-C)) - 1),
\label{eqn:tanh}
\end{equation}
where $\epsilon(s)$ gives the neutron capture efficiency at source
radius $s$.

	To determine the uncertainty on the simulation's prediction of
capture efficiency, we first calculated the mean capture efficiency in
the \dto volume, given the two functions shown in
Fig.~\ref{fig:funcomps}, as follows:
\begin{equation}
  \epsilon = \frac{\int_{0}^{600.5}s^{2}\epsilon(s) ds
  }{\int_{0}^{600.5}s^{2}ds}.
\end{equation}
We took the difference of 0.8\% between data and simulation as a
baseline uncertainty.  (The mean detection efficiency measured this
way was 35.6\%).

	The normalization of the curves shown in
Fig.~\ref{fig:funcomps} depends on the strength of the $^{252}$Cf
source, which we know to 0.7\% based on {\it ex-situ} measurements.
An overall shift in reconstructed event positions, discussed in
Sec.~\ref{sec:hitcal}, also changed the measured efficiency in data
relative to the simulation results.  By varying the value of this
shift within its range of uncertainty we found it resulted in an
additional 0.3\% uncertainty in capture efficiency.  The uncertainty
in the fit parameters of the neutron detection efficiency function was
included conservatively by taking the entire statistical uncertainty
on the data efficiency measurements of Fig.~\ref{fig:funcomps}, which
yields another 0.9\%.  Lastly, we included a 0.1\% uncertainty to
account for the fraction of $^{250}$Cf in the $^{252}$Cf source (only
$^{252}$Cf is simulated by the Monte Carlo code).  The overall
uncertainty on the neutron capture efficiency, calculated by adding
these in quadrature, was 1.4\%.

	We checked these results by performing an independent time
series analysis, in which we fit directly for the efficiency at each
source deployment point based on the rates of neutron capture and
$\gamma$ fission events (the source strength is not an input
parameter).  The fit included parameters associated with the overall
fission rate, backgrounds from accidental coincidences, and the mean
capture time for neutrons.  We obtained the efficiency as a function
of source radial position, to which we fit the same efficiency
function from Eq.~\ref{eqn:tanh}, and extracted the volume-weighted
capture efficiency directly (rather than by comparison to Monte
Carlo).  The mean efficiency calculated this way was $35.3\pm 0.6$\%,
in excellent agreement with the value of 35.6\% from the direct
counting method, and well within the uncertainties on both
measurements.

\subsection{Phase~I Neutron Capture Efficiency Uncertainties}

	The measurement of neutron capture efficiency uncertainty for
Phase~I is more difficult than for Phase~II, primarily because the
lower capture cross section in Phase~I made identification of neutron
events from the $^{252}$Cf source difficult. The number of detected
neutrons per fission was small (less than one on average), and the
long capture time (roughly 50~ms) made coincidences more likely to be
accidental pile-up of prompt fission $\gamma$s than neutrons following
the $\gamma$s.

	Instead of using the burst algorithm, we separated neutron
events from fission \gam s based on their differing energies and mean
free paths in \dtonosp.  Events were required to be more than 150~cm
from the source position and to have energies above the mean energy
expected for a neutron capture event, for both data and Monte Carlo
events.  The detected rate of events after these cuts was used for the
data and Monte Carlo simulation comparison.

\begin{figure}
\begin{center}
\includegraphics[width=0.48\textwidth]{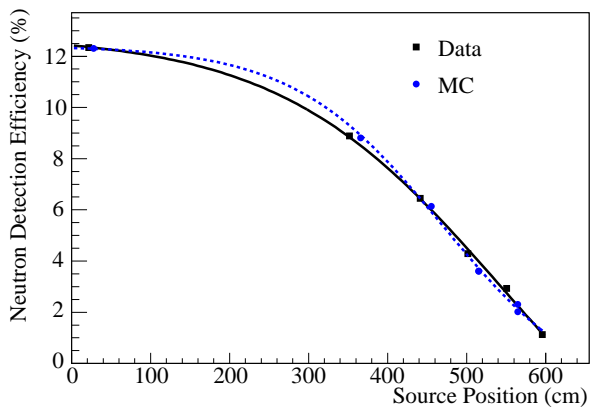}
\caption{\label{fig:cfmcdata_d2o}(Color online) Comparison of the fit
functions to the data and Monte Carlo in Phase~I.}
\end{center}
\vspace{-4ex}
\end{figure}

	An additional parameter was added to the neutron detection
efficiency function for these data, as follows:
\begin{equation}
\epsilon(s) = A ({\rm tanh}(B(s-C)) - D),
\label{eqn:tanhprime}
\end{equation}
and the resulting fits to data and Monte Carlo are shown in
Figure~\ref{fig:cfmcdata_d2o}.

	The difference of the volume-weighted integrals of the two
curves is just 0.9\%, but the small value is clearly due to
cancellation differences at different radii.  The shape difference is
driven by small differences between the data and Monte Carlo fits at
large radii, which are likely due to unassessed systematic errors on
the data points themselves.  We included additional uncertainties to
account for these.  In particular, we included a 0.6\% uncertainty
associated with the statistical uncertainties of the data and Monte
Carlo neutron detection efficiency function parameters, and an
additional 0.6\% uncertainty associated with knowledge of the source
position.  We also included a further uncertainty of 0.9\% to account
for data and Monte Carlo differences in the energy cut applied to
select neutrons.

	We applied the same source-strength uncertainties as for the
Phase~II analysis, namely the 0.7\% absolute source strength
calibration, and 0.1\% from the (unmodeled) contamination of
$^{250}$Cf in the $^{252}$Cf source.  The total uncertainty on the
neutron capture efficiency for Phase~I comes to 2\%.

\begin{figure}
\begin{center}
\includegraphics[width=0.48\textwidth]{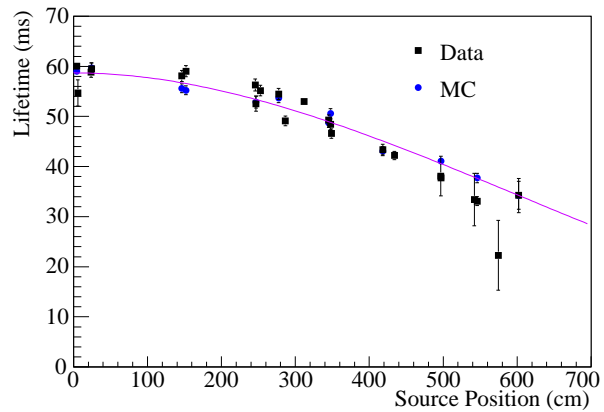}
\caption{\label{fig:d2otau}(Color online) Mean neutron capture time
from the time series analysis in Phase~I as a function of source
position.  The line shows the best fit to the simulation using a cubic
polynomial.}
\end{center}
\vspace{-4ex}
\end{figure}

	To check our estimates, we also performed a time series
analysis of the $^{252}$Cf data.  Unlike Phase~II, for Phase~I we
cannot extract the absolute efficiency to compare with that derived
from the direct counting method because of the 150~cm reconstruction
cut.  Instead, we performed the time series analysis on both Monte
Carlo and source data runs, and compared them. We found the fractional
difference between the source-derived and Monte Carlo-derived
efficiencies to be just 0.3\%, well within the 2\% uncertainty
obtained from the direct counting method.  One output of the time
series analysis is the neutron capture time: the time between neutron
emission from the $^{252}$Cf source and capture on a deuteron.
Figure~\ref{fig:d2otau} shows the neutron capture time as a function
of source radial position for both data and Monte Carlo.  As the
$^{252}$Cf source approaches the acrylic vessel and light water
region, the capture time decreases significantly.  The overall
agreement between the measured capture times in data and Monte Carlo
is very good throughout most of the volume.

\section{Backgrounds \label{sec:backgrounds}} 

Lowering the energy threshold opened the analysis window to additional
background contamination, predominantly from radioactive decays of
\bisp and \tlsp in the $^{238}$U and $^{232}$Th chains, respectively.
In Phase~II, neutron capture on $^{23}$Na produced a low level of
$^{24}$Na in the detector which, in its decay to $^{24}$Mg, produced a
low energy $\beta$ and two $\gamma$s. One of these \gam s has an
energy of 2.75~MeV, which could photodisintegrate a deuteron.  The
result was some additional electron-like and neutron capture
background events.  In addition, radon progeny that accumulated on the
surface of the AV during construction could have created neutrons
through ($\alpha$,$n$) reactions on isotopes of carbon and oxygen
within the acrylic.

	In the past, most of these backgrounds were estimated using
separate self-contained analyses and then subtracted from the measured
neutrino fluxes.  In this analysis, the Monte Carlo simulation was
used to create PDFs for each of 17 sources of background events
(except for PMT $\beta$-\gamsp events, for which an analytic PDF was
used in each phase, as described in Sec.~\ref{s:pmtpdf}), and the
numbers of events of each type were parameters in the signal
extraction fits.  Table~\ref{t:bkgs} lists the sources of
physics-related backgrounds that were included in the fits.

\begin{table}[!ht]
\begin{center}
\begin{tabular}{lccc}
\hline \hline Detector Region & Phase~I & Phase~II\\ \hline \hline
\dto volume & Internal \bi & Internal \bi \\ & Internal \tl & Internal
\tl \\ & & $^{24}$Na \\ \hline Acrylic vessel & Bulk \bi & Bulk \bi \\
& Bulk \tl & Bulk \tl \\ & Surface ($\alpha$,$n$) $n$s & Surface
($\alpha$,$n$) $n$s \\ \hline \hto volume & External \bi & External
\bi \\ & External \tl & External \tl \\ & PMT $\beta$-\gam s & PMT
$\beta$-\gam s \\ \hline \hline
\end{tabular}
\caption[Sources of background events in the LETA analysis.]{The
sources of physics-related background events in the LETA analysis.}
\label{t:bkgs}
\end{center}
\end{table}

All of the Monte Carlo-generated PDFs were verified using calibration
sources.  {\it Ex-situ} measurements~\cite{htio, mnox} of background
levels in the \dto and \hto provided {\it a priori} information for
several of them, which were used as constraints in the signal
extraction fits.  In addition, corrections were applied after the
signal extraction fits to account for a number of background event
types that contributed much smaller levels of contamination.  The
following sections describe these procedures.

\subsection{Background PDFs}

Most of the PDFs used in the signal extraction were created from Monte
Carlo simulations of the specific event types.  However, because of
the limited number of simulated PMT $\beta$-\gamsp events available in
the radial range of interest, an analytic parameterization of the PDF
was used, as described in Sec.~\ref{s:pmtpdf}.  This was verified by
comparison to the simulation and uncertainties associated with the
value of each parameter were propagated in the signal extraction fits.

The remainder of the background PDFs were verified by comparison of
calibration data to simulated events.  The \dto and \hto backgrounds
were verified using the \dto- and \htonosp-region radon spikes in
Phase~II and calibration sources deployed in these regions.  Bulk AV
backgrounds were verified using the $^{238}$U and $^{232}$Th sources,
and surface ($\alpha$,$n$) neutrons using the \cf source deployed near
the AV.

\begin{figure}[!ht]
\begin{center}
\includegraphics[width=0.48\textwidth]{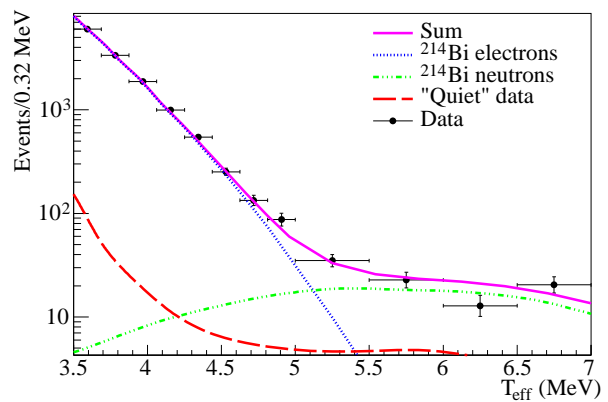}
\caption{\label{f:spikefits}(Color online) One dimensional projection
of the fit to the internal radon spike data.}
\end{center} 
\end{figure}

\begin{figure}[!ht]
\begin{center}
\includegraphics[width=0.48\textwidth]{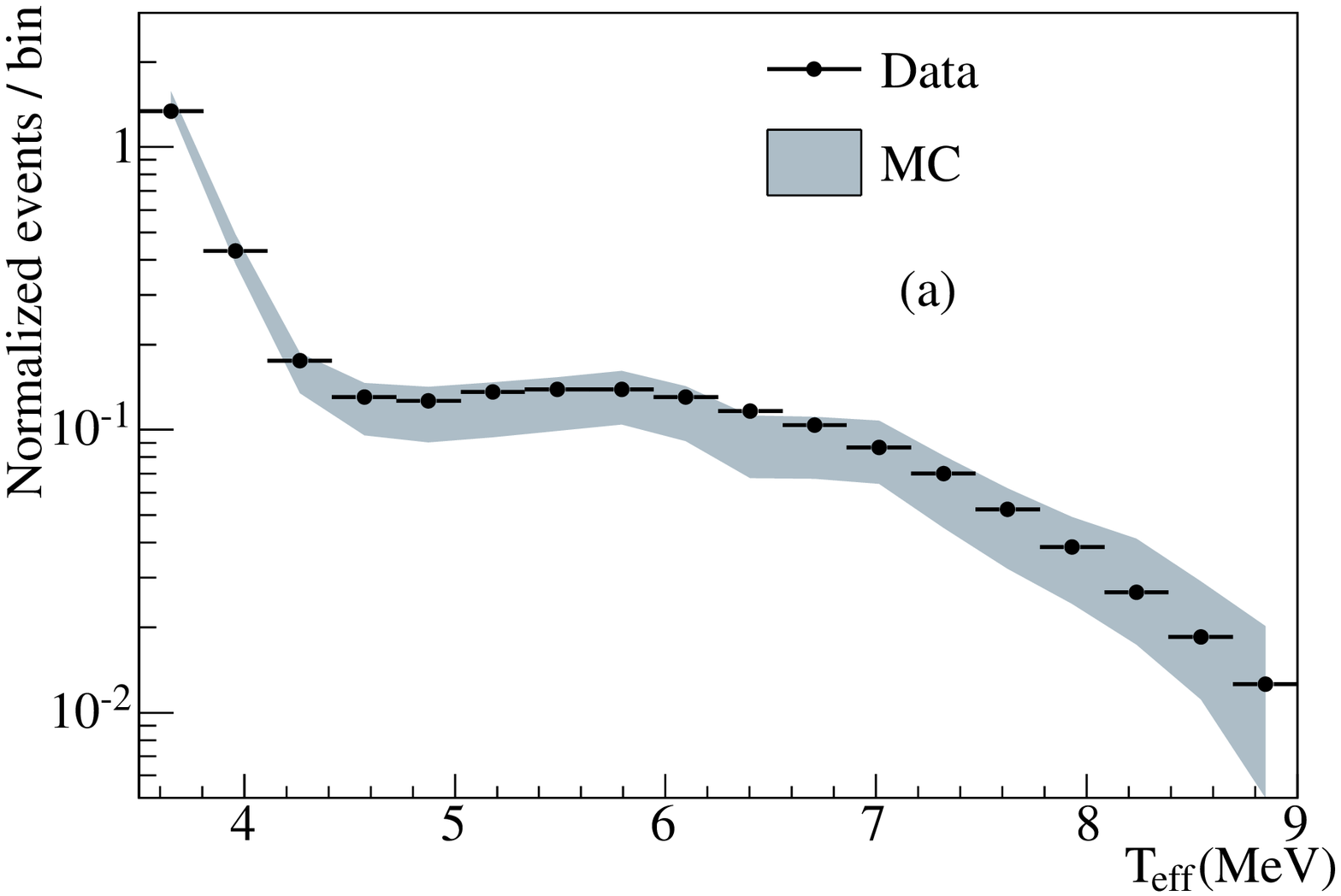}
\includegraphics[width=0.48\textwidth]{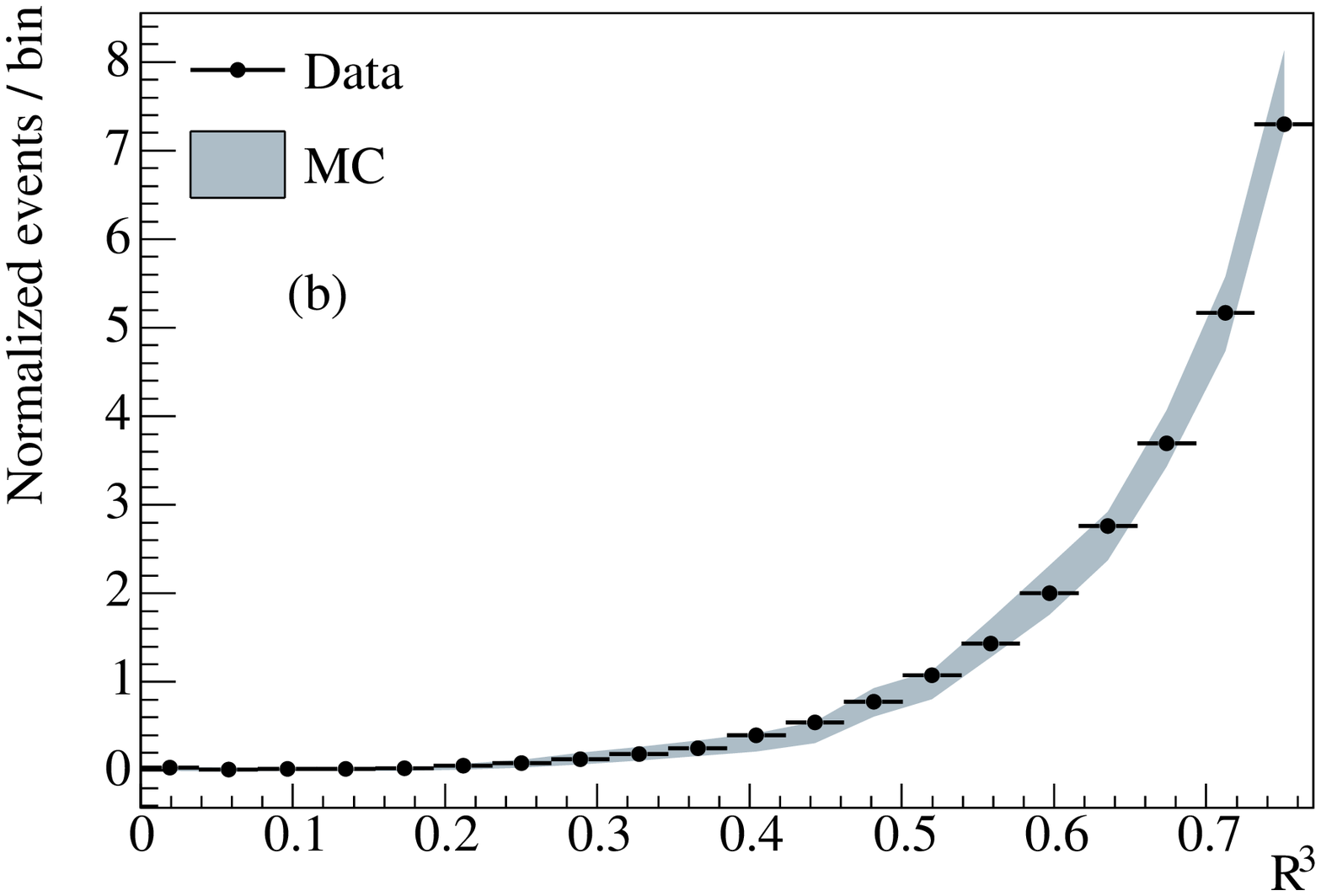}
\includegraphics[width=0.48\textwidth]{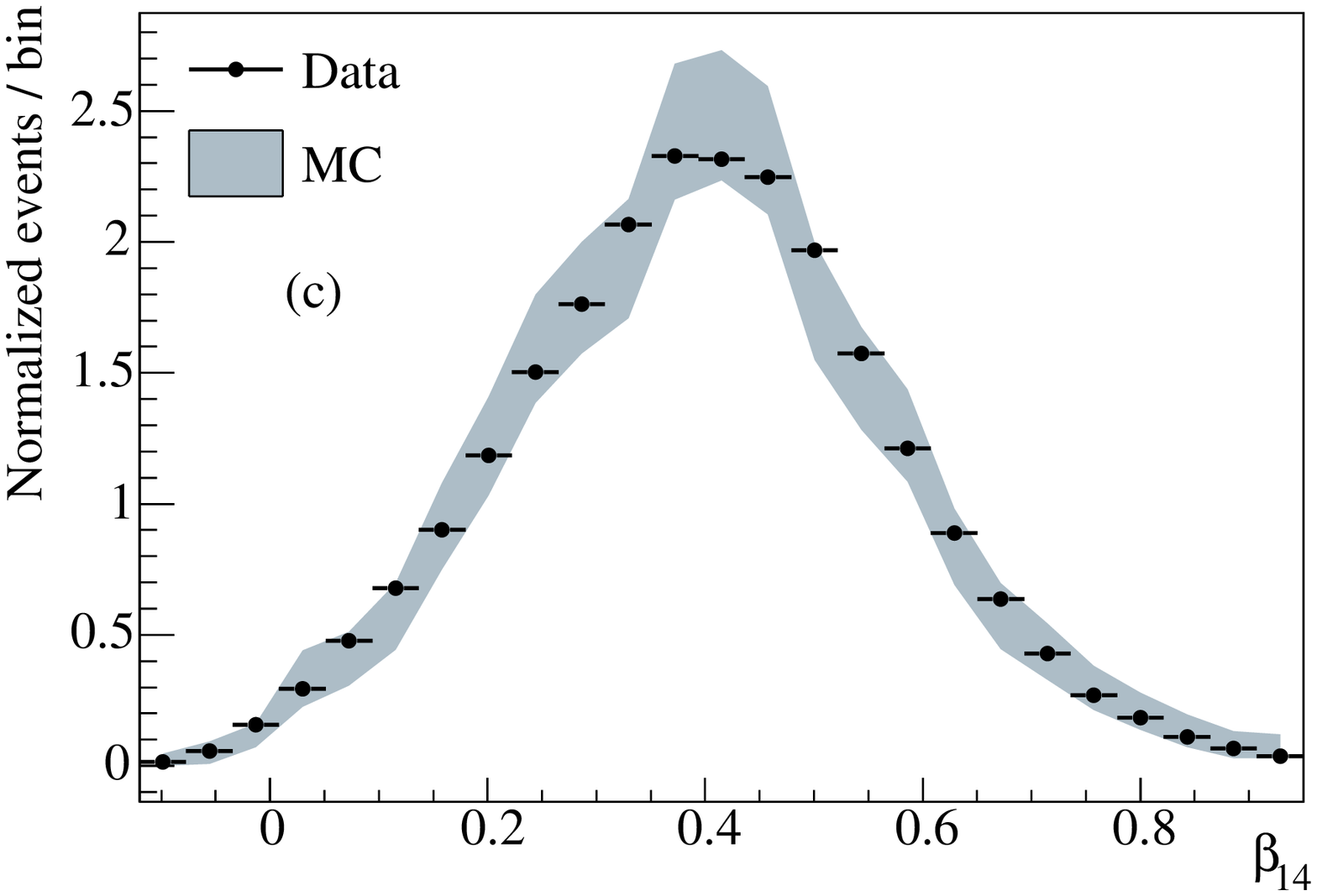}
\caption{\label{f:avtl}(Color online) Comparison of data to simulation
for $^{232}$Th source runs near the AV in Phase~II, in (a) $T_{\rm
eff}$, (b) $R^3$, and (c) \betnosp.  The band represents the 1$\sigma$
uncertainty on the Monte Carlo-prediction, taking the quadrature sum
of the statistical uncertainties with the effect of applying the
dominant systematic uncertainties.}
\end{center} 
\end{figure}

In all cases, the data and Monte Carlo event distributions agreed to
within the systematic uncertainties already defined for the PDFs.
Figure~\ref{f:spikefits} shows the energy dimension of a fit to the
internal radon spike.  The fit was performed using the unbinned signal
extraction code (see Sec.~\ref{s:kernel}) in a simplified
configuration, as described in Sec.~\ref{s:eres:saltelec}.  The result
is a good fit to the data, in particular at low energy.
Figure~\ref{f:avtl} shows a comparison of data to simulation for the
$^{232}$Th source deployed near the AV.  A band is shown for the
simulated events, representing the quadrature sum of the statistical
uncertainties with the effect of applying the dominant systematic
uncertainties.  The distributions in $T_{\rm eff}$, $R^3$ and \bof
show good agreement within the 1$\sigma$ uncertainties.

The cross section for photodisintegration affects the relative
normalization of the neutron and electron parts of the background
PDFs.  The simulation used a theoretical value for the cross section
and the associated 2\% uncertainty was propagated in the signal
extraction fits.

The simulation of $^{24}$Na events used to generate a PDF was done
under the assumption of a uniform distribution of events within the
detector, since a primary source of $^{24}$Na was the capture of
neutrons produced by deployed calibration sources on $^{23}$Na.
$^{24}$Na was also introduced via the neck, and via the water systems,
which connected near the top and bottom of the AV.  Therefore, the
signal extraction fits were redone with different spatial
distributions, in which the events originated either at the neck of
the AV or at the bottom, with a conservatively chosen 10\% linear
gradient along the $z$-axis.  The difference from the baseline
(uniform distribution) fit was taken as a systematic uncertainty.

\subsection{Low Energy Background Constraints}
\label{s:bkgconst}

Several radioassays were performed during data taking to measure the
concentrations of radon and radium in the \dto and \hto regions, as
described in previous publications \cite{longd2o, htio, mnox}.
Although equilibrium was broken in the decay chains, the results are
expressed in terms of equivalent amounts of $^{238}$U and $^{232}$Th
assuming equilibrium for ease of comparison with other measurements.
The results were used to place constraints on the expected number of
background events in the analysis window.  During Phase II, there was
a leak in the assay system used to measure the $^{238}$U chain
contamination that was not discovered until after data taking had
ended, so there is no accurate constraint on the $^{238}$U level in
the \dto during that phase.  Other limits based on secondary assay
techniques were found to be too loose to have any impact on the signal
extraction results and so were disregarded.  The results of the assays
are given in tables \ref{t:exsitud} and \ref{t:exsituh}.

\begin{table}[!ht]
\begin{center}
\begin{tabular}{lcc}
\hline \hline Phase & Isotope & Concentration ($\times$ 10$^{-15}$ g/g
of \dto)\\ \hline I & $^{238}$U & 10.1$\,^{+3.4}_{-2.0} $ \\ &
$^{232}$Th & 2.09$\,\pm\,$0.21(stat)$\,^{+0.96}_{-0.91}$(syst)\\
\hline II & $^{238}$U & --- \\ & $^{232}$Th &
1.76$\,\pm\,$0.44(stat)$\,^{+0.70}_{-0.94}$(syst) \\ \hline \hline
\end{tabular}
\caption[{\it Ex-situ} constraints on background events.]{$^{238}$U
and $^{232}$Th concentrations in the \dto volume, determined from {\it
ex-situ} radioassays in Phases I and II.}
\label{t:exsitud}
\end{center}
\end{table}

\begin{table}[!ht]
\begin{center}
\begin{tabular}{lcc}
\hline \hline Phase & Isotope & Concentration (g/g of \hto)\\ \hline I
& $^{238}$U & 29.5$\,\pm\,$ 5.1$\times$ 10$^{-14}$\\ & $^{232}$Th &
8.1$\,^{+2.7}_{-2.3} $$\times$ 10$^{-14}$\\ \hline II & $^{238}$U &
20.6$\,\pm\,$ 5.0$\times$ 10$^{-14}$\\ & $^{232}$Th & 5.2$\,\pm\,$
1.6$\times$ 10$^{-14}$\\ \hline \hline
\end{tabular}
\caption[{\it Ex-situ} constraints on background events.]{$^{238}$U
and $^{232}$Th concentrations in the \hto volume, determined from {\it
ex-situ} radioassays in Phases I and II.}
\label{t:exsituh}
\end{center}
\end{table}

These concentrations were converted into an expected number of events
and were applied as constraints in the signal extraction fits, as
described in Sec.~\ref{s:penaltyterms}.

{\it In-situ} analyses \cite{simsthesis} were used to predict the
number of background events from $^{24}$Na decays in Phase II.  The
predicted value of $392\pm117.6$ events was applied as a constraint in
the signal extraction fits.

\subsection{PMT $\beta$-\gamsp PDF}
\label{s:pmtpdf}

We use the term ``PMT events'' to refer to all radioactive decays in
the spherical shell region encompassing the PMTs and the PSUP.  These
events were primarily $^{208}$Tl decays originating from $^{232}$Th
contamination in the PMT/PSUP components.

PMT events occurred at a high rate, but only a tiny fraction of them
reconstructed inside the signal box and within the fiducial volume: in
Phase~I, the acceptance was only $1.7\times 10^{-8}$ and in Phase~II
it was $5.9\times 10^{-8}$.  Therefore, an enormous amount of computer
time would be needed to generate enough events to create a PDF.
Creation of a multi-dimensional PDF based entirely on simulation was
therefore deemed to be impractical.

A high rate thorium source was deployed near the PSUP in both phases
to help model these events.  However, interpretation of this data was
complicated by the fact that a point source with a sufficiently high
rate tends to produce significant `pile-up' of multiple events that
trigger in the same time window.  This pile-up changes the topology of
the events to the extent that they are not characteristic of PMT
$\beta$-\gam s, so they cannot be used directly as a model.

Therefore, an analytic parameterization of the PDF, given in
Eq.~\eqref{e:pmtpdf}, was used.  For this, the \cts dimension was
assumed to be flat; the remaining three-dimensional PDF was of the
form:

\begin{eqnarray}
	\label{e:pmtpdf}
P_{PMT}(T_{\rm eff},& \beta_{14}&, R^3) = e^{A\,T_{\rm eff}} \times
 (e^{B\,R^3} + C)\nonumber \\ &\times& \mathcal{N}(\beta_{14}\, |\,
 \bar{\beta}_{14}=D + ER^3, \sigma = F),
\end{eqnarray}
where $\mathcal{N}(x|\bar{x}, \sigma)$ is a Gaussian distribution in
$x$ with mean $\bar{x}$ and standard deviation $\sigma$.  The \bet
dimension was determined from a Gaussian fit to Monte Carlo events, in
which $\bar{\beta}_{14}$ was allowed a linear dependence on $R^3$.

The source location of the PMT events, their large number, and the
fact that they must reconstruct nearly 3~m from their origin to appear
inside the fiducial volume means that they have features that
distinguish them from other sources of backgrounds.  Therefore, we
were able to extract a prediction for the total number of PMT events,
as well as for the shape of the energy and radial dimensions of the
PDF, from the data itself, by performing a bifurcated analysis.

In a bifurcated analysis, two independent cuts are selected that
discriminate signal from background.  The behavior of these cuts when
applied both separately and in combination is used to assess the
number of signal and background events in the analysis window.  We
assume that the data set consists of $\nu$ signal events and $\beta$
background events, so that the total number of events is $S=\beta +
\nu$.  The background contamination in the final signal sample is just
the fraction of $\beta$ that passes both cuts.  If the acceptances for
background and signal events by cut $i$ are $y_i$ and $x_i$,
respectively, the contamination is $y_1 y_2 \beta$ and the number of
signal events is $x_1 x_2 \nu$.

Given the number, $a$, of events that pass both cuts, the number, $b$,
that fail cut 1 but pass cut 2, and the number, $c$, that pass cut 1
but fail cut 2, we then relate these with a system of equations:
\begin{eqnarray}
a+c&=&x_1 \nu+y_1 \beta,\\
\label{eq1}
a+b&=&x_2 \nu+y_2 \beta,\\
\label{eq2}
a&=&x_1 x_2\nu+y_1 y_2 \beta,\\
\label{eq3}
\beta+\nu&=&S,
\label{eq3prime}
\end{eqnarray}
\noindent
which we solve analytically, using Monte Carlo-predictions for the cut
acceptances, to determine the contamination, $K= y_1 y_2 \beta$, in
the signal sample.  A feature of this method is that it produces a
contamination estimate without including events from the signal box
(those that pass both cuts) in the analysis.

In this analysis, the `background' comprised the PMT events and the
`signal' all other events, including both neutrino interactions and
non-PMT radioactive decays.  The cuts chosen were the in-time ratio
(ITR) cut, because it selected events that were reconstructed far from
their true origin, and the early charge (EQ) cut because it selected
events in which a large amount of light produced hits early in time in
a small number of tubes.  These tend to be characteristics of PMT
events (see Sec.~\ref{s:cutdeschlc}).

For a bifurcated analysis to work, the probabilities of passing the
cuts must be statistically independent. To demonstrate this, we
loosened the cuts, and found that the increase in the number of
background events agreed well with what would be expected if they were
independent.

	 One result of the bifurcated analysis is a prediction for the
number of PMT events in the analysis window, which was used as a
constraint in the binned likelihood signal extraction fits, as
described in Sec.~\ref{s:penaltyterms}.

The acceptance of signal events ($x_1 x_2$) $\neq 1.0$ and therefore
some non-PMT events were also removed by the cuts. Such events falsely
increase the count of background events in the three `background
boxes'.  We limited the impact of this effect by restricting the
analysis to the 3.5--4.5~MeV region, which was overwhelmingly
dominated by PMT events. We also included a correction for the number
of non-PMT events in each of the background boxes by using estimates
from the Monte Carlo simulation for the acceptance of all other
signals and backgrounds, and verifying these predictions with radon
spike data.  ($^{214}$Bi, a radon daughter, is the dominant background
other than the PMT events in this region).

	To estimate the number of non-PMT events in each of the three
background boxes, we multiplied the Monte Carlo-predicted acceptances
of non-PMT events by the expected total number of these events in the
data set.  The procedure was therefore iterative: a PMT PDF was
created using initial estimates for the total number of non-PMT events
in the data set and their acceptances; the bifurcated analysis was
used to predict the number of PMT events in the signal box; the data
were re-fit with this new PMT constraint; the total number of non-PMT
events in the data set, based upon the new fit, was then used to
update the non-PMT event correction in the background boxes in the
bifurcated analysis, and so on.  In practice, the bifurcated analysis
itself was simply included within the signal extraction fit, so the
prediction for the number of PMT events could be recalculated as the
fit progressed, and the penalty factor in the likelihood calculation
from the resulting constraint could be varied accordingly.  To
determine systematic uncertainties on this overall procedure, we
tested the analysis on sets of fake data and compared the prediction
of the bifurcated analysis to the known true number of PMT
$\beta$-$\gamma$ events in the signal box.

	We verified the bifurcated analysis results by comparing the
prediction of the total number of PMT $\beta$-$\gamma$ events in the
signal box to an estimate made with an independent analysis performed
outside the fiducial volume.  This independent analysis looked for
events that occurred at high radius and were inward-pointing, which
are characteristics of PMT $\beta$-$\gamma$ events, and extrapolated
that count into the fiducial volume.  The measurements agreed with the
bifurcated analysis to well within the uncertainties on the two
methods.

To predict the shape of the PMT PDF, the bifurcated analysis was
performed in discrete bins in $T_{\rm eff}$ and $R^3$.  Unlike the
prediction for the total number of PMT events in the data set, this
calculation was not included in the signal extraction, so a fixed
estimate of the contamination of non-PMT events in the three
background boxes was applied.  This estimate was derived from a signal
extraction fit performed on a small subset of the data.  To take
uncertainties into account, bifurcated analyses were performed on
Monte Carlo-generated `fake' data sets with the dominant systematic
and statistical uncertainties applied in turn, to determine the effect
of each on the extracted shape for the PMT PDF.  The differences of
the results from the unshifted version were added in quadrature to
obtain an additional uncertainty on the shape.

\begin{figure}[!ht]
\begin{center}
\includegraphics[width=0.48\textwidth]{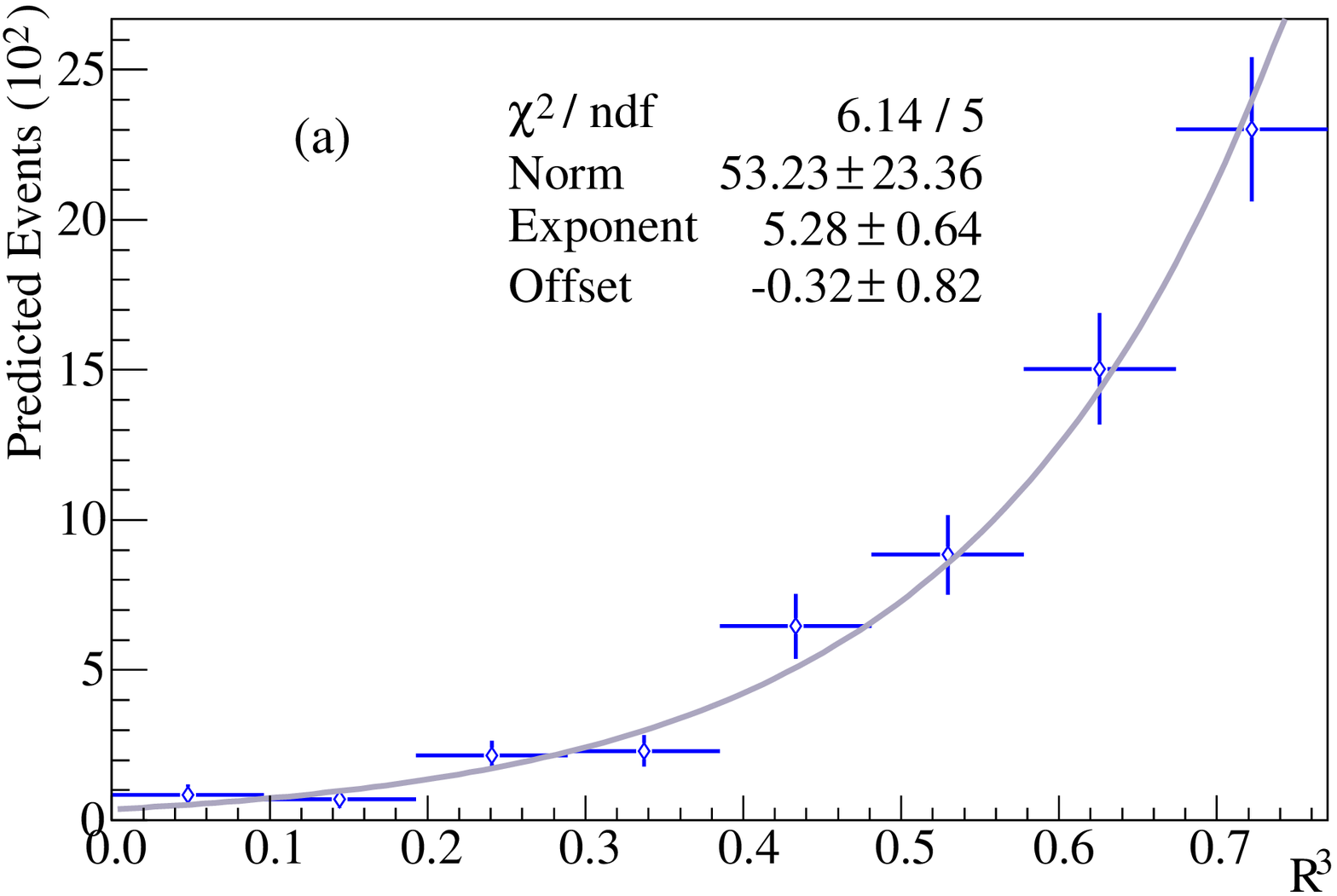}
\includegraphics[width=0.48\textwidth]{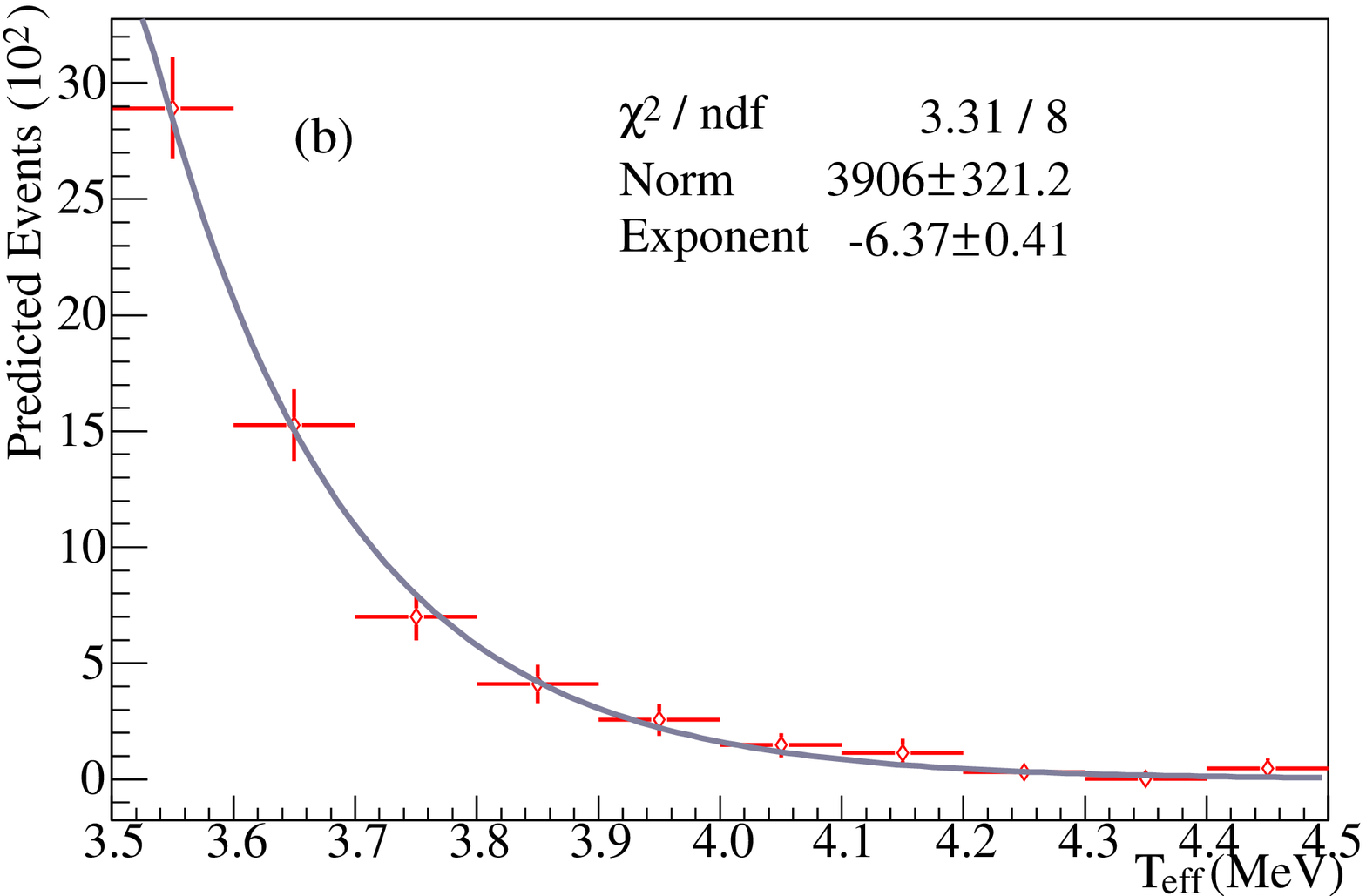}
\caption{\label{f:saltpmtpdf}(Color online) Predicted shapes for the
PMT PDF in (a) $R^3$ and (b) $T_{\rm eff}$ in Phase~II.}
\end{center} 
\end{figure}

A number of functional forms were fit to the $T_{\rm eff}$ and $R^3$
distributions to determine the best parameterizations for the shapes.
An exponential was found to be a good fit to the energy profile and an
exponential plus a constant offset to the radial distribution (see
Eq.~\eqref{e:pmtpdf}).  The fit results for Phase~II are shown in
Figure~\ref{f:saltpmtpdf}.

The parameters from the fits shown in Fig.~\ref{f:saltpmtpdf} were
varied in the signal extraction by applying a Gaussian penalty factor
to the likelihood function, as described in Sec.~\ref{s:penaltyterms}.
The mean of the Gaussian was the central fit value from
Fig.~\ref{f:saltpmtpdf} and the standard deviation was taken as the
total uncertainty in this value, including both the fit uncertainty
from Fig.~\ref{f:saltpmtpdf} and the additional systematic
uncertainties described above.  Results for both phases are shown in
Table~\ref{t:bafin}.  The fits to the bifurcated analysis prediction
for the $R^3$ distribution showed a significant correlation between
the exponent and the offset, with correlation coefficients of 0.846
and 0.883 in Phases~I and~II, respectively.  This correlation was
included in the Gaussian penalty factor in the signal extraction fits.

\begin{table}[!ht]
\begin{center}
\begin{tabular}{lrr}
\hline \hline Parameter& \multicolumn{1}{c}{Phase~I} &
\multicolumn{1}{c}{Phase~II}\\ \hline Energy exponent, A (/MeV) &
$-$5.94 $\pm$ 0.96 & $-$6.37 $\pm$ 0.81\\ $R^3$ exponent, B & 5.83
$\pm$ 0.96 & 5.28 $\pm$ 0.79\\ $R^3$ offset, C & $-$0.40 $\pm$ 1.43 &
$-$0.32 $\pm$ 1.16\\ \hline \hline
\end{tabular}
\caption{Parameters defining the PMT PDF shape, as defined in
Eq.~\eqref{e:pmtpdf}.}
\label{t:bafin}
\end{center}
\end{table}

\subsection{Limits on Instrumental Backgrounds}

Because instrumental background events were not modeled by the
simulation, their contamination in the analysis window was determined
directly from the data.  A bifurcated analysis was used, similar to
that described in Sec.~\ref{s:pmtpdf}.  In this instance, two sets of
cuts were used to define the analysis: the instrumental cuts and the
high-level cuts, described in Sec.~\ref{s:cuts}.  The numbers of
events in the data set failing each and both sets of cuts were used to
estimate the contamination by instrumental backgrounds.

As was done in Sec.~\ref{s:pmtpdf}, a prediction of the number of good
(physics) events that failed the instrumental cuts was used to correct
the number of events in each of the background boxes.  We obtained
this prediction using the cut acceptances given in Sec.~\ref{s:inssac}
and an estimate of the numbers of signal and radioactive background
events in the data set.  The analysis was performed at two energy
thresholds in order to study the energy dependence of the
contamination.  Results are given in Table~\ref{t:instcon}.

\begin{table}[!ht]
\begin{center}
\begin{tabular}{lcc}
\hline \hline & \multicolumn{2}{c}{Threshold} \\ Phase & 3.5$\,$MeV &
4.0$\,$MeV \\ \hline I & 2.64 $\pm$ 0.22 & 0.09 $\pm$ 0.42 \\ II &
4.48 $\pm$ 0.27 & 0.52 $\pm$ 0.23 \\ \hline \hline
\end{tabular}
\caption[Instrumental contamination.]{Estimated number of instrumental
contamination events in the full data set at different analysis
thresholds.}
\label{t:instcon}
\end{center}
\end{table}

Since these events were not modeled in the simulation, it is difficult
to directly predict their effect on the signal extraction fit results.
However, because virtually all of them fall into the lowest energy
bin, they are unlikely to appear like neutron events.  Since the
$T_{\rm eff}$ distributions of CC and ES signals were unconstrained in
the signal extraction fit, they could mimic these event types.
Therefore, a conservative approach was taken, in which the estimated
contamination from the 3.5$\,$MeV analysis was applied as an
additional uncertainty in the lowest energy bin for both the CC and ES
signals.

\subsection{Atmospheric Backgrounds}\label{s:atmbkg}

The NUANCE neutrino Monte Carlo simulation package~\cite{nuance} was
used to determine the contribution of atmospheric neutrino events to
the data set.  The estimated number of atmospheric neutrino events was
not large enough to merit introducing an additional event type into
the already complex signal extraction procedure.  Instead, 15
artificial data sets were created that closely represented the best
estimate for the real data set, including all neutrino signals and
radioactive backgrounds in their expected proportions.  The NUANCE
simulation was used to predict the distribution of atmospheric
neutrino events in each of the four observable parameters used to
distinguish events in the signal extraction fit (see
Sec.~\ref{sec:sigex}), and a number of such events were included in
each artificial data set, drawn from the estimate for the number in
the true data.  Signal extraction was performed on these sets to
determine which signal the events would mimic in the extraction.  This
resulted in a small correction to the NC flux of $4.66\pm0.76$ and
$17.27\pm2.83$ events to be subtracted in Phases I and II,
respectively, and small additional uncertainties for the CC and ES
rates, mostly at the sub-percent level.

Atmospheric events were often characterized by a high-energy primary
followed by several neutrons.  Therefore, there was significant
overlap with events identified by the `coincidence cut', which removed
events that occurred within a fixed time period of each other.  This
overlap was exploited to verify the predicted number of atmospheric
events.  Without application of the coincidence cut, a total of
$28.2\pm5.4$ and $83.9\pm15.9$ atmospheric neutrino events were
predicted in Phases I and II, respectively.  The coincidence cut
reduced these numbers to $21.3\pm4.0$ and $29.8\pm5.7$ events, which
were the numbers used in the creation of the initial artificial data
sets.  A second group of sets was created, using the pre-coincidence
cut estimates for the number of events, to determine the change in the
NC flux due to the additional events.  The signal extraction was then
performed on a subset of the real data, both with and without the
application of the coincidence cut, and the observed difference in the
NC flux was entirely consistent with the predictions, thus verifying
the method used to derive the NC flux correction.

\subsection{Isotropic Acrylic Vessel Background (IAVB)}

Early in the SNO analyses, a type of instrumental background was
discovered that reconstructed near the AV and was characterized by
very isotropic events ($\beta_{14}<0.15$).  At higher energies
($N_{\rm hit}>60$), these events form a distinct peak in a histogram
of \bo, and they are easily removed from the data by a combination of
the fiducial volume and isotropy cuts.  However, at lower energies,
position reconstruction errors increase and the isotropy distributions
of the IAVB and other events broaden and join, so that removal of the
IAVB events by these cuts is no longer assured.

Accurate simulation of these events is difficult because the physical
mechanism that produces the IAVB events has not been identified and
crucial IAVB event characteristics cannot be predicted.  These include
the light spectrum, photon timing distribution, location, and
effective event energy.  To circumvent this problem, simulated events
were generated that covered a wide range of possibilities.  Three
event locations were modeled: on the exterior and interior AV
surfaces, and uniformly distributed within the AV acrylic.  Events
were generated at three different photon wavelengths that cover the
range of SNO detector sensitivity: 335, 400, and 500~nm.  The photons
were generated isotropically, with the number of photons in an event
chosen from a uniform distribution with a maximum above the energy
range used in the neutrino analysis.  The photon time distribution was
a negative exponential, with the time constant for an event chosen
from a truncated Gaussian with mean and standard deviation of 5~ns.

Using PDFs derived from the simulated event samples, maximum
likelihood signal extraction code was used to estimate the number of
IAVB events in the data in the vicinity of the AV, between 570 and
630~cm from the detector center, in accompaniment with the CC, ES, and
NC neutrino event types and \tl and \bi backgrounds in the \dto, AV,
\hto, and PMTs.  This was done separately for each of the nine
simulated photon wavelength/event location combinations.  Because the
energy distribution of the IAVB events was unknown, the IAVB
extractions were done as a function of \nhits in 11 bins.  The ratio
of the number of IAVB events that passed all the neutrino cuts to
those that fit near the AV in each \nhits bin was calculated for each
simulated IAVB case as a function of event energy.  These ratios were
used, together with the estimated numbers of such events near the AV,
to estimate the IAVB contamination in the neutrino sample as a
function of energy for each of the simulated IAVB cases.

The polar-angle distributions of hit PMTs in the simulated IAVB events
were studied in a coordinate system centered on the middle of the AV,
with its $z$-axis along the radial vector through the fitted event
location.  There are marked differences in these distributions among
the different simulated cases due to optical effects of the AV.
Comparisons of these distributions were made between simulated events
and high \nhitsnosp, high isotropy events in the data that reconstruct
near the AV (presumed to be IAVB events).  A fit was made to find the
weighted combination of the simulated cases that best fit the high
\nhits data.  The resulting weights were assumed to be valid at all
energies, and were used together with the contamination ratios
discussed above: first, to estimate the total IAVB background expected
in the neutrino analysis data set as a function of energy (totaling 27
and 32 events above 3.5~MeV in Phases I and II, respectively) and,
second, to generate a set of simulated IAVB events representative of
those expected to contaminate the neutrino data.

A test similar to that described in Sec.~\ref{s:atmbkg} was performed.
Fifteen artificial data sets were created that also contained
simulated IAVB events based on estimates of the weighted contributions
of the simulated cases and their energy distributions.  It was found
that the majority of the IAVB events fit out as other background event
types, so that the result of adding the simulated IAVB background was
only small additional uncertainties for each of the neutrino flux
parameters, with no required corrections.  The increase in uncertainty
for the NC flux was evaluated at $0.26\%$.  The increases of the CC
uncertainties were also mostly at the sub-percent level, and the
increase in uncertainties on the ES rates were so small as to be
negligible ($< 0.01\%$).

\subsection{Additional Neutron Backgrounds}

A full study of other possible sources of neutron background events,
such as from events such as ($\alpha$,$n$) reactions and terrestrial
and reactor antineutrino interactions, was presented in previous
publications~\cite{longd2o, nsp}.  The full set of simulated NC events
was used to adjust these numbers for the lowered energy threshold and
for the live times and detection efficiencies in the two phases to
give a final correction to the NC flux of $3.2\pm0.8$ and $12.0\pm3.1$
neutron capture events in Phases~I and II, respectively.

\section{Signal Extraction Methods \label{sec:sigex}}

An extended maximum likelihood method was used to separate event types
based on four observable parameters: the effective electron kinetic
energy, $T_{\rm eff}$; the angle of the event direction with respect
to the vector from the Sun, \ctsnosp; the normalized cube of the
radial position in the detector, $R^3$; and the isotropy of the PMT
hits, \betnosp.  Two independent techniques were used, as described in
sections~\ref{s:mxf} and~\ref{s:kernel}. One method used binned PDFs
and the other an unbinned, ``kernel estimation'' approach.

We performed two distinct types of fit.  The first extracted the
detected electron energy spectra for CC and ES events in individual
$T_{\rm eff}$ bins, without any model constraints on the shape of the
underlying neutrino spectrum. We refer to this as an `unconstrained'
fit.  The second fit exploited the unique capabilities of the SNO
detector to directly extract the energy-dependent $\nu_e$ survival
probability (Sec.~\ref{s:kerpoly}).  The survival probability was
parameterized as a polynomial function and applied as a distortion to
the $^8$B neutrino energy spectrum (taken from~\cite{winter}).  The
shapes of the CC and ES $T_{\rm eff}$ spectra were recomputed from the
distorted $^8$B spectrum as the fit progressed, allowing the
polynomial parameters to vary in the fit.  The overall fluxes were
also constrained in this fit through the requirement of unitarity.
The features in common for the two signal extraction approaches are
described below.

The types of events included in the fit were the three neutrino
interaction types (CC, ES and NC) and 17 background event types across
the two phases of data, as defined in Table~\ref{t:bkgs}.  The
likelihood was maximized with respect to the number of events of each
signal type, and several systematic parameters affecting the shapes of
the PDFs, as described in Sections~\ref{s:mxf} and~\ref{s:kernel}.

To extract energy spectra for the CC and ES neutrino signals in the
unconstrained fits, CC and ES PDFs were created in discrete $T_{\rm
eff}$ intervals and the fitted numbers of events in these intervals
were allowed to vary independently.  The energy spectra for events
from the NC interaction and from radioactive backgrounds have no
dependence on the neutrino oscillation model, and so the shapes of
these spectra were fixed within their systematic uncertainties.

The flux of solar neutrinos was assumed to be constant, so a single
set of neutrino-related fit parameters was applied to both phases.
Therefore, the neutrino signal parameters varied in the fit were an NC
rate and a number of CC and ES rates in discrete energy intervals, as
defined in Sections~\ref{s:mxf} and~\ref{s:kernel}.  Although SNO was
primarily sensitive to the $^8$B chain of solar neutrinos, we included
a fixed contribution of solar hep neutrinos, which was not varied in
the fit.  Based on results from a previous SNO analysis~\cite{nsp}, we
used 0.35, 0.47, and 1.0 times the Solar Standard Model (SSM) prediction 
for CC, ES, and NC hep
neutrinos, respectively. Taken together, these correspond to 16.4
events in Phase I and 33.3 events in Phase II.

To take into account correlations between parameters,
multi-dimensional PDFs were used for all signals.  In the
unconstrained fits, CC and ES were already divided into discrete
energy bins, and three-dimensional PDFs were created in each bin for
the other observables: $P(\beta_{14}, R^3, \cos \theta_{\odot})$.  In
the survival probability fits, fully four-dimensional PDFs were used
for CC and ES events.  For the NC and background PDFs the \cts
distribution is expected to be flat, since there should be no
dependence of event direction on the Sun's position, but correlations
exist between the other observables. For these event types, the PDFs
were factorized as $P(T_{\rm eff}, \beta_{14}, R^3)\times P(\cos
\theta_{\odot})$.

Uncertainties in the distributions of the observables were treated as
parameterized distortions of the Monte Carlo PDF shapes.  The dominant
systematic uncertainties were allowed to vary in the fit in both
signal extraction methods.  Less significant systematics were treated
as in previous SNO analyses~\cite{longd2o}, using a `shift-and-refit'
approach: the data were refit twice for each systematic uncertainty,
with the model PDFs perturbed by the estimated positive and negative
1~$\sigma$ values for the uncertainty in a given parameter.  The
differences between the nominal flux values and those obtained with
the shifted PDFs were taken to represent the 68\% C.L. uncertainties,
and the individual systematic uncertainties were then combined in
quadrature to obtain total uncertainties for the fluxes.

\subsection{Systematic Uncertainties: Phase Correlations}
\label{s:correl}

Uncertainties related to theoretical quantities that are unaffected by
detector conditions (such as the photodisintegration cross section
uncertainty) were applied to both phases equally. Uncertainties in
quantities dependent on detector conditions (such as energy
resolution) were treated independently in each phase. Uncertainties in
quantities that partly depend on the operational phase (such as
neutron capture efficiency, which depends both on a common knowledge
of the $^{252}$Cf source strength and on the current detector
conditions) were treated as partially correlated. For the latter, the
overall uncertainty associated with each phase thus involved a common
contribution in addition to a phase-specific uncertainty.  Since
neutron capture events were more similar to electron-like events in
Phase~I than in Phase~II, several of the neutron-related uncertainties
applied to Phase~II only.  The correlations are summarized in
Table~\ref{tab:systcorr}.

\begin{table}[!h]
\begin{center}
\begin{tabular}{lcc}
\hline \hline Systematic uncertainty & Correlation \\ \hline Energy
scale & Both \\ Electron energy resolution & Uncorrelated \\ Neutron
energy resolution & Phase~II only \\ Energy linearity & Correlated \\
\bof electron scale & Correlated \\ \bof neutron scale & Phase~II only
\\ \bof electron width & Correlated \\ \bof neutron width & Phase~II
only \\ \bof energy dependence & Correlated \\ Axial scaling &
Uncorrelated \\ $z$ scaling & Uncorrelated \\ $x$, $y$, $z$ offsets &
Uncorrelated \\ $x$, $y$, $z$ resolutions & Uncorrelated \\ Energy
dependent fiducial volume & Uncorrelated \\ \cts resolution &
Uncorrelated \\ PMT $T_{\rm eff}$ exponent & Uncorrelated \\ PMT $R^3$
exponent & Uncorrelated \\ PMT $R^3$ offset & Uncorrelated \\ PMT \bof
intercept & Uncorrelated \\ PMT \bof radial slope & Uncorrelated \\
PMT \bof width & Uncorrelated \\ Neutron capture & Both \\
Photodisintegration & Correlated \\ $^{24}$Na distribution & Phase~II
only \\ Sacrifice & Uncorrelated \\ IAVB & Uncorrelated \\
Atmospherics backgrounds & Uncorrelated \\ Instrumental contamination
& Uncorrelated \\ Other neutrons & Uncorrelated \\ \hline \hline
\end{tabular}
\caption{Phase correlations of the systematic
uncertainties. ``Correlated'' refers to a correlation coefficient of
1.0 between the phases and ``uncorrelated'' refers to a coefficient of
0.0.  ``Both'' means an uncertainty was treated as partially
correlated between the phases.}
\label{tab:systcorr}
\end{center}
\end{table}

\subsection{Binned-Histogram Unconstrained Fit}
\label{s:mxf}

In this approach, the PDFs were created as three-dimensional
histograms binned in each observable dimension, as summarized in
Table~\ref{t:mxfbins}.  For CC and ES, three-dimensional PDFs were
created in each $T_{\rm eff}$ interval, to fully account for
correlations between all four observable dimensions.  Fifty rate
parameters were fitted: the CC and ES rates in each of 16 spectral
bins, the NC normalization and 17 background PDF normalizations.
Dominant systematic uncertainties were allowed to vary within their
uncertainties, or `floated', by performing one-dimensional scans of
the likelihood in the value of each systematic parameter.  This
involved performing the fit multiple times at defined intervals in
each systematic parameter and extracting the value of the likelihood,
which included a Gaussian factor whose width was defined by the
independently estimated uncertainty on that parameter, as described in
Sec.~\ref{s:penaltyterms}.  This combined \textit{a priori} knowledge
from the calibration data and Monte Carlo studies used to parameterize
systematic uncertainties with information inherent in the data itself.
If a new likelihood maximum was found at an offset from the existing
best estimate of a particular systematic parameter, then the offset
point was defined as the new best estimate.  An iterative procedure
was used to take into account possible correlations between
parameters.  The final uncertainties on each parameter were defined by
where the log likelihood was 0.5 less than at the best-fit point, and
the differences in each fitted flux parameter between these points and
the best-fit point were taken as the associated systematic
uncertainties for that parameter.  For more details of this approach,
see~\cite{orebithesis}.

\begin{table}[!ht]
\begin{center}
\begin{tabular}{lccccc} 
\hline \hline Observable && Min & Max & Bins & Bin width\\ \hline CC,
ES $T_{\rm eff}$ && 3.5~MeV & 11.5~MeV & 16 & 0.5~MeV \\ 
\multirow{2}{*}{Other $T_{\rm eff}$} && 3.5~MeV & 5.0~MeV & 6 & 0.25~MeV\\
 && 5.0~MeV & 11.5~MeV & 13
& 0.5~MeV\\ $\cos \theta_{\odot}$ && $-$1.0 & 1.0 & 8& 0.25\\ $R^3$ &&
0.0 & 0.77025 & 5& 0.15405\\ $\beta_{14}$ && $-$0.12 & 0.95& 15 &
0.0713\\ \hline \hline
\end{tabular}
\caption{\label{t:mxfbins}PDF configurations used for the
binned-histogram signal extraction approach.}
\end{center}
\end{table}

The parameters floated using this approach, along with their relevant
correlations, as described in Sec.~\ref{s:correl}, were:
\begin{itemize}
 \item Energy scale (both correlated and uncorrelated in each phase)
 \item Energy resolution (uncorrelated in each phase)
 \item \bet scale for electron-like events (correlated between phases)
 \item PMT $\beta$-$\gamma$ $R^3$ exponent (uncorrelated in each
phase, see Sec.~\ref{s:pmtpdf})
 \item PMT $\beta$-$\gamma$ $R^3$ offset (uncorrelated in each phase,
 see Sec.~\ref{s:pmtpdf})
 \item PMT $\beta$-$\gamma$ $T_{\rm eff}$ exponent (uncorrelated in
each phase, see Sec.~\ref{s:pmtpdf})
\end{itemize}
The remaining systematic uncertainties were applied using the
`shift-and-refit' approach.

\subsection{Unbinned Unconstrained Fit Using Kernel Estimation}
\label{s:kernel}

In this approach, the PDFs were created by kernel estimation.  Like
standard histogramming techniques, kernel estimation starts with a
sample of event values, $t_i$, drawn from an unknown distribution,
$P(x)$.  Based on this finite sample, the parent distribution is
approximated by $\hat{P}(x)$, which is a sum of kernel functions,
$K_i(x)$, each centered at an event value from the sample:
\begin{equation}
	\label{eq:kernel_basic}
	\hat{P}(x) = \frac{1}{n}\sum_{i=1}^n K_i(x - t_i).
\end{equation}
The most common choice of form of kernel functions is the normalized
Gaussian distribution,
\begin{equation}
	K(x/h) = \frac{1}{h\sqrt{2\pi}}e^{-(x/h)^2/2},
\end{equation}
where $h$ is called the \emph{bandwidth} of the kernel.  One can pick
a different bandwidth, $h_i$, for the kernel centered over each event.

Kernel-estimated density functions have many useful properties.  If
the kernel functions are continuous, then the density function will
also be continuous.  In one dimension, kernel estimation can also be
shown to converge to the true distribution slightly more quickly than
a histogram with bin size the same as the kernel bandwidth.
Generalizing the kernel estimation method to multiple dimensions is
done by selecting a kernel with the same dimensionality as the PDF.
We used a multi-dimensional Gaussian kernel that was simply the
product of one-dimensional Gaussians.  We followed the prescription
given in~\cite{cranmer} for the selection of bandwidths for each event
in each dimension.

By varying the values associated with the events in the PDF sample
individually, kernel estimation can very naturally be extended to
incorporate systematic variation of PDF shapes.  For example, energy
scale is incorporated by a transformation of the simulated event
values, $t_i\rightarrow (1 + \alpha) \times t_i$, where $\alpha$ is a
continuously variable parameter.  Such transformations preserve the
continuity and analyticity of the PDF.  We can then add these
systematic distortion parameters to the likelihood function, and also
optimize with respect to them using a gradient descent method.  This
allows correlations between systematics and neutrino signal
parameters, as well as between systematics themselves, to be naturally
handled by the optimization algorithm.  In addition, the information
in the neutrino data set itself helps to improve knowledge of detector
systematics.

Three kinds of systematic distortions can be represented within this
formalism.  Transformations like energy scale and position offset have
already been mentioned.  A Gaussian resolution systematic can be
floated by transforming the bandwidth, $h$, through analytic
convolution.  Finally, re-weighting systematics, such as the neutron
capture efficiency, are represented by varying the weight of events in
the sum.

The main challenge in using kernel estimation with large data sets is
the computational overhead associated with repeatedly re-evaluating
the PDFs as the parameters associated with detector response vary.  We
made several algorithmic improvements to make kernel estimation more
efficient and did much of the calculation on off-the-shelf 3D graphics
processors.  For more detail on the implementation of the fit on the
graphics processors, see~\cite{seibertthesis}.

The kernel-estimated PDFs had the same dimensionality over the same
ranges of the observables as the binned fit, except with an upper
energy limit of 20~MeV instead of 11.5~MeV.  CC rates were extracted
in 0.5~MeV intervals up to 12~MeV, with a large 12--20~MeV interval at
the end of the spectrum.  To reduce the number of free parameters in
the fit, ES rates were extracted in a 3.5--4.0~MeV interval, in 1~MeV
intervals from 4~MeV to 12~MeV, and in a final 12--20~MeV interval.
The CC and ES PDFs were fixed to be flat in the $T_{\rm eff}$
dimension within each $T_{\rm eff}$ interval.  During fitting, the
following parameters, corresponding to the dominant systematic
uncertainties, were allowed to vary continuously:
\begin{itemize}
	\item Energy scale (both correlated and uncorrelated in each
	phase)
	\item Energy resolution (uncorrelated in each phase)
	\item \bet electron and neutron scales
 	\item PMT $\beta$-$\gamma$ $R^3$ exponent (uncorrelated in
 	each phase)
	\item PMT $\beta$-$\gamma$ $R^3$ offset (uncorrelated in each
	phase)
	\item PMT $\beta$-$\gamma$ $T_{\rm eff}$ exponent
	(uncorrelated in each phase)
\end{itemize}
Altogether there were 18 CC parameters, 10 ES parameters, 1 NC
parameter, 17 background normalization parameters, and 16 detector
systematic parameters.  The remaining systematic uncertainties were
applied using the `shift-and-refit' approach.

\subsection{Energy-Dependent $\nu_e$ Survival Probability Fit Using Kernel
Estimation}
\label{s:kerpoly}

	The unique combination of CC, ES, and NC reactions detected by
SNO allowed us to fit directly for the energy-dependent $\nu_e$
survival probability without any reference to flux models or other
experiments.  Such a fit has several advantages over fitting for the
neutrino mixing parameters using the NC rate and the `unconstrained'
CC and ES spectra described in the previous sections.

	The unconstrained fits described in Secs.~\ref{s:mxf}
and~\ref{s:kernel} produce neutrino signal rates for CC and ES in
intervals of reconstructed energy, $T_{\rm eff}$, with the free
parameters in the fit directly related to event counts in each $T_{\rm
eff}$ interval.  Although this simplifies implementation of the signal
extraction fit, physically-relevant quantities, such as total $^8$B
neutrino flux and neutrino energy spectra, are entangled with the
energy response of the SNO detector.  Comparing the unconstrained fit
to a particular model therefore requires convolving a distorted $^8$B
neutrino spectrum with the differential cross sections for the CC and
ES interactions, and then further convolving the resulting electron
energy spectra with the energy response of the SNO detector to obtain
predictions for the $T_{\rm eff}$ spectra.

Moreover, the unconstrained fits of Secs.~\ref{s:mxf}
and~\ref{s:kernel} have more degrees of freedom than are necessary to
describe the class of MSW distortions that are observable in the SNO
detector.  For example, the RMS width of $T_{\rm eff}$ for a 10~MeV
neutrino interacting via the CC process is nearly 1.5~MeV.  Therefore,
adjacent $T_{\rm eff}$ bins in the unconstrained fit are correlated,
but this information is not available to the minimization routine to
constrain the space of possible spectra.  By fitting for an
energy-dependent survival probability, we enforce continuity of the
energy spectrum and thereby reduce covariances with backgrounds, most
notably $^{214}$Bi events.  Events from the CC reaction can no longer
easily mimic the steep exponential shape of the background energy
distribution.  In addition, systematic uncertainties that are
correlated between the CC and NC events will naturally cancel in this
approach within the fit itself.

We therefore performed a signal extraction fit in which the free
parameters directly described the total $^8$B neutrino flux and the
energy-dependent $\nu_e$ survival probabilities.  We made the
following assumptions:
\begin{itemize}
	\item The observed CC and ES $T_{\rm eff}$ spectra come from a
	fixed distribution of neutrino energies, $E_{\nu}$, with the
	standard differential cross sections;
	\item The $\nu_e$ survival probability can be described by
	a smooth, slowly varying function of $E_\nu$ over the range of 
	neutrino energies to which the SNO detector is sensitive;
	\item The CC, ES and NC rates are directly related
	through unitarity of the neutrino mixing matrix;
	\item $\nu_e$ regeneration in the Earth at night 
        can be modeled as a linear perturbation
	to the daytime $\nu_e$ survival probability.
\end{itemize}

\begin{figure}[!ht]
\begin{center}
\includegraphics[width=0.48\textwidth]{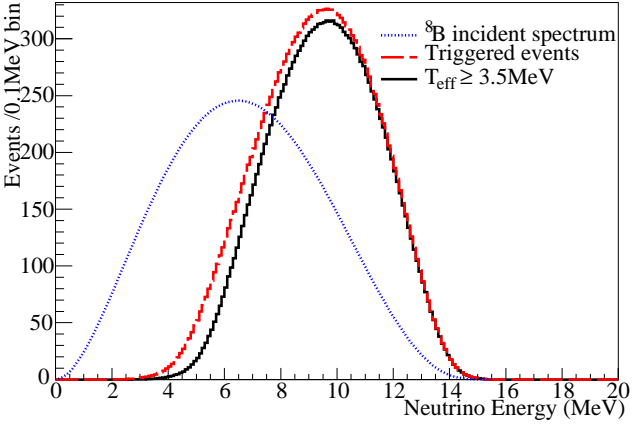}
\caption{\label{f:b8spec}(Color online) Simulation of the undistorted
energy spectrum of $^8$B neutrinos that trigger the detector, before
the application of the $T_{\rm eff}$ threshold, and after a $T_{\rm
eff} > 3.5$~MeV cut is applied, normalized to the SSM prediction.  The
sharp cut in $T_{\rm eff}$ results in a smooth roll-off in detection
efficiency for energies less than the peak energy.  Also shown is the
spectrum of incident neutrinos predicted by~\cite{winter}, arbitrarily
normalized, to illustrate the effect of detector sensitivity.}
\end{center} 
\end{figure}

Given these assumptions, we performed a fit in which the neutrino
signal was described by six parameters:
\begin{itemize}
	\item $\Phi_{^8{\rm B}}$ - the total $^8$B neutrino flux;
	\item $c_0$, $c_1$, $c_2$ - coefficients in a quadratic expansion 
	of the daytime $\nu_e$ survival probability around $E_\nu = 10$~MeV;
	\item $a_0$, $a_1$ - coefficients in a linear expansion of the 
	day/night asymmetry around $E_\nu = 10$~MeV.
\end{itemize}
The day/night asymmetry, $A$, daytime $\nu_e$ survival probability,
$P_{ee}^{\rm day}$, and nighttime $\nu_e$ survival probability,
$P_{ee}^{\rm night}$, that correspond to these parameters are:
\begin{eqnarray}
	A(E_\nu) & = & a_0 + a_1(E_\nu - 10\;{\rm MeV}) \label{eq:dn}
	\\ P_{ee}^{\rm day}(E_\nu) & = & c_0 + c_1 (E_\nu - 10\;{\rm
	MeV}) \label{eq:poly} \nonumber \\ & & \; + c_2 (E_\nu -
	10\;{\rm MeV})^2 \\ P_{ee}^{\rm night}(E_\nu) & = &
	P_{ee}^{\rm day} \times \frac{1 + A(E_\nu)/2}{1 - A(E_\nu)/2}
\end{eqnarray}
The survival probabilities were parameterized in this way to reduce
correlations between $c_0$ and the higher order terms by expanding all
functions around the detected $^8$B spectrum peak near 10~MeV.  The
simulated neutrino energy spectrum after application of the analysis
cuts, shown in Figure~\ref{f:b8spec}, rapidly drops in intensity away
from 10~MeV.  The broad $T_{\rm eff}$ resolution of the detector in
combination with the limited range of detectable neutrino energies
limits our sensitivity to sharp distortions.  For this reason, we
chose to fit for a smooth, polynomial expansion of the survival
probability.  By using a generic form, we allow arbitrary models of
neutrino propagation and interaction to be tested, including standard
MSW effects, as long as they meet the assumptions described above.
Monte Carlo studies demonstrated that this analytical form was
sufficient to model the class of MSW distortions to which the SNO
detector was sensitive.  We propagated the uncertainty in the shape of
the undistorted $^8$B energy spectrum as an additional
`shift-and-refit' systematic uncertainty to ensure the extracted
survival probability incorporated this model dependence.

To implement this fit, we performed a `four phase' signal extraction,
dividing the data and the PDFs into Phase~I-day, Phase~I-night,
Phase~II-day, and Phase~II-night groups.  Background decay rates from
solid media, such as the acrylic vessel and the PMTs, were constrained
to be identical day and night.  Decay rates in the \dto and \hto
regions were free to vary between day and night to allow for day/night
variations in the water circulation and filtration schedules.  We
floated the same detector systematics as in the unconstrained fit
described in Sec.~\ref{s:kernel}.  The fit has 6 neutrino parameters,
26 background normalization parameters, and 16 detector systematic
parameters, for a total of 48 free parameters.

We constructed the PDFs in the same way as described in
Sec.~\ref{s:kernel}, with the exception of the CC and ES signals.
Instead of creating a 3D PDF ($\beta_{14}$, $R^3$, $\cos
\theta_\odot$) for intervals in $T_{\rm eff}$ in the undistorted
spectrum, we created 4D PDFs ($T_{\rm eff}$, $\beta_{14}$, $R^3$,
$\cos \theta_\odot$) for separate $E_\nu$ intervals in the undistorted
spectrum.  There were 9 CC and 9 ES PDFs in each of the 4 day/night
phases, with $E_\nu$ boundaries at $4, 6, 7, 8, 9, 10, 11, 12, 13,$
and $15$~MeV.

During optimization, the signal rates associated with the 76 CC, ES
and NC PDFs were not allowed to vary freely, but were determined by
the 6 neutrino parameters.  We defined an `ES survival probability':
\begin{eqnarray}
	P_{\rm ES}^{\rm day}(E_\nu) & = & P_{ee}^{\rm day} + \epsilon
	(1 - P_{ee}^{\rm day}(E_\nu)) \\ P_{\rm ES}^{\rm night}(E_\nu)
	& = & P_{ee}^{\rm night} + \epsilon (1 - P_{ee}^{\rm
	night}(E_\nu))
\end{eqnarray}
where $\epsilon = 0.156$ is the approximate ratio between the
$\nu_{\mu,\tau}$ and $\nu_e$ ES cross sections.  The ES cross-section
ratio is not constant as a function of neutrino energy, so we took the
variation with energy as an additional systematic uncertainty.  The
signal rates were defined in terms of $\Phi_{^8{\rm B}}$, $P_{ee}$ and
$P_{\rm ES}$ to be:
\begin{eqnarray}
R_{\mathrm{NC}} & = & \Phi_{^8{\rm B}} \\ R_{\mathrm{CC}, i}^{\rm day}
& = & \frac{\Phi_{^8{\rm B}}}{E_{i} -
E_{i-1}}\int_{E_{i-1}}^{E_{i}}dE_\nu \; P_{ee}^{\rm day}(E_\nu) \\
R_{\mathrm{CC}, i}^{\rm night} & = & \frac{\Phi_{^8{\rm B}}}{E_{i} -
E_{i-1}}\int_{E_{i-1}}^{E_{i}}dE_\nu \; P_{ee}^{\rm night}(E_\nu) \\
R_{\mathrm{ES}, i}^{\rm day} & = & \frac{\Phi_{^8{\rm B}}}{E_{i} -
E_{i-1}}\int_{E_{i-1}}^{E_{i}}\; dE_\nu \; P_{\rm ES}^{\rm
day}(E_\nu)\\ R_{\mathrm{ES}, i}^{\rm night} & = & \frac{\Phi_{^8{\rm
B}}}{E_{i} - E_{i-1}}\int_{E_{i-1}}^{E_{i}} dE_\nu \; P_{\rm ES}^{\rm
night}(E_\nu)
\end{eqnarray}
where $E_0$ is 4~MeV and $E_i$ is the upper energy boundary of the
$i$-th $E_\nu$ interval.

The survival probability fit included the same `shift-and-refit'
systematics as the unconstrained fit, along with all of the day/night
systematics used in previous analyses~\cite{longd2o,nsp}.  These
systematics accounted for diurnal variations in reconstructed
quantities, such as energy scale and vertex resolution, as well as
long-term variation in detector response which could alias into a
day/night asymmetry.  In addition, the non-uniformity of the $\cos
\theta_\odot$ distributions of CC and ES events can also alias into a
day/night asymmetry, so we incorporated additional day/night
systematic uncertainties on all observables in the CC and ES PDFs.

\subsection{Application of Constraints}
\label{s:penaltyterms}

\textit{A priori} information from calibrations and background
measurements was included in the fits to constrain some of the fit
parameters, in particular several of the radioactive backgrounds
(discussed in Sec.~\ref{s:bkgconst}) and any systematic parameters
floated in the fit.

The extended likelihood function had the form: 
\begin{equation}
\mathcal{L}(\vec{\alpha},\vec{\beta}) =
\mathcal{L}_{data}(\vec{\alpha} | \vec{\beta})
\mathcal{L}_{calib}(\vec{\beta})
\end{equation}
where $\vec{\alpha}$ represents the set of signal parameters being fit
for, $\vec{\beta}$ represents the nuisance parameters for the
systematic uncertainties that were floated in the fits,
$\mathcal{L}_{data}(\vec{\alpha} | \vec{\beta})$ is the extended
likelihood function for the neutrino data given the values of those
parameters, and $\mathcal{L}_{calib}(\vec{\beta})$ is a constraint
term representing prior information on the systematic parameters,
obtained from calibration data and {\it ex-situ} measurements.  The
contribution to $\mathcal{L}_{calib}(\vec{\beta})$ for each systematic
parameter had the form:
\begin{equation}
\mathcal{L}_{calib}({\beta_i}) =
e^{\frac{-(\beta_i-\mu_i)^2}{2\sigma_i^2} }
\end{equation}
where $x_i$ is the value of parameter $i$, and $\mu_i$ and $\sigma_i$
are the estimated value and uncertainty determined from external
measurements (with asymmetric upper and lower values for $\sigma_i$
where required).  This results in a reduction of the likelihood as the
parameter value moves away from the \textit{a priori} estimate.

\subsection{Bias Testing}

To verify that the signal extraction methods were unbiased, we used
half the Monte Carlo events to create `fake data' sets, and the
remaining events to create PDFs used in fits to the fake data sets.  A
fit was performed for each set and the results were averaged to
evaluate bias and pull in the fit results.

We created 100 sets containing only neutrino events, 45 sets also
containing internal background events, and 15 sets containing the full
complement of neutrino events and internal and external backgrounds.
The numbers of fake data sets were limited by the available computing
resources.

The two signal extraction methods gave results that were in excellent
agreement for every set.  The biases for the neutrino fluxes were
consistent with zero, and the Gaussian pull distributions were
consistent with a mean of zero and standard deviation of 1.

Additional tests were performed in which one or more systematic shifts
were applied to the event observables in the fake data sets, and the
corresponding systematic parameters were floated in the fit, using
\textit{a priori} inputs as in the final signal extraction fits, to
verify that the two independent methods for propagating systematic
uncertainties were also unbiased.  In all cases, the true values for
the neutrino fluxes were recovered with biases consistent with zero.

\subsection{Corrections to PDFs}
\label{sec:correc}
A number of corrections were required to account for residual
differences between data and PDFs derived by simulation.  An offset of
the laserball position along the $z$-axis during calibration of PMT
timing introduced an offset to reconstructed positions along this axis
in the data.  A correction was therefore applied to all data events,
as described in Sec.~\ref{sec:hitcal}.  In addition, a number of
corrections were applied to the reconstructed energy and isotropy of
events (see Sections~\ref{sec:ecorr} and~\ref{sec:beta14},
respectively).

The Monte Carlo simulation was used to link the neutrino rates between
the two phases, thus taking into account variations in detector
efficiency and livetime.  Several corrections were applied to the
Monte Carlo flux predictions, as described below.

The predicted number of events for signal type $i$ per unit of
incident flux, including all correction factors, is:

\begin{eqnarray}
\label{e:se:corr}
N_i & = & N^{\rm MC}_{i} \, \delta^{\rm sim} \, \delta^{\rm acc}_{i}\,
N^{\rm iso}_{i}\, N^{\rm D}_{i} \, N^e_{i}\, R_i \, \tau,
\end{eqnarray}

\noindent where:

\begin{itemize}
\item $N^{\rm MC}_i$ is the number of events predicted by the Monte
Carlo simulation for signal $i$ per unit incident flux.  This is
recalculated as needed to account for any systematic shifts applied to
the PDFs.
\item $\delta^{\rm sim}$ corrects for events aborted in the simulation
due to photon tracking errors.  This correction increases with the
number of photons in an event.
\item $\delta^{\rm acc}_{i}$ corrects for differences in the acceptances 
of the instrumental and high level cuts for data and Monte Carlo 
events (Sec.~\ref{s:cutacc}).
\item $N^{\rm iso}_{i}$ is a correction to account for CC interactions
on chlorine and sodium nuclei in the \dto volume that are not
modeled in the simulation.  This correction is relevant only to the CC
signal in Phase~II.
\item $N^{\rm D}_{i}$ is a correction to the number of target
deuterons and hence is relevant to CC and NC only.
\item $N^e_{i}$ is a correction to the number of target electrons and
hence is relevant to ES only.
\item $R_{i}$ accounts for radiative corrections to the
neutrino-deuteron interaction cross section for NC.  Radiative
corrections relevant to the CC and ES interactions were included in
the simulation.
\item $\tau$ corrects for deadtime introduced into the data set by the
instrumental cuts.
\end{itemize}

These corrections are summarized in Table~\ref{t:se:fluxc}.

\begin{table}[!ht]
\begin{center}
\begin{tabular}{lccccc}
\hline Correction & Phase && CC & ES & NC \\ \hline \hline
$\delta^{\rm sim}$ & I, II&& \multicolumn{3}{c}{(1.0 -
0.0006238$\times T_{\rm eff}$)$^{-1}$} \\ $\delta^{\rm acc}_{i}$ &I &&
0.9924 & 0.9924 & 0.9924 \\ $\delta^{\rm acc}_{i}$ &II && 0.9930 &
0.9930 & 0.9954 \\ $N^{\rm iso}_{i}$ &II && 1.0002 & --- & --- \\
$N^{\rm D}_{i}$ & I, II&&1.0129 & --- & 1.0129 \\ $N^e_{i}$ &I, II&&
--- & 1.0131 & --- \\ $R_{i}$ &I, II&& --- & --- & 0.977 \\ $\tau$ &I
&& 0.979 & 0.979 & 0.979 \\ $\tau$ &II && 0.982 & 0.982 & 0.982 \\
\hline
\end{tabular}
\caption[Flux corrections.]{Corrections applied to the expected number
of CC, ES and NC events used in the signal extraction fits.}
\label{t:se:fluxc}
\end{center}
\end{table}

\section{ Results \label{sec:results}}

The detailed improvements made to this analysis, as described in
previous sections, allow a more precise extraction of the neutrino
flux parameters and, as a result, of the MSW oscillation parameters.
Results from the unconstrained fit are given in Sec.~\ref{res:uncon}
and from the energy-dependent fit to the $\nu_e$ survival probability
in Sec.~\ref{res:sprob}.  This new method for directly extracting the
form of the $\nu_e$ survival probability from the signal extraction
fit produces results that are straightforward to interpret.  A direct
comparison can be made of the shape of the extracted survival
probability to model predictions, such as the LMA-predicted low-energy
rise.

Sec.~\ref{res:mixp} describes the measurements of the neutrino
oscillation parameters.  As has been observed in a number of recent
publications~\cite{t131,t132,t133}, the different dependence of the
$\nu_e$ survival probability on the mixing parameters $\theta_{12}$
and $\theta_{13}$ between solar and reactor neutrino experiments means
that a comparison of solar data to reactor antineutrino data from the
KamLAND experiment allows a limit to be placed on the value of
$\sin^2\theta_{13}$.  The new precision achieved with the LETA
analysis in the measurement of $\tan^2\theta_{12}$ results in a better
handle on the value of $\sin^2\theta_{13}$ in such a three-flavor
oscillation analysis.  Results of this analysis are presented in
Sec.~\ref{res:mixp}, including a constraint on the value of
$\sin^2\theta_{13}$.

\subsection{Unconstrained Fit}
\label{res:uncon}
Our measurement of the total flux of active $^8$B solar neutrinos,
using the NC reaction ($\Phi_{\textrm{NC}}$) is found to be:
\begin{itemize}
\item Binned-histogram method
\end{itemize}
$\Phi_{\textrm{NC}}^{binned} = 5.140 \,^{+0.160}_{-0.158}
\textrm{(stat)} \,^{+0.132}_{-0.117} \textrm{(syst)} \times 10^6\,\rm
cm^{-2}\,s^{-1} $
\begin{itemize}
\item Kernel estimation method
\end{itemize}
$\Phi_{\textrm{NC}}^{kernel} = 5.171 \,^{+0.159}_{-0.158}
\textrm{(stat)} \,^{+0.132}_{-0.114} \textrm{(syst)} \times 10^6\,\rm
cm^{-2}\,s^{-1} $
\newline

This represents $^{+4.0}_{-3.8}$\% total uncertainty on the flux,
which is more than a factor of two smaller than the best of previous
SNO results.  The statistical uncertainty has been reduced by nearly
$\sqrt2$, to 3.1\%.  However, the largest improvement is in the
magnitude of the systematic uncertainty, which has been reduced from
7.3\% and 6.3\% in previous analyses of Phase~II~\cite{nsp} and
Phase~III~\cite{snoncd} data, respectively, to 2.4\% (taking the
average of the upper and lower values).

\begin{figure}[!ht]
\begin{center}
\includegraphics[width=0.48\textwidth]{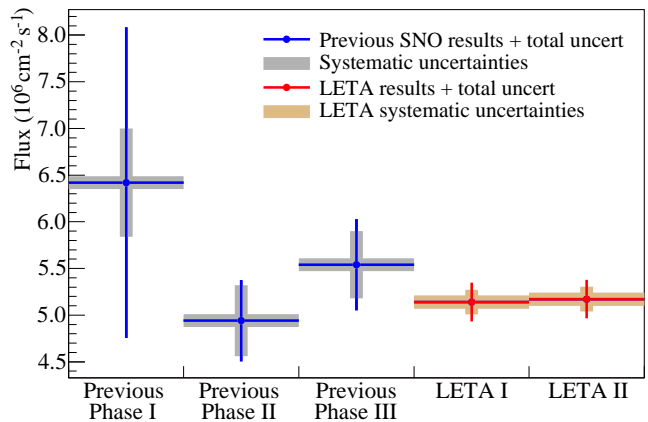}
\caption{\label{f:nccomp}(Color online) Total $^8$B neutrino flux
results using the NC reaction from both unconstrained signal
extraction fits in comparison to unconstrained fit results from
previous SNO analyses.  `LETA I' refers to the binned-histogram method
and `LETA II' to the kernel estimation method.}
\end{center} 
\end{figure}

Figure~\ref{f:nccomp} shows a comparison of these results to those
from previous analyses of SNO data.  Note that the $^8$B spectral
shape used in the previous Phase~I and Phase~II
analyses~\cite{oldb8spec} differs from that used here~\cite{winter}.
The bands represent the size of the systematic uncertainties on each
measurement, thus illustrating the improvements achieved with this
analysis.

Throughout this analysis, the quoted `statistical' uncertainties
represent the uncertainty due to statistics of all signals and
backgrounds in the fit, with correlations between event types taken
into account.  Therefore, they include uncertainties in the separation
of signal events from backgrounds in the fits.  For example, the
statistical uncertainties on the quoted results for
$\Phi_{\textrm{NC}}$ include both the Poisson uncertainty in the
number of NC events, and covariances with other event types.  This is
different from previous SNO analyses, in which the background events
were not included in the signal extraction fits and any uncertainty in
the level of background events was propagated as an additional
systematic uncertainty.

The two independent signal extraction fit techniques are in excellent
agreement, both in the central NC flux value and in the magnitude of
the uncertainties.  The result from the binned-histogram method is
quoted as the final unconstrained fit result for ease of comparison to
previous analyses, which used a similar method for PDF creation.

This result is in good agreement with the prediction from the BS05(OP)
SSM of 5.69$\times 10^6\,\rm cm^{-2}\,s^{-1} $~\cite{bs05}, to within
the theoretical uncertainty of $\pm16$\%.  It is also in good
agreement with the BS05(AGS,OP) model prediction of 4.51$\times
10^6\,\rm cm^{-2}\,s^{-1} \pm16$\%~\cite{bs05}, which was constructed
assuming a lower heavy-element abundance in the Sun's surface.

\begin{figure}[!ht]
\begin{center}
\includegraphics[width=0.48\textwidth]{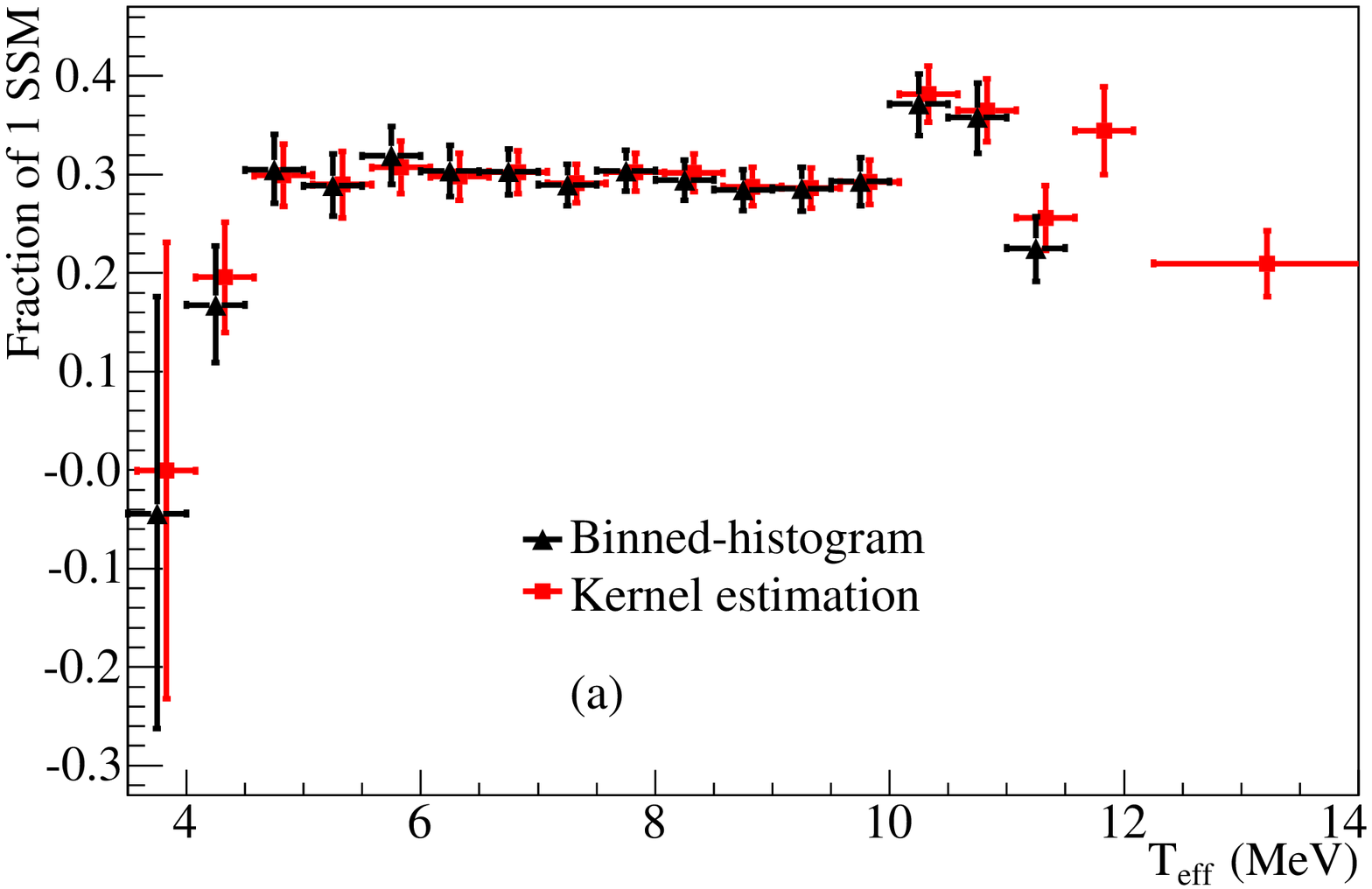}
\includegraphics[width=0.48\textwidth]{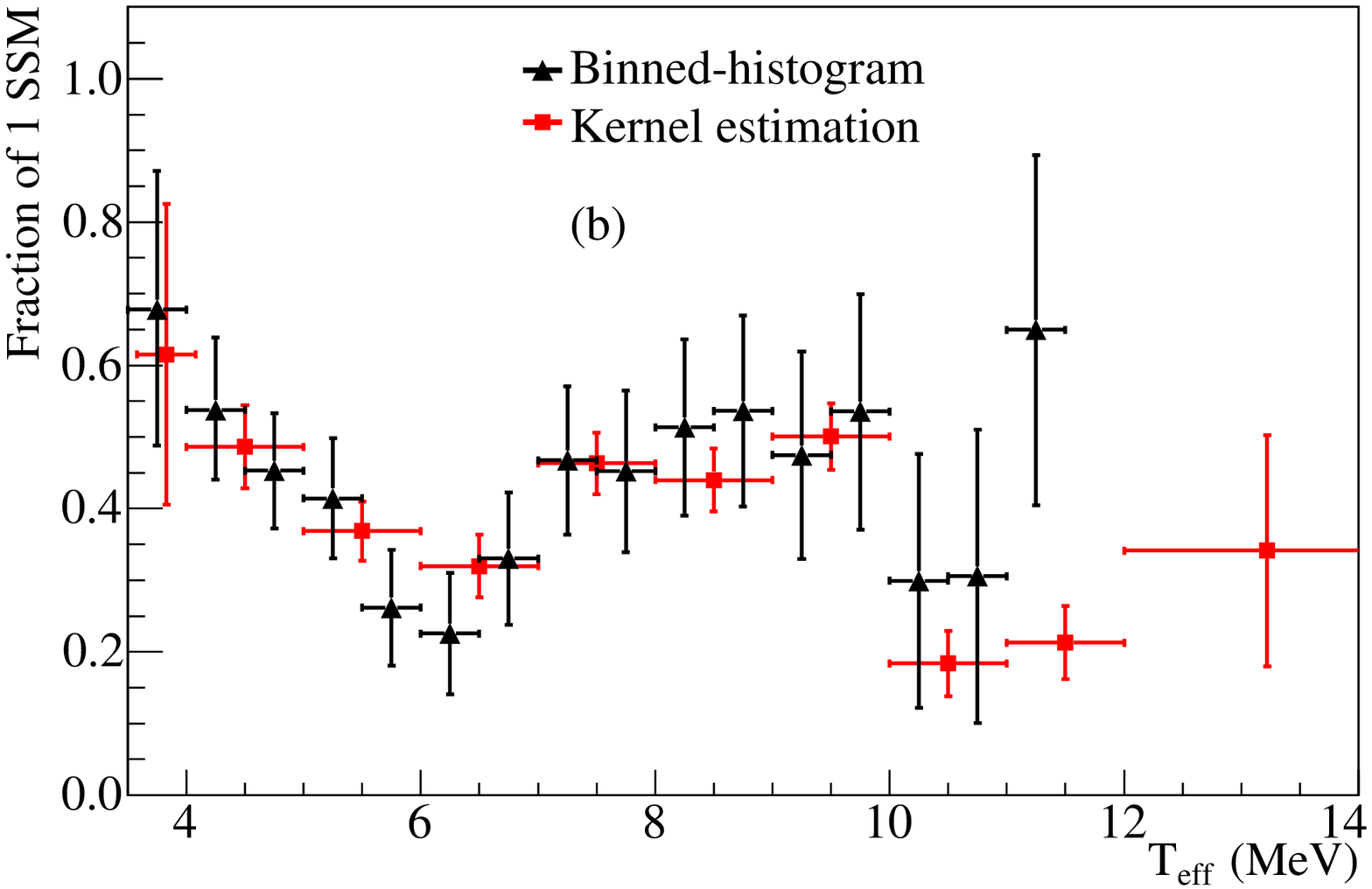}
\caption{\label{f:overlayspec}(Color online) Extracted a) CC and b) ES
electron spectra as a fraction of one unoscillated SSM (BS05(OP)),
from both signal extraction fits, with total uncertainties.  The final
12--20~MeV bin in the kernel estimation fit is plotted at the mean of
the spectrum in that range.  Both spectra are consistent with the
hypothesis of no distortion (a flat line).}
\end{center} 
\end{figure}

The extracted CC and ES electron spectra from both signal extraction
fits, in terms of the fraction of one unoscillated SSM, using the
BS05(OP) model flux of 5.69$\times 10^6\,\rm cm^{-2}\,s^{-1}
$~\cite{bs05}, are shown in Figure~\ref{f:overlayspec}.  An
unsuppressed, undistorted spectrum would correspond to a flat line at
1.0.  A greater suppression is observed for CC events than ES, since
the ES spectrum includes some contribution from $\nu_{\mu}$ and
$\nu_{\tau}$ whereas CC is sensitive only to $\nu_e$.  Both spectra
are consistent with the hypothesis of no distortion.  The results from
the two independent signal extraction fits are again in excellent
agreement for both the central fit values and the uncertainties.

\begin{figure}[!ht]
\begin{center}
\includegraphics[width=0.48\textwidth]{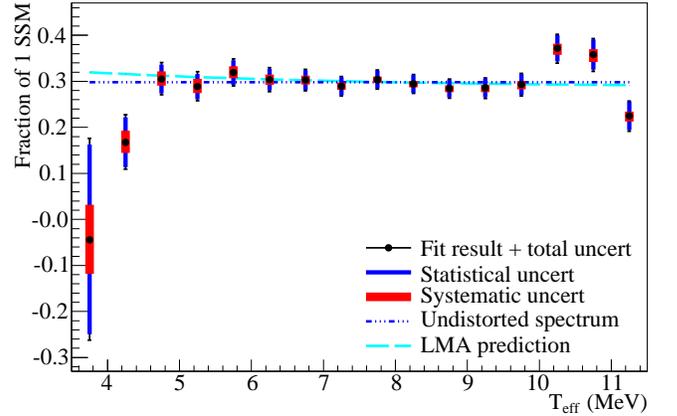}
\caption{\label{f:mxfspec}(Color online) Extracted CC electron
spectrum as a fraction of one unoscillated SSM (BS05(OP)) from the
binned-histogram signal extraction fit, with the uncertainties
separated into statistical (blue bars) and systematic (red band)
contributions.  The predictions for an undistorted spectrum, and for
the LMA point $\Delta m^2_{21} = 7.59\times 10^{-5}\,{\rm eV}^2$ and
$\tan^2 \theta_{12} = 0.468$ (taken from a previous global
solar+KamLAND fit~\cite{snoncd} and floating the $^8$B flux scale) are
overlaid for comparison.  }
\end{center} 
\end{figure}

Figure~\ref{f:mxfspec} shows the CC electron spectrum extracted from
the binned-histogram signal extraction fit with the errors separated
into the contributions from statistical and systematic uncertainties.
As for the NC flux result, the uncertainties are dominated by those
due to statistics (which includes the ability to distinguish signal
from background).  This demonstrates the effect of the significant
improvements made both in the determination of the individual
systematic uncertainties, as presented in previous sections, and in
the improved treatment of the dominant systematic uncertainties,
whereby the self-consistency of the data itself was used to further
constrain the allowed ranges of these parameters.  It is worth noting
that correlations between bins, which are not shown, tend to reduce
the significance of any observed shape.  Fitting to an undistorted
spectrum (the flat line on Fig.~\ref{f:mxfspec}) gives a $\chi^2$
value of 21.52 for 15 degrees of freedom, which is consistent with the
hypothesis of no distortion.  The prediction for the $T_{\rm eff}$
spectrum for CC events taken from the best fit LMA point from a
previous global analysis of solar data~\cite{snoncd} is also overlaid
on Fig.~\ref{f:mxfspec}.  The $\chi^2$ value of the fit of the
extracted spectrum to this prediction is 22.56 for 15 degrees of
freedom, demonstrating that the data are also consistent with the LMA
prediction.

\begin{figure}[!ht]
\begin{center}
\includegraphics[width=0.35\textwidth]{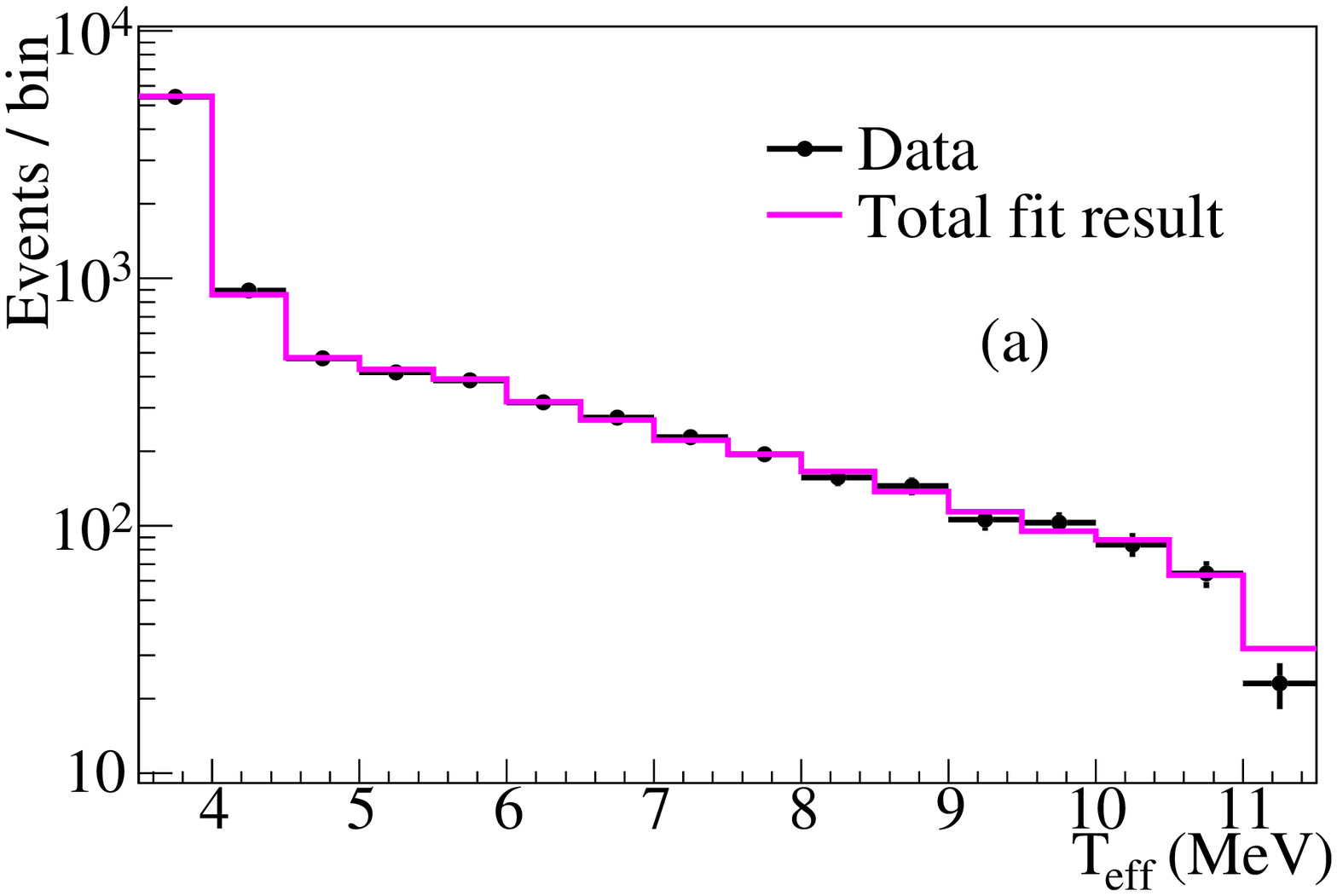}
\includegraphics[width=0.35\textwidth]{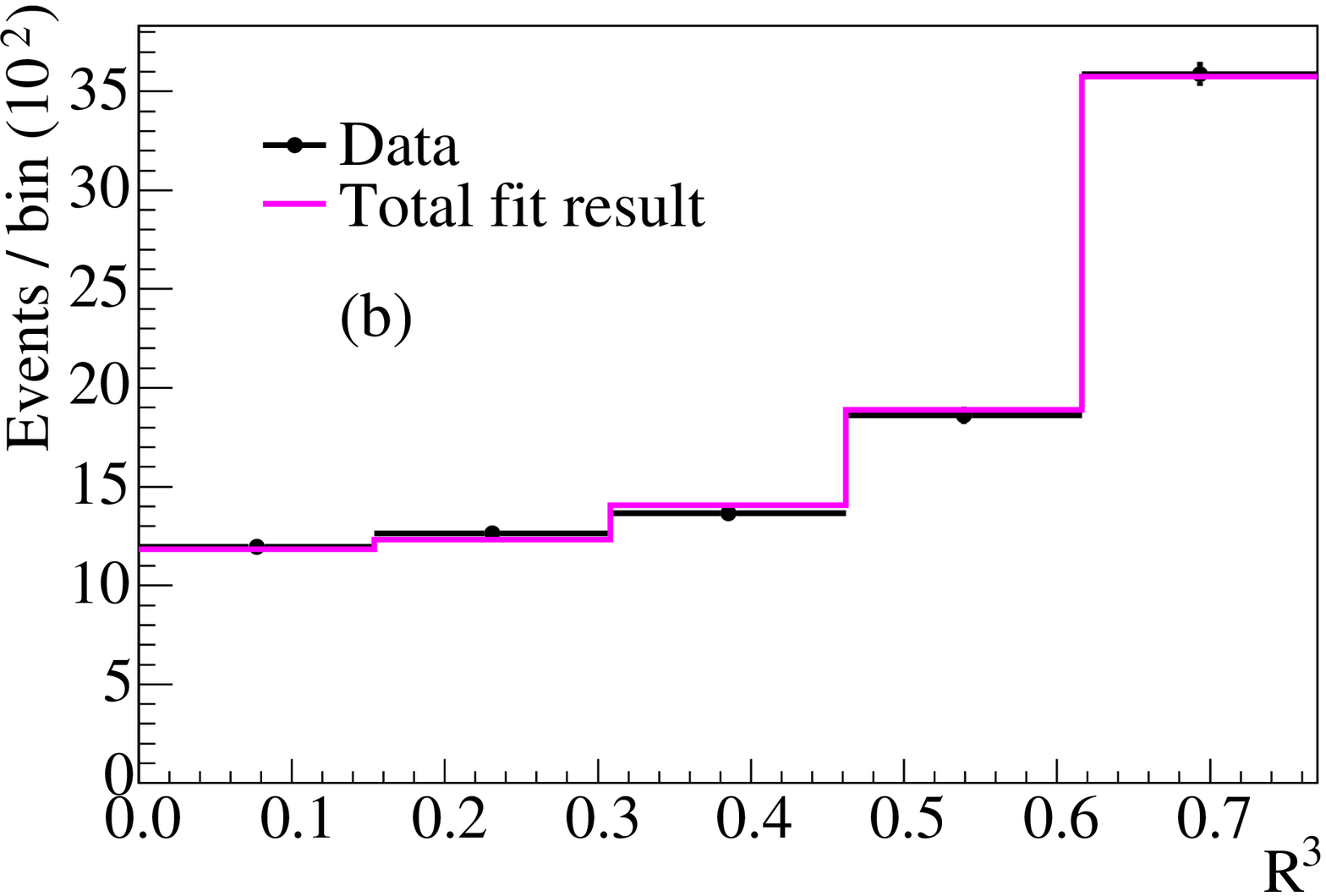}
\includegraphics[width=0.35\textwidth]{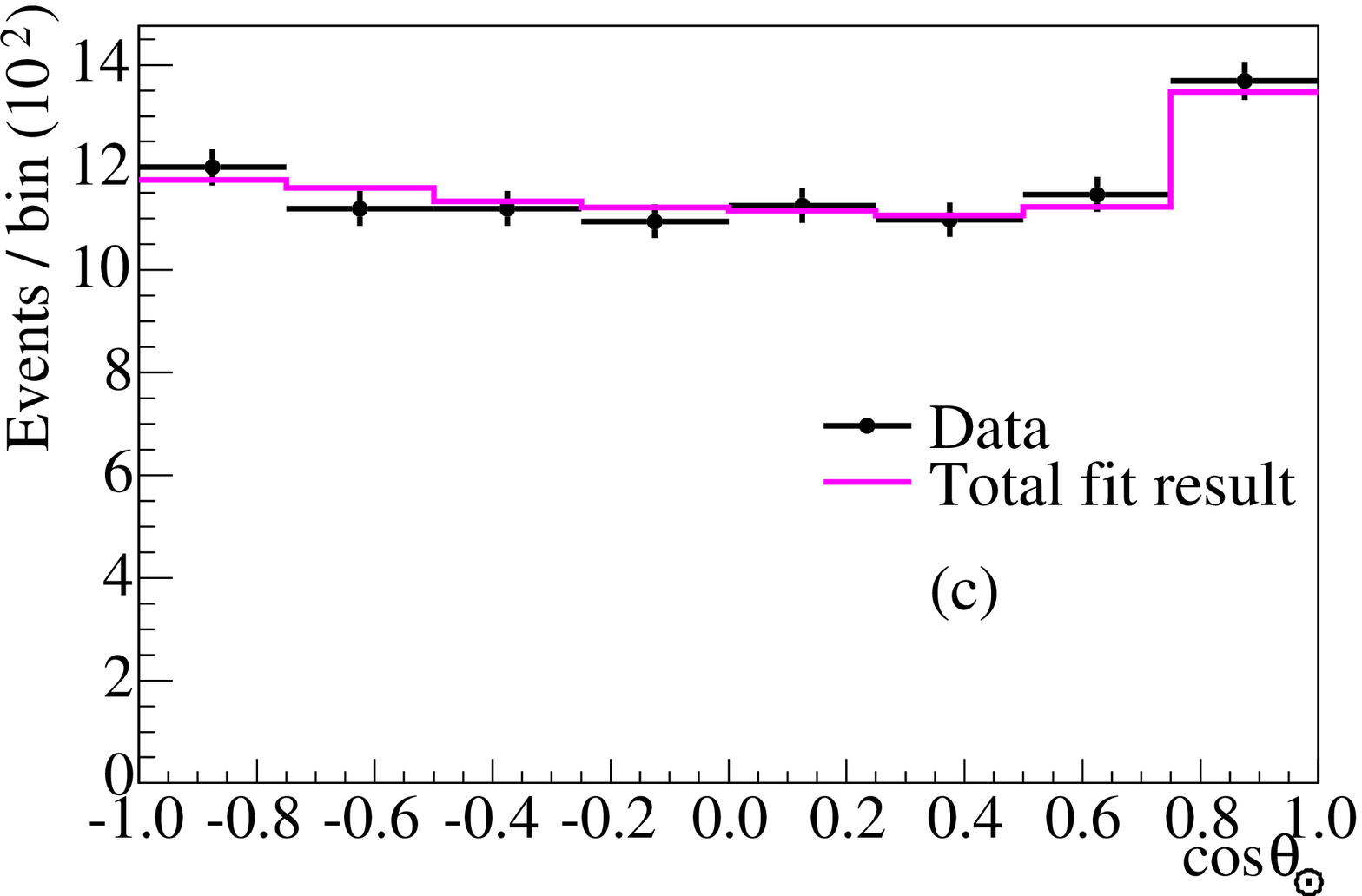}
\includegraphics[width=0.35\textwidth]{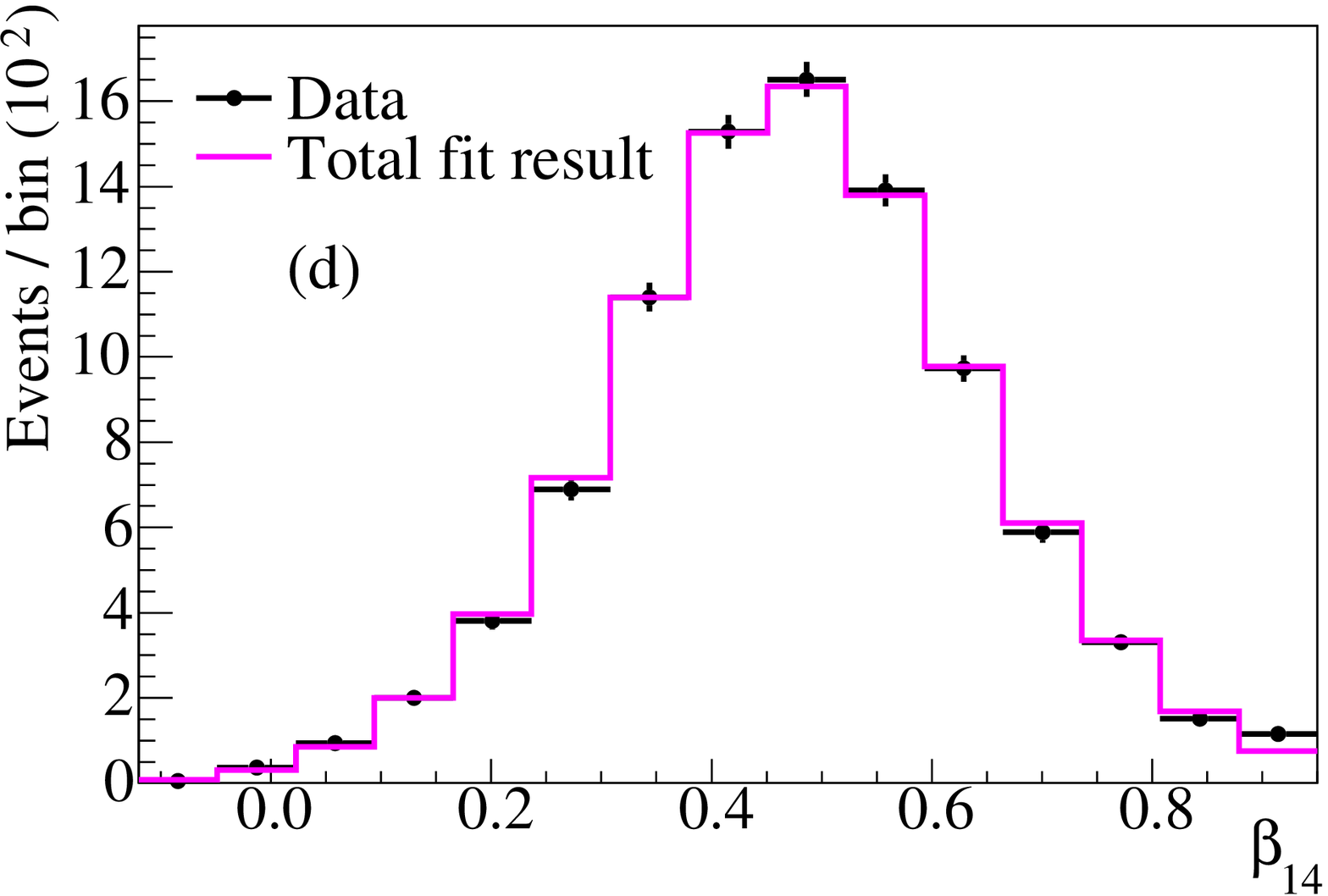}
\caption{\label{f:mxffitsd}(Color online) One dimensional projections
of the fit in each observable parameter in Phase~I, from the
binned-histogram signal extraction.  The panels show the fit projected
onto (a) energy ($T_{\rm eff}$), (b) radius cubed ($R^3$), (c)
direction (\ctsnosp), and (d) isotropy (\betnosp).}
\end{center} 
\end{figure}

\begin{figure}[!ht]
\begin{center}
\includegraphics[width=0.35\textwidth]{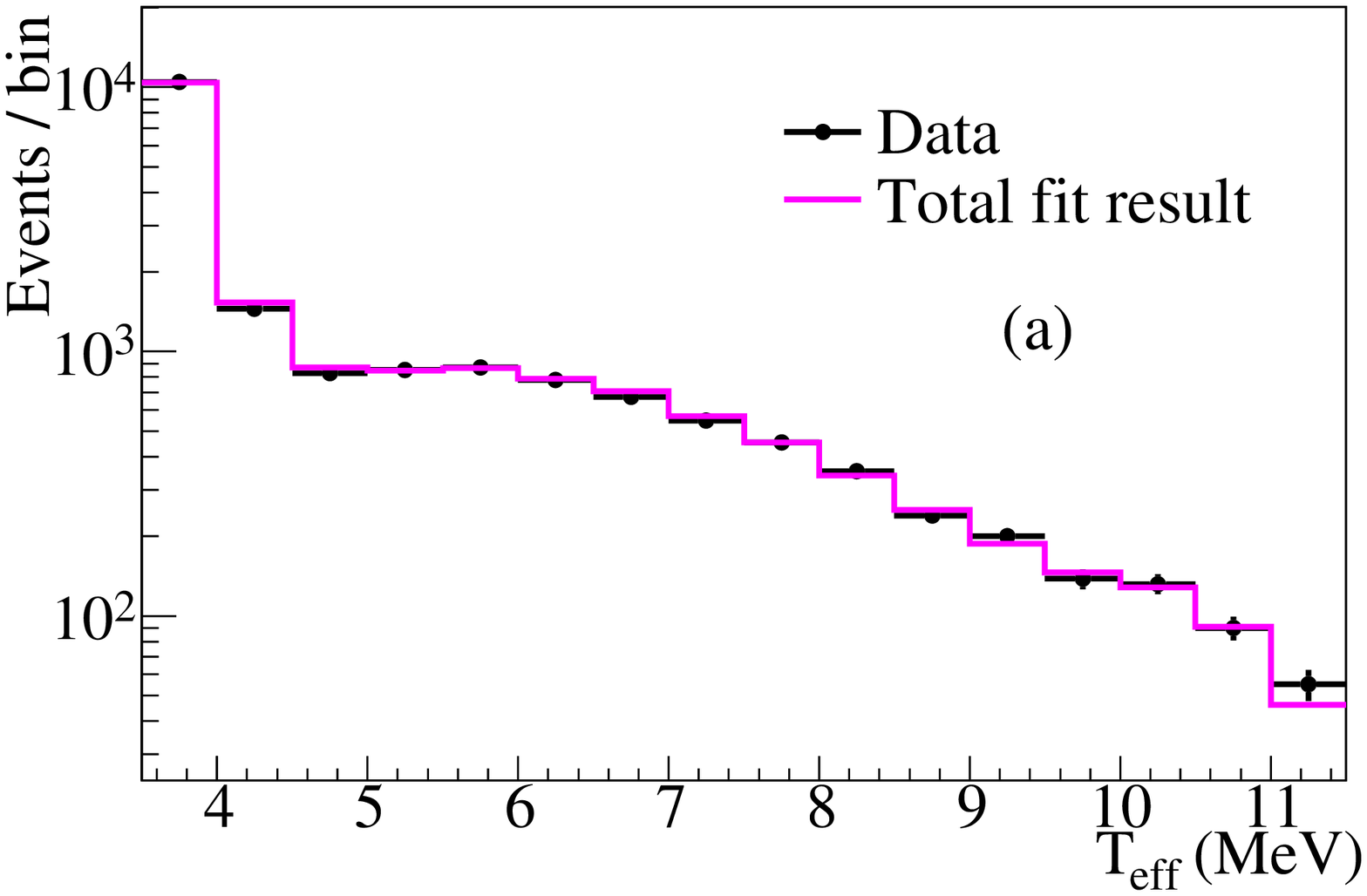}
\includegraphics[width=0.35\textwidth]{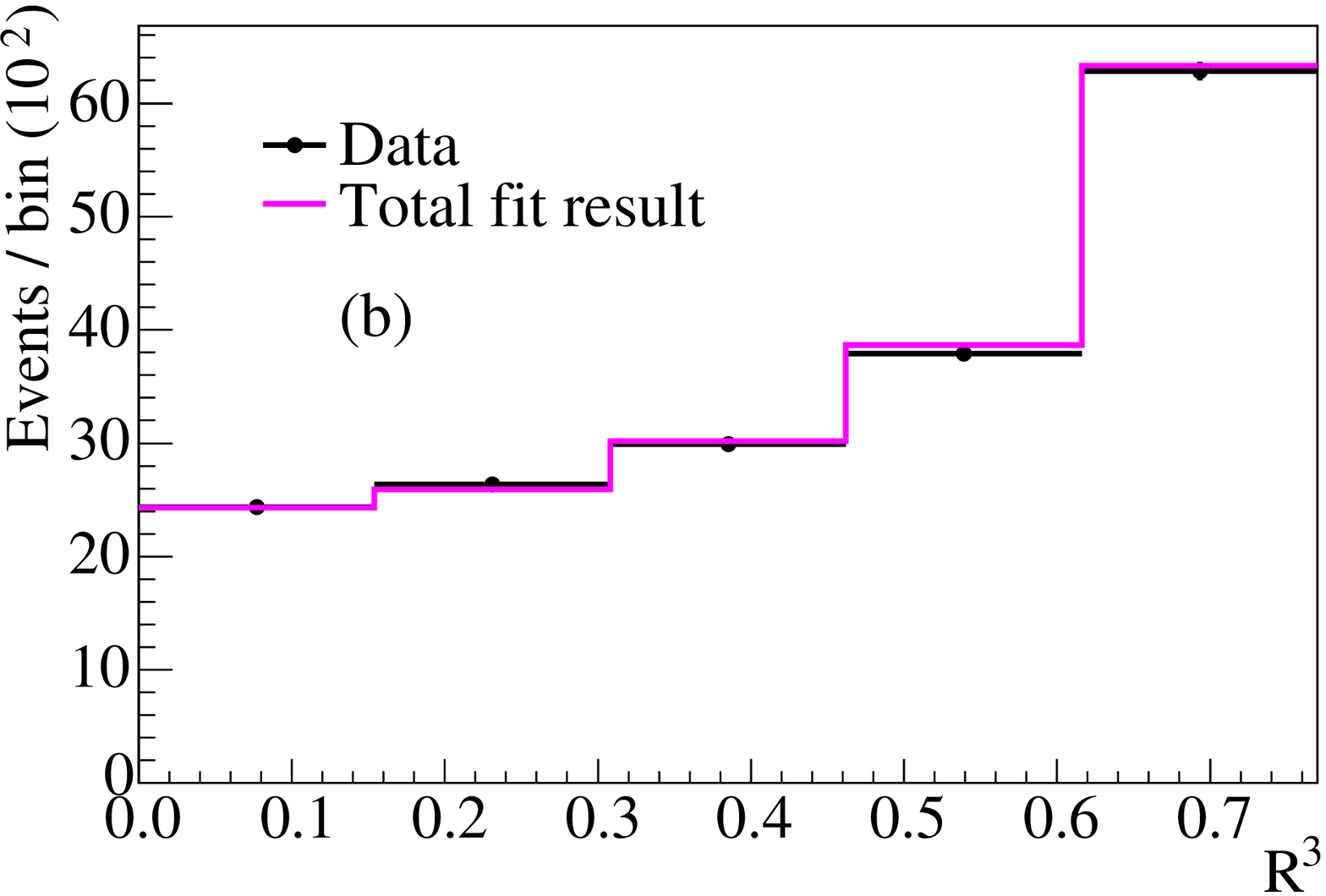}
\includegraphics[width=0.35\textwidth]{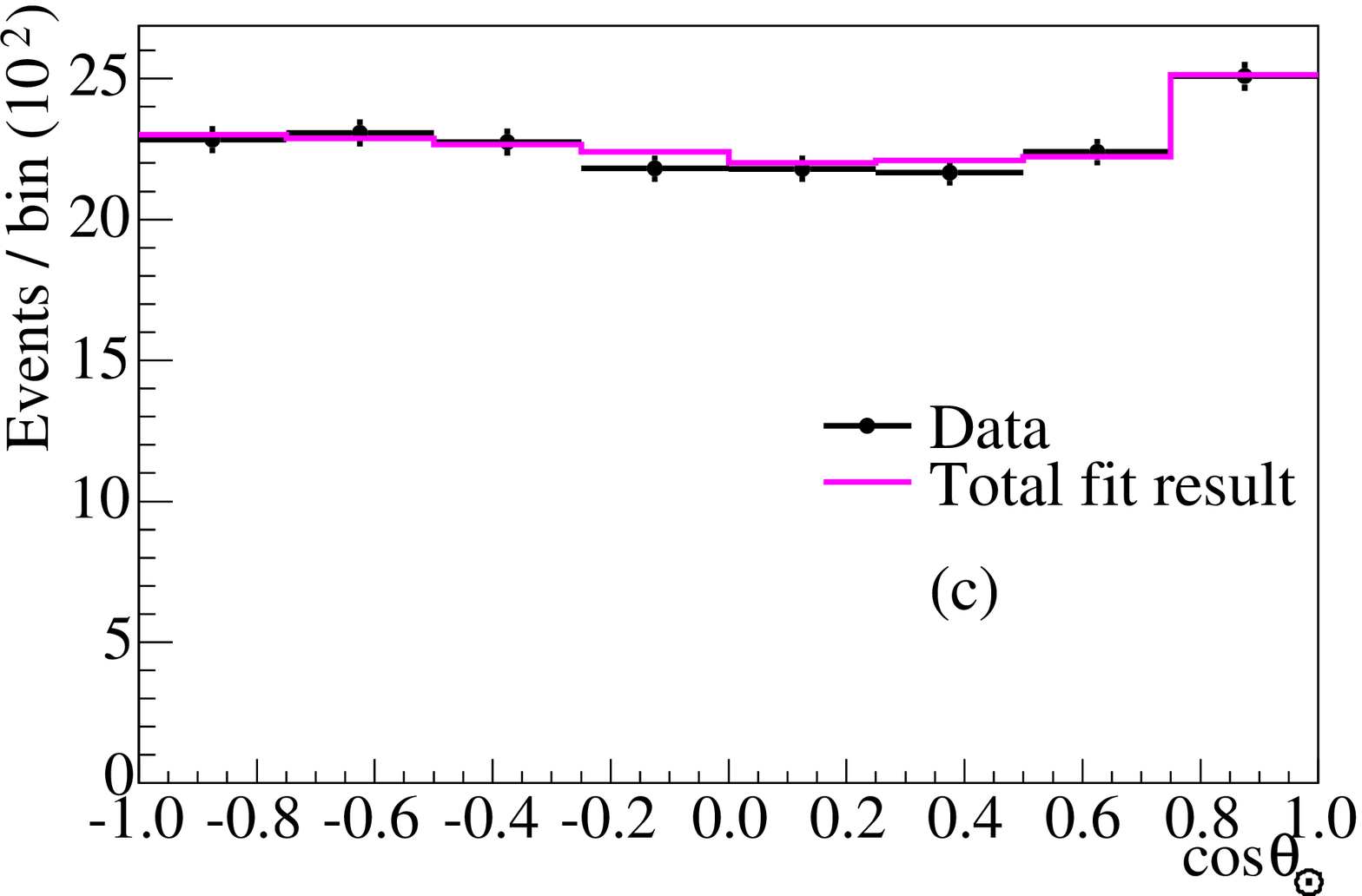}
\includegraphics[width=0.35\textwidth]{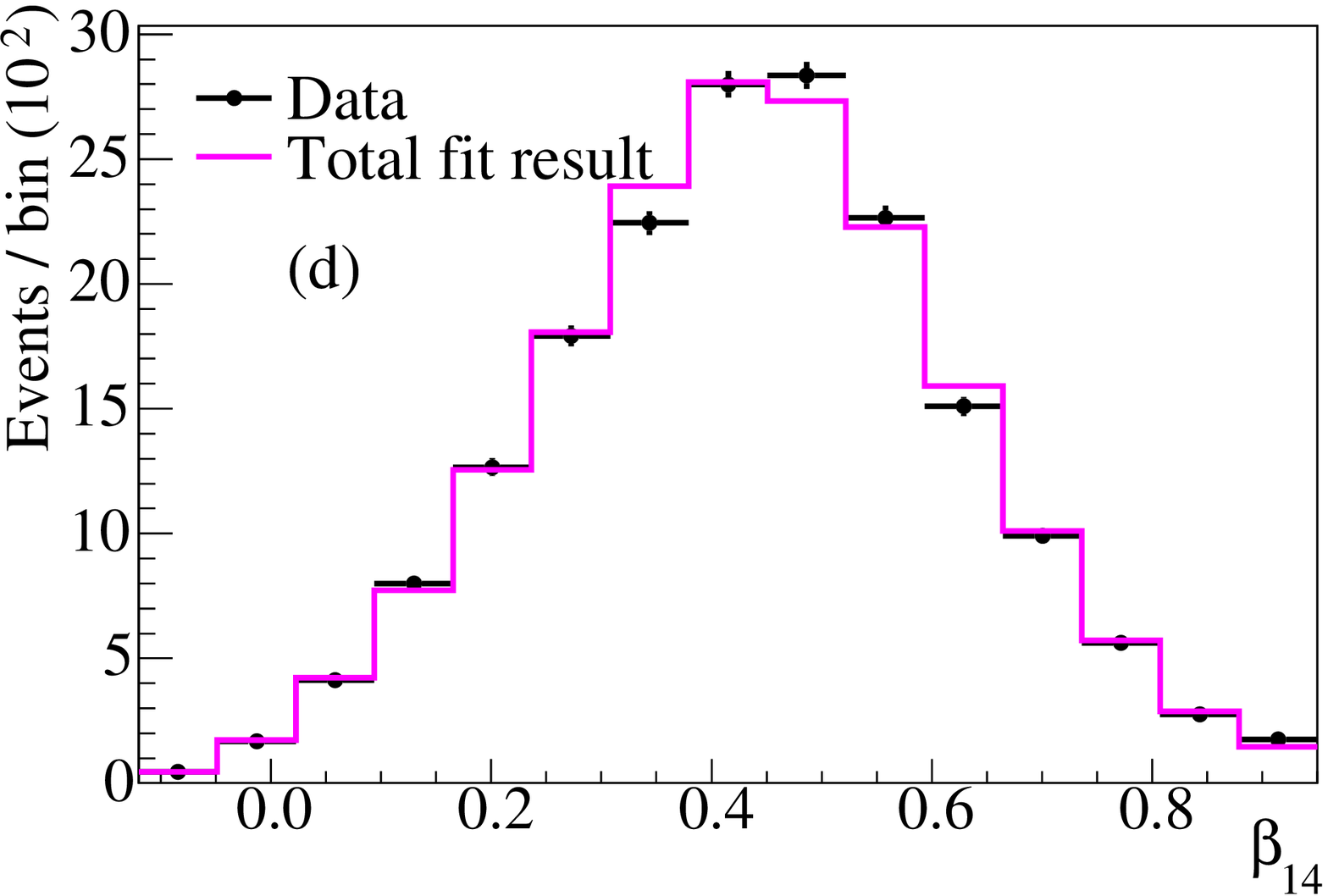}
\caption{\label{f:mxffitss}(Color online) One dimensional projections
of the fit in each observable parameter in Phase~II, from the
binned-histogram signal extraction.  The panels show the fit projected
onto (a) energy ($T_{\rm eff}$), (b) radius cubed ($R^3$), (c)
direction (\ctsnosp), and (d) isotropy (\betnosp).}
\end{center} 
\end{figure}

\begin{figure}[!ht]
\begin{center}
\includegraphics[width=0.48\textwidth]{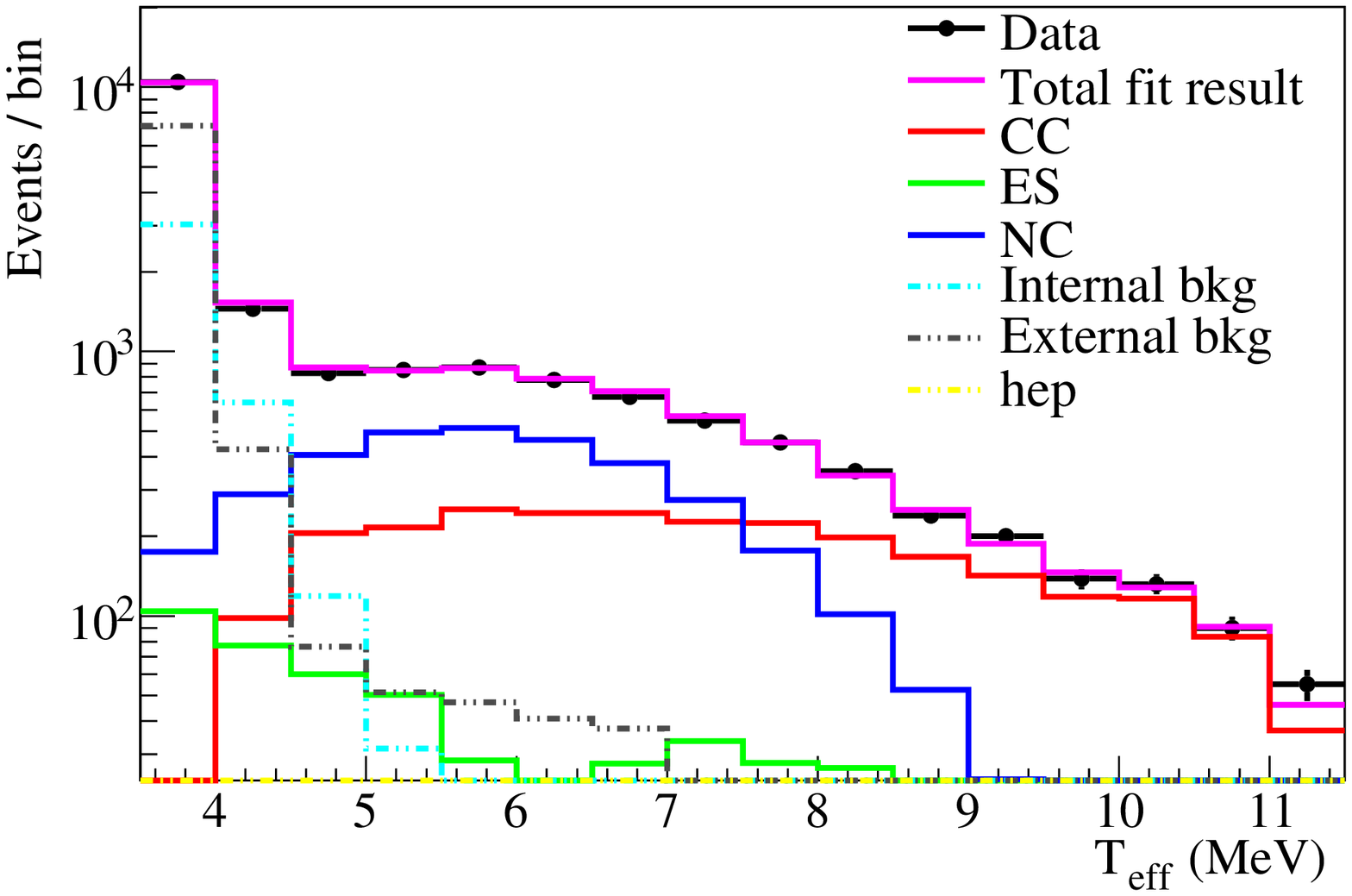}
\caption{\label{f:mxffits2}(Color) One dimensional projection of the
fit in $T_{\rm eff}$ in Phase~II from the binned-histogram signal
extraction, with the individual signals separated into the three
neutrino interactions, internal backgrounds (within the \dto volume),
external backgrounds (in the AV, \htonosp, and PMTs) and hep neutrino
events.}
\end{center} 
\end{figure}

The one-dimensional projections of the fits in each observable
parameter from the binned-histogram signal extraction are shown for
each phase in Figures~\ref{f:mxffitsd} and~\ref{f:mxffitss}.  Of
particular note is the clear ES peak observed in the \cts fits for
both phases (Figs.~\ref{f:mxffitsd}(c) and~\ref{f:mxffitss}(c)),
demonstrating the extraction of ES events over the integrated energy
spectrum, even with the low 3.5~MeV threshold.  The error bars
represent statistical uncertainties; systematic uncertainties are not
shown.  Figure~\ref{f:mxffits2} shows the one-dimensional projection
in $T_{\rm eff}$ from Phase~II (as in Fig.~\ref{f:mxffitss}(a)) but
with the fitted contributions from individual signal types separated
into six categories: CC, ES, and NC neutrino events, internal
backgrounds (within the \dto volume), external backgrounds (in the AV,
\htonosp, and PMTs) and hep neutrino events.

\noindent The $\chi^2$ for the one-dimensional projections of the fit
are given in Table~\ref{t:mxfchi}.  These were evaluated using
statistical uncertainties only and are, therefore, a conservative test
of goodness-of-fit in the one-dimensional projections.  In all
dimensions, the final result is a good fit to the data.

\begin{table}[!h]
\begin{center}
\begin{tabular}{lcccc}
\hline \hline Phase & Observable &&& $\chi^2$ (data points) \\ \hline I
& $T_{\rm eff}$ &&& 8.17 (16) \\ & \cts &&& 3.69 (8) \\ & $\rho$ &&& 2.61
(5) \\ & \bet &&& 20.99 (15) \\ \hline II & $T_{\rm eff}$ &&& 13.64 (16)
\\ & \cts &&& 3.07 (8) \\ & $\rho$ &&& 2.98 (5) \\ & \bet &&& 26.25 (15)
\\ \hline \hline
\end{tabular}
\caption[$\chi^2$ values for the fit of extracted signals to the
data.]{$\chi^2$ values for the fit of the extracted signals from the
binned-histogram signal extraction to the data set for one-dimensional
projections in each of the four observables, in each phase. These were
evaluated using statistical uncertainties only.  The number of data
points used for the $\chi^2$ calculations are given afterwards in
parentheses.  Because these are one-dimensional projections of a fit
in four observables, the probability of obtaining these $\chi^2$
values cannot be simply evaluated; these are simply quoted as a
qualitative demonstration of goodness-of-fit.}
\label{t:mxfchi}
\end{center}
\end{table}

Table~\ref{t:neutflux} in Appendix~\ref{a:tables} shows the extracted
number of events for the neutrino fit parameters from the
binned-histogram signal extraction fit, with total statistical plus
systematic uncertainties.

Table~\ref{t:bkgcomp} shows the total number of background events
extracted by each signal extraction in each phase, and a breakdown of
the number of background neutron events occurring within each region
of the detector.  The two methods are in good agreement based on
expectations from studies of Monte Carlo-generated `fake' data sets.
For comparison, the total number of events in each data set is also
given (taken from Table~\ref{t:cuts}).  Due to the exponential shape
of the energy spectra of most sources of background in this fit, the
majority of the background events fit out in the lowest two bins in
$T_{\rm eff}$, illustrating one of the major challenges of the low
energy analysis.

\begin{table}[!h]
\begin{center}
\begin{tabular}{lcccc}
\hline \hline & \multicolumn{2}{c}{Phase~I} &
 \multicolumn{2}{c}{Phase~II} \\ Background & LETA~I & LETA~II &
 LETA~I & LETA~II \\ \hline Total background events & 6148.9 & 6129.8
 & 11735.0 & 11724.6 \\ \hline \dto neutrons & 29.7 & 34.0 & 122.4 &
 133.5 \\ AV neutrons & 214.9 & 191.4 & 295.7 & 303.4 \\ \hto neutrons
 & 9.9 & 8.4 & 27.7 & 26.3 \\ \hline Total data events &
 \multicolumn{2}{c}{9337} & \multicolumn{2}{c}{18228} \\ \hline \hline
\end{tabular}
\caption{Number of background events extracted from the signal
extraction fits for each method.  `LETA~I' refers to the
binned-histogram signal extraction, and `LETA~II' refers to the kernel
estimation method.  The total number of events in each data set is
also given, taken from Table~\ref{t:cuts}.}
\label{t:bkgcomp}
\end{center}
\end{table}

Tables~\ref{t:sigsyst}--\ref{t:sigsyst1} in Appendix~\ref{a:tables}
show the effects of the individual systematic uncertainties on the
extracted NC rate, the CC rate in two energy intervals (4.0--4.5~MeV
and 9.5--10.0~MeV) and the ES rate in the 3.5--4.0~MeV interval, all
taken from the binned-histogram fit.  The dominant source of
uncertainty on the total neutrino flux measured with the NC reaction
is the neutron capture uncertainty.  Further significant contributions
come from the Phase~II energy resolution, the \bof scale for neutron
capture events, the energy-dependent fiducial volume, and the
cut-acceptance uncertainties.

\begin{figure}[!ht]
\begin{center}
\includegraphics[width=0.48\textwidth]{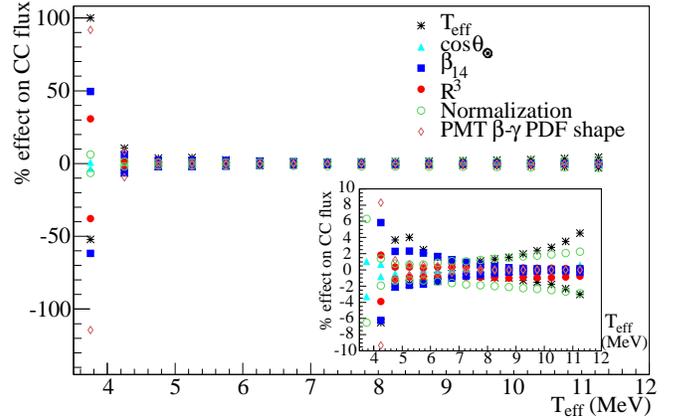}
\caption{\label{f:ccsyst}(Color) Effect of systematic uncertainties on
the extracted CC electron spectrum.  The inset shows the same plot on
a larger scale.}
\end{center} 
\end{figure}

Figure~\ref{f:ccsyst} shows the effects of several groups of
systematic uncertainties on the extracted CC electron spectrum, taken
from the binned-histogram fit.  Four groups cover systematic effects
that apply to the observables ($T_{\rm eff}$, \ctsnosp, $R^3$ and
\betnosp), in which the individual contributions are summed in
quadrature (for example the $T_{\rm eff}$ group includes the effect of
energy scale, resolution and linearity); `normalization' uncertainties
include neutron capture, cut-acceptance, energy-dependent fiducial
volume and photodisintegration uncertainties; the final group consists
of uncertainties in the shape of the PMT $\beta$-$\gamma$ PDF.  The
dominant sources of the systematic uncertainties on the shape of the
CC electron spectrum are energy resolution and the shape of the PMT
$\beta$-$\gamma$ PDF, particularly as a function of $T_{\rm eff}$.
The \bof scale for electron-like events is also a significant
contributor.  It is worth noting that the contribution from the
fiducial volume uncertainty, which was significant in previous
analyses~\cite{nsp}, is now relatively small.

The two signal extraction methods are in excellent agreement for all
the neutrino flux parameters, as well as the sources of background
events.  This is a stringent test of the result, since the two methods
differed in several fundamental ways:
\begin{itemize}
\item Formation of the PDFs

The methods used to create the PDFs were entirely independent: one
using binned histograms, and the other using smooth, analytic,
kernel-estimated PDFs.
\item Treatment of systematic uncertainties

The dominant systematics in the fits were `floated' using different
approaches: in the kernel method they were floated directly, whereas
an iterative likelihood scan was used in the binned-histogram
approach.
\item PMT $\beta$-$\gamma$ constraint

In the binned-histogram method, a constraint on the total number of
PMT events was implemented using a bifurcated analysis of the data
(Sec.~\ref{s:pmtpdf}), whereas no constraint was applied in the kernel
method.
\end{itemize}
That these independent approaches give such similar results
demonstrates the robust nature of the analysis and the final results.

\subsection{Survival Probability Fit}
\label{res:sprob}

Under the assumption of unitarity (for example, no oscillations
between active and sterile neutrinos), the NC, CC, and ES rates can be
directly related.  Based on this premise, a signal extraction fit was
performed in which the free parameters directly described the total
$^8$B neutrino flux and the $\nu_e$ survival probability.  This fit
therefore produces a measure of the total flux of $^8$B neutrinos that
naturally includes information from all three interaction types.
Applying this approach, the uncertainty on the flux was reduced in
comparison to that from the unconstrained fit (Sec.~\ref{res:uncon}).
The total flux measured in this way ($\Phi_{^8{\rm B}}$) is found to
be:
\begin{displaymath}
	\Phi_{^8{\rm B}} = 5.046 \,^{+0.159}_{-0.152} \textrm{(stat)}
	\,^{+0.107}_{-0.123} \textrm{(syst)} \times 10^6\,\rm
	cm^{-2}\,s^{-1},
\end{displaymath}
which represents $^{+3.8}_{-3.9}$\% total uncertainty.  This is the
most precise measurement of the total flux of $^8$B neutrinos from the
Sun ever reported.

The survival probability was parameterized as a quadratic function in
$E_{\nu}$, representing $P_{ee}^{\rm day}$, and a linear day/night
asymmetry, as defined in Eqs.~\eqref{eq:dn} and~\eqref{eq:poly} of
Sec.~\ref{s:kerpoly}.  The best-fit polynomial parameter values and
uncertainties are shown in Table~\ref{t:poly_pars}, and the
correlation matrix is shown in Table~\ref{t:poly_corr}, both presented
in Appendix~\ref{a:tables}.  For all the extracted parameters, the
total uncertainty is dominated by that due to statistics.

\begin{figure}[!ht]
\begin{center}
\includegraphics[width=0.48\textwidth]{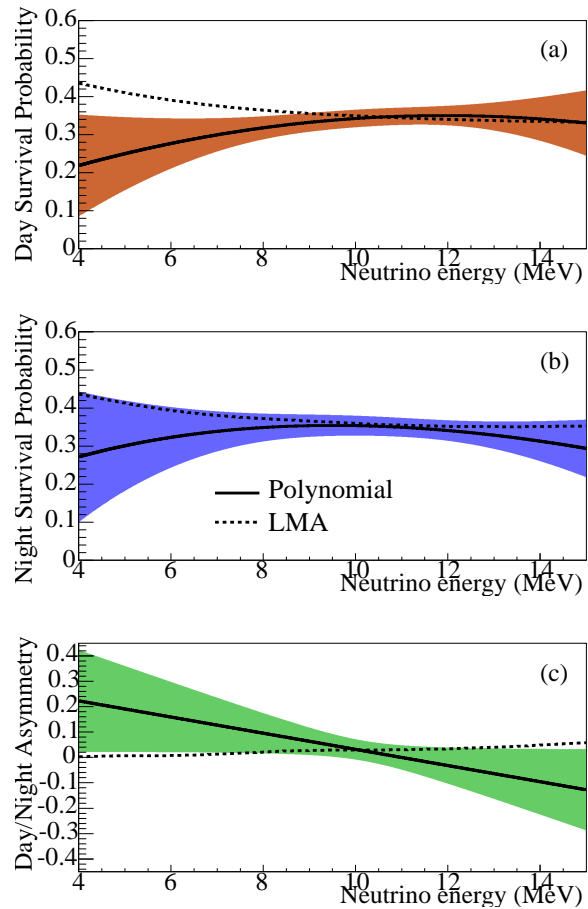}
\caption{\label{f:poly_band}(Color online) Best fit and RMS spread in
the (a) $P_{ee}^{\rm day}(E_\nu)$, (b) $P_{ee}^{\rm night}(E_\nu)$,
and (c) $A(E_\nu)$ functions.  The survival probabilities and
day/night asymmetry for the LMA point $\Delta m^2_{21} = 7.59\times
10^{-5}\,{\rm eV}^2$ and $\tan^2 \theta_{12} = 0.468$, taken from a
previous global solar+KamLAND fit~\cite{snoncd}, are shown for
comparison.}
\end{center} 
\end{figure}

Figure~\ref{f:poly_band} shows the RMS spread in the best fit survival
probabilities, $P_{ee}^{\rm day}(E_\nu)$ and $P_{ee}^{\rm
night}(E_\nu)$, and day/night asymmetry, $A(E_\nu)$.  The bands were
computed by sampling the parameter space 1000 times, taking into
account the parameter uncertainties and correlations.  Overlaid on
Fig.~\ref{f:poly_band} are the predicted shapes of the day and night
survival probabilities and the day/night asymmetry for the best-fit
point from a previous global analysis of solar data~\cite{snoncd}.

The advantage of this direct parameterization for the survival
probability is that model testing becomes straightforward.  We can
test the goodness-of-fit to an undistorted spectrum by setting $c_1 =
c_2 = 0.0$ in Eq.~\ref{eq:poly}, and we can test the goodness-of-fit
to a model with no day/night asymmetry by setting $a_0 = a_1 = 0.0$ in
Eq.~\ref{eq:dn}.  Requiring both simultaneously, we find a $\Delta
\chi^2 = 1.94$ for 4 degrees of freedom, demonstrating that the
extracted survival probabilities and day/night asymmetry are
consistent with the hypothesis of no spectral distortion and no
day/night asymmetry.  For comparison, the $\Delta \chi^2$ value of the
fit to the LMA point shown in Fig.~\ref{f:poly_band} is 3.9 for 4
degrees of freedom, showing that the data are also consistent with
LMA.

\begin{figure}[!ht]
\begin{center}
\includegraphics[width=0.35\textwidth]{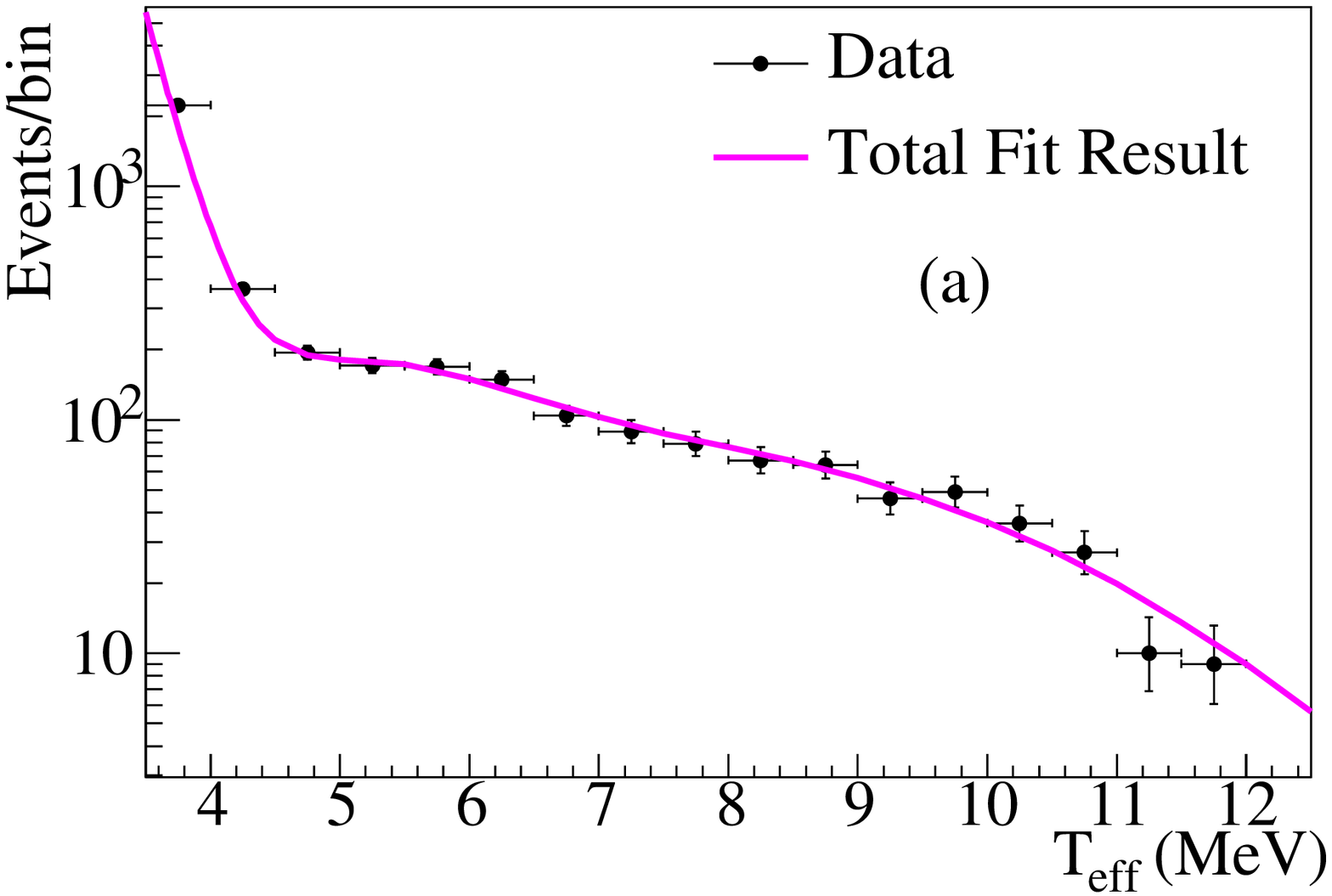}
\includegraphics[width=0.35\textwidth]{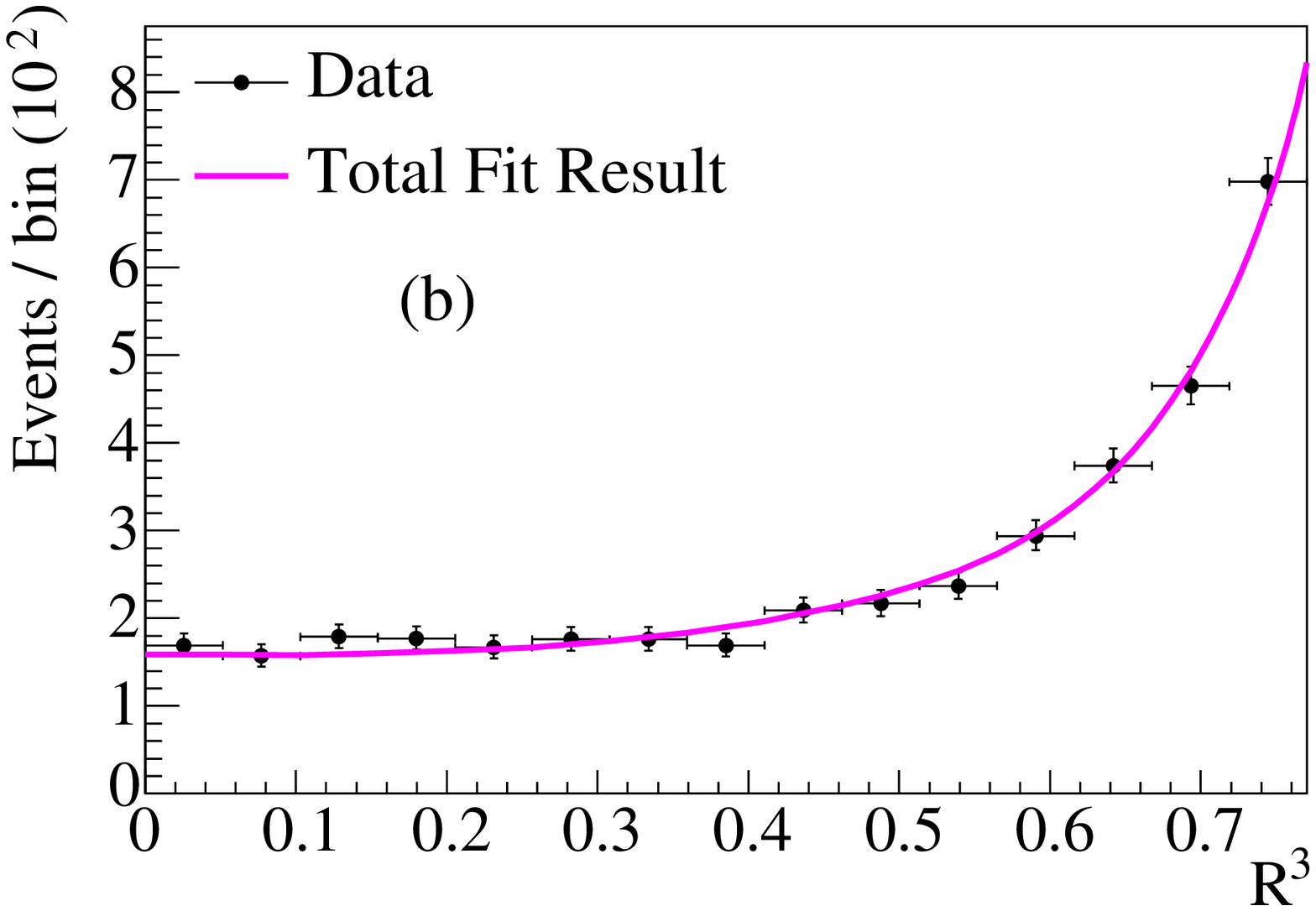}
\includegraphics[width=0.35\textwidth]{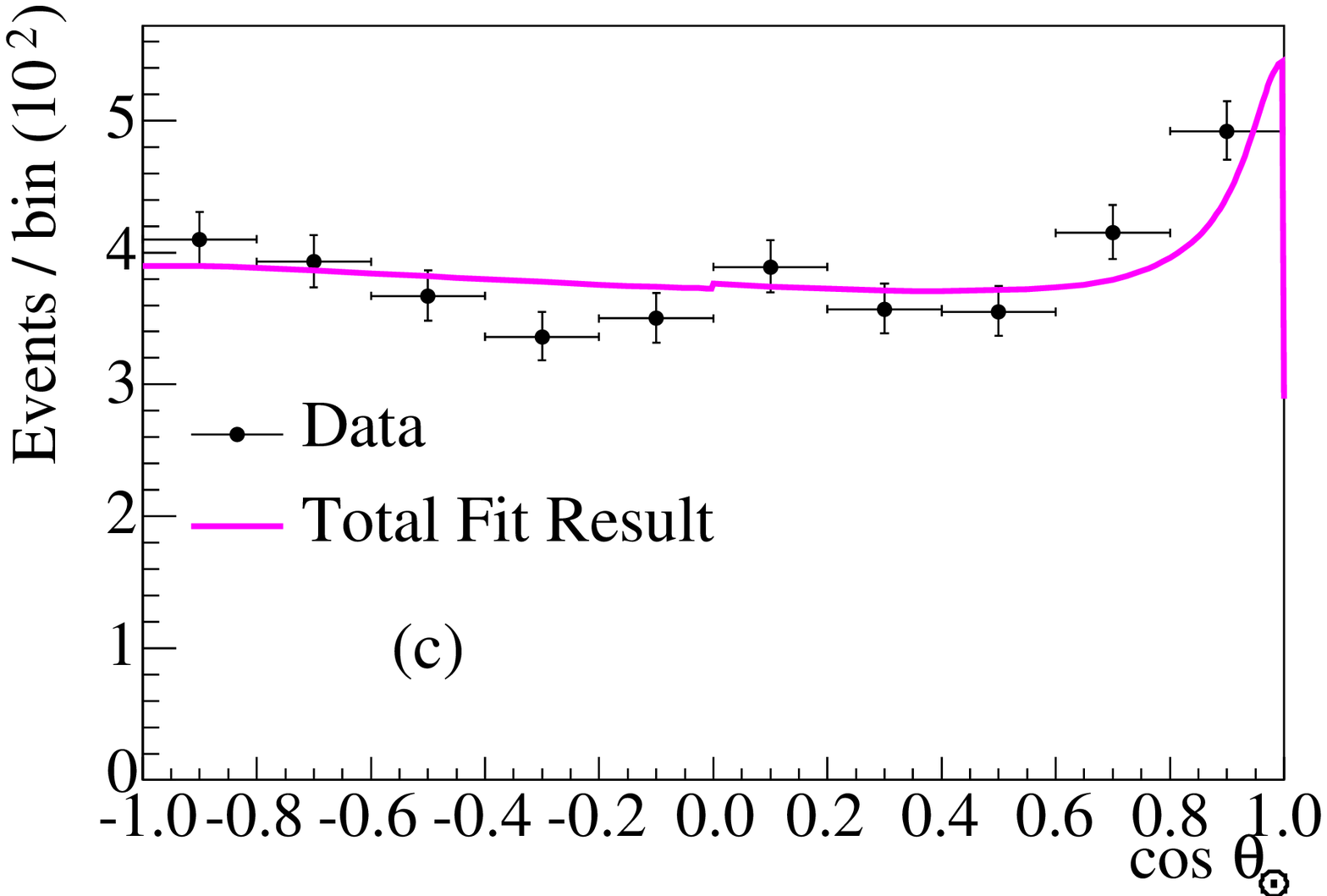}
\includegraphics[width=0.35\textwidth]{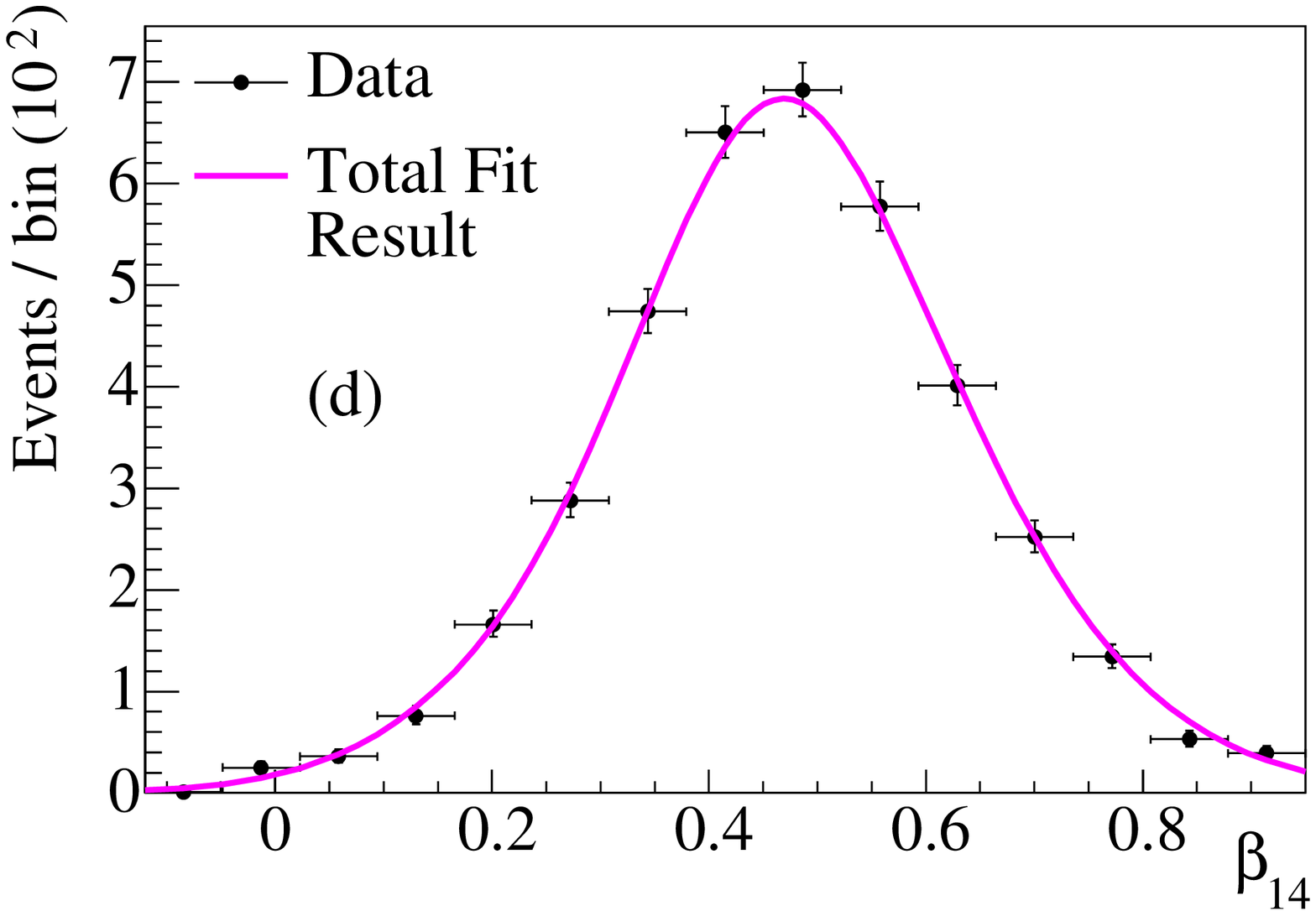}
\caption{\label{f:poly_1d_d2o}(Color online) One dimensional
projections of the fit in Phase~I-day, from the polynomial survival
probability fit.  The panels show the fit projected onto (a) energy
($T_{\rm eff}$), (b) radius cubed ($R^3$), (c) direction (\ctsnosp),
and (d) isotropy (\betnosp).  The binning of data is purely for
display purposes; the fits were performed unbinned.}
\end{center} 
\end{figure}
\begin{figure}[!ht]
\begin{center}
\includegraphics[width=0.35\textwidth]{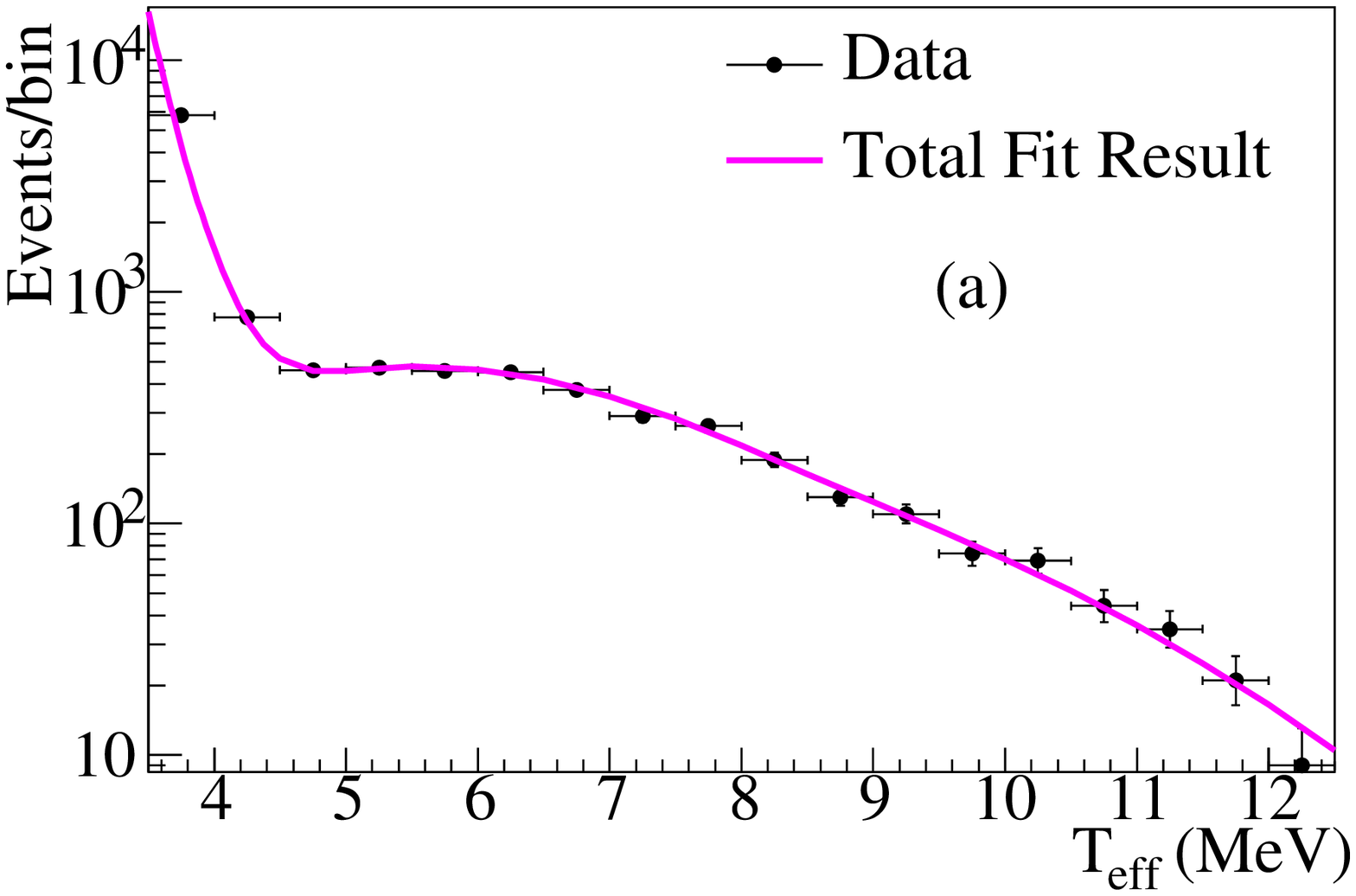}
\includegraphics[width=0.35\textwidth]{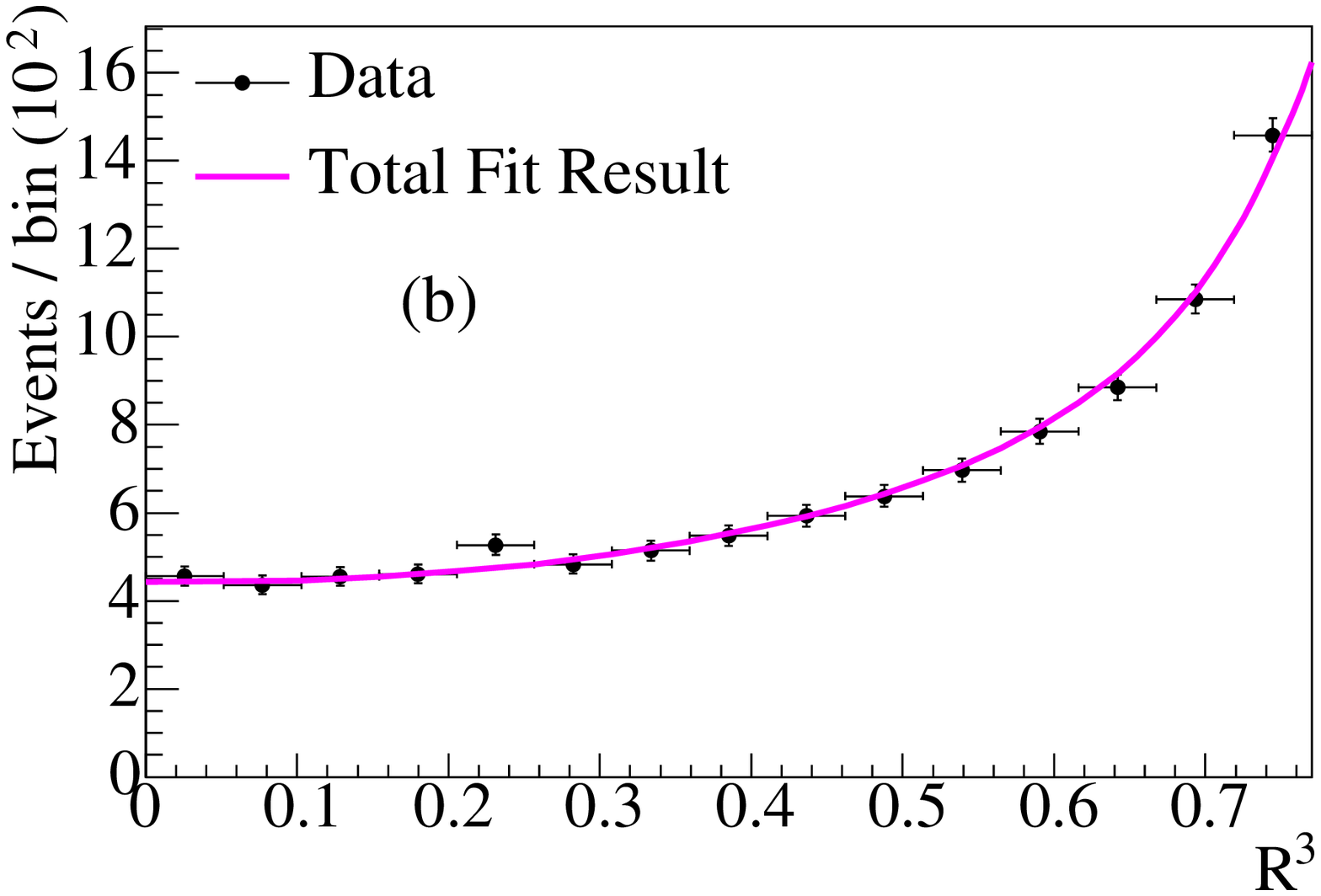}
\includegraphics[width=0.35\textwidth]{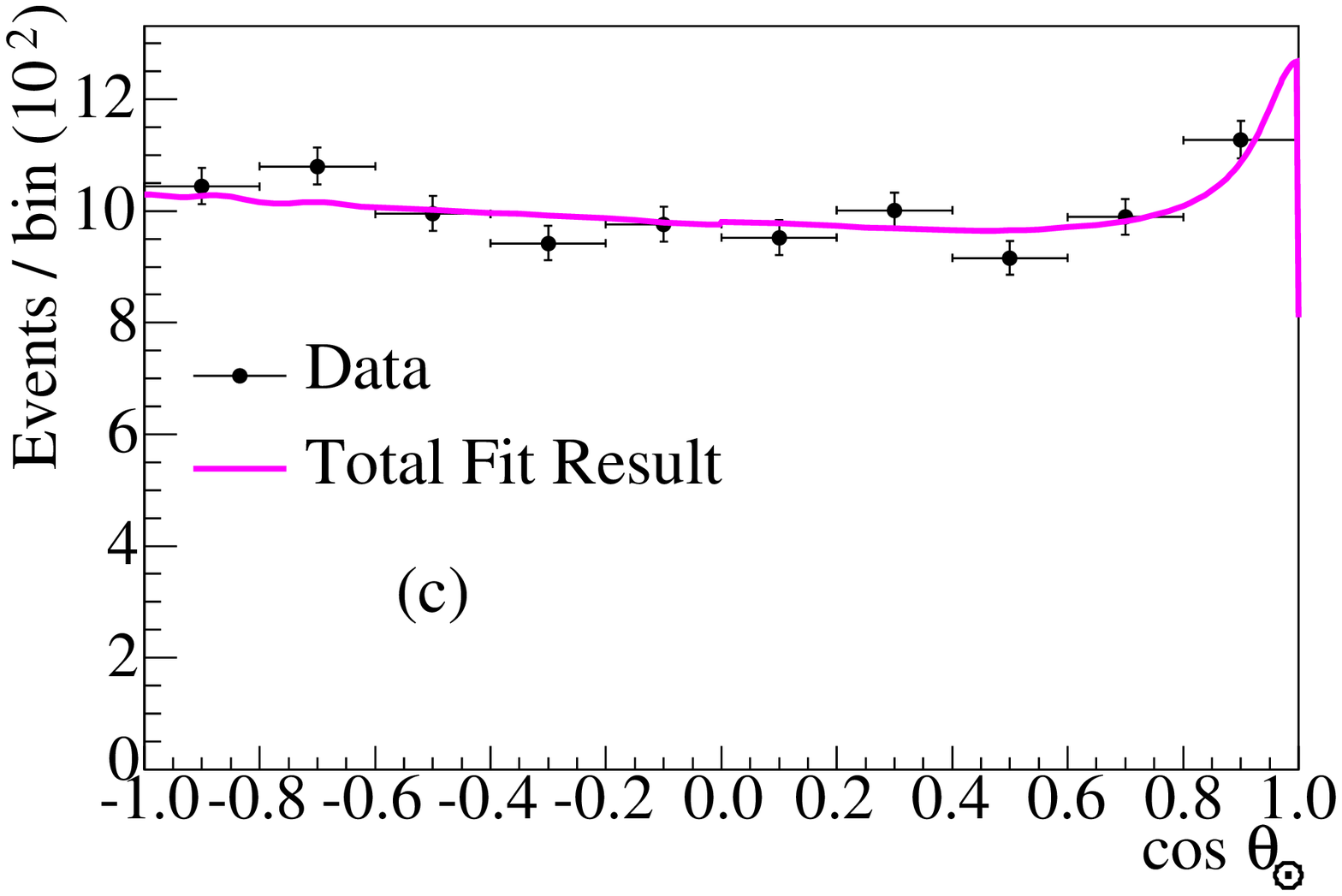}
\includegraphics[width=0.35\textwidth]{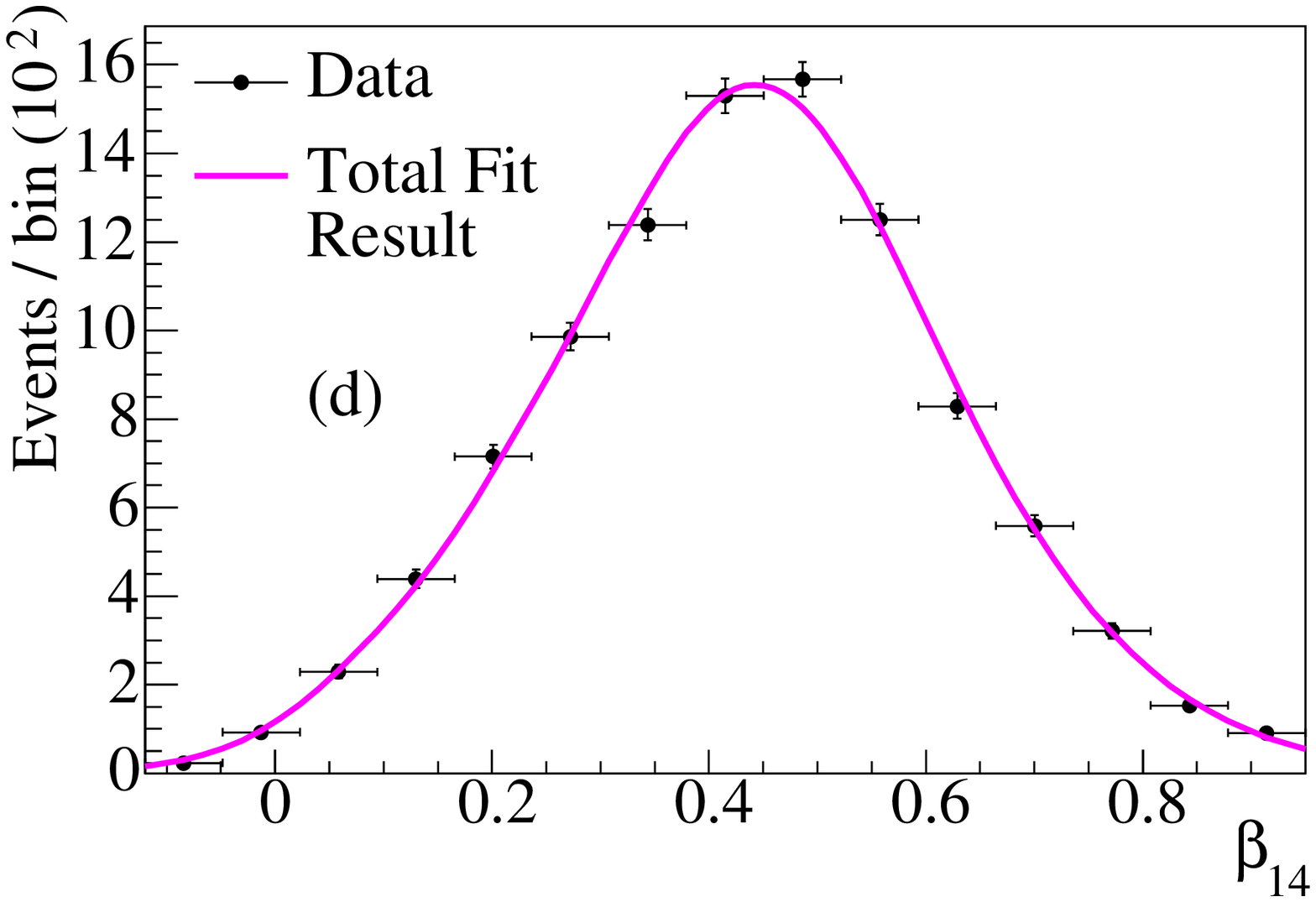}
\caption{\label{f:poly_1d_salt}(Color online) One dimensional
projections of the fit in Phase~II-night, from the polynomial survival
probability fit.  The panels show the fit projected onto (a) energy
($T_{\rm eff}$), (b) radius cubed ($R^3$), (c) direction (\ctsnosp),
and (d) isotropy (\betnosp).  The binning of data is purely for
display purposes; the fits were performed unbinned.}
\end{center} 
\end{figure}

This method for parameterizing the day/night asymmetry differs from
previous SNO analyses, which quoted an asymmetry for each interaction
type:
\begin{eqnarray}
	A = 2\frac{(\phi_N - \phi_D)}{(\phi_N + \phi_D)},
\end{eqnarray}
where $\phi_D$ and $\phi_N$ are the interaction rates measured for the
day and night data sets, respectively.  A combined analysis of the
results from Phase~I and Phase~II, assuming an undistorted neutrino
spectrum, gave a result of $A = 0.037 \pm 0.040$~\cite{nsp}.  For
comparison, the current analysis made no assumption about the shape of
the underlying neutrino spectrum, except that it is a smooth, slowly
varying function of $E_\nu$ over the range of neutrino energies to
which the SNO detector is sensitive.  The value of $a_0$ extracted
under this assumption was $a_0 = 0.032 \pm 0.040$.  Uncertainty on the
day/night asymmetry measurement has always been dominated by
statistics, so the improvements made to systematic uncertainties in
this analysis have a small effect.  The effect of the additional
statistics gained by going lower in energy appears to be balanced by
the additional degrees of freedom allowed in the shape of the neutrino
energy spectrum.

The one-dimensional projections of the fits in the observable
parameters for Phase~I-day and Phase~II-night are shown in Figures
\ref{f:poly_1d_d2o} and \ref{f:poly_1d_salt}.

\subsection{Mixing Parameters}
\label{res:mixp}

A three-flavor, active solar neutrino oscillation model has four
parameters: $\theta_{12}$ and $\theta_{13}$, which quantify the
strength of the mixing between flavor and mass eigenstates, and
$\Delta m^2_{21}$ and $\Delta m^2_{31}$, the differences between the
squares of the masses of the neutrino propagation eigenstates.  The
approximation of $\Delta m^2_{31} \sim \Delta m^2_{32}$ can be made
because $|\Delta m^2_{32}| \gg |\Delta m^2_{21}|$, while the remaining
mixing angle, $\theta_{23}$, and the CP-violating phase, $\delta$, are
irrelevant for the oscillation analysis of solar neutrino data.

For the sake of comparison with other oscillation analyses, this work
employed $\tan^2\theta_{12}$ to quantify the leading effects of the
mixing angles for solar neutrino oscillations.  Smaller effects due to
$\theta_{13}$ are quantified with $\sin^2\theta_{13}$.  The value of
$\Delta m^2_{31}$ was fixed to $+2.3\times 10^{-3}\
\mathrm{eV^2}$~\cite{pdg08}, an assumption that was necessary for the
numerical determination of the three-flavor survival probabilities,
but whose precise value had very little impact on our calculation.

The parameters describing the $P_{ee}(E_\nu)$ function for solar
neutrinos are, in order of importance, $\theta_{12}$, $\Delta
m^2_{21}$, $\theta_{13}$, and $\Delta m^2_{31}$. For experiments
sensitive to neutrinos from terrestrial sources, near the detector,
the survival probabilities were accurately calculated using a formula
without the effect of matter. The inclusion of matter effects in the
survival probability calculation for solar neutrino experiments
involves the numerical integration of a system of coupled differential
equations:
\begin{equation}
i \, \frac{d}{dx} \psi_{\alpha}(x) = H_f \, \psi_{\alpha}(x)  \, ,
\end{equation}
where $H_f$ is the Hamiltonian in flavor space, including matter
effects in both the Sun and the Earth, $x$ is the position along the
propagation direction, and $\psi_{\alpha}(x)$ is a vector containing
the real and imaginary coefficients of the wave function, where
$\alpha = $ (e, $\mu$, $\tau$). The system was solved for each new
value of $x$ as the wave function was propagated from the Sun to a
given detector on the Earth.  The probabilities were then calculated
from the magnitudes of the wave function coefficients.  The
integration was performed with the adaptative Runge-Kutta algorithm.
Radial profiles of the electron density and neutrino production in the
Sun were taken from the BS05(OP) model~\cite{bs05}.  The matter
density inside the Earth was taken from the Preliminary Reference
Earth Model (PREM)~\cite{prem}, which is the most widely accepted data
since the density profile is inferred from seismological
considerations.  For more details on the survival probability
calculation, see~\cite{olivier}.

Constraints on neutrino mixing parameters can be derived by comparing
neutrino oscillation model predictions with experimental data, as has
been done in previous SNO analyses~\cite{longd2o,nsp,snoncd}.  The
approach for the interpretation of the solar and reactor neutrino data
used the covariance $\chi^2$ method.  From a series of observables
with an associated set of measured parameters from a number of
experiments, the corresponding theoretical expectations were
calculated for a given neutrino oscillation parameter hypothesis.  In
order to calculate the model prediction for the neutrino yield at a
given detector, each of the neutrino fluxes that the detector was
sensitive to was weighted with the neutrino survival probabilities,
convolved with the cross-sections for the neutrino-target interactions
as well as with the detector response function, and then considered
above the experiment's energy threshold.  The $\chi^2$ function
quantifies the difference between the experimental data and
theoretical model expectation for the observable under study.

In the results presented here, the free parameters were the neutrino
mixing parameters and the total flux of the $^8$B and hep
neutrinos. The survival probabilities and, hence, the fluxes and
spectra of solar neutrinos and reactor antineutrinos were fully
constrained by the mixing parameters.  The $\chi^2$ function in each
case was minimized over a fine grid of points with respect to
$\tan^2\theta_{12}$, $\sin^2\theta_{13}$, and $\Delta m^2_{21}$.  The
$\Delta \chi^2 = \chi^2 - \chi^2_{\rm{min}}$ differences were the
indicators of the confidence levels (C.L.) in the one- and
two-dimensional projections.  The 68\%, 95\%, and 99.78\% C.L. regions
in two-dimensional parameter projections were drawn following the
standard definitions: $\Delta \chi^2 =$ 2.279, 5.99, and 11.83,
respectively. For one-dimensional projections the errors on the
parameter were the standard $1 \sigma$ C.L. at $\Delta \chi^2 =
1$. For all projections shown in this section, the $\chi^2$ was
minimized with respect to the undisplayed parameters at each point in
the MSW space.

The information from the LETA survival probability measurement was
included by evaluating the polynomial survival probability and
day/night asymmetry (as defined in Eqs.~\eqref{eq:dn}
and~\eqref{eq:poly} of Sec.~\ref{s:kerpoly}) that best represented the
model prediction at each point in the MSW plane.  To do this, it was
necessary to take into account the sensitivity of the SNO detector
(including effects such as the energy dependence of the cross
sections, reaction thresholds, and analysis cuts) so that the
parameterization of the model prediction at each point in the MSW
plane sampled the neutrino energy spectrum in the same manner and over
the same range as the data.  We calculated the number of detected
events that passed all the cuts as a function of neutrino energy using
the Monte Carlo simulation, and what was thus equivalent to a
`detected neutrino energy spectrum' (given in Table~\ref{t:enu_spec}
in Appendix~\ref{a:tables}) was distorted by the model-predicted
survival probability at each point in the MSW plane.  This was fit to
a similarly obtained spectrum, now distorted by the polynomial
parameterization, allowing the five polynomial parameters to vary in
the fit.  At each point in the plane, we then calculated the $\chi^2$
value of the fit of the model-predicted polynomial parameters ($c_0$,
$c_1$, $c_2$, $a_0$, and $a_1$) to the result from the signal
extraction, taking into account all uncertainties and correlations as
output by the signal extraction fit.  The SNO rates from
Phase~III~\cite{snoncd} were treated as a separate data set.

\begin{figure}[!ht]
\begin{center}
\includegraphics[width=0.48\textwidth]{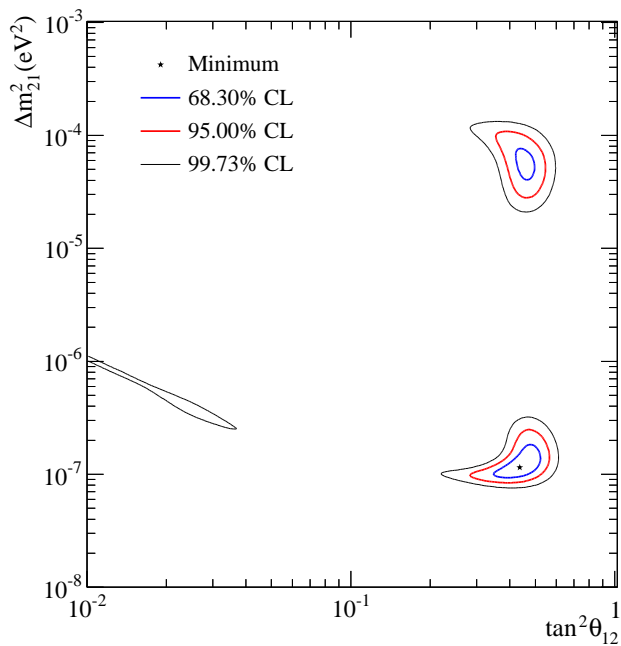}
\caption{(Color) SNO (LETA~+~Phase~III) two-flavor oscillation
parameter analysis.\label{f:contour-12-snoleta}}
\end{center} 
\end{figure}

Figure~\ref{f:contour-12-snoleta} shows the allowed regions of the
$(\tan^2\theta_{12},\Delta m^2_{21})$ parameter space when the LETA
data were analyzed in combination with the rates from
Phase~III~\cite{snoncd}.  The $2\nu$ contours were projected from the
parameter space at a constant value of $\sin^2\theta_{13}=0.0$, making
them equivalent to an effective two-flavor analysis.  While the best
fit point falls in the so-called `LOW' region, with $\Delta m^2_{21} =
1.15\,^{+0.38}_{-0.18}\times 10\,^{-7}(\mathrm{eV}^2)$ and
$\tan^2\theta_{12} =0.437\,^{+0.058}_{-0.058}$, the significance
levels of the LOW and the higher mass Large Mixing Angle (LMA) regions
are very similar.  The predicted shape for the survival probability is
very flat in both regions, and the day/night asymmetry is expected to
be small, so the SNO-only analysis has little handle on distinguishing
the two regions.  A notable difference between LOW and LMA is in the
predicted sign of the slope of the energy dependence of the day/night
asymmetry, with LOW predicting a negative slope, as was extracted in
the polynomial survival probability signal extraction fit reported in
Sec.~\ref{res:sprob}.

As described above, the observables from the SNO LETA fit used in the
hypothesis testing were the polynomial parameters of the survival
probability.  In a full global analysis, event yields were used for
the other solar neutrino experiments, including the SNO Phase~III
results.  For each set of parameters, the oscillation model was used
to predict the rates in the Chlorine~\cite{home2},
Gallium~\cite{sage}, and Borexino~\cite{bor} experiments, the
Super-Kamiokande Phase~I zenith spectra~\cite{sk1} and Phase~II
day/night spectra~\cite{sksol}, and the KamLAND rates and
spectrum~\cite{kam}, as well as the SNO rates~\cite{snoncd} and
spectra.  The expected rates and spectra were divided by the
respective predictions, calculated without oscillations, to remove the
effects of the model scaling factors.  The unitless rates were then
used in the global $\chi^2$ calculation.

Although the $\Phi_{^8{\rm B}}$ scale was determined in the LETA
signal extraction, we re-introduced it as a free parameter in the
$\chi^2$ minimization at each point in the parameter space to
constrain it with all solar data.  The uncertainty of the scale was
retrieved from its marginal distribution, as was done for the
oscillation parameters.

The SNO LETA covariance matrix was taken from the signal extraction
output given in Table~\ref{t:poly_corr}, as before.  For other
experiments, the total covariance matrix was assembled from the
individual statistical and systematic components, as described
in~\cite{nsp}.  Correlations between SNO's LETA and other solar
experimental results were allowed via the floated $\Phi_{^8{\rm B}}$
scale parameter.

The KamLAND rates and spectrum were predicted using three-flavor
vacuum oscillations.  Publicly available information about the KamLAND
detector and nearby reactors were included in our calculation, which
reproduced the unoscillated spectrum of Fig.~1 of Ref.~\cite{kam} with
good accuracy.  To include the effects of three-flavor oscillations,
we then compared the $\chi^2$ obtained with non-zero values of
$\theta_{13}$ with those obtained with $\theta_{13}=0$, for each set
of ($\tan^{2}\theta_{12}$,\,$\Delta m^2_{21}$) values.  In this way,
we built a $\Delta\chi^2$ function to parameterize the change of the
$\chi^2$ map in Fig.~2 of Ref.~\cite{kam} due to a non-zero value of
$\theta_{13}$.  This allowed us to include the KamLAND experiment in
our three-flavor neutrino oscillation analysis and to precisely
reproduce KamLAND's two-flavor neutrino contours.  When including the
KamLAND antineutrino spectrum we assumed CPT invariance, and we used
the KamLAND data only to constrain the oscillation parameters (as
opposed to the $^8$B flux scale), whereas all other solar neutrino
rates were used to collectively determine the absolute scale of the
${}^{8}\mathrm{B}$ neutrino flux as well as the oscillation
parameters.

\begin{figure}[!ht]
\begin{center}
\includegraphics[width=0.42\textwidth]{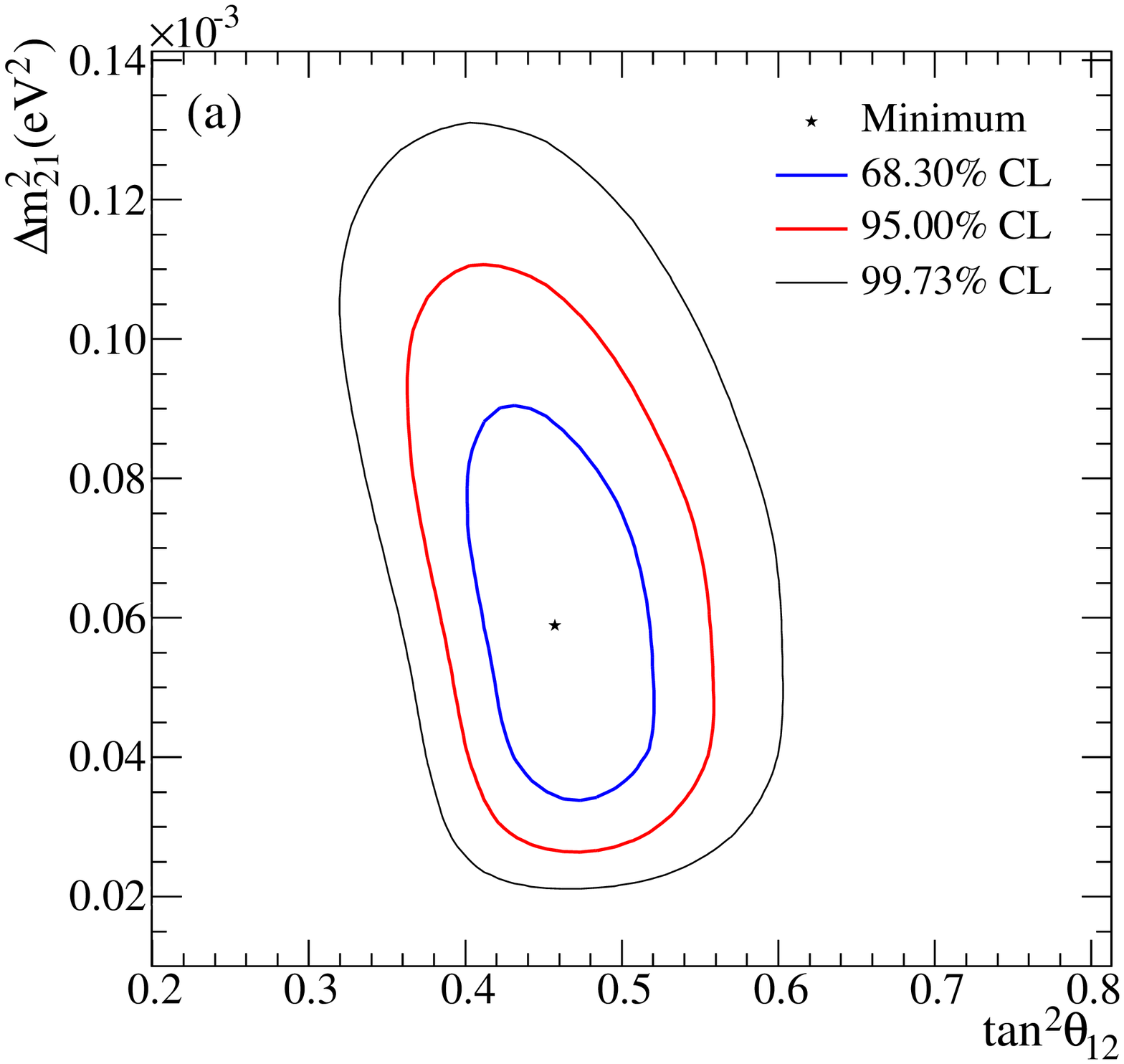}
\includegraphics[width=0.42\textwidth]{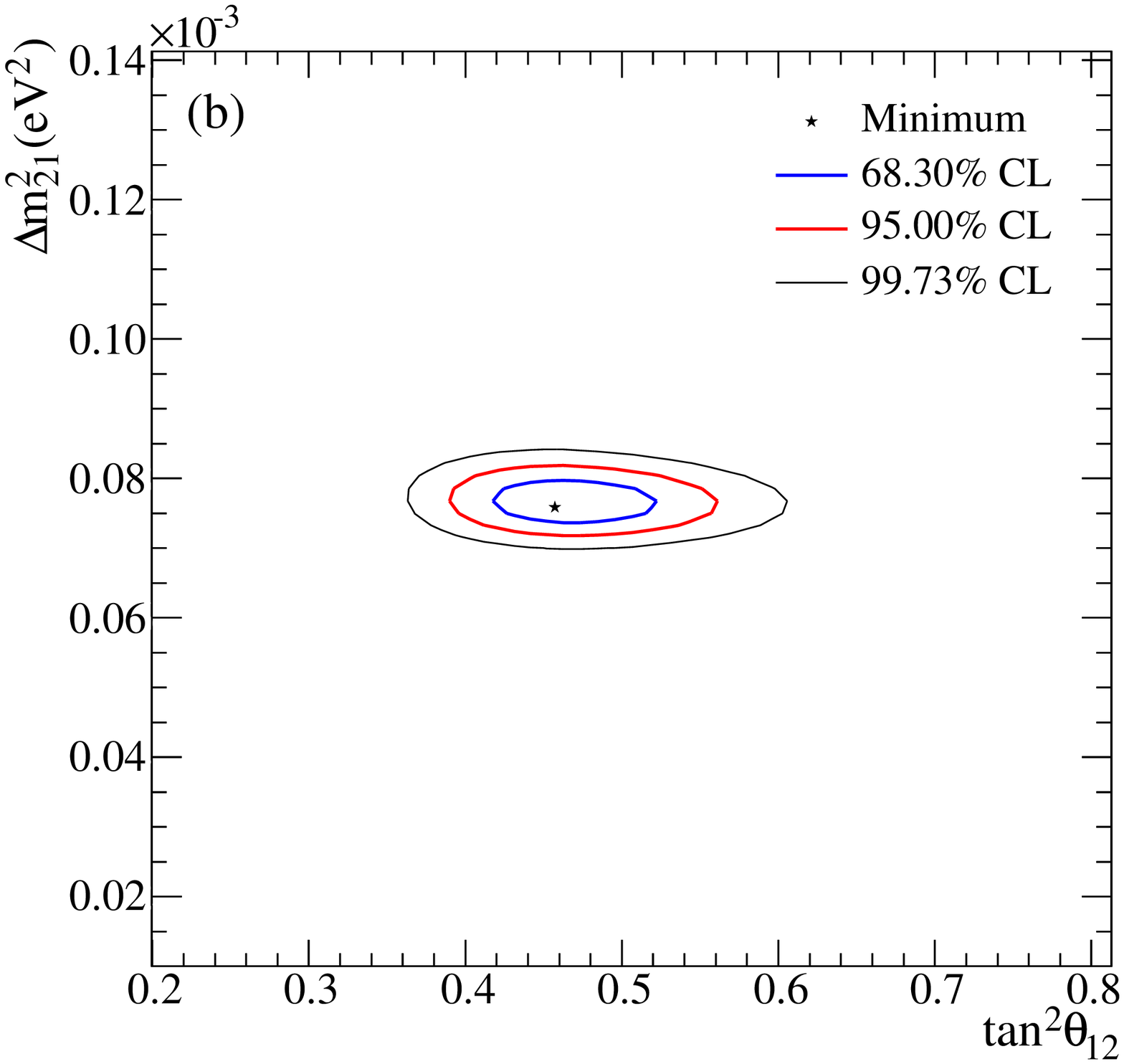}
\caption{(Color) Two-flavor oscillation parameter analysis for a)
global solar data and b) global solar + KamLAND data.  The solar data
includes: SNO's LETA survival probability day/night curves; SNO
Phase~III integral rates; Cl; SAGE; Gallex/GNO; Borexino; SK-I zenith
and SK-II day/night spectra.
\label{f:contour-12-solar}} 
\end{center} 
\end{figure}

Figure~\ref{f:contour-12-solar} shows the allowed regions of the
$(\tan^2\theta_{12},\Delta m^2_{21})$ parameter space when the global
solar data and the KamLAND data were analyzed, both separately and
together, in a two-flavor analysis.  It is interesting to note that
the global solar analysis does not significantly alter the constraints
in the LMA region relative to the SNO-only analysis.

\begin{figure}[!ht]
\begin{center}
\includegraphics[width=0.42\textwidth]{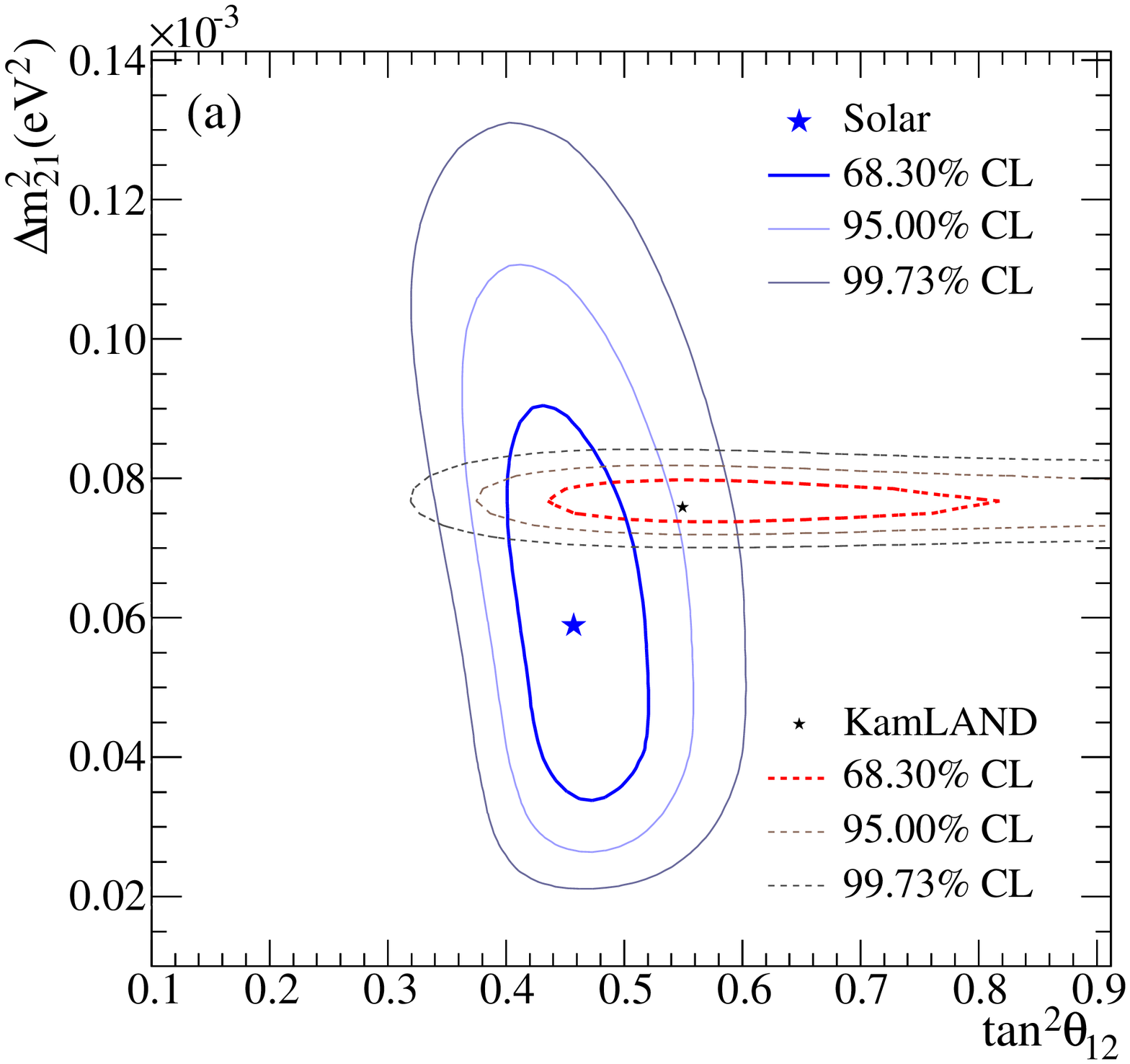}
\includegraphics[width=0.42\textwidth]{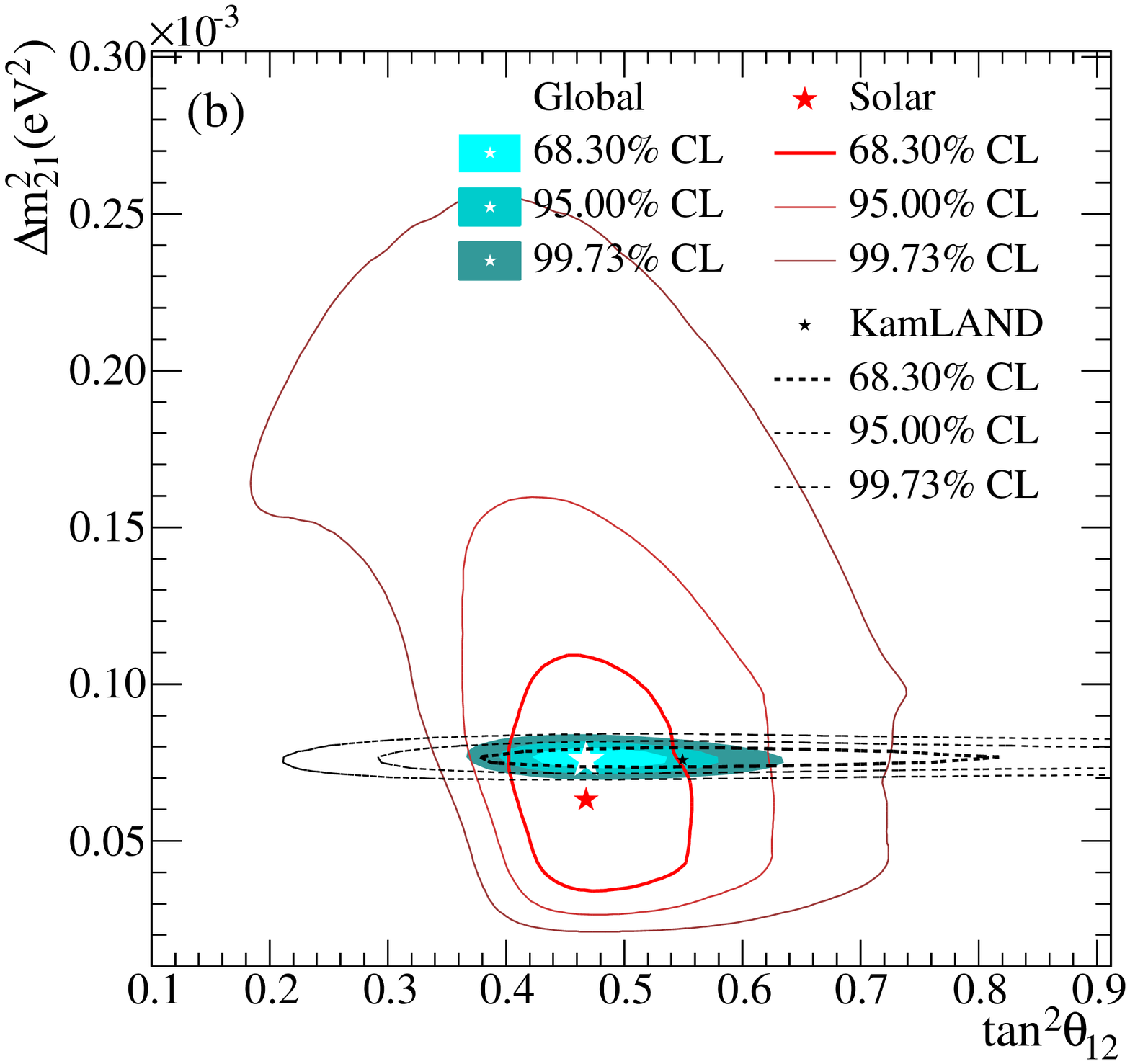}
\caption{(Color) Solar and KamLAND oscillation parameter analysis for
a) a two-flavor oscillation hypothesis and b) a three-flavor
hypothesis.  The solar data includes SNO's LETA survival probability
day/night curves, SNO Phase~III integral rates, Cl, SAGE, Gallex/GNO,
Borexino, SK-I zenith and SK-II day/night spectra.  The $\chi^2$ is
minimized with respect to all undisplayed parameters, including
$\sin^2\theta_{13}$ and $\Phi_{^8{\rm B}}$.\label{f:contour-3nu} }
\end{center} 
\end{figure}

Figure~\ref{f:contour-3nu} shows the results of a three-flavor
oscillation analysis.  Fig.~\ref{f:contour-3nu}(a) shows an overlay of
the global solar and the KamLAND allowed regions in
$(\tan^2\theta_{12},\Delta m^2_{21})$ parameter space, under a
two-flavor hypothesis.  Fig.~\ref{f:contour-3nu}(b) shows the same
overlay for the three-flavor hypothesis.  Allowing the value of
$\sin^2\theta_{13}$ to be non-zero clearly brings the two regions into
much better agreement.  The three-flavor contours show the effect of
allowing both $\Phi_{^8{\rm B}}$ and $\sin^2\theta_{13}$ to float at
each point in space.  Allowing these extra degrees of freedom worsens
the uncertainties on the two dominant oscillation parameters,
$\tan^2\theta_{12}$ and $\Delta m^2_{21}$.  The regions obtained with
all solar data are consistent with the SNO-only data and show an
extension of the space towards larger values of $\tan^2\theta_{12}$
when $\sin^2\theta_{13}$ is allowed to vary.  In contrast, the
three-flavor KamLAND contours show an extension towards smaller values
of $\tan^2\theta_{12}$.

\begin{figure}[!ht]
\begin{center}
\includegraphics[width=0.42\textwidth]{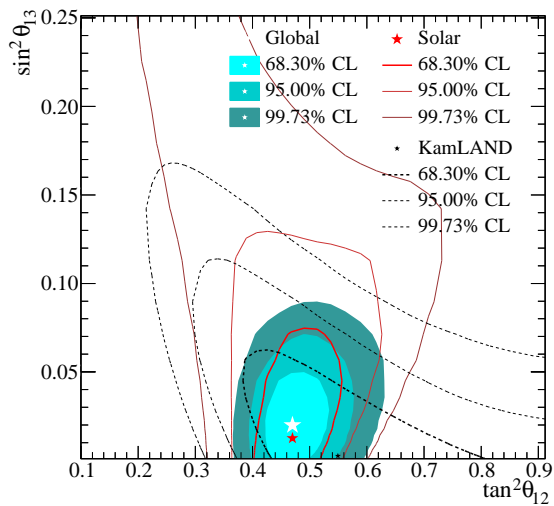}
\caption{(Color) Solar oscillation parameter analysis, identical to
Fig.~\ref{f:contour-3nu}(b), but projected in the mixing angle space.
The $\chi^2$ is minimized with respect to all undisplayed parameters,
including $\Delta m^2_{21}$ and $\Phi_{^8{\rm
B}}$.\label{f:contour-13-solarkam}}
\end{center} 
\end{figure}

Figure~\ref{f:contour-13-solarkam} shows the confidence regions in the
$(\tan^2\theta_{12},\sin^2\theta_{13})$ space.  The directionality of
the contours explains the excellent agreement of $\tan^2\theta_{12}$
between the solar and KamLAND experiments when $\sin^2\theta_{13}$ is
allowed to vary in the fit.

Tables~\ref{t:oscpars} and~\ref{t:oscpars3} summarize the oscillation
parameter results from the various two- and three-flavor oscillation
analyses, respectively.  When all solar experiments are combined with
data from the KamLAND reactor antineutrino experiment in a two-flavor
fit, the best fit point is found to be at
$\theta_{12}=34.06\,^{+1.16}_{-0.84}$ degrees and $\Delta
m^2_{21}=7.59\,^{+0.20}_{-0.21}\times 10^{-5}$~eV$^2$.  The
uncertainty on the mixing angle has been noticeably reduced in
comparison to SNO's previous analyses, resulting in the world's best
measurement of $\theta_{12}$ to date.  The global value of
$\Phi_{^8{\rm B}}$ from this fit is extracted to a precision of
$^{+2.38}_{-2.95}$\%.  The combination with KamLAND in a three-flavor
fit has allowed us to constrain $\sin^2\theta_{13}$, giving a value of
$\sin^2\theta_{13}=2.00^{+2.09}_{-1.63}\times 10^{-2}$.  This implies
an upper bound of $\sin^2\theta_{13}< 0.057$ (95\% C.L.).

\begin{table}[!h]
\begin{center}
\begin{tabular}{lcc}
\hline \hline Oscillation analysis & $\tan^2\theta_{12}$ & $\Delta
m^2_{21}(\mathrm{eV}^2)$\\ \hline SNO (LOW) &
$0.437\,^{+0.058}_{-0.058}$ & $1.15\,^{+0.38}_{-0.18}\times
10\,^{-7}$\\ SNO (LMA) & $0.457\,^{+0.038}_{-0.042}$ &
$5.50\,^{+2.21}_{-1.62}\times 10\,^{-5}$\\ Solar &
$0.457\,^{+0.038}_{-0.041}$ & $5.89\,^{+2.13}_{-2.16}\times
10\,^{-5}$\\ Solar+KamLAND & $0.457\,^{+0.040}_{-0.029}$ &
$7.59\,^{+0.20}_{-0.21}\times 10\,^{-5}$\\ \hline &
$\chi^2_{\mathrm{min}}/\mathrm{ndf}$ & $\Phi_{^8{\rm B}}$ ($\times
10^6\,\rm cm^{-2}\,s^{-1} $)\\ \hline SNO (LOW) & $6.80/9$ &
$5.013\,^{+0.176}_{-0.199}$ \\ SNO (LMA) & $8.20/9$ &
$4.984\,^{+0.205}_{-0.182}$ \\ Solar & $67.5/89$ &
$5.104\,^{+0.199}_{-0.148}$\\ Solar+KamLAND & $82.8/106$ &
$5.013\,^{+0.119}_{-0.148}$\\ \hline \hline
\end{tabular}
\caption{Best-fit neutrino oscillation parameters and extracted $^8$B
flux from a two-flavor oscillation analysis.  The `SNO' results are
from the combined LETA + Phase~III oscillation analysis.
Uncertainties listed are $\pm 1\sigma$ after the $\chi^2$ was
minimized with respect to all other parameters.}
\label{t:oscpars}
\end{center}
\end{table}

\begin{table}[!h]
\begin{center}
\begin{tabular}{lcc}
\hline \hline Oscillation analysis & $\tan^2\theta_{12}$ & $\Delta
m^2_{21}(\mathrm{eV}^2)$\\ \hline Solar & $0.468\,^{+0.052}_{-0.050}$
& $6.31\,^{+2.49}_{-2.58}\times 10\,^{-5}$\\ Solar+KamLAND &
$0.468\,^{+0.042}_{-0.033}$ & $7.59\,^{+0.21}_{-0.21}\times
10\,^{-5}$\\ \hline & $\chi^2_{\mathrm{min}}/\mathrm{ndf}$ &
$\Phi_{^8{\rm B}}$ ($\times 10^6\,\rm cm^{-2}\,s^{-1} $)\\ \hline
Solar & $67.4/89$ & $5.115\,^{+0.159}_{-0.193}$\\ Solar+KamLAND &
$81.4/106$ & $5.087\,^{+0.171}_{-0.159}$\\ \hline
&\multicolumn{2}{c}{$\sin^2\theta_{13}(\times 10\,^{-2})$} \\ \hline
Solar & \multicolumn{2}{c}{$< 8.10\, (95\%\, {\rm\, C.L.)}$}\\
Solar+KamLAND& \multicolumn{2}{c}{$2.00\,^{+2.09}_{-1.63}$}\\ \hline
\hline
\end{tabular}
\caption{Best-fit neutrino oscillation parameters and extracted $^8$B
flux from a three-flavor oscillation analysis.  Uncertainties listed
are $\pm 1\sigma$ after the $\chi^2$ was minimized with respect to all
other parameters.}
\label{t:oscpars3}
\end{center}
\end{table}

\section{Summary and Conclusions}
\label{s:summary}
	We have described here a joint low energy threshold analysis
of SNO's Phase~I and Phase~II data sets down to an effective kinetic
energy threshold of $T_{\rm eff}=3.5$~MeV.  The low threshold
increased the statistics of the CC and ES events by roughly 30\%, and
of NC events by $\sim$70\%.  A new energy estimator improved the
energy resolution by 6\%, thus reducing the number of background
events reconstructing above threshold by $\sim$60\%.  Separation of
electron-like and neutron capture events was improved by the joint fit
of data from Phases~I and~II, due to the difference in neutron
detection sensitivity in the two phases.  In addition, use of
calibration data to correct the Monte Carlo-generated PDF shapes, and
reduction of systematic uncertainties, have all contributed to
increased precision on both the total $^8$B solar neutrino flux and
the derived neutrino mixing parameters.  Fitting our data without
constraints on the shape of the underlying neutrino energy spectrum or
the unitarity of the mixing matrix gives a total $^8$B neutrino flux
of $\phi_{\rm NC} =
5.14^{+0.21}_{-0.20}\mbox{\,(stat\,$\oplus$\,syst)} \times
10^6$cm$^{-2}$ s$^{-1}$, measured by the NC reaction only, where
$\oplus$ refers to the quadrature sum.  This is in good agreement with
the predictions of recent Standard Solar Models.  The uncertainties on
this result are more than a factor of two better than in our previous
publications.  The CC and ES reconstructed electron spectra for this
fit are consistent with the hypothesis of no spectral distortion, and
with the best fit LMA point.

	We have also used the unique capabilities of the SNO detector
to perform the first direct fit to data for the energy-dependent
$\nu_e$ survival probability, without any reference to flux models or
other experiments.  The fit for the survival probability assumes
unitarity of the neutrino mixing matrix, and that the underlying
neutrino spectrum follows a smoothly-distorted $^8$B shape. We have
parameterized the survival probability as a second-order polynomial,
allowing for a linear energy-dependent asymmetry between day and night
spectra.  The fit gives us a total $^8$B neutrino flux of
$\Phi_{^8{\rm B}} =
5.05^{+0.19}_{-0.20}\mbox{\,(stat\,$\oplus$\,syst)} \times
10^6$cm$^{-2}$ s$^{-1}$.  No evidence for either a significant
spectral distortion or a day/night asymmetry was found.

	With the results of the survival probability fit, we have
created contours that show the allowed regions of the mixing
parameters, finding that for SNO data alone the best fit point is in
the LOW region of parameter space, but consistent with the LMA region
at the 68.3\% confidence level.  Combining all solar experiments and
the KamLAND reactor antineutrino experiment in a two-flavor fit, we
find the best fit point is at $\theta_{12}=34.06\,^{+1.16}_{-0.84}$
degrees and $\Delta m^2_{21}=7.59\,^{+0.20}_{-0.21}\times
10^{-5}$~eV$^2$.  The uncertainty on the mixing angle has been
noticeably reduced from SNO's previous analyses, resulting in the
world's best measurement of $\theta_{12}$.  The global value of
$\Phi_{^8{\rm B}}$ from this fit was extracted to a precision of
$^{+2.38}_{-2.95}$\%.  In a three-flavor fit, we find
$\sin^2\theta_{13}=2.00^{+2.09}_{-1.63}\times 10^{-2}$.  This
implies an upper bound of $\sin^2\theta_{13}< 0.057$
at the 95\% confidence level.

\section{Acknowledgments}
 This research was supported by: Canada: Natural Sciences and
Engineering Research Council, Industry Canada, National Research
Council, Northern Ontario Heritage Fund, Atomic Energy of Canada,
Ltd., Ontario Power Generation, High Performance Computing Virtual
Laboratory, Canada Foundation for Innovation, Canada Research Chairs;
US: Department of Energy, National Energy Research Scientific
Computing Center, Alfred P. Sloan Foundation; UK: Science and
Technology Facilities Council; Portugal: Funda\c{c}\={a}o para a
Ci\^{e}ncia e a Technologia.  We thank the SNO technical staff for
their strong contributions. We thank the University of Liverpool and
the Texas Advanced Computing Center for their grants of CPU time. We
thank NVIDIA for the donation of a Tesla graphics card.  We thank Vale
Inco, Ltd. for hosting this project.

\bibliographystyle{apsrev}

\appendix

\section{SNO Results: Data Tables}
\label{a:tables}

Table~\ref{t:neutflux} shows the extracted numbers of neutrino events
from the binned-histogram signal extraction fit in each phase.  The
total statistical plus systematic uncertainties are given.

\begin{table}[!h]
\begin{center}
\begin{tabular}{lcc}
\hline \hline Parameter & \multicolumn{2}{c}{Number of Events} \\ &
\multicolumn{1}{c}{Phase~I} & \multicolumn{1}{c}{Phase~II} \\ \hline
CC 3.5--4.0~MeV & $-$15.47$\,^{+76.96}_{-76.06}$ &
$-$21.78$\,^{+108.36}_{-107.09}$ \\ CC 4.0--4.5~MeV &
69.98$\,^{+25.19}_{-24.44}$ & 98.39$\,^{+35.41}_{-34.36}$ \\ CC
4.5--5.0~MeV & 147.00$\,^{+17.26}_{-16.42}$ &
205.70$\,^{+24.16}_{-22.98}$ \\ CC 5.0--5.5~MeV &
154.37$\,^{+17.35}_{-16.53}$ & 215.89$\,^{+24.27}_{-23.11}$ \\ CC
5.5--6.0~MeV & 180.52$\,^{+16.93}_{-16.59}$ &
252.94$\,^{+23.72}_{-23.25}$ \\ CC 6.0--6.5~MeV &
174.63$\,^{+14.99}_{-14.94}$ & 244.55$\,^{+21.00}_{-20.92}$ \\ CC
6.5--7.0~MeV & 175.52$\,^{+13.53}_{-13.63}$ &
245.02$\,^{+18.88}_{-19.03}$ \\ CC 7.0--7.5~MeV &
163.28$\,^{+11.90}_{-12.02}$ & 227.47$\,^{+16.57}_{-16.75}$ \\ CC
7.5--8.0~MeV & 161.09$\,^{+10.93}_{-11.10}$ &
224.83$\,^{+15.26}_{-15.50}$ \\ CC 8.0--8.5~MeV &
142.23$\,^{+9.73}_{-9.98}$ & 198.41$\,^{+13.57}_{-13.92}$ \\ CC
8.5--9.0~MeV & 119.69$\,^{+8.61}_{-8.86}$ &
167.51$\,^{+12.05}_{-12.40}$ \\ CC 9.0--9.5~MeV &
101.34$\,^{+7.75}_{-8.04}$ & 142.44$\,^{+10.89}_{-11.29}$ \\ CC
9.5--10.0~MeV & 84.03$\,^{+6.90}_{-7.16}$ &
118.39$\,^{+9.71}_{-10.09}$ \\ CC 10.0--10.5~MeV &
82.49$\,^{+6.72}_{-7.08}$ & 116.39$\,^{+9.49}_{-9.99}$ \\ CC
10.5--11.0~MeV & 58.75$\,^{+5.69}_{-5.98}$ & 83.36$\,^{+8.07}_{-8.48}$
\\ CC 11.0--11.5~MeV & 25.90$\,^{+3.71}_{-3.83}$ &
36.88$\,^{+5.28}_{-5.46}$ \\ ES 3.5--4.0~MeV &
74.10$\,^{+21.21}_{-20.76}$ & 104.30$\,^{+29.85}_{-29.22}$ \\ ES
4.0--4.5~MeV & 55.00$\,^{+10.34}_{-9.98}$ &
77.34$\,^{+14.54}_{-14.04}$ \\ ES 4.5--5.0~MeV &
42.92$\,^{+7.63}_{-7.63}$ & 60.32$\,^{+10.72}_{-10.72}$ \\ ES
5.0--5.5~MeV & 35.90$\,^{+7.28}_{-7.28}$ & 50.37$\,^{+10.22}_{-10.21}$
\\ ES 5.5--6.0~MeV & 20.25$\,^{+6.27}_{-6.27}$ &
28.33$\,^{+8.78}_{-8.78}$ \\ ES 6.0--6.5~MeV &
15.25$\,^{+5.73}_{-5.73}$ & 21.33$\,^{+8.02}_{-8.01}$ \\ ES
6.5--7.0~MeV & 19.73$\,^{+5.51}_{-5.51}$ & 27.58$\,^{+7.70}_{-7.71}$
\\ ES 7.0--7.5~MeV & 23.97$\,^{+5.31}_{-5.32}$ &
33.69$\,^{+7.46}_{-7.47}$ \\ ES 7.5--8.0~MeV &
19.72$\,^{+4.91}_{-4.92}$ & 27.79$\,^{+6.92}_{-6.93}$ \\ ES
8.0--8.5~MeV & 18.75$\,^{+4.49}_{-4.51}$ & 26.54$\,^{+6.36}_{-6.39}$
\\ ES 8.5--9.0~MeV & 16.16$\,^{+4.01}_{-4.02}$ &
22.65$\,^{+5.61}_{-5.63}$ \\ ES 9.0--9.5~MeV &
11.47$\,^{+3.49}_{-3.49}$ & 16.38$\,^{+4.98}_{-4.99}$ \\ ES
9.5--10.0~MeV & 10.23$\,^{+3.14}_{-3.15}$ & 14.64$\,^{+4.49}_{-4.50}$
\\ ES 10.0--10.5~MeV & 4.38$\,^{+2.60}_{-2.60}$ &
6.27$\,^{+3.72}_{-3.72}$ \\ ES 10.5--11.0~MeV &
3.37$\,^{+2.26}_{-2.26}$ & 4.83$\,^{+3.24}_{-3.24}$ \\ ES
11.0--11.5~MeV & 5.18$\,^{+1.94}_{-1.95}$ & 7.44$\,^{+2.79}_{-2.80}$
\\ NC & 870.17$\,^{+35.07}_{-33.29}$ & 3257.04$\,^{+131.26}_{-124.61}$
\\ \hline \hline
\end{tabular}
\caption{Extracted number of events for each neutrino parameter from
the binned-histogram signal extraction fit, in each phase, with total
uncertainties.}
\label{t:neutflux}
\end{center}
\end{table}

Tables~\ref{t:sigsyst}--\ref{t:sigsyst2} show the effects of the
individual systematic uncertainties on the extracted NC rate, the CC
rate in two energy intervals (4.0--4.5~MeV and 9.5--10.0~MeV) and the
ES rate in the 3.5--4.0~MeV interval, taken from the binned-histogram
unconstrained signal extraction fit.

\begin{table}[!h]
\begin{center}
\begin{tabular}{lcrrrr}
\hline \hline Systematic &Phase & \multicolumn{4}{c}{Effect on rate
/\%} \\ && NC & CC1 & CC12 & ES0 \\ \hline $T_{\rm eff}$ scale (+) &
I, II & $-$0.293 & $-$2.037 & $-$2.144 & $-$0.156 \\ $T_{\rm eff}$
scale ($-$) & I, II & 0.137 & 0.475 & 0.913 & 0.035 \\ $T_{\rm eff}$
scale (+) & I & 0.030 & $-$0.956 & $-$0.337 & $-$0.148 \\ $T_{\rm
eff}$ scale ($-$) & I & $-$0.084 & 1.659 & 0.652 & 0.236 \\ $T_{\rm
eff}$ scale (+) & II & $-$0.307 & 0.317 & $-$1.094 & 0.105 \\ $T_{\rm
eff}$ scale ($-$) & II & 0.177 & $-$0.493 & 0.584 & $-$0.133 \\
$T_{\rm eff}$ resn (elec) (+) & I & 0.008 & $-$3.999 & $-$0.013 &
$-$0.439 \\ $T_{\rm eff}$ resn (elec) ($-$) & I & $-$0.030 & 7.656 &
0.017 & 1.399 \\ $T_{\rm eff}$ resn (elec) (+) & II & 0.653 & $-$5.005
& $-$0.006 & $-$0.531 \\ $T_{\rm eff}$ resn (elec) ($-$) & II &
$-$0.716 & 6.597 & 0.027 & 0.480 \\ $T_{\rm eff}$ resn (neut) (+) & I,
II & 0.065 & $-$0.054 & $-$0.023 & $-$0.006 \\ $T_{\rm eff}$ resn
(neut) ($-$) & I, II & $-$0.041 & $-$0.058 & 0.046 & 0.013 \\ $T_{\rm
eff}$ linearity (+) & I, II & 0.130 & $-$0.160 & 0.379 & $-$0.125 \\
$T_{\rm eff}$ linearity ($-$) & I, II & $-$0.132 & 0.287 & $-$0.372 &
0.301 \\ \bet elec scale (+) & I, II & 0.634 & $-$5.064 & $-$0.082 &
$-$0.648 \\ \bet elec scale ($-$) & I, II & $-$0.622 & 5.559 & 0.086 &
0.607 \\ \bet neut scale (+) & I, II & 0.719 & $-$1.962 & $-$0.040 &
$-$0.068 \\ \bet neut scale ($-$) & I, II & $-$0.411 & 1.204 & 0.029 &
0.048 \\ \bet elec width (+) & I, II & 0.306 & $-$1.263 & $-$0.079 &
$-$0.027 \\ \bet elec width ($-$) & I, II & $-$0.286 & 2.342 & 0.058 &
0.099 \\ \bet neut width (+) & I, II & 0.067 & $-$0.240 & $-$0.002 &
$-$0.014 \\ \bet neut width ($-$) & I, II & $-$0.054 & 0.217 & 0.012 &
0.017 \\ \bet E$-$dep (+) & I, II & 0.227 & 1.661 & $-$0.054 & 0.299
\\ \bet E$-$dep ($-$) & I, II & $-$0.246 & $-$0.999 & 0.068 & $-$0.228
\\ \hline \hline
\end{tabular}
\caption{Effect of systematic uncertainties in $T_{\rm eff}$ and \bet
on the NC rate, the CC rate in the intervals 4.0--4.5~MeV (``CC1'')
and 9.5--10.0~MeV (``CC12''), and the ES rate in the interval
3.5--4.0~MeV (``ES0'').  Systematics shown as applying to both phases
were treated as 100\% correlated between the phases.  The `(+)' and
`($-$)' labels refer to the result of applying the positive and
negative side of each double-sided uncertainty.}
\label{t:sigsyst}
\end{center}
\end{table}

\begin{table}[!h]
\begin{center}
\begin{tabular}{lcrrrr}
\hline \hline Systematic &Phase & \multicolumn{4}{c}{Effect on rate
/\%} \\ && NC & CC1 & CC12 & ES0 \\ \hline Angular resn (+) & I &
$-$0.032 & $-$0.688 & $-$0.075 & 1.176 \\ Angular resn ($-$) & I &
0.039 & 0.648 & 0.128 & $-$1.477 \\ Angular resn (+) & II & $-$0.058 &
$-$0.458 & $-$0.172 & 3.219 \\ Angular resn ($-$) & II & 0.065 & 0.298
& 0.194 & $-$3.488 \\ Axial scale (+) & I & $-$0.030 & 0.261 & 0.128 &
0.047 \\ Axial scale ($-$) & I & 0.188 & $-$2.377 & $-$0.746 &
$-$1.344 \\ Axial scale (+) & II & 0.030 & $-$0.366 & 0.079 & $-$0.037
\\ Axial scale ($-$) & II & $-$0.320 & $-$1.981 & $-$0.493 & $-$0.892
\\ Z scale (+) & I & $-$0.052 & 0.377 & 0.151 & 0.018 \\ Z scale ($-$)
& I & 0.000 & 0.000 & 0.000 & 0.000 \\ Z scale (+) & II & 0.004 &
0.007 & 0.018 & 0.044 \\ Z scale ($-$) & II & $-$0.070 & $-$0.906 &
$-$0.130 & $-$0.112 \\ X offset (+) & I & $-$0.002 & $-$0.075 &
$-$0.010 & $-$0.444 \\ X offset ($-$) & I & 0.004 & $-$0.103 &
$-$0.000 & 0.032 \\ X offset (+) & II & 0.009 & $-$0.538 & 0.009 &
$-$0.075 \\ X offset ($-$) & II & $-$0.007 & 0.002 & 0.003 & 0.022 \\
Y offset (+) & I & $-$0.035 & $-$0.034 & 0.000 & 0.009 \\ Y offset
($-$) & I & 0.005 & $-$0.084 & 0.002 & $-$0.101 \\ Y offset (+) & II &
$-$0.029 & $-$0.695 & 0.035 & $-$0.279 \\ Y offset ($-$) & II & 0.003
& $-$0.146 & 0.007 & 0.046 \\ Z offset (+) & I & 0.011 & $-$0.275 &
$-$0.032 & $-$0.642 \\ Z offset ($-$) & I & $-$0.003 & $-$0.060 &
0.002 & 0.112 \\ Z offset (+) & II & $-$0.168 & $-$1.009 & 0.006 &
$-$0.317 \\ Z offset ($-$) & II & $-$0.013 & 0.027 & 0.005 & 0.132 \\
X resn & I & $-$0.002 & $-$0.206 & $-$0.004 & $-$0.216 \\ X resn & II
& 0.052 & $-$0.732 & 0.003 & $-$0.020 \\ Y resn & I & $-$0.007 & 0.079
& $-$0.002 & $-$0.109 \\ Y resn & II & 0.038 & $-$0.417 & 0.019 &
$-$0.201 \\ Z resn & I & $-$0.003 & 0.173 & $-$0.002 & $-$0.224 \\ Z
resn & II & 0.115 & $-$1.354 & 0.023 & $-$0.418 \\ \hline \hline
\end{tabular}
\caption{Effect of systematic uncertainties in \cts and $R^3$ on the
NC rate, the CC rate in the intervals 4.0--4.5~MeV (``CC1'') and
9.5--10.0~MeV (``CC12''), and the ES rate in the interval 3.5--4.0~MeV
(``ES0'').  The `(+)' and `($-$)' labels refer to the result of
applying the positive and negative side of each double-sided
uncertainty.}
\label{t:sigsyst1}
\end{center}
\end{table}

\begin{table}[!h]
\begin{center}
\begin{tabular}{lcrrrr}
\hline \hline Systematic &Phase & \multicolumn{4}{c}{Effect on rate
/\%} \\ && NC & CC1 & CC12 & ES0 \\ \hline E$-$dep fid vol (+) & I &
0.397 & $-$0.277 & $-$1.735 & 0.378 \\ E$-$dep fid vol ($-$) & I &
$-$0.230 & 0.119 & 1.027 & $-$0.233 \\ E$-$dep fid vol (+) & II &
$-$0.698 & 0.794 & $-$1.144 & 0.322 \\ E$-$dep fid vol ($-$) & II &
0.825 & $-$0.994 & 1.376 & $-$0.389 \\ Cut acceptance (+) & I, II &
$-$0.357 & $-$0.519 & $-$0.434 & $-$0.451 \\ Cut acceptance ($-$) & I,
II & 1.039 & 1.299 & 1.136 & 1.171 \\ Photodisint.n (+) & I, II &
$-$0.180 & 0.134 & $-$0.002 & 0.026 \\ Photodisint.n ($-$) & I, II &
0.183 & $-$0.100 & 0.004 & $-$0.023 \\ neut cap (+) & I & $-$0.049 &
$-$0.797 & 0.003 & $-$0.074 \\ neut cap ($-$) & I & 0.044 & 0.829 &
$-$0.001 & 0.084 \\ neut cap (+) & II & $-$1.306 & 0.616 & $-$0.001 &
0.062 \\ neut cap ($-$) & II & 1.338 & $-$0.612 & 0.003 & $-$0.060 \\
neut cap (+) & I, II & $-$0.759 & 0.040 & $-$0.000 & $-$0.001 \\ neut
cap ($-$) & I, II & 0.770 & $-$0.053 & 0.001 & $-$0.011 \\ $^{24}$Na
model (+) & II & 0.028 & $-$0.751 & 0.008 & $-$0.056 \\ $^{24}$Na
model ($-$) & II & 0.067 & $-$0.463 & 0.003 & $-$0.182 \\ PMT $T_{\rm
eff}$ exponent (+) & I & 0.009 & $-$6.482 & $-$0.003 & $-$1.469 \\ PMT
$T_{\rm eff}$ exponent ($-$) & I & 0.002 & 3.217 & 0.004 & 0.821 \\
PMT $T_{\rm eff}$ exponent (+) & II & 0.046 & $-$0.814 & 0.001 &
$-$0.196 \\ PMT $T_{\rm eff}$ exponent ($-$) & II & 0.011 & $-$0.328 &
0.003 & 0.010 \\ PMT $R^3$ exponent (+) & I & $-$0.048 & $-$2.875 &
0.003 & $-$0.402 \\ PMT $R^3$ exponent ($-$) & I & 0.035 & 1.746 &
0.000 & 0.238 \\ PMT $R^3$ exponent (+) & II & 0.023 & $-$2.371 &
0.002 & $-$0.185 \\ PMT $R^3$ exponent ($-$) & II & 0.004 & 0.870 &
$-$0.000 & 0.440 \\ PMT $R^3$ offset (+) & I & 0.053 & 5.674 &
$-$0.004 & 0.774 \\ PMT $R^3$ offset ($-$) & I & $-$0.016 & $-$2.113 &
0.003 & $-$0.203 \\ PMT $R^3$ offset (+) & II & $-$0.005 & 0.735 &
$-$0.000 & 0.370 \\ PMT $R^3$ offset ($-$) & II & 0.001 & $-$1.014 &
0.003 & $-$0.111 \\ PMT \bet mean (+) & I & $-$0.042 & $-$2.271 &
0.002 & $-$0.714 \\ PMT \bet mean ($-$) & I & 0.062 & 0.559 & 0.000 &
0.509 \\ PMT \bet mean (+) & II & $-$0.516 & 4.456 & 0.029 & 0.396 \\
PMT \bet mean ($-$) & II & 0.524 & $-$4.102 & $-$0.027 & $-$0.802 \\
PMT \bet width (+) & I & 0.075 & $-$1.388 & $-$0.001 & $-$0.008 \\ PMT
\bet width ($-$) & I & $-$0.070 & 0.192 & 0.005 & 0.060 \\ PMT \bet
width (+) & II & 0.357 & $-$1.054 & $-$0.006 & 0.257 \\ PMT \bet width
($-$) & II & $-$0.365 & 1.394 & 0.009 & $-$0.459 \\ \hline \hline
\end{tabular}
\caption{Effect of relative normalization uncertainties and systematic
uncertainties in background PDFs on the NC rate, the CC rate in the
intervals 4.0--4.5~MeV (``CC1'') and 9.5--10.0~MeV (``CC12''), and the
ES rate in the interval 3.5--4.0~MeV (``ES0'').  Systematics shown as
applying to both phases were treated as 100\% correlated between the
phases.  The `(+)' and `($-$)' labels refer to the result of applying
the positive and negative side of each double-sided uncertainty.}
\label{t:sigsyst2}
\end{center}
\end{table}

The direct signal extraction fit to the $\nu_e$ survival probability
parameterized the neutrino fluxes as:
\begin{itemize}
	\item $\Phi_{^8{\rm B}}$ - the total $^8$B neutrino flux;
	\item $c_0$, $c_1$, $c_2$ - coefficients in a quadratic
	expansion of the daytime $\nu_e$ survival probability around
	$E_\nu = 10$~MeV;
	\item $a_0$, $a_1$ - coefficients in a linear expansion of the
	day/night asymmetry around $E_\nu = 10$~MeV.
\end{itemize}
Where the day/night asymmetry, $A$, daytime $\nu_e$ survival
probability, $P_{ee}^{\rm day}$, and nighttime $\nu_e$ survival
probability, $P_{ee}^{\rm night}$, that correspond to these parameters
are:
\begin{eqnarray}
	A(E_\nu) & = & a_0 + a_1(E_\nu - 10\;{\rm MeV}) \\ P_{ee}^{\rm
	day}(E_\nu) & = & c_0 + c_1 (E_\nu - 10\;{\rm MeV}) \nonumber
	\\ & & \; + c_2 (E_\nu - 10\;{\rm MeV})^2 \\ P_{ee}^{\rm
	night}(E_\nu) & = & P_{ee}^{\rm day} \times \frac{1 +
	A(E_\nu)/2}{1 - A(E_\nu)/2}
\end{eqnarray}

The best-fit polynomial parameter values and uncertainties are shown
in Table~\ref{t:poly_pars}, and the correlation matrix is shown in
Table~\ref{t:poly_corr}.

\begin{table}[!htb]
\begin{center}
\begin{tabular}{lrrrr}
\hline \hline Parameter & Value & \multicolumn{1}{c}{Stat} &
\multicolumn{1}{c}{Syst} & \multicolumn{1}{c}{D/N Syst} \\ \hline
$a_0$ & 0.0325 & $^{+0.0366}_{-0.0360}$ & $^{+0.0059}_{-0.0092}$ &
$^{+0.0145}_{-0.0148}$ \\ $a_1$ & $-$0.0311 & $^{+0.0279}_{-0.0292}$ &
$^{+0.0104}_{-0.0056}$ & $^{+0.0140}_{-0.0129}$ \\ $c_0$ & 0.3435 &
$^{+0.0205}_{-0.0197}$ & $^{+0.0111}_{-0.0066}$ &
$^{+0.0050}_{-0.0059}$ \\ $c_1$ & 0.00795 & $^{+0.00780}_{-0.00745}$ &
$^{+0.00308}_{-0.00335}$ & $^{+0.00236}_{-0.00240}$ \\ $c_2$ &
$-$0.00206 & $^{+0.00302}_{-0.00311}$ & $^{+0.00148}_{-0.00128}$ &
$^{+0.00057}_{-0.00074}$ \\ \hline \hline
\end{tabular}
\caption{Extracted polynomial parameter values, statistical
uncertainties, average systematic uncertainties, and day/night
systematic uncertainties from the survival probability fit.}
\label{t:poly_pars}
\end{center}
\end{table}

\begin{table}[!htb]
\begin{center}
\begin{tabular}{c|rrrrrr}
\hline \hline & \multicolumn{1}{c}{$\Phi_{^8{\rm B}}$} &
 \multicolumn{1}{c}{$a_0$} & \multicolumn{1}{c}{$a_1$} &
 \multicolumn{1}{c}{$c_0$} & \multicolumn{1}{c}{$c_1$} &
 \multicolumn{1}{c}{$c_2$} \\ \hline $\Phi_{^8{\rm B}}$ & 1.000 &
 $-$0.166 & 0.051 & $-$0.408 & 0.103 & $-$0.246 \\ $a_0$ & $-$0.166 &
 1.000 & $-$0.109 & $-$0.263 & 0.019 & $-$0.123 \\ $a_1$ & 0.051 &
 $-$0.109 & 1.000 & $-$0.005 & $-$0.499 & $-$0.031 \\ $c_0$ & $-$0.408
 & $-$0.263 & $-$0.005 & 1.000 & $-$0.101 & $-$0.321 \\ $c_1$ & 0.103
 & 0.019 & $-$0.499 & $-$0.101 & 1.000 & $-$0.067 \\ $c_2$ & $-$0.246
 & $-$0.123 & $-$0.031 & $-$0.321 & $-$0.067 & 1.000 \\ \hline \hline
\end{tabular}
\caption{Correlation matrix for the polynomial survival probability fit.}
\label{t:poly_corr}
\end{center}
\end{table}

\newpage

Table~\ref{t:enu_spec} lists the Monte Carlo-generated neutrino energy
spectrum for events that passed all the standard analysis cuts (the
``detected neutrino energy spectrum'').  Events are separated into
those occurring during the daytime and during the nighttime.

\clearpage

\begin{table*}[t]
\begin{center}
\begin{tabular}{c|c|c||c|c|c}
\hline \hline Energy (MeV) & Day & Night & Energy (MeV) & Day & Night
\\ \hline 2.2 & 7.717 $\times 10^{-6}$ & 7.726 $\times 10^{-6}$ & 9.0
& 7.675 $\times 10^{-1}$ & 9.699 $\times 10^{-1}$ \\ 2.6 & 7.211
$\times 10^{-5}$ & 8.505 $\times 10^{-5}$ & 9.4 & 7.858 $\times
10^{-1}$ & 9.970 $\times 10^{-1}$ \\ 3.0 & 5.074 $\times 10^{-4}$ &
6.592 $\times 10^{-4}$ & 9.8 & 7.882 $\times 10^{-1}$ & 1.000 \\ 3.4 &
2.168 $\times 10^{-3}$ & 2.992 $\times 10^{-3}$ & 10.2 & 7.666 $\times
10^{-1}$ & 9.723 $\times 10^{-1}$ \\ 3.8 & 7.339 $\times 10^{-3}$ &
8.796 $\times 10^{-3}$ & 10.6 & 7.298 $\times 10^{-1}$ & 9.251 $\times
10^{-1}$ \\ 4.2 & 1.599 $\times 10^{-2}$ & 1.971 $\times 10^{-2}$ &
11.0 & 6.725 $\times 10^{-1}$ & 8.524 $\times 10^{-1}$ \\ 4.6 & 3.165
$\times 10^{-2}$ & 3.948 $\times 10^{-2}$ & 11.4 & 5.974 $\times
10^{-1}$ & 7.573 $\times 10^{-1}$ \\ 5.0 & 6.130 $\times 10^{-2}$ &
7.632 $\times 10^{-2}$ & 11.8 & 5.117 $\times 10^{-1}$ & 6.485 $\times
10^{-1}$ \\ 5.4 & 1.099 $\times 10^{-1}$ & 1.375 $\times 10^{-1}$ &
12.2 & 4.137 $\times 10^{-1}$ & 5.256 $\times 10^{-1}$ \\ 5.8 & 1.768
$\times 10^{-1}$ & 2.221 $\times 10^{-1}$ & 12.6 & 3.167 $\times
10^{-1}$ & 4.000 $\times 10^{-1}$ \\ 6.2 & 2.595 $\times 10^{-1}$ &
3.266 $\times 10^{-1}$ & 13.0 & 2.211 $\times 10^{-1}$ & 2.807 $\times
10^{-1}$ \\ 6.6 & 3.491 $\times 10^{-1}$ & 4.403 $\times 10^{-1}$ &
13.4 & 1.368 $\times 10^{-1}$ & 1.748 $\times 10^{-1}$ \\ 7.0 & 4.398
$\times 10^{-1}$ & 5.560 $\times 10^{-1}$ & 13.8 & 7.208 $\times
10^{-2}$ & 9.023 $\times 10^{-2}$ \\ 7.4 & 5.260 $\times 10^{-1}$ &
6.667 $\times 10^{-1}$ & 14.2 & 2.965 $\times 10^{-2}$ & 3.786 $\times
10^{-2}$ \\ 7.8 & 6.061 $\times 10^{-1}$ & 7.713 $\times 10^{-1}$ &
14.6 & 9.843 $\times 10^{-3}$ & 1.248 $\times 10^{-2}$ \\ 8.2 & 6.761
$\times 10^{-1}$ & 8.508 $\times 10^{-1}$ & 15.0 & 2.799 $\times
10^{-3}$ & 3.578 $\times 10^{-3}$ \\ 8.6 & 7.275 $\times 10^{-1}$ &
9.243 $\times 10^{-1}$ & 15.4 & 2.008 $\times 10^{-4}$ & 2.086 $\times
10^{-4}$ \\ \hline \hline
\end{tabular}
\caption{ Monte Carlo-generated undistorted $^8$B neutrino energy
spectrum for events that passed all the applied analysis cuts, divided
into those occurring during the daytime and during the nighttime.  The
spectra have been normalized to the peak nighttime response, and the
relative scales of the day and night spectra reflect the livetime and
detector acceptance differences between day and night.  The quoted
energies are the central values of 0.4~MeV intervals.  The spectrum is
zero outside the displayed range.}
\label{t:enu_spec}
\end{center}
\end{table*}

\end{document}